\documentclass[sigconf]{acmart}

\usepackage{algorithm, algorithmicx}
\usepackage{algpseudocode}
\usepackage[american]{babel}
\usepackage{microtype}
\usepackage{times}
\usepackage{latexsym}
\usepackage{graphicx}
\usepackage{epsfig}
\usepackage[tight,footnotesize]{subfigure}

\usepackage{fancybox}
\usepackage{url}
\usepackage{color, soul}
\usepackage{xcolor}
\usepackage{bm}
\usepackage{multirow}
\usepackage{bbold}
\usepackage{mwe}
\usepackage{colortbl}

\usepackage{enumitem}

\usepackage{diagbox}
\DeclareMathOperator{\E}{\mathbb{E}}
\newtheorem{theorem}{Theorem}

\newtheorem{definition}[theorem]{Definition}

\usepackage{xspace}

\newcommand{\paragraphbe}[1]{\smallskip\noindent{\bf {#1}.}~}

\newcommand{\reffig}[1]{Figure~\ref{#1}}

\newcommand{\refsec}[1]{Section~\ref{#1}}

\newcommand{\refeq}[1]{Eq.~(\ref{#1})}

\newcommand{\refalg}[1]{Algorithm~\ref{#1}}

\newcommand{\ie}{{\em i.e.}}
\newcommand{\eg}{{\em e.g.}}
\newcommand{\subbb}[1]{{\smallskip \noindent{\bf #1}}}

\definecolor{RoyalBlue}{RGB}{58,83,164}

\newcommand{\ccsrev}[1]{#1}
\makeatletter
\newcommand*{\rom}[1]{\expandafter\@slowromancap\romannumeral #1@}
\makeatother

\copyrightyear{2024}
\acmYear{2024}
\setcopyright{acmlicensed}\acmConference[CCS '24]{Proceedings of the 2024 ACM SIGSAC Conference on Computer and Communications Security}{October 14--18, 2024}{Salt Lake City, UT, USA}
\acmBooktitle{Proceedings of the 2024 ACM SIGSAC Conference on Computer and Communications Security (CCS '24), October 14--18, 2024, Salt Lake City, UT, USA}
\acmDOI{10.1145/3658644.3670298}
\acmISBN{979-8-4007-0636-3/24/10}

\begin{document}
\title{Data Poisoning Attacks to Locally Differentially Private Frequent Itemset Mining Protocols}



\author{Wei Tong}
\affiliation{%
  \institution{State Key Laboratory of Novel Software\\ Technology, Nanjing University}
  \city{Nanjing}
  \country{China}
}
\email{weitong@outlook.com}

\author{Haoyu Chen}
\affiliation{%
  \institution{State Key Laboratory of Novel Software\\ Technology, Nanjing University}
  \city{Nanjing}
  \country{China}
}
\email{corheyc@gmail.com}

\author{Jiacheng Niu}
\affiliation{%
  \institution{State Key Laboratory of Novel Software\\ Technology, Nanjing University}
  \city{Nanjing}
  \country{China}
}
\email{lingdangsmoke@gmail.com}

\author{Sheng Zhong}
\authornote{Corresponding author}
\affiliation{%
  \institution{State Key Laboratory of Novel Software\\ Technology, Nanjing University}
  \city{Nanjing}
  \country{China}
}
\email{zhongsheng@nju.edu.cn}

\renewcommand{\shortauthors}{Tong et al.}

\begin{abstract}
Local differential privacy (LDP) provides a way for an untrusted data collector to aggregate users' data without violating their privacy. Various privacy-preserving data analysis tasks have been studied under the protection of LDP, such as frequency estimation, frequent itemset mining, and machine learning. Despite its privacy-preserving properties, recent research has demonstrated the vulnerability of certain LDP protocols to data poisoning attacks. However, existing data poisoning attacks are focused on basic statistics under LDP, such as frequency estimation and mean/variance estimation. As an important data analysis task, the security of LDP frequent itemset mining has yet to be thoroughly examined. In this paper, we aim to address this issue by presenting novel and practical data poisoning attacks against LDP frequent itemset mining protocols. By introducing a unified attack framework with composable attack operations, our data poisoning attack can successfully manipulate the state-of-the-art LDP frequent itemset mining protocols and has the potential to be adapted to other protocols with similar structures. We conduct extensive experiments on three datasets to compare the proposed attack with four baseline attacks. The results demonstrate the severity of the threat and the effectiveness of the proposed attack.
\end{abstract}




\maketitle

\section{Introduction}
The increasing popularity of data analysis using user data has raised concerns over potential violations of privacy. As a promising solution for privacy-preserving data collection and analysis, Differential Privacy (DP)~\cite{DBLP:journals/fttcs/DworkR14,DBLP:conf/tcc/DworkMNS06} and its local version, Local Differential Privacy (LDP)~\cite{DBLP:conf/focs/KasiviswanathanLNRS08,DBLP:conf/ccs/ErlingssonPK14} have been widely studied. By adding perturbations to the data prior to reporting it, LDP enables data collection and analysis without compromising user privacy. Many LDP protocols have been proposed by both the academic and industrial communities. 
Various data analysis tasks, \eg, frequency estimation~\cite{DBLP:conf/ccs/ErlingssonPK14,DBLP:conf/uss/WangBLJ17}, heavy hitters~\cite{DBLP:journals/tdsc/0001LJ21,DBLP:conf/nips/BassilyNST17}, and frequent itemset mining~\cite{DBLP:conf/sp/WangLJ18,DBLP:conf/ccs/QinYYKXR16}, have been studied under the notion of local differential privacy. 
Major technology companies, including Google, Apple, and Microsoft, have used LDP to collect user data in widely used systems and applications, such as Chrome~\cite{DBLP:conf/ccs/ErlingssonPK14}, iOS~\cite{iosdp}, and Windows~\cite{DBLP:conf/nips/DingKY17}. 

Despite the promising applications of local differential privacy, recent work~\cite{DBLP:conf/sp/CheuSU21,DBLP:conf/uss/CaoJG21,DBLP:journals/corr/abs-2111-11534} has demonstrated the vulnerability of LDP protocols to data poisoning attacks. This is because the data collection in LDP is performed in a distributed manner, and some users may be malicious in real-world scenarios, allowing them to submit carefully crafted values to manipulate the aggregation results. In addition, some LDP protocols employ encoding schemes to spread the influence of a reported value and reduce communication complexity, making the data poisoning attacks more powerful. The local perturbation nature of LDP protocols also makes the data poisoning attacks more concealed, making it difficult for the data collector to distinguish between the poisoned data and the normal data. Previous studies have shown that poisoning attacks can significantly impact the performance of frequency estimation~\cite{DBLP:conf/sp/CheuSU21, DBLP:conf/uss/CaoJG21}, heavy hitters~\cite{DBLP:conf/uss/CaoJG21}, and key-value data aggregation~\cite{DBLP:journals/corr/abs-2111-11534} in LDP protocols. However, frequent itemset mining protocols, which aim to discover frequently occurring events, patterns, or associations, have not been investigated in terms of their vulnerability to data poisoning attacks. Despite being an indispensable and essential component of many data analysis tasks, as well as being widely used in various applications, the security of LDP frequent itemset mining protocols remains unexplored. 

In this work, we investigate the security of LDP frequent itemset mining. Our focus is to examine the susceptibility of LDP frequent itemset mining protocols to data poisoning attacks. We propose a novel framework to demonstrate how malicious users can manipulate the results of frequent itemset mining by submitting carefully crafted data to the collector. These attacks can result in the identification of incorrect frequent itemsets and can have a significant impact on pattern discovery and association rule mining tasks. For instance, an attacker could undermine an advertising campaign strategy or promote a harmful product combination by exploiting the vulnerabilities in LDP frequent itemset mining protocols. 

Although previous studies~\cite{DBLP:conf/sp/CheuSU21, DBLP:conf/uss/CaoJG21} have shown the feasibility of attacking the LDP frequency estimation, which is a basic component for the LDP frequent itemset mining protocols~\cite{DBLP:conf/sp/WangLJ18,DBLP:conf/ccs/QinYYKXR16}, or attacking the LDP heavy hitter identification, which shares a similar goal of identifying the most frequent items, it is important to note that successfully attacking these protocols does not necessarily imply the ability to successfully attack LDP frequent itemset mining protocols. Compared with the existing attacks on other LDP protocols, there are three unique challenges to highlight for attacking LDP frequent itemset mining protocols. (1) \textit{The user data is in a more varied format.} In frequent itemset mining, each user is associated with a set of items, while in frequency estimation/heavy hitter identification and key-value aggregation, each user is associated with a single item or a pair of key-value data, respectively. This makes it challenging to craft user data as the domain of set-valued data grows exponentially with the number of items. 
(2) \textit{The goal of the attacker is more difficult to achieve.} Unlike in frequency estimation and key-value aggregation, where the goal is to estimate the frequencies, the goal in frequent itemset mining is to rank the itemsets and output the top-$k$ itemsets as results. Simply changing the frequencies of some itemsets may not guarantee a change in the ranking. 
(3) \textit{The protocols are generally more sophisticated.} An LDP frequent itemset mining protocol usually consists of multiple phases of interactions with users, which creates the opportunity for multiple rounds of poisoning. \ccsrev{While the multiple phases of interactions indeed present a larger attack surface for the attacker, they also create a more complicated task for the attacker to exploit and manage.} Specifically, it is challenging for an attacker to coordinate the different phases in a manner that maximizes the attack performance with a limited number of malicious users. 

To address the challenges, we leverage a general framework for LDP frequent itemset mining protocols and identify the possible attack surfaces by analyzing the phases that are most susceptible to be manipulated. We observe that the majority of current LDP frequent itemset mining protocols have two crucial phases: (a) pruning the domain by constructing a candidate set of itemsets, and (b) estimating the frequencies of candidate itemsets and selecting the top-$k$ as the final results. As both phases rely on users' data for aggregation, attackers can leverage these phases to manipulate the results of the mining process. Additionally, some protocols employ an adaptive approach to determine parameters during execution, offering another avenue for attackers to exploit. In particular, we propose our novel \textit{Adaptive Orchestration Attack} (AOA), which is built upon three types of composable attack operations. Given an LDP frequent itemset mining protocol, these attack operations enable the attacker to estimate the attack resource available for each phase, select a refined set of targets, and allocate resources effectively to generate poisoned data to achieve a satisfactory attack outcome. 

We apply the proposed attack to two state-of-the-art set-valued frequent item mining protocols, LDPMiner~\cite{DBLP:conf/ccs/QinYYKXR16} and SVIM~\cite{DBLP:conf/sp/WangLJ18}, and a leading frequent itemset mining protocol, SVSM~\cite{DBLP:conf/sp/WangLJ18}.  Set-valued frequent item mining is a special case of frequent itemset mining, where each user still possesses a set of items, \ie, an itemset, but its goal is to identify the frequent itemsets with only one item. We first consider the scenario where the attack is well-informed. Then, we extend our attack to the cases with partial-knowledge attackers and and a man-in-the-middle attack scenario. In addition, we compare the performance of our attack with four baseline attacks, including two random attacks and two improved variants of a state-of-the-art attack against LDP mechanisms to demonstrate the superior effectiveness of our proposed attack. 

The contributions of this paper are summarized as follows:
\begin{itemize}[leftmargin=10pt]
\item To the best of our knowledge, we are the first to study data poisoning attacks to LDP frequent itemset mining protocols. 
    
\item We propose an effective attack, which is feasible under various practical threat models and can successfully compromise major state-of-the-art LDP frequent itemset mining protocols under a unified framework, with the help of composable attack operations. 

\item We demonstrate the effectiveness of our attacks against two set-valued frequent item mining protocols and a frequent itemset mining protocol on a synthetic dataset and two real-world datasets, compared with four baselines. Moreover, some potential defenses for countering the proposed attacks have been discussed and evaluated.

\end{itemize}
\section{Preliminaries}
\subsection{Local Differential Privacy}
In the local setting of differential privacy, each \textit{user} perturbs its input $x$ by using a perturbation mechanism $\mathcal{M}$ and sends $\mathcal{M}(x)$ to the \textit{aggregator}. Below we review the formal definition of local differential privacy (LDP).

\begin{definition}[Local Differential Privacy]
	A randomized mechanism $\mathcal{M}: \mathcal{X} \rightarrow \mathcal{Y}$ satisfies $\epsilon$-local differential privacy, if and only if for any two inputs $x$ and $x'$ and for any output $y$ of $\mathcal{M}$, 
    \[
     \Pr[\mathcal{M}(x) = y] \leq \exp(\epsilon) \cdot\Pr[\mathcal{M}(x') = y].
    \]
\end{definition}

\paragraphbe{Frequency oracles} One of the major parts that are used in the LDP frequent itemset mining is frequency oracle ($\rm{FO}$)~\cite{DBLP:conf/sp/WangLJ18}, which is a basic component that can estimate the frequencies of items in a domain. Four commonly used frequency estimation protocols are Generalized Randomized Response (GRR)~\cite{DBLP:conf/ccs/ErlingssonPK14}, Succinct Histogram (SH)~\cite{DBLP:conf/stoc/BassilyS15}, Optimized Local Hashing (OLH)~\cite{DBLP:conf/uss/WangBLJ17}, and RAPPOR~\cite{DBLP:conf/ccs/ErlingssonPK14}. For a frequency oracle $\rm{FO}$, let $\Psi_{\rm{FO}(\epsilon)}(x_j)$ be the function of perturbing the value $x_j$ of user $j$, $\langle x_j\rangle$ be the encoded value of $x_j$ under the FO, and $\Phi_{\rm{FO}(\epsilon)}(R, t)$ be the function of aggregating the reports $R$ and getting the frequency of item $t$. The framework in~\cite{DBLP:conf/uss/WangBLJ17} shows that the aggregated \ccsrev{relative} frequency of item $t$ can be computed as:
\begin{equation}\label{eq:freq}
  \tilde{f}_t = \frac{\tilde{c}_t}{n} = \frac{\sum_{j=1}^{n} \mathbb{1}_{\mathrm{supp}(y_j)}(t) - q}{n(p-q)},
\end{equation}
where $y_j$ is the perturbed value reported by user $j$; \ccsrev{$n$ is the number of users; $p$ and $q$ are the perturbation parameters;} and $\mathbb{1}_{\mathrm{supp}(y_j)}(t)$ is an indicator function that returns 1 if $t\in \mathrm{supp}(y_j)$, otherwise 0. The support function $\mathrm{supp}(y_j)$ returns all the input values that $y_j$ supports (\ie, the input values that can produce $y_j$ in the LDP protocol). 

\paragraphbe{Attacks to LDP FOs} For the sake of completeness, we review the study of attacks to the LDP frequency estimation. In~\cite{DBLP:conf/uss/CaoJG21}, \ccsrev{Cao et al.} investigate how to poison LDP frequency estimation. In their paper, the concept of frequency gain is introduced to formulate the changes that can be achieved by falsely reported values. Formally, the frequency gain on an item $t$ is 
\begin{equation}\label{eq:freqgain}
\Delta \tilde{f_t} = \frac{\sum_{\hat{y}\in \hat{Y}}({\mathbb{1}}_{\mathrm{supp}(\hat{y})}(t)-\sum_{y\in Y}{\mathbb{1}}_{\mathrm{supp}(y)}(t)/n)}{(n+m)(p-q)}
\end{equation}
Based on the frequency gain, they have proposed the maximal gain attack (MGA)~\cite{DBLP:conf/uss/CaoJG21} against LDP frequency estimation, which can promote the frequencies of chosen items. Specifically, given a set of items $T$, MGA tries to maximize the total frequency gain on $T$. 
The total frequency gain that can be achieved by the attacker can be considered as the attack resource for targeting the protocols.
Although a theoretical bound of the maximum value of the attack resource has been given by Cao et al.~\cite{DBLP:conf/uss/CaoJG21}, the attacker is not always able to obtain this maximum value in practice. For example, in OLH and SH, finding the maximum frequency gain contributed by a single reported value will cost an exponential search complexity. 

\subsection{Set-Valued Frequent Itemset Mining}
We consider two tasks, \ie, the set-valued frequent item mining~\cite{DBLP:conf/ccs/QinYYKXR16,DBLP:conf/sp/WangLJ18} and frequent itemset mining~\cite{DBLP:conf/sp/WangLJ18}, under the definition of local differential privacy. We assume that there are $n$ users, and each user $j \in [n]$ holds a subset of a set of $d$ items $\mathcal{I} = \{1,2,\ldots, d\}$. Denote the item subset, \ie, \textit{transaction}, of user $j$ by $S_j \subseteq \mathcal{I}$. The frequency of an item $t$ is defined as $f_t = \sum_{S_j} \mathbb{1}(t \in S_j)/n$; and the frequency of the itemset $I$ is defined as $f_I = \sum_{S_j} \mathbb{1}(I \subseteq S_j)/n$. The frequent item/itemset mining task is to find the top-$k$ most frequent items/itemsets in the corresponding domains.

Then we provide an overview of the family of locally differentially private frequent itemset mining protocols. To illustrate the basic idea of the differentially private frequent itemset mining, we consider two widely used mechanisms in this family: SVIM/SVSM~\cite{DBLP:conf/sp/WangLJ18} and LDPMiner~\cite{DBLP:conf/ccs/QinYYKXR16}. 
Based on the analysis from~\cite{DBLP:conf/sp/WangLJ18}, a frequent itemset mining protocol consists of three major parts: (1) FO choosing and parameter setting ($\mathsf{ConfigPro}$); (2) Domain pruning ($\mathsf{PruneDom}$); and (3) Estimation and top-$k$ selection ($\mathsf{SelectTop}$). For the sake of completeness, we review LDPMiner~\cite{DBLP:conf/ccs/QinYYKXR16}, SVIM~\cite{DBLP:conf/sp/WangLJ18}, SVSM~\cite{DBLP:conf/sp/WangLJ18}, \ccsrev{ FIML~\cite{DBLP:conf/cikm/LiGGWY22}, and PrivSet~\cite{DBLP:conf/infocom/0003HNWXY18}} \ccsrev{with respect to} the above frequent itemset mining framework. 

\paragraphbe{LDPMiner} The three parts of LDPMiner are
\begin{description}[leftmargin=10pt]
    \item[$\mathsf{ConfigPro}$.] LDPMiner sets the parameters: privacy budget $\epsilon_1$ and $\epsilon_2$ for $\mathsf{PruneDom}$ and $\mathsf{SelectTop}$, respectively; padding size $l_1 = L$ for $\mathsf{PruneDom}$ and $l_2=2k$ for $\mathsf{SelectTop}$. The SH mechanism is chosen as FO in  $\mathsf{PruneDom}$ and the RAPPOR mechanism~\cite{DBLP:conf/ccs/ErlingssonPK14} is chosen as FO in $\mathsf{SelectTop}$. 
    
    \item[$\mathsf{PruneDom}$.] The protocol narrows down the item domain by finding the top $2k$ items as a candidate set $\mathcal{C}$ by using SH with privacy budget $\epsilon_1$. \ccsrev{This step lets user} $j$ pad its set of items $S_j$ to length $l_1$ if $|S_j| < l_1$ before she/he selects an item $t_j$ randomly from $S_j$ and applies $\Psi_{\mathrm{SH}(\epsilon_1)}(t_j)$. 
    
    \item[$\mathsf{SelectTop}$.] LDPMiner lets each user report $\Psi_{\mathrm{RAPPOR}(\epsilon_2)}(t_j)$, where $t_j$ \ccsrev{is randomly} selected  from $S_j \cap \mathcal{C}$ (if $|S_j \cap S| < l_2$, first pads the intersection to size of $l_2$). Then LDPMiner estimates the frequencies of items in $\mathcal{C}$ by applying $\Phi_{\rm{RAPPOR}(\epsilon_2)}(R,t)$. Then LDPMiner selects $k$ items in $\mathcal{C}$ with the highest estimated frequencies as the top-$k$ frequent items. 
\end{description}

\paragraphbe{SVIM} The three parts of the SVIM protocol are
\begin{description}[leftmargin=10pt]
    \item[$\mathsf{ConfigPro}$.] The protocol first partitions the users into three groups and sets the initial system parameters: privacy budget $\epsilon$; padding size $l_1 = 1$ for $\mathsf{PruneDom}$ and $l_2 = L$ for $\mathsf{SelectTop}$, 
    where $L$ is the $90$-th percentile of all the cardinalities of the intersection between the value of each user and the candidate set. GRR will be chosen as the $\rm{FO}$ if $d<l(4l-1)e^\epsilon+1$ (for $l = l_1, l_2$), where $d=|I|$ for $\mathsf{PruneDom}$ and $d=2k$ for $\mathsf{SelectTop}$; otherwise, OLH will be chosen as the $\rm{FO}$.
    
    \item[$\mathsf{PruneDom}$.] The protocol uses the first group of users to narrow down the item domain by finding the top $2k$ items and sets them as the candidate set $\mathcal{C}$. In this part, each user first pads the set of items to length $l_1$ with dummy items and reports only one item $t_j$ from her/his set of items by using $\Psi_{\rm{FO}(\epsilon)}(t_j)$. 
    
    \item[$\mathsf{SelectTop}$.] SVIM estimates the frequencies of $2k$ candidate items by using the FO under settings decided by the former steps as $\Psi_{\rm{FO}(\epsilon)}(t_j)$, where $t_j$ is sampled from a padded set with length $l_2$ of $S_j \cap \mathcal{C}$. 
    Then, SVIM estimates the frequencies of items in $\mathcal{C}$ by applying $\Phi_{\rm{FO}(\epsilon)}(R,t)$ for $t\in \mathcal{C}$, and updates the estimation with a factor to fix the bias induced by padding. Then SVIM selects $k$ items in $\mathcal{C}$ with the highest estimated frequencies as the top-$k$ frequent items. 
\end{description}

\paragraphbe{SVSM} The three parts of the SVSM protocol are similar to those of SVIM. SVSM can be regarded as a modified version of SVIM, where the \textit{items} are replaced by the itemsets. However, the domain of all itemsets is exponentially larger than the domain of items, making it infeasible to estimate the frequencies of all itemsets. SVSM uses the estimated item frequencies from SVIM to help itself prune the size of the domain of itemsets 
in the $\mathsf{PruneDom}$ phase. 

\begin{description}[leftmargin=10pt]
    \item[$\mathsf{ConfigPro}$.] Users are partitioned into three groups, similar to SVIM, and the protocol decides the system parameters: privacy budget $\epsilon$; padding size $l$ for $\mathsf{SelectTop}$. GRR is chosen as the $\rm{FO}$ if $d<l(4l-1)e^\epsilon+1$, where $d=2k$ for $\mathsf{SelectTop}$; otherwise, OLH is chosen as the $\rm{FO}$.
    
    \item[$\mathsf{PruneDom}$.] After calling SVIM with the first group of users, SVSM gets the top-$k$ items $\mathcal{K}$ along with their estimated frequencies: $\{ \tilde{f}_x | x\in \mathcal{K} \}$. The candidate itemsets are selected by guessing the frequencies of itemsets as
    \(
    \tilde{f}_{I} = \prod_{t \in I} \frac{0.9\tilde{f}_t}{\max_{x\in \mathcal{K}} \tilde{f}_x}
    \) from the power set of $\mathcal{K}$. 
    The $2k$ itemsets with the highest guessed frequencies are selected as the candidate itemsets $\mathcal{S}$.

    \item[$\mathsf{SelectTop}$.] SVSM estimates the frequencies of $2k$ candidate itemsets by having user $j$ report $\Psi_{\rm{FO}(\epsilon)}(I_j)$, where $I_j$ is sampled from the set of padded itemsets of length $l$. Then, SVSM estimates the frequencies of the itemsets in $\mathcal{S}$ by applying $\Phi_{\rm{FO}(\epsilon)}(R,I)$ for $I \in \mathcal{S}$, and updates the estimate with a factor to fix the bias induced by the padding.
    Then SVSM selects $k$ itemsets in $\mathcal{S}$ with the highest estimated frequencies as the top-$k$ frequent itemsets. 
\end{description}

\paragraphbe{FIML} The FIML protocol is able to find both the frequent items and frequent itemsets. We denote these two functions as FIML-I and FIML-IS in this paper, and most parts of them share similar steps. The three parts of the FIML protocol are
\begin{description}[leftmargin=10pt]
    \item[$\mathsf{ConfigPro}$.] The protocol first divides the users into three groups and sets the privacy budget $\epsilon$. The OLH mechanism is chosen as FO in $\mathsf{PruneDom}$, and the GRR mechanism with a binary domain is chosen as FO in $\mathsf{SelectTop}$.
    
    \item[$\mathsf{PruneDom}$.] For frequent items, the protocol prunes the domain by finding the top $1.5k$ items using OLH and sets them as the candidate set $\mathcal{C}$. In this part, each user samples and reports only one item $t_j$ from their set of items using $\Psi_{\rm{OLH}(\epsilon)}(t_j)$. For frequent itemsets, the protocol constructs the candidate itemsets from the mined top-$k$ items and prunes the set of candidate itemsets to a size of $1.5k$.

    \item[$\mathsf{SelectTop}$.] In this part, FIML allows the aggregator to choose an item/itemset randomly for each user from the candidate set and send it to the user. Each user responds by using function $\Psi_{\rm{GRR}(\epsilon)}(\cdot)$ to indicate whether the received item/itemset is one of her/his top-$k$ items/itemsets. Then, the aggregator uses $\Phi_{\rm{GRR}(\epsilon)}(\cdot)$ to estimate the frequency of each candidate item/itemset and identify the top-$k$ items/itemsets.
\end{description}

\paragraphbe{PrivSet} The PrivSet protocol~\cite{DBLP:conf/infocom/0003HNWXY18} is a set-valued data aggregation mechanism, and we can wrap it into a naive top-$k$ item mining protocol by simply selecting $k$ items with the highest estimated frequency as the final result. 
\begin{description}[leftmargin=10pt]
    \item[$\mathsf{ConfigPro}$.] The PrivSet protocol sets the parameters: privacy budget $\epsilon$, padding size $l$, and domain size $d$. Then, the report set size $\kappa$ is determined based on these three parameters to minimize the square error of the estimation.
    \item[$\mathsf{PruneDom}$.] The PrivSet protocol does not include such a phase. 
    \item[$\mathsf{SelectTop}$.]  In PrivSet, reports from all users are collected, and the protocol identifies the top-$k$ items from the aggregated result. 
\end{description}

\subsection{Threat Model}
\paragraphbe{Attacker's capability}
We follow the assumption made in previous work that the attacker can inject/corrupt users and make them report arbitrary values to the aggregator that conform to the formats of the LDP protocols~\cite{DBLP:conf/uss/CaoJG21, DBLP:conf/sp/CheuSU21}. The corrupted users can create the original values, the encrypted values, and the corrupted values. In addition, we assume that the attacker can provide poisoned data in a man-in-the-middle (MITM) fashion by eavesdropping, crafting, and replaying perturbed values from benign users to the aggregator. We assume that the attacker can inject/corrupt or intercept reports from at most $m \leq n$ users. If we assume that the attacker will inject/corrupt or intercept as many users as possible, then we have that the number of honest users is $n - m$. The set of honest users is denoted by $U^b$ and the set of corrupted users by $U^c$. 

\paragraphbe{Attacker's background knowledge} We assume that the attacker knows the predefined parameters of the LDP protocols. Since the aggregator has no idea which user is malicious or benign, it will provide each user with the implementation/setup for running the LDP protocols. However, the attacker has no prior knowledge of the parameters that are adaptively determined by the protocols. We also consider two major types of attackers who have different levels of knowledge about the information regarding the frequent itemset mining process: \textit{full-knowledge} and \textit{partial-knowledge} attackers. 

\textit{Full-knowledge:} The attacker knows the true frequencies of all items/itemsets. Although the assumption is somewhat strong, there are some scenarios where it is applicable. For example, there are two aggregators who are competitors. One of them has completed the data aggregation and wants to prevent the other from mining the (genuine) frequent items. In addition, we can use this type of attacker to estimate the maximal severity of such threats. 

\textit{Partial-knowledge:} The attacker only knows the items/itemsets on the injected/corrupted users. The attacker can estimate the true frequencies of all items/itemsets based on the held data by the injected/corrupted users. A variant of the partial-knowledge scenario is the \textit{man-in-the-middle (MITM) attack}, where the attacker can only estimate the frequencies of items/itemsets based on intercepted perturbed reports instead of the true values of the items/itemsets.

\paragraphbe{Attacker's goal}
Suppose an attacker tries to manipulate top-$k$ frequent itemset mining such that the aggregator will get false results. To poison the results of frequent itemset mining, the attacker carefully selects the responses to the aggregator in the interactive process of the LDP protocol. Denoted by the set of responses of all users $\mathcal{R}$, the set of responses of all benign users $\mathcal{R}^b$, and the set of responses of all corrupted users $\mathcal{R}^c$. For each user $j$, the response $R_j \in \mathcal{R}$ is a sequence of perturbed values $y_{j,1}, y_{j,2}, \ldots$ generated by the LDP protocol. If the users have only a single interaction with the aggregator, then $R_j = y_j$. Specifically, the goal of the attacker is to lower the accuracy of the aggregation results. Formally, the attacker's goal is 
\ccsrev{
      \begin{equation}\label{eq:goal3}
        \min \quad \sum_{t\in \mathcal{I}_k}\mathbb{1}(\tilde{f}^*_t > \tilde{f}^*_{k+1})
      \end{equation}
      }
where $\mathcal{I}_k$ is the set of true top-$k$ items before the attack and $\mathbb{1}(cond)$ is an indicator function that returns 1 if $cond$ is true; otherwise 0.

    \begin{table}[]
    \caption{\ccsrev{Notation}}
    \label{tab:notation}
    \ccsrev{
    \begin{tabular}{c|l}
    \hline
    Notation                            & Description               \\ \hline\hline
    $x_j$, $\langle x_j\rangle$         & value and encoded value of user $j$\\
    $\mathcal{R}$, $\mathcal{R}_j$      & the responses of all users and the response of user $j$\\
    $\mathcal{R}^b$, $\mathcal{R}^c$    & the set of responses of benign and corrupted users\\
    $\Psi_{\rm{FO}(\epsilon)}(\cdot)$   & function for perturbing values   \\ 
    $\Phi_{\rm{FO}(\epsilon)}(\cdot, \cdot)$  & function for estimating frequencies \\ 
    $y_j$, $\mathrm{supp}(y_j)$     & perturbed value of user $j$ and its support   \\
    $S(y)$ & set of items/itemsets supported by report $y$\\
    $\mathcal{I}$, $d$ & domain of items and its size                               \\
    $S_j$                             & set of items of user $j$                    \\ 
    $f_t$, $\tilde{f}_t$              & frequency of item $t$ and its estimation    \\ 
    $f_I$, $\tilde{f}_I$              & frequency of itemset $I$ and its estimation \\ 
    $n$, $m$                          & size of benign and corrupted users          \\
    $U^b$, $U^c$                 & sets benign and corrupted users           \\
    $\mathcal{C}$, $\mathcal{S}$      & candidate item/itemset set\\
    $\mathcal{K}$                    & estimated top-k items\\
    $T$, $L$                          & attacker's target set and its size\\
    \hline
    \end{tabular}}
    \end{table}
\section{Attack Framework and Operations}
\label{sec:blocks}
The major building blocks of LDP frequent itemset mining protocols are FOs. The $\mathsf{PruneDom}$ and $\mathsf{SelectTop}$ of a LDP frequent itemset mining protocol use FOs as tools to select the candidates, set configurations, and choose top-$k$ results based on users' reports. 
Therefore, the basic idea for the attacks is to leverage the manipulation of the FOs with respect to the upper-level LDP frequent itemset mining protocols. 
Specifically, we introduce an attack toolkit including three operations for attacking against LDP frequent itemset mining protocols. 
The first operation is \textit{attack resource estimation}, which can estimate the amount of resources that can be used for the attack given a certain number of corrupted users. The second operation is \textit{target set refinement}, which allows the attacker to select a candidate set of target itemsets for the attack. The third operation is \textit{poisoned data generation}. Given the amount of attack resources and the set of targets, the attacker tries to allocate the resources and generate the poisoned responses strategically to maximize the effect of the attack. 

\subsection{Overview}
\paragraphbe{Attack surfaces} We define a poisoning attack based on a given LDP frequent itemset mining protocol with $\mathsf{ConfigPro}$, $\mathsf{PruneDom}$, and $\mathsf{SelectTop}$. By choosing a set of responses $\mathcal{R}^c$ of the corrupted users, the attacker can potentially manipulate the set of candidate items/itemsets generated by $\mathsf{PruneDom}$, the estimated frequencies of the items/itemsets in the candidate set via  $\mathsf{SelectTop}$, and the parameters chosen in $\mathsf{ConfigPro}$.

$\mathsf{ConfigPro}$: An LDP frequent itemset mining protocol tries to choose a type of encoding format, which will be used in the following $\mathsf{SelectTop}$ step. The goal of the attacker is to trigger the protocol to choose a configuration of encoding format that can be leveraged to amplify the attack ability in the following step. Specifically, the padding length in SVIM and SVSM is decided by all users, where the attacker can greedily increase support for larger length values or smaller ones to influence the padding length choice.
    
$\mathsf{PruneDom}$: The protocol tries to select a set of candidates $\mathcal{C}$ containing the top-$k$ results ($|\mathcal{C}| \geq k$) from the item domain $\mathcal{I}$ or itemset domain $2^\mathcal{I}$, respectively. Therefore, the goal of the attacker is to manipulate the protocol in this phase so that as many non-top-$|\mathcal{C}|$ items/itemsets as possible can be pushed into $\mathcal{C}$. The attacker can estimate the itemset frequencies of the benign users and therefore identify its targets as those itemsets with high frequency but not in the genuine top-$k$ results. Then the attacker refines its target set until the attack resources meet the requirement. Finally, the attacker will craft the poisoned reports and submit them to the aggregator.

$\mathsf{SelectTop}$: The attacker tries to make as many non-top-$k$ items or itemsets in $\mathcal{C}$ as top-$k$ results by crafting the reports of corrupted users, which is similar to the goal of manipulating $\mathsf{PruneDom}$. However, there are three differences when trying to manipulate $\mathsf{SelectTop}$: (1) $\mathsf{SelectTop}$ uses different parameters (\eg, FO, padding size, privacy budget) compared to $\mathsf{PruneDom}$; (2) the item/itemsets domain is $\mathcal{C}$ in $\mathsf{SelectTop}$ instead of $\mathcal{I}$ or $2^\mathcal{I}$ in $\mathsf{PruneDom}$ for frequent item mining and frequent itemset mining, respectively; (3) true top-$k$ items/itemsets remaining in the candidate set are usually with high ranks, meaning that making more progress in this phase will cost more attack resources on each target. 

\paragraphbe{Attack operations} To make the above attack framework possible, we introduce three attack operations: \textit{Attack Resource Estimation}, \textit{Target Set Refinement}, and \textit{Poisoned Data Generation}. For each part of the frequent itemset mining protocol, the attacker uses these three operations to perform an effective attack. Specifically, the attack resource estimation operation is employed to estimate the available attack resources for data poisoning, given a specified number of injected/corrupted users or maliciously modified reports. Then the target set refinement operation refines the target set based on the frequencies of items or itemsets. This operation aims to select targets from the domain of items or itemsets with the highest likelihood of being promoted to the top-$k$ results through the impact of poisoned data. Subsequently, the poisoned data generation operation allocates the estimated attack resources to the selected target itemsets. This operation determines the poisoned reports needed to be crafted in the LDP frequent itemset mining protocol. 

These three attack operations can be applied to any of the aforementioned attack surfaces within an LDP frequent item mining protocol. We design the attack resource estimation and poisoned data generation operations in terms of different LDP FOs because different frequent itemset mining protocols use different LDP FOs as the primitives in their design: SVSM and SVIM use GRR and OLH; LDPMiner uses SH and RAPPOR, and propose a unified target set refinement operation. We use $\mathsf{AREst}$, $\mathsf{TSRef}$, and $\mathsf{PDGen}$ to denote these three attack operations, respectively. In addition, we use $\mathsf{Operation}$-$\mathsf{FO}$ to denote the specified versions of an attack operation $\mathsf{Operation}$ with respect to a given LDP frequency oracle $\mathsf{FO}$, \eg, $\mathsf{AREst}$-$\mathsf{GRR}$, $\mathsf{PDGen}$-$\mathsf{OLH}$. The details of these three operations are shown in \refsec{sec:ops:arest}, \refsec{sec:ops:tsref}, and \refsec{sec:ops:pdgen}, respectively. 

\subsection{Attack Resource Estimation} 
\label{sec:ops:arest}
We define the attack resources as the supports provided by the reports of users. 
Based on the frequency gain formulation (\refeq{eq:freqgain}) in~\cite{DBLP:conf/uss/CaoJG21}, 
given a set of corrupted users, the amount of attack resources can be maximized by maximizing the frequency gain of each corrupted user with crafted reports. We consider four FO primitives: GRR, OLH, SH, and RAPPOR. 

\paragraphbe{GRR} Given the perturbation function of GRR:
    \[
        y = \Psi_{\mathrm{GRR}(\epsilon)}(t)=\langle v\rangle,
    \]
     we have that item $t$ is supported by the reported value if and only if the reported item $v$ is the same as item $t$ based on the analysis in~\cite{DBLP:conf/uss/CaoJG21}. Therefore, each reported value $y$ can support only one item. Then we have the amount of resources produced by corrupted users $U^c$ as 
        $
        \Omega_{\mathrm{GRR}}(U^c) = |U^c| = m.
        $

\paragraphbe{OLH} According to the analysis in~\cite{DBLP:conf/uss/CaoJG21}, the attacker needs to find an appropriate seed and hash value to try to support more items in the target set. Specifically, the attacker's goal is to find a seed $r$ and a hash value $v$ such that as many items in the target set as possible are hashed to the reported value $v$.  
\ifx\lncsshort\undefined
    MGA~\cite{DBLP:conf/uss/CaoJG21} adopts a method of finding the best seed by enumerating $h$ (say 1,000) random hash functions, without estimating the amount of attack resources that can be used.
\fi
Below we characterize the amount of attack resources. Let $\alpha$ denote the probability that any given item is supported by a randomly reported value. We assume that there are $h$ times of random sampling from the hash function space, and the corresponding reported values are $\hat{y}^{(1)}, \hat{y}^{(2)}, \ldots, \hat{y}^{(h)}$. For any given set of target items $T$, let $\mathrm{supp}(\hat{y}^{(i)}; T)$ be the amount of items supported in $T$ by the reported value $\hat{y}^{(i)}$. With $h$ times of random sampling, we have the maximum number of supported items among them is 
\begin{equation}
\label{eq:sstar}
    s^* =\max_{i=1,2,\ldots, h}\{\mathrm{supp}(\hat{y}^{(i)}; T)\},
\end{equation}
and the cumulative probability distribution of $s^*$ is
\begin{equation}
   F_{s^*}(l)=\prod\limits_{i=1}^{h} \left[\sum_{j=0}^{l} \binom{b}{j} \alpha^j\beta^{b-j}\right], ~~\text{for}~~l = 0, 1, \ldots, b.  
\end{equation}
  where $\beta = 1-\alpha$ and $b = |T|$. Thus the expectation of $s^*$ is
\begin{equation}
\label{eq:esstar}
  E_{s^*}(h, b, \alpha) \triangleq \mathbb{E}[s^*]=\sum\limits_{i=0}^{b}(1-F_{s^*}(i)), 
\end{equation}
  which is a function with parameters: the times of random sampling $h$, the size $b$ of the target set, and the probability $\alpha$.
  
  For OLH, we have that a randomly chosen reported value will support any item with a probability of $1/d$ (\ie, $\alpha = 1/d$), where $d$ is the size of item domain considered in the aggregation, and then we have that the expectation of the amount of resources produced by corrupted users $U^c$ is
  \begin{equation*}
     \mathbb{E}\left[\Omega_{\mathrm{OLH}}(U^c)\right]  = \mathbb{E}\left[\mathrm{supp}(\mathcal{R}^c; T)\right]
     = \sum_{j=1}^m  E_{s^*}(h_j, |T|, 1/d), 
  \end{equation*}
  where $h_j$ is the times of random sampling for the $j$-th corrupted user to produce the reported value, and we can set all the $h_j = h$ if we assume all the corrupted users choose the same times of random sampling.

\paragraphbe{SH} SH is a special case of OLH with the size of the encoded domain as 2. Therefore, we can use the same attack resource estimation operation as OLH for SH. 

\paragraphbe{RAPPOR} The RAPPOR protocol allows a report to support multiple itemsets. \ccsrev{This means that, given a set of corrupted users $U^c$ with a size of $m$,} the amount of resources required for manipulating RAPPOR is $m \times L$, where $L$ represents the size of the target set. Alternatively, if the attacker chooses all the itemsets as targets, the amount of resources allowed could be $m \times d$. 

\subsection{Target Set Refinement} 
\label{sec:ops:tsref}
Since the item domain is large and the amount of attack resources is limited, it is important to select a set of target items to be the focus of the attack. Basically, there are two ways to manipulate the results of frequent item mining: reduce the frequencies of items in $I_k$ (i.e., the true top-$k$ items), or increase the frequencies of non-top-$k$ items and make them the top-$k$ result. 

However, due to the property of LDP mechanisms, if other users' reports remain unchanged, a user can only increase the frequency of a particular item, not decrease it. Thus, our goal is to push as many non-top-$k$ items as possible into the final top-$k$ result, and we denote the top-$k$ result after attack by $\mathcal{I}^*_k$. Intuitively, the attacker should always pick those items with high frequency but not in $\mathcal{I}_k$. \ccsrev{If the attacker chooses a set of items $\{\hat{t}_1, \hat{t}_2, \ldots, \hat{t}_{L}\}$ and wants to push all of them into the top-$k$ results,} the frequency of any chosen item should be larger than the $(k+1-L)$-th item before the attack, and we call this item a \textit{pivot item}. Denote by $x_1, x_2, \ldots, x_{d}$ the sorted items based on their frequencies that are contributed by the benign users from high to low and the corresponding frequencies contributed by the benign users are denoted by $\tilde{f}^b_1, \tilde{f}^b_2, \ldots, \tilde{f}^b_d$. The gap between any item $t$ and the pivot item is defined as 
$
    \mathrm{Gap}_{t}=\max\{\tilde f^b_{x_{k+1-L}}-\tilde f^b_{t}, 0\}. 
$

Denote by the frequencies contributed by the corrupted users as $\tilde{f}^c_1, \tilde{f}^c_2, \ldots, \tilde{f}^c_d$. Clearly, item $t$ can be pushed into the top-$k$ results if the attacker can make a frequency gain on $t$ such that 
\[
f^*(t) = \tilde f^c_{t} - \tilde f^c_{x_{k+1-L}} \geq \mathrm{Gap}_{t}. 
\]
After the attack, the estimation of item $t$'s total frequency is $\tilde f^b_{t} + \tilde f^c_{t}$. Below we specify how to determine the number of items to attack and refine the target set. The attacker sorts all the items based on their frequencies that are contributed by the benign users from high to low and picks items from ranked $k+1$ as the candidates in the target set. The frequency increment required to push $L$ items into the top-$k$ result is$\sum\limits_{i=1}^{L} (\tilde{f}^b_{x_{k+1-L}}-\tilde{f}^b_{x_{k+i}})$.

We can choose the largest $L$ as long as the attack resources are sufficient. Formally, the size of the target set is determined as 
\begin{align}
    & L^* = \arg\max_L \sum\limits_{i=1}^{L} (\tilde{f}^b_{x_{k+1-L}}-\tilde{f}^b_{x_{k+i}}), \\
    \text{subject to} ~~& \sum\limits_{i=1}^{L} (\tilde{f}^b_{x_{k+1-L}}-\tilde{f}^b_{x_{k+i}}) \leq \mathbb{E}[\Omega_{\mathrm{FO}}(U^c)].
\end{align}
and the target set is $T^*=\{x_{k+1},...,x_{k+L^*}\}$.

\subsection{Poisoned Data Generation} 
\label{sec:ops:pdgen}
For simplicity, we let the refined target set as $T^*=\{\hat{x}_1,\hat{x}_2,...,\hat{x}_{L^*}\}$. 
Our goal is to construct a set of poisoned responses $\mathcal{R}^c$ for the corrupted users $U^c$ such that 
\begin{equation}
    f^*(\hat{x}) \geq \mathrm{Gap}_{\hat{x}}, ~~\forall \hat{x}\in T^*.
\end{equation}
Below we present how to allocate the attack resources with respect to the resource constraint and $T^*$ to generate poisoned reports for the different FOs used in the LDP frequent itemset mining protocols.

\paragraphbe{GRR} Each reported value $\hat{y}$ can only support at most one item, \ie, $|S(\hat y)| \leq 1$, and increase the frequency of supported item by 
    \begin{equation}
    \label{eq:freq_inc}
        \mathrm{INC}_\mathrm{GRR} = \frac{1}{n(p-q)} = \frac{e^\epsilon + 1}{(n+m)(e^\epsilon-1)}, 
    \end{equation}
\ccsrev{where $p$ is the probability of reporting the true value of a user in GRR, and $q$ is the probability of reporting a perturbed value.}
Therefore, for each $\hat{x}\in T^*$, it requires $\lceil \mathrm{Gap}_{\hat{x}}/{\mathrm{INC}_{\mathrm{GRR}}} \rceil$ corrupted users to report $\langle \hat{x} \rangle$ such that item $\hat{x}$ can get enough increase of the frequency, and the total number of corrupted users required for the set of target items $T^*$ is 
\[
    \left\lceil \frac{\sum_{\hat{x}\in T^*}\mathrm{Gap}_{\hat{x}}}{\mathrm{INC}_\mathsf{GRR}} \right\rceil.
\]
    
\ifx\lncsshort\undefined
\begin{algorithm}
    \renewcommand{\algorithmicrequire}{\textbf{Input:}}  
    \renewcommand{\algorithmicensure}{\textbf{Output:}}  
    \caption{Attack resource allocation for OLH}
    \label{alg:alloc_olh}
    \begin{algorithmic}[1]
        \Require $T^*$,  $\mathrm{Gap}_{\hat{x}}$ (for $\hat{x}\in T^*$), $h$
        \Ensure $\mathcal{R}^c$
        \Function{Sample}{$\hat{x}_\mathsf{axis}$}
            \State $sup_{\max} \gets 0$
            \For{$i=0 \to h$}
                \State $r \gets \Call{RandomInt}$; $sup \gets 0$
                \For{$\hat{x} \in T^*$}
                    \If{$H_r(\hat{x})=H_r(\hat{x}_{\mathsf{axis}})$}
                        \State $sup \gets sup+1$
                    \EndIf
                \EndFor
                \If{$sup > sup_{\max}$}
                    \State $y \gets \langle r,H_r(\hat{x}_{\mathsf{axis}})\rangle$
                \EndIf
            \EndFor
            \State \Return{$y$}
        \EndFunction
        \State $ index \gets \arg\max_{\hat{x}} {\mathrm{Gap}_{\hat{x}}}$; $ \mathcal{R}^c \gets \emptyset$
        \While{$\mathrm{Gap}_{index} > 0$}
            \State $ \hat{y} \gets \Call{Sample}{index}$; $ \mathcal{R}^c \gets \mathcal{R}^c \cup \hat{y}$
            \For{$\hat{x}\in T^*$}
                \State $\mathrm{Gap}_{\hat{x}} \gets \mathrm{Gap}_{\hat{x}}-\tilde f^*_{\hat{x}}(\hat{y})$
            \EndFor
            \State $ index \gets \arg\max_{\hat{x}} {\mathrm{Gap}_{\hat{x}}}$
        \EndWhile
        \State \Return{$\mathcal{R}^c$}
    \end{algorithmic}  
\end{algorithm}
\else
\begin{algorithm}
    \renewcommand{\algorithmicrequire}{\textbf{Input:}}  
    \renewcommand{\algorithmicensure}{\textbf{Output:}}  
    \caption{Attack resource allocation for OLH}
    \label{alg:alloc_olh}
    \begin{algorithmic}[1]
        \Require $T^*$,  $\mathrm{Gap}_{\hat{x}}$ (for $\hat{x}\in T^*$), $h$
        \Ensure $\mathcal{R}^c$
        \State $ index \gets \arg\max_{\hat{x}} {\mathrm{Gap}_{\hat{x}}}$
        \While{$\mathrm{Gap}_{index} > 0$}
            \State $ \hat{y} \gets \Call{Sample}{index}$
            \For{$\hat{x}\in T^*$}
                \State $\mathrm{Gap}_{\hat{x}} \gets \mathrm{Gap}_{\hat{x}}-\tilde f^*_{\hat{x}}(\hat{y})$
            \EndFor
            \State $ index \gets \arg\max_{\hat{x}} {\mathrm{Gap}_{\hat{x}}}$
        \EndWhile
    \end{algorithmic}  
\end{algorithm}
\fi

\paragraphbe{OLH} Hash functions are used in OLH to map the items to an encoded domain, which makes a reported value can support multiple different items. For an item $\hat{x}\in T^*$, let $\hat{y} = \langle r, H_r(\hat{x}) \rangle$ to support it, where $H_r$ is a hash function with a random seed $r$. The reported value $\hat{y}$ can support a chosen item $\hat{x}_\mathsf{axis}$ and other items in $T^*$ as many as possible and the total number items $\hat{y}$ supports is 
\[
    1 + E_{s^*}(h, |T^*|-1, 1/d),
\]
where $E_{s^*}(h, b, \alpha)$ is the function defined in~\refeq{eq:esstar}. 
    
    For the sampling sequence $\hat{y}^{(1)}, \hat{y}^{(2)}, \ldots, \hat{y}^{(h)}$, the maximum support $s^*$, which is defined in~\refeq{eq:sstar}, and the corresponding reported value $\hat{y}$, the amount of resources, denoted by $S(\hat{y};t)$, allocated to any item $t \in \mathcal{I}$ satisfies 
    \begin{equation}
    \mathbb{E}[S(\hat{y};t)] = \left\{
    \begin{aligned}
        & 1, && t = \hat{x}_\mathsf{axis}, \\
        & \frac{E_{s^*}}{|T^*|-1}, && t\in T^*\ \text{and} \ t\neq  \hat{x}_\mathsf{axis}, \\
        & 1/d, && t \notin T^*.
    \end{aligned}
    \right.
  \end{equation}
    
    Then the actual increase of the frequency on item $t$ by reporting poisoned response $\hat{y}$ is 
    \begin{equation}
        \Delta \tilde{f}_t^*(\hat{y}) = \mathrm{INC}_{\mathrm{OLH}}\cdot \left( S(\hat{y};t) - \frac{1}{d}\right),
    \end{equation}
    where $\mathrm{INC}_{\mathrm{OLH}} = \frac{1}{(n+m)(p-1/d)}$. 
    
Based on the above formulation, we can find that given a fixed total amount of resources, the support for each target item in $T^*$ is divided equally on expectation. For the items that are not in $T^*$, a certain amount of resources will still be allocated to them due to the randomness of hash functions. In addition, the total amount of resources may be not able to adequately support all the items in $T^*$. We have proposed an algorithm based on the water-filling 
technique for allocating the attack resources. Details of the algorithm are shown in~\refalg{alg:alloc_olh}.

\paragraphbe{SH} Similar to OLH, SH is also a kind of local hashing protocol. It is a special case of OLH with the size of the encoded domain as 2. Therefore, we can use the same poisoned data generation operation as OLH for SH.

\paragraphbe{RAPPOR} The attack resource estimation of RAPPOR in \refsec{sec:ops:arest} shows that the attack resources $\Omega_{\mathrm{RAPPOR}}(U^c) = m\times L$, given the set of corrupted users $U^c$. 

However, we note that if the attacker uses all $mL$ attack resources, every $m$ attack resources allocated for attacking RAPPOR are coupled and can only be allocated to a single target. Each target receives its $m$ supports and these supports cannot be allocated to other targets. This implies that any target cannot receive more than $m$ supports when the attacker tries to promote its frequency. Therefore, a succinct way to generate poisoned reports with respect to RAPPOR FO is to set the bits representing the itemsets in the target set $T^*$ to $1$ and the other bits to $0$. 

\section{Attacking LDP Frequent Itemset Mining Protocols}
\label{sec:framework}
We specify the attacks against two well-adopted set-valued frequent item mining protocols: LDPMiner~\cite{DBLP:conf/ccs/QinYYKXR16} and SVIM~\cite{DBLP:conf/sp/WangLJ18}. In addition, we describe how to apply the attacks against a state-of-the-art frequent itemset mining protocol: SVSM~\cite{DBLP:conf/sp/WangLJ18} and discuss how to adopt the proposed attack for an attacker with limited knowledge.  

\subsection{Attacking LDPMiner}
\subbb{Manipulating} $\mathsf{PruneDom}$.
Given a fixed padding length $l$, \ccsrev{each benign user pads} its value (\ie, a set of items) to the padding length and submits a randomly picked item from the set by using SH. The attacker can estimate the genuine frequency and expected frequency gain to acquire the attack resource requirements and amounts for every combination of target items by the attack operation $\mathsf{AREst}$-$\mathsf{SH}$. The attacker thus picks out an available target set where the attack resource amount can meet the requirement by $\mathsf{TSRef}$. 

By submitting values supporting these target items, the attacker can promote the target items into a candidate set by allocating the attack resources with $\mathsf{PDGen}$-$\mathsf{SH}$. Note that the attacker needs to push at least $k+1$ non-top-$k$ items into the $2k$ candidate set $\mathcal{C}$ to squeeze out genuine top-$k$ results.

\subbb{Manipulating} $\mathsf{SelectTop}$.
To manipulate $\mathsf{SelectTop}$ of LDPMiner, the attacker first considers the item domain that shrinks from $\mathcal{I}$ to $\mathcal{C}$ and the aggregator picks top-$k$ rather than top-$2k$ frequent items as the result compared with manipulating $\mathsf{PruneDom}$. Usually, the attacker has squeezed out some top-$k$ items in the $\mathsf{PruneDom}$ phase. The left genuine top-$k$ items have a higher frequency than those squeezed out, which means the frequency bound becomes higher, and squeezing out more true top-$k$ items becomes harder. Specifically, the attacker employs $\mathsf{AREst}$-$\mathsf{RAPPOR}$ and $\mathsf{PDGen}$-$\mathsf{RAPPOR}$ to estimate the attack resources and generate poisoned data. 

\subsection{Attacking SVIM}
\label{sec:attacksvim}
\subbb{Manipulating} $\mathsf{PruneDom}$.
In the $\mathsf{PruneDom}$ phase of SVIM, the padding length $l$ is set to 1, which is equivalent to no padding at all. Since users have different sizes of values (sets of items), the estimation can be biased. Another difference from LDPMiner is that SVIM chooses an adapted FO instead of SH, but the choice is out of the attacker's control since the FO is chosen based on the item domain and padding length, which are set in advance instead of being determined by the collected reports from users. Similar to LDPMiner, the attacker can let the corrupted users submit poisoned reports to increase the frequency estimation of target items by using $\mathsf{AREst}$-$\mathsf{FO}$, $\mathsf{TSRef}$, and $\mathsf{PDGen}$-$\mathsf{FO}$, where $\mathsf{FO}$ is $\mathsf{GRR}$ or $\mathsf{OLH}$ based on the setting of the SVIM protocol. In addition, we will also show how to manipulate SVIM to choose a certain FO when the protocol uses an adaptive and data-driven way to determine the padding length. 

\subbb{Manipulating} $\mathsf{ConfigPro}$ and $\mathsf{SelectTop}$.
A unique challenge for SVIM is that the protocol may employ an adaptive method to choose the FO from GRR or OLH, and the attacker does not know in advance which FO will be chosen. In SVIM, OLH is chosen if 
\begin{equation}\label{eq:adpfo}
   d\ge l(4l-1)e^\epsilon+1, 
\end{equation}
otherwise GRR is chosen as the FO~\cite{DBLP:conf/sp/WangLJ18}. In the above equation, $d$ is the domain size, which is fixed for each phase, and $l$ is the padding size. The aggregator uses OLH to estimate the distribution of the length of users' values and sets $l$ as the 90th percentile of the value length among all users for the $\mathsf{SelectTop}$ phase. This allows the attacker to manipulate the value of $l$ by having the corrupted users falsely report the value length. 

Therefore, the first step to manipulate $\mathsf{SelectTop}$ of SVIM is to craft corrupted users' reports to trick the aggregator into choosing a certain FO. In~\refeq{eq:adpfo} we can notice that OLH is chosen if $l$ is small enough, otherwise GRR is chosen. Hence, the attacker has two options: 1) let the corrupted users submit reports to support as many target items as possible while having as small a value length as possible to decrease the padding length $l$ so that the aggregator will choose OLH as the FO; 2) increase the padding length $l$ to make GRR be chosen as the FO by letting the corrupted users report larger value lengths. 

Intuitively, both options seem viable. However, we find that the former strategy hardly works because the item domain size $d$ already shrinks to $2k$ in $\mathsf{SelectTop}$. Since $d$ is quite small and the right-hand side of~\refeq{eq:adpfo} grows asymptotically proportional to the square of $l$, the padding size must be very small for {OLH} to be chosen. 
In our experiment, we find that even if the attacker controls up to 10\% of the users to submit crafted reports, the padding length is still too large to get the OLH to be chosen as the FO. Moreover, it may also limit the attack capability to lower the padding length. 
Therefore, we choose the later option of manipulating the aggregator to set a longer padding length and choose GRR as the FO. Let's explain why this is the better strategy with an example. For the first option, suppose all the reports of the benign users provide $100$ supports for different length values, the attacker must provide $900$ supports for lower values to make the $90$\textit{th} percentile in the attacker's control. In contrast, if the attacker tries to increase the estimate of the $90$\textit{th} percentile by providing reports that support higher length values, only $11$ supports are needed to make the $90$\textit{th} percentile under the control of the attacker. In addition, since the attacker has manipulated the aggregator to set a longer padding length, the benign users will pad their value with more dummy items, giving the attacker a greater opportunity to push more target items into the final top-$k$ results.

Given that the SVIM protocol employs OLH to estimate the padding size $l$, the attacker utilizes $\mathsf{AREst}$-$\mathsf{OLH}$, $\mathsf{TSRef}$, and $\mathsf{PDGen}$-$\mathsf{OLH}$ to manipulate the aggregator to choose GRR as the FO. The attacker manipulates SVIM by using $\mathsf{AREst}$-$\mathsf{GRR}$ to estimate the supports that can be allocated for the attack, $\mathsf{TSRef}$ to select targets, and $\mathsf{PDGen}$-$\mathsf{GRR}$ to generate poisoned reports to affect the estimation of top-$k$ results. 

\subsection{Attacking SVSM}
\label{sec:attacksvsm}
\subbb{Manipulating} $\mathsf{PruneDom}$. SVSM performs the SVIM protocol as the first step of its $\mathsf{PruneDom}$. 
After the top-$k$ frequent items are mined by the SVIM, SVSM uses the result to pruned the domain of itemsets. We note that the aggregator has no further interaction with users after the SVIM protocol is executed in the $\mathsf{PruneDom}$ of SVIM. Therefore, the attacker can only manipulate the first step of SVSM's $\mathsf{PruneDom}$ phase by the attack in~\refsec{sec:attacksvim} (including manipulating $\mathsf{PruneDom}$ and $\mathsf{SelectTop}$ phases of SVIM), and cannot influence the following parts of this phase. Since the determination of the set of candidate itemsets is based on the results of SVIM (\ie, the top-$k$ frequent items), if we can successfully attack SVIM, the performance of SVSM will also be significantly degraded. 

\subbb{Manipulating} $\mathsf{ConfigPro}$ and $\mathsf{SelectTop}$. We use a similar way as in SVIM to manipulate these two parts of SVSM by treating the itemsets in SVSM as items in SVIM. First, we manipulate the server to choose a longer padding length by committing reports that support long padding length values. Then we estimate the true itemset frequencies from our knowledge and choose target itemsets to be supported in the final estimation part.

\subsection{Attacking FIML}
\label{sec:attackfiml}
\subbb{Manipulating} $\mathsf{PruneDom}$. FIML utilizes OLH to estimate the frequencies of items and construct candidate sets. Consequently, an attacker can estimate the genuine frequency and obtain an attack resource estimation using the AREst-OLH attack operation. The attacker selects targets using $\mathsf{TSRef}$ and generates poisoned reports using $\mathsf{PDGen}$-$\mathsf{OLH}$. To manipulate the results and undermine the accuracy of the top-$k$ results, the attacker must include at least $k+1$ non-top-$k$ items in the candidate set $\mathcal{C}$, which has a size of $1.5k$.

\subbb{Manipulating} $\mathsf{SelectTop}$. In the protocol's $\mathsf{SelectTop}$ phase, a membership query is performed by selecting an random item/itemset from the candidate set for each user and asking the user whether the chosen item/itemset is one of her/his top-$k$ items/itemsets. In FIML, the GRR mechanism with a binary domain is used as FO for this step. When responding to a query, the attacker follows a specific strategy: if the received item/itemset is in the attacker's target set, the attacker reports $1$; otherwise, the attacker reports $0$.

\subsection{Adapting to Scenarios with Less Information}
\label{sec:lessknow}
The major assumption of the full-knowledge attack is that the attacker knows the true frequencies of all items/itemsets, which could be a restrictive assumption in some cases. We try to remove this assumption and adapt our \textit{Adaptive Orchestration Attack} (AOA) to more practical scenarios. Specifically, we introduce two attacks, which have limited information about the prior knowledge, against LDP frequent itemset mining protocols: 
    
\subbb{Partial-knowledge attack.} Because the attacker has compromised a group of users, she/he can use the values of those users to estimate the frequencies of items. Specifically, in the full-knowledge attack, we assume that the attacker knows the true frequencies of items or itemsets, $\{f_t\}_{t\in \mathcal{I}}$ or $\{f_I\}_{I \subseteq \mathcal{I}}$, respectively. The partial-knowledge attack only requires that the attacker knows the transactions possessed by the corrupted users, and the attacker uses these transactions to estimate the frequencies of items. For the partial-knowledge attack, due to the limit of the observation, it may be difficult for the attacker to accurately estimate the frequencies of items. Previous work~\cite{DBLP:conf/www/FangSLGT021} shows that some statistical techniques, \eg, Bootstrapping, can be used to refine the estimated frequencies in this case. 

\subbb{Man-in-the-middle (MITM) attack.} In the case where the attacker can eavesdrop on the messages between the users and the collector, the attacker can capture the perturbed reports from the users. Thus, the attacker can also use this knowledge to estimate the frequencies from benign users.  The attacker can exploit the aggregation part of the LDP protocol and obtain the estimated frequencies of items/itemsets. Unlike the partial-knowledge attack, the MITM attacker can only obtain the perturbed data of the LDP protocol.  When the privacy budget is low, the perturbed reports of the users may be inaccurate, which leads to a poor target set determination for the attacker. However, the low privacy budget does not affect the attacker's simulation of the data collector's frequency estimation, because the attacker received exactly the same perturbed data from users as the collector. 

We note that the attack resource estimation and the poisoned data generation operations do not rely on the knowledge of the frequencies of items/itemsets and such scenarios with less information only affect the target set refinement operation. 

\begin{figure}
	\centering
	\subfigure[IBM, $\mathrm{ACC}$]{
		\begin{minipage}[c]{0.2\textwidth}
		\centering
        \includegraphics[width=1\textwidth]{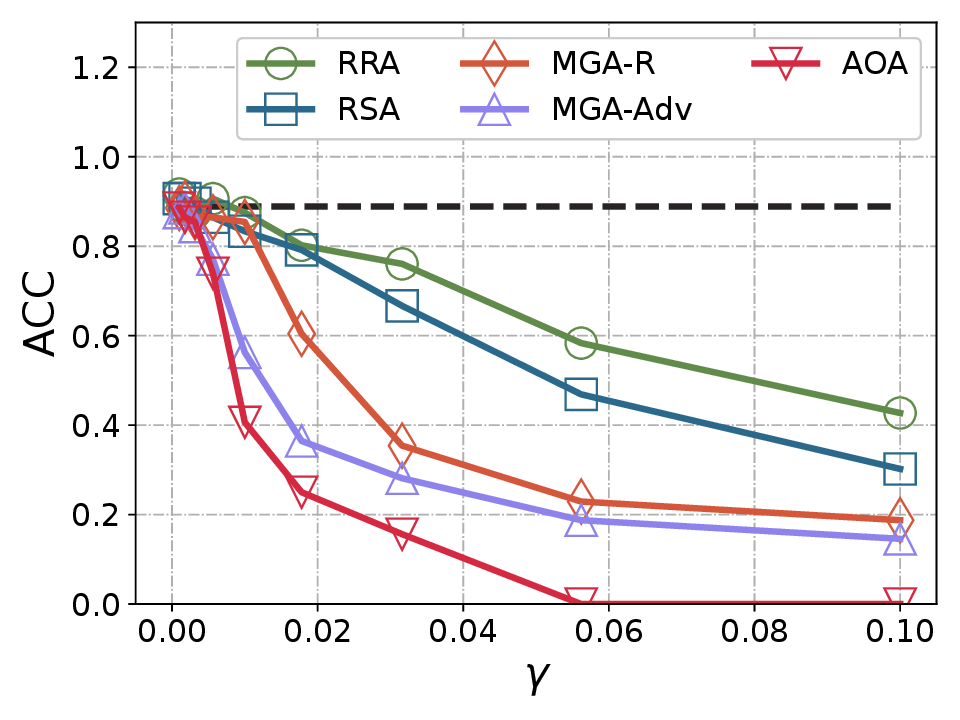}
		\end{minipage}%
	}
        	\subfigure[IBM, $\mathrm{NCR}$]{
		\begin{minipage}[c]{0.2\textwidth}
		\centering
        \includegraphics[width=1\textwidth]{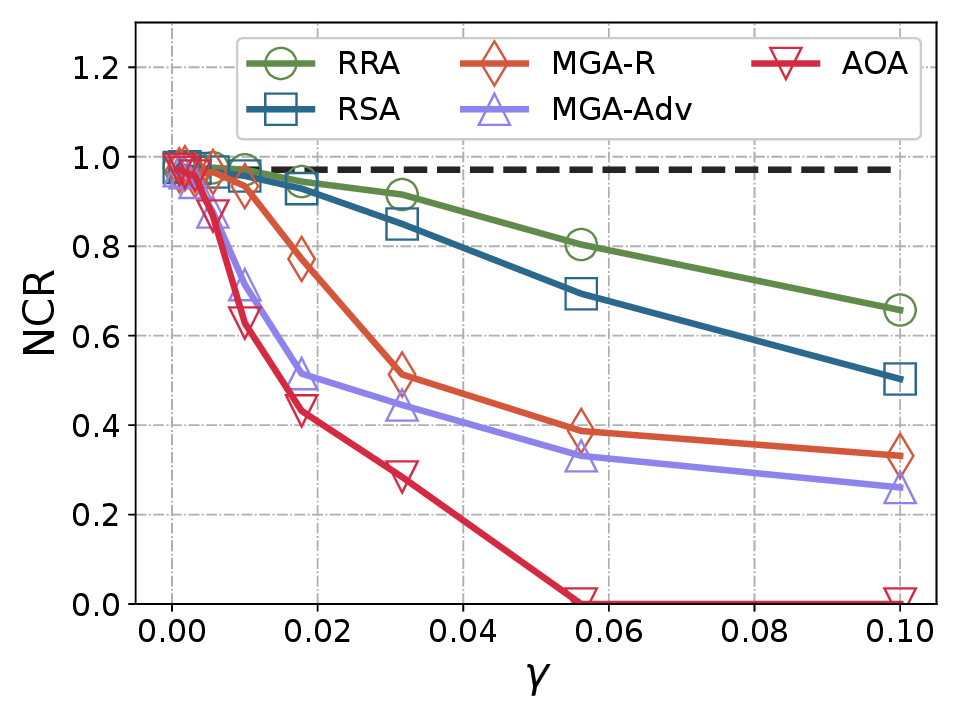}
		\end{minipage}%
    }
    
      \subfigure[Kosarak, $\mathrm{ACC}$]{
		\begin{minipage}[c]{0.2\textwidth}
		\centering
        \includegraphics[width=1\textwidth]{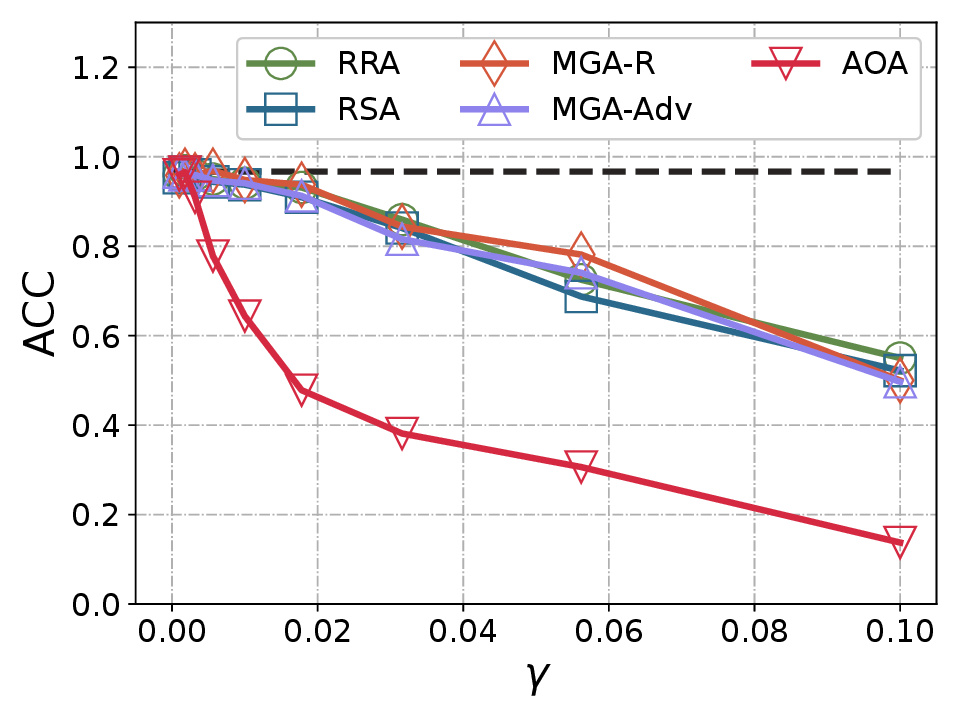}
		\end{minipage}%
	}
     	\subfigure[Kosarak, $\mathrm{NCR}$]{
		\begin{minipage}[c]{0.2\textwidth}
		\centering
        \includegraphics[width=1\textwidth]{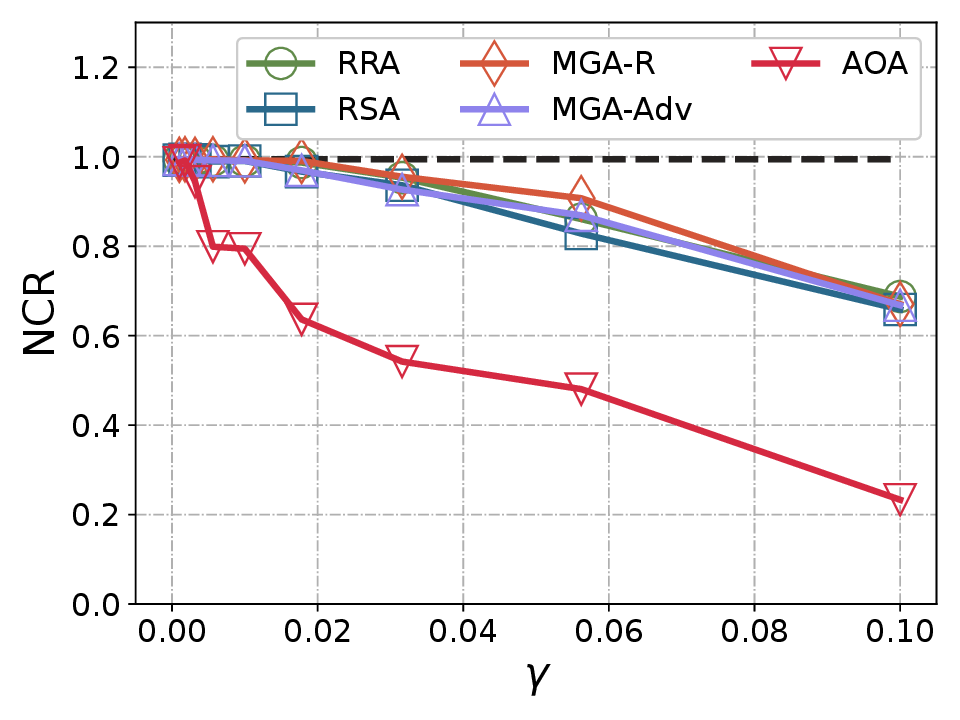}
		\end{minipage}%
    }
    
     \subfigure[BMS-POS, $\mathrm{ACC}$]{
		\begin{minipage}[c]{0.2\textwidth}
		\centering
        \includegraphics[width=1\textwidth]{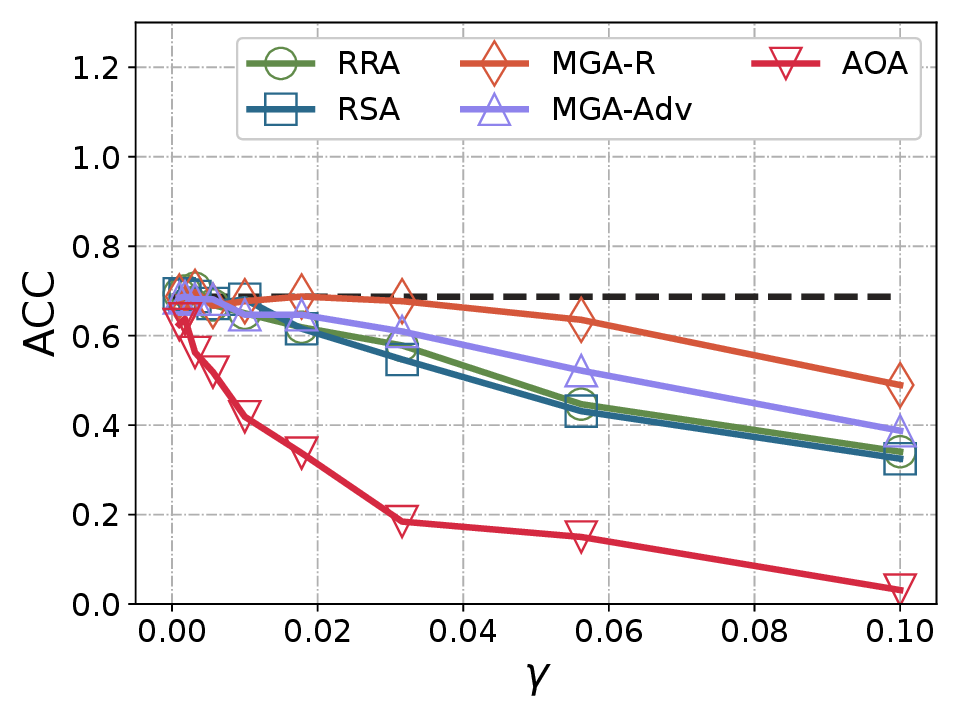}
		\end{minipage}%
	}
	\subfigure[BMS-POS, $\mathrm{NCR}$]{
		\begin{minipage}[c]{0.2\textwidth}
		\centering
        \includegraphics[width=1\textwidth]{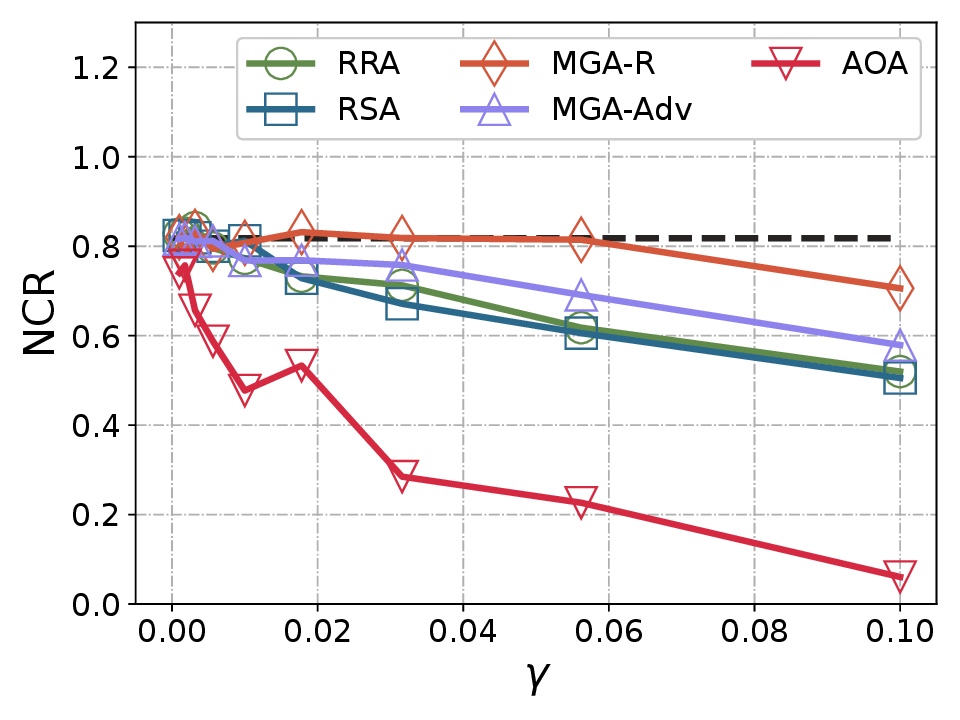}
		\end{minipage}%
    }

    \centering
	\caption{Attacking SVSM, with $k = 32$, $\epsilon=4.0$ (black dashed line - no attack).}
    \Description[Attacking SVSM ($\epsilon=4.0$)]{Attacking SVSM, with $k = 32$, $\epsilon=4.0$ (black dashed line - no attack).}
	\label{fig:gamma:svsm}
\end{figure}

\begin{figure}
	\centering
	\subfigure[\ccsrev{IBM, $\mathrm{ACC}$}]{
		\begin{minipage}[c]{0.2\textwidth}
		\centering
        \includegraphics[width=1\textwidth]{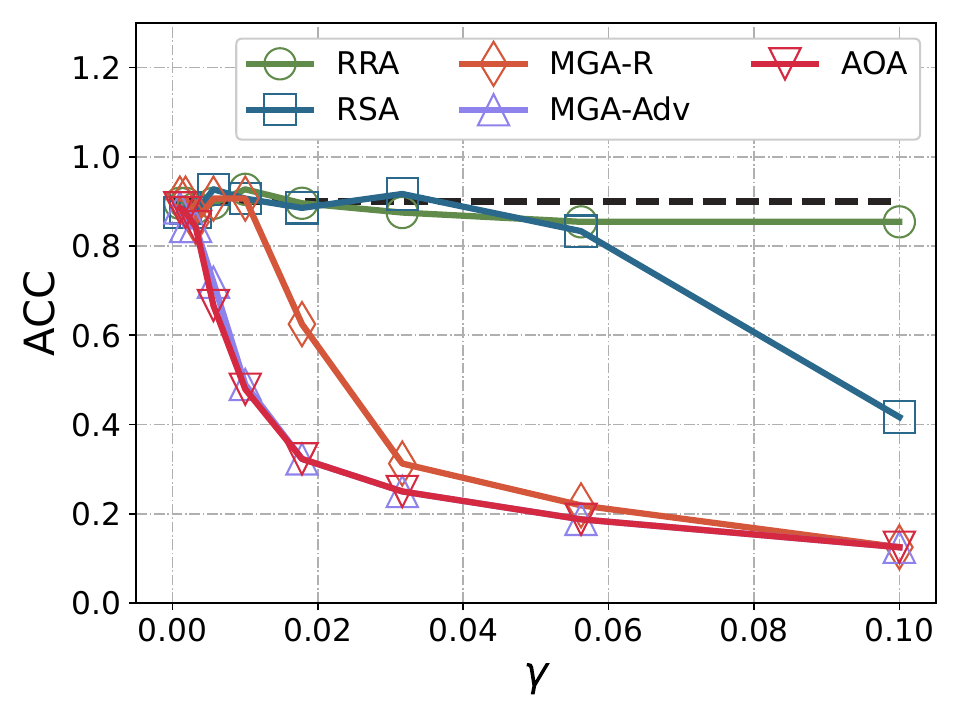}
		\end{minipage}%
	}
 	\subfigure[\ccsrev{IBM, $\mathrm{NCR}$}]{
		\begin{minipage}[c]{0.2\textwidth}
		\centering
        \includegraphics[width=1\textwidth]{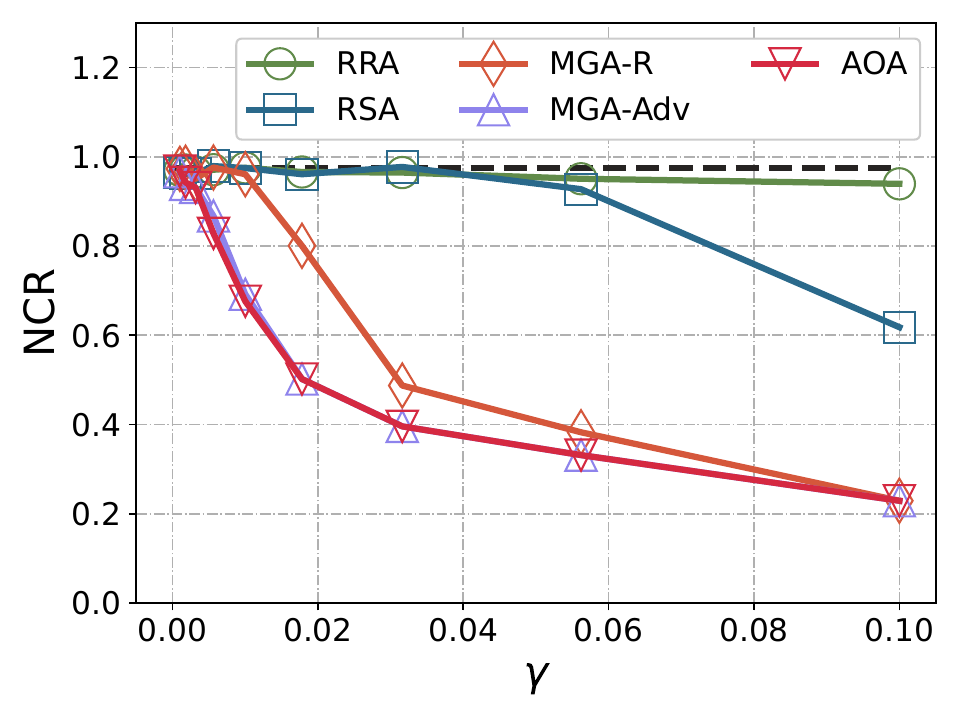}
		\end{minipage}%
    }
 
      \subfigure[\ccsrev{Kosarak, $\mathrm{ACC}$}]{
		\begin{minipage}[c]{0.2\textwidth}
		\centering
        \includegraphics[width=1\textwidth]{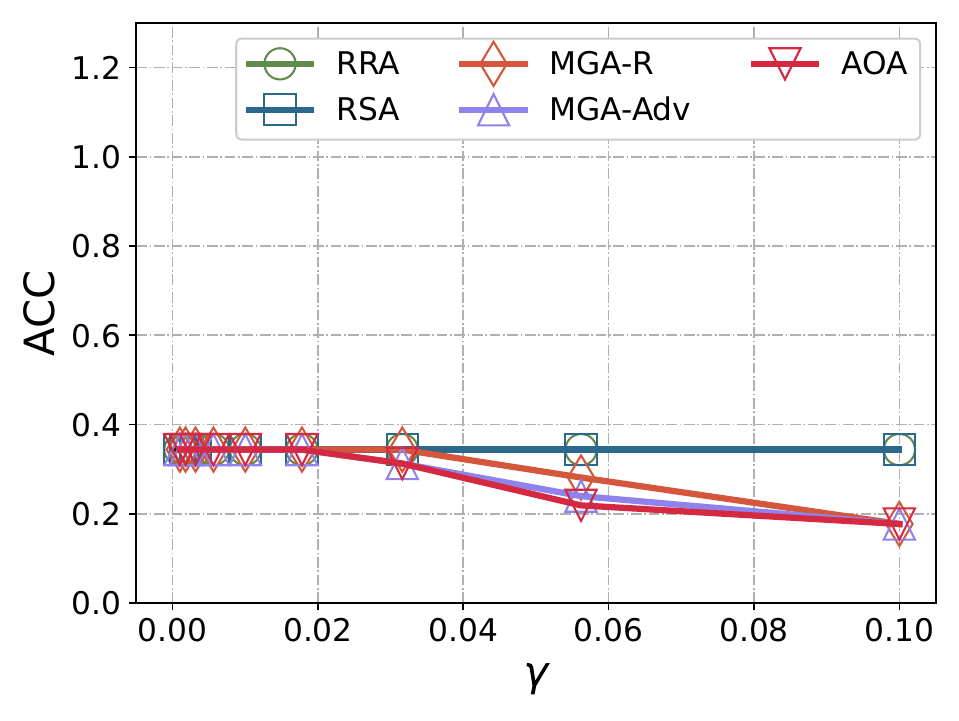}
		\end{minipage}%
	}
     	\subfigure[\ccsrev{Kosarak, $\mathrm{NCR}$}]{
		\begin{minipage}[c]{0.2\textwidth}
		\centering
        \includegraphics[width=1\textwidth]{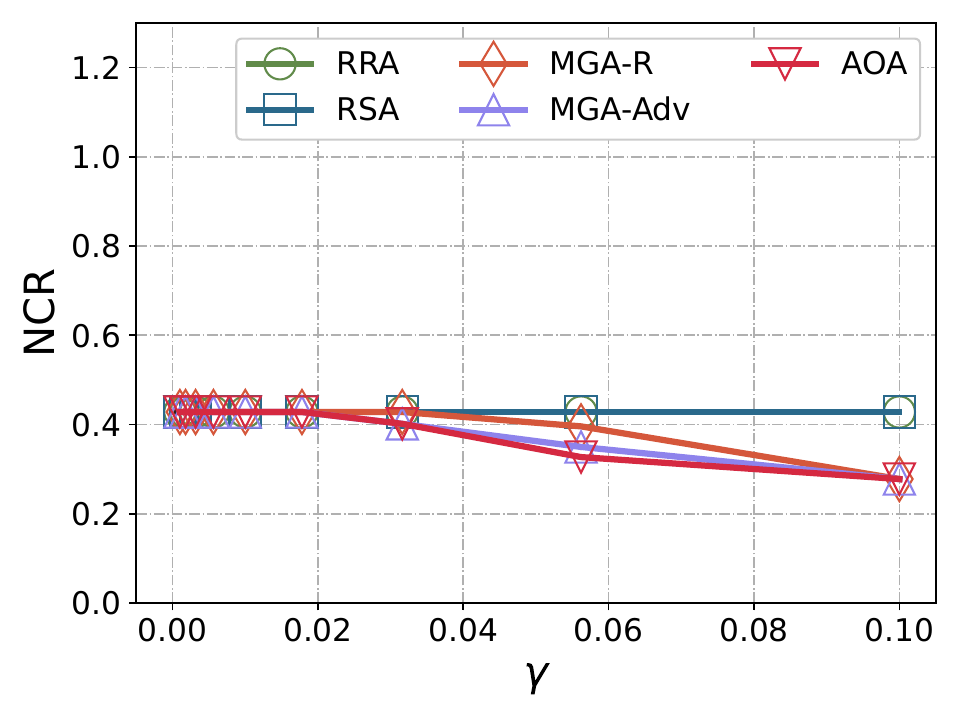}
		\end{minipage}%
    }
 
     \subfigure[\ccsrev{BMS-POS, $\mathrm{ACC}$}]{
		\begin{minipage}[c]{0.2\textwidth}
		\centering
        \includegraphics[width=1\textwidth]{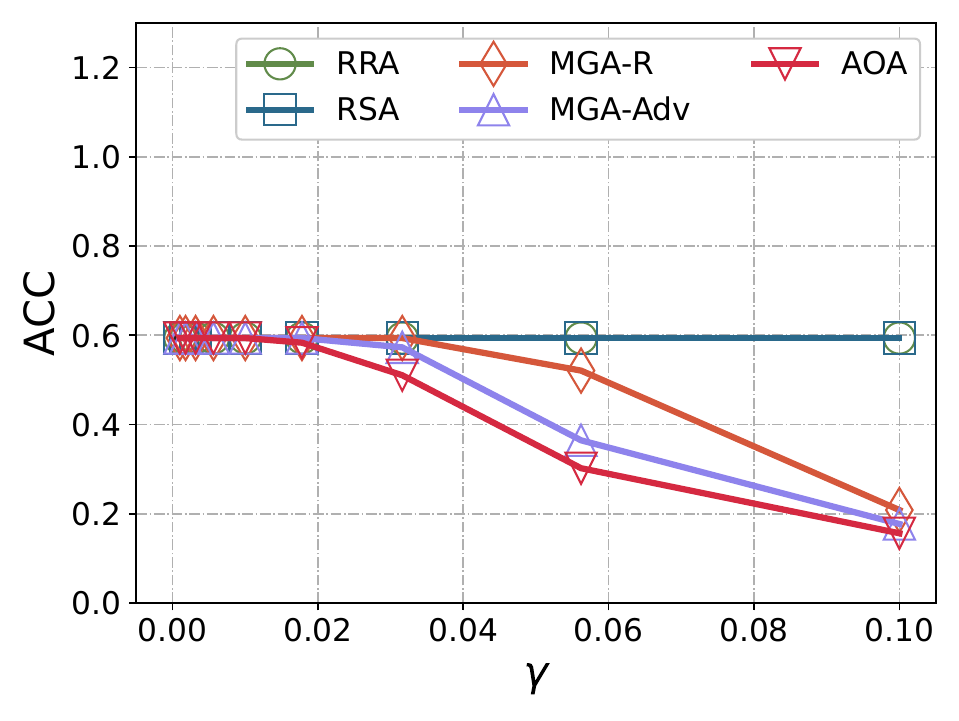}
		\end{minipage}%
	}
	\subfigure[\ccsrev{BMS-POS, $\mathrm{NCR}$}]{
		\begin{minipage}[c]{0.2\textwidth}
		\centering
        \includegraphics[width=1\textwidth]{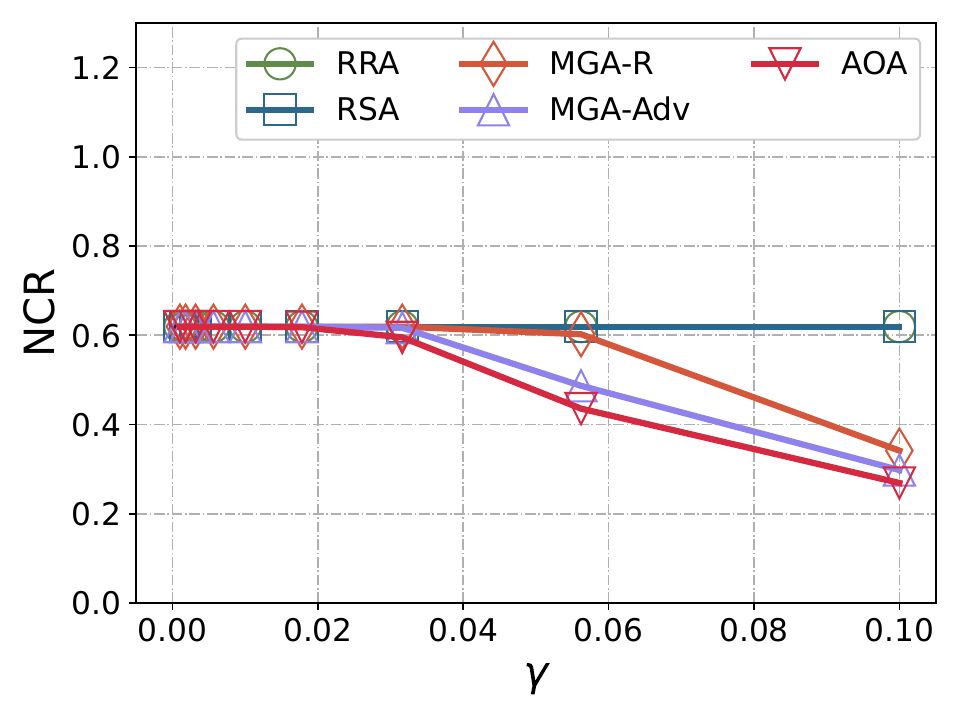}
		\end{minipage}%
    }

    \centering
	\caption{\ccsrev{Attacking FIML-IS, with $k = 32$, $\epsilon=4.0$ (black dashed line - no attack).}}
 \Description[Attacking FIML-IS ($\epsilon=4.0$)]{Attacking FIML-IS, with $k = 32$, $\epsilon=4.0$ (black dashed line - no attack).}
	\label{fig:gamma:fimlitemset:eps4}
\end{figure}

\begin{figure}
	\centering
	\subfigure[IBM, $\mathrm{ACC}$]{
		\begin{minipage}[c]{0.2\textwidth}
		\centering
        \includegraphics[width=1\textwidth]{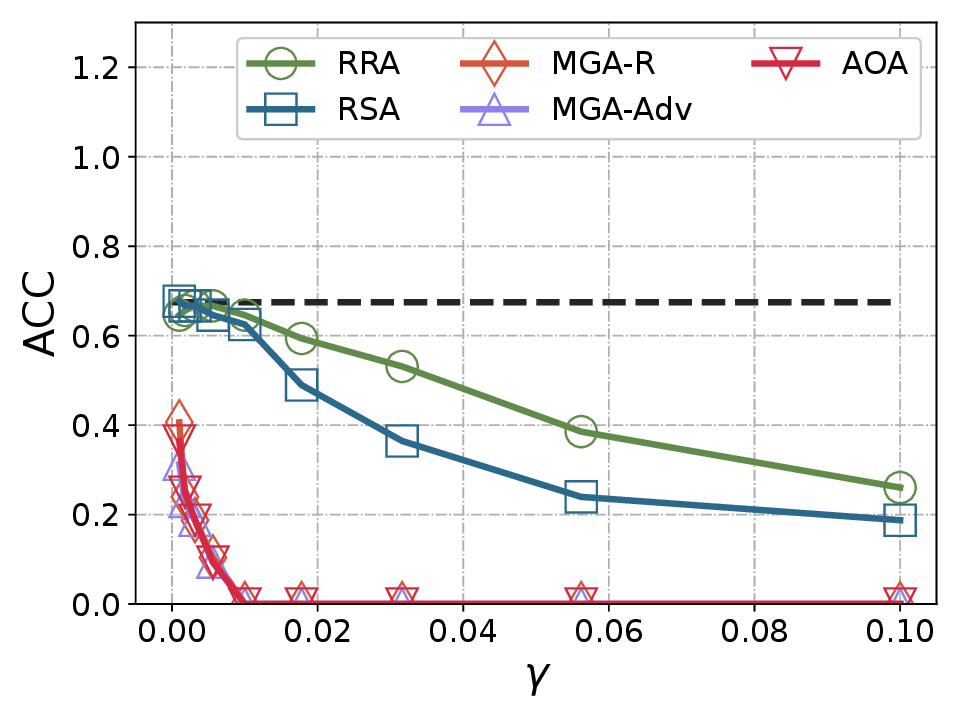}
		\end{minipage}%
	}
 	\subfigure[IBM, $\mathrm{NCR}$]{
		\begin{minipage}[c]{0.2\textwidth}
		\centering
        \includegraphics[width=1\textwidth]{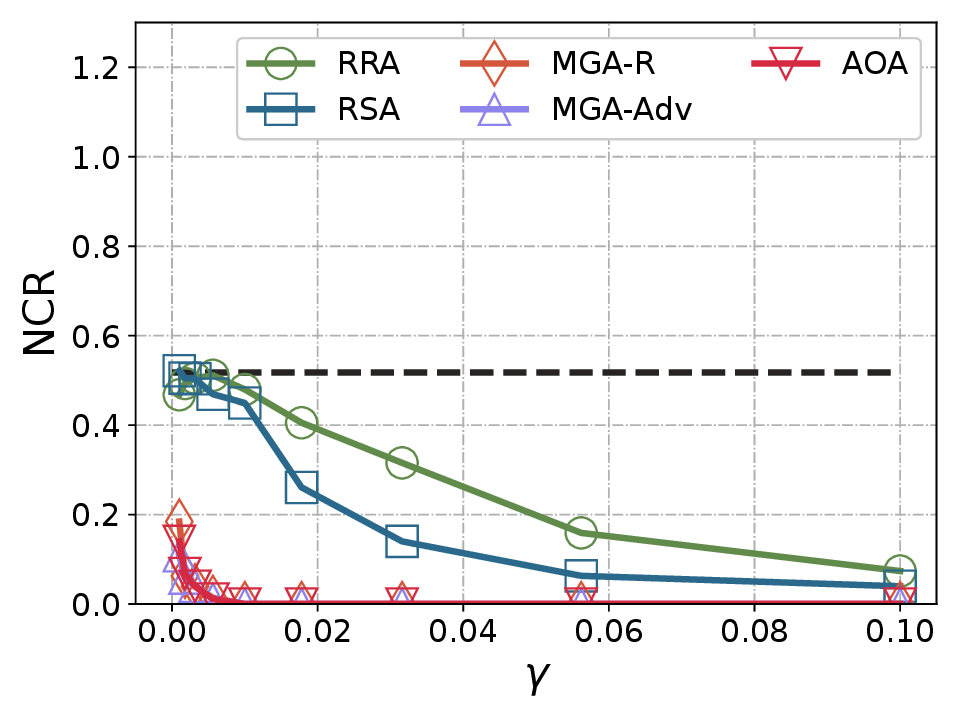}
		\end{minipage}%
    }
    
     \subfigure[Kosarak, $\mathrm{ACC}$]{
		\begin{minipage}[c]{0.2\textwidth}
		\centering
        \includegraphics[width=1\textwidth]{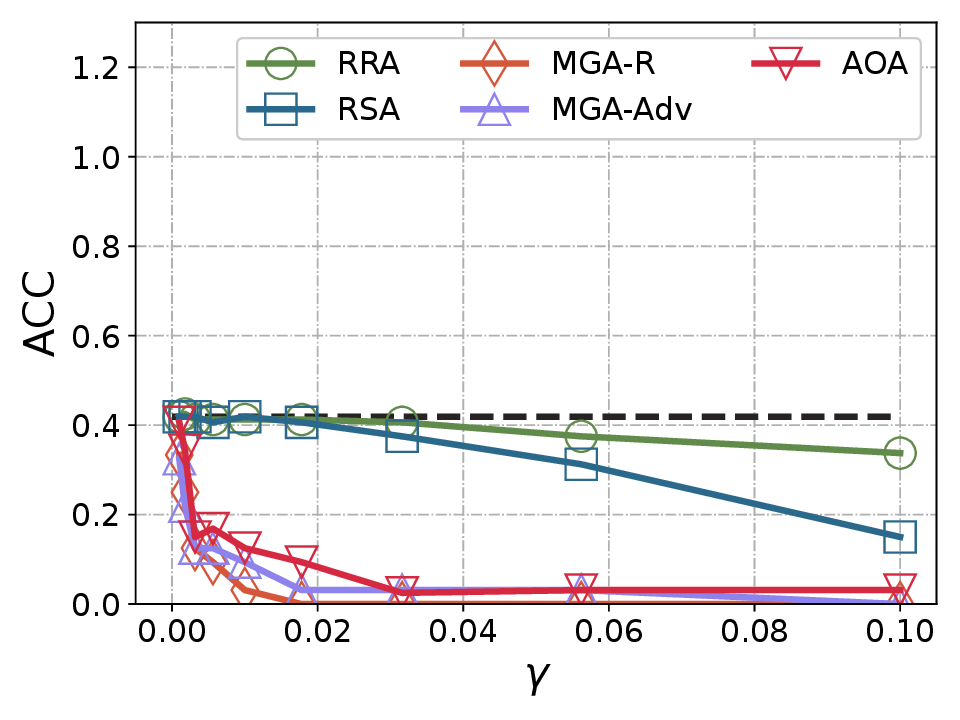}
		\end{minipage}%
	}
 	\subfigure[Kosarak, $\mathrm{NCR}$]{
		\begin{minipage}[c]{0.2\textwidth}
		\centering
        \includegraphics[width=1\textwidth]{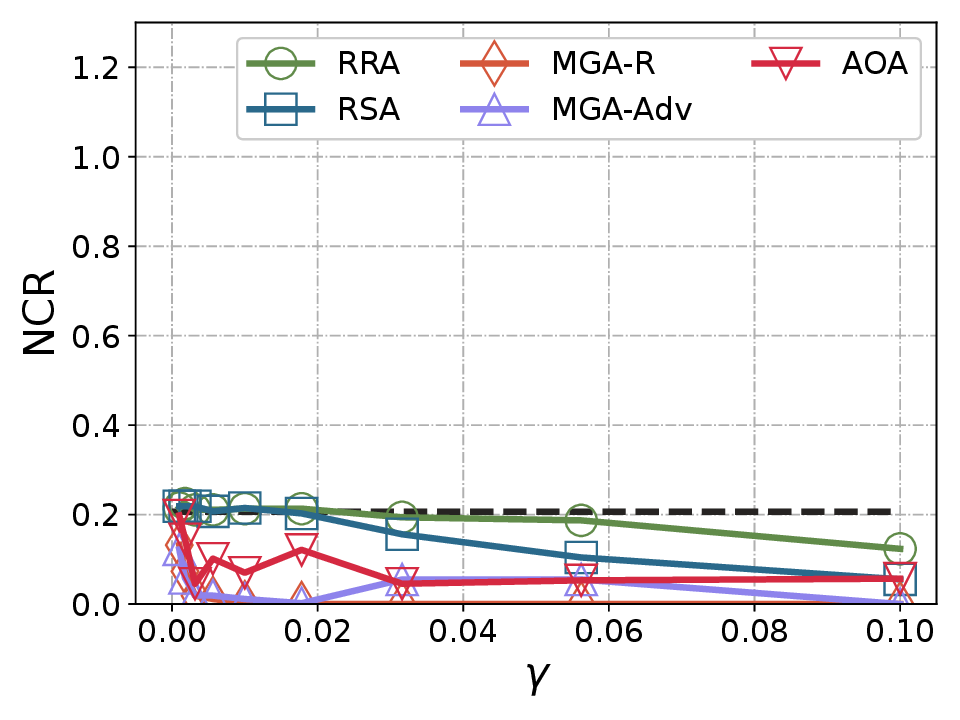}
		\end{minipage}%
    }
    
     \subfigure[BMS-POS, $\mathrm{ACC}$]{
		\begin{minipage}[c]{0.2\textwidth}
		\centering
        \includegraphics[width=1\textwidth]{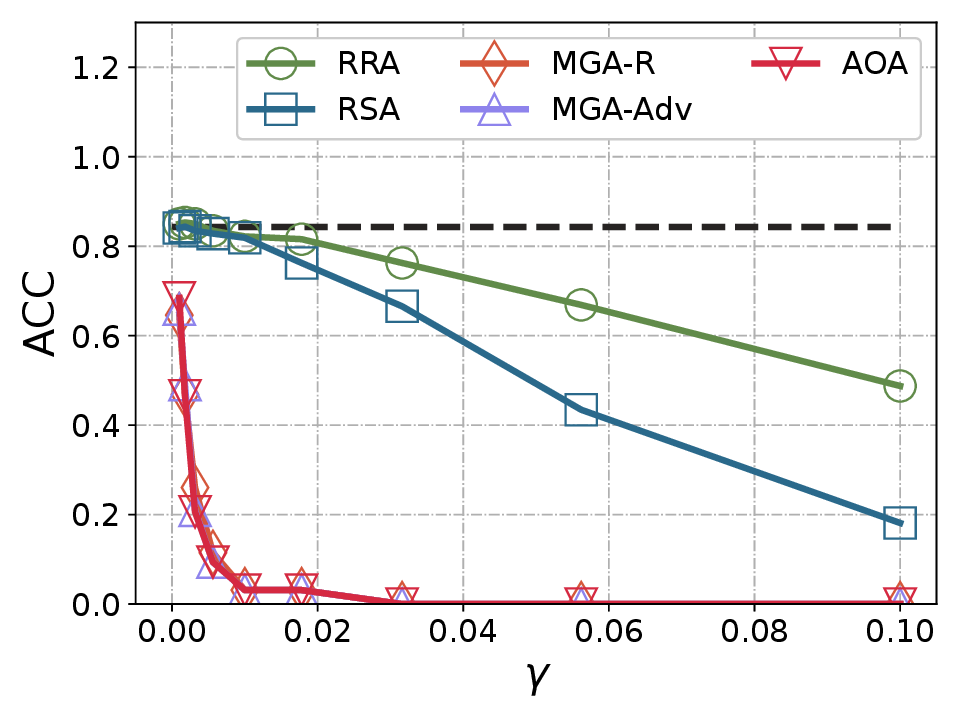}
		\end{minipage}%
	}
	\subfigure[BMS-POS, $\mathrm{NCR}$]{
		\begin{minipage}[c]{0.2\textwidth}
		\centering
        \includegraphics[width=1\textwidth]{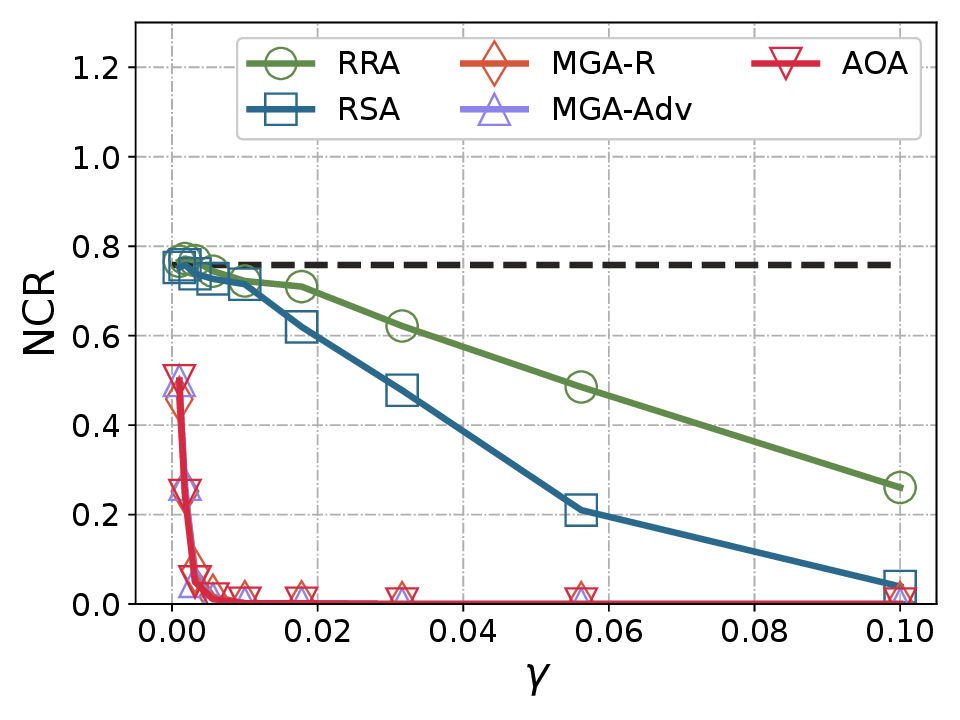}
		\end{minipage}%
    }
    \centering
	\caption{Attacking LDPMiner, with $k = 32$, $\epsilon=4.0$ (black dashed line - no attack).} 
  \Description[Attacking LDPMiner ($\epsilon=4.0$)]{Attacking LDPMiner, with $k = 32$, $\epsilon=4.0$ (black dashed line - no attack).}
	\label{fig:gamma:ldpminer}
\end{figure}

\section{Evaluation}
In this section, we introduce the setup for experiments. Next, we present the experimental results for attacking LDP frequent itemset mining protocol, SVSM, LDP frequent item mining protocols, LDPMiner and SVIM. We \ccsrev{also} evaluate the impact of different parameters of the LDP protocols and the attacks with limited knowledge. The codes are available in \url{https://github.com/CorneyHeY/Poison-Attack-LDP-Frequent-Itemset-Mining-CCS2024}. 

\subsection{Setup}
We compare the proposed attack with four other attacks:
\begin{itemize}[leftmargin=10pt]
    \item \textit{Random Report Attack} (RRA): For each corrupted user $j$, RRA randomly selects a sequence of perturbed values (a report) for it to answer the aggregator's requests in the LDP frequent itemset mining protocols.
    
    \item \textit{Random Set-Value Attack} (RSA): RSA randomly selects a subset of items for each corrupted user. The size of the chosen subset is equal to the cardinality of the original set of items held by the corresponding user.
    
    \item Variants of \textit{Maximal Gain Attack} (MGA)~\cite{DBLP:conf/uss/CaoJG21}: MGA is a data poisoning attack against LDP frequency estimation, which cannot be directly applied to attack against LDP frequent itemset mining because it can only promote the frequencies of items when the targets have been determined. Therefore, we consider two baseline attacks that incorporate the proposed attack operations into MGA for attacking LDP frequent itemset mining protocols. 1) MGA-R: this attack first estimates the frequencies of items, uniformly randomly selects $k$ items from the non-top-$k$ items as the targets, and then maximizes the frequency gains of these items by MGA; 2) MGA-Adv: this attack uses $\mathsf{AREst}$ to estimate the attack resources, uses $\mathsf{TSRef}$ to get the set of target items, and then promotes the frequencies of items in the target set by MGA. 
\end{itemize}

\paragraphbe{Datasets} We evaluate the proposed attacks on three datasets. 
\begin{itemize}[leftmargin=10pt]
    \item IBM Synthesize~\cite{DBLP:books/mit/fayyadPSU96/AgrawalMSTV96}: A dataset generated by IBM Synthetic Data Generator~\footnote{\url{https://github.com/zakimjz/IBMGenerator}, provided by Mohammed J. Zaki.}, containing 1.8 million transactions with 1,000 items. 
    \item BMS-POS~\cite{KDDCup-2000}: A dataset of points of sale data, containing more than half a million users and 1,657 items~\footnote{\url{https://github.com/cpearce/HARM}, prepossessed by Chris Pearce.}.
    \item Kosarak~\cite{DBLP:conf/wsdm/Benson0T18}: A dataset of clickstreams on a Hungarian news portal, containing more than one million users and the domain of items is 42,000.
\end{itemize}

\begin{figure}
	\centering
	\subfigure[IBM, $\mathrm{ACC}$]{
		\begin{minipage}[c]{0.2\textwidth}
		\centering
        \includegraphics[width=1\textwidth]{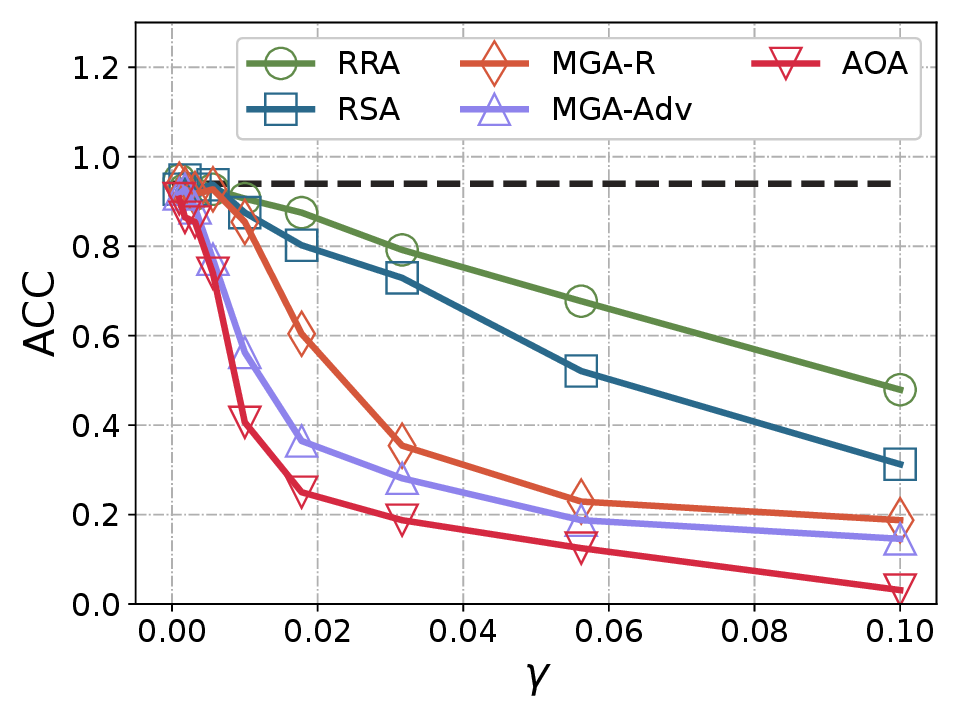}
		\end{minipage}%
	}
 	\subfigure[IBM, $\mathrm{NCR}$]{
		\begin{minipage}[c]{0.2\textwidth}
		\centering
        \includegraphics[width=1\textwidth]{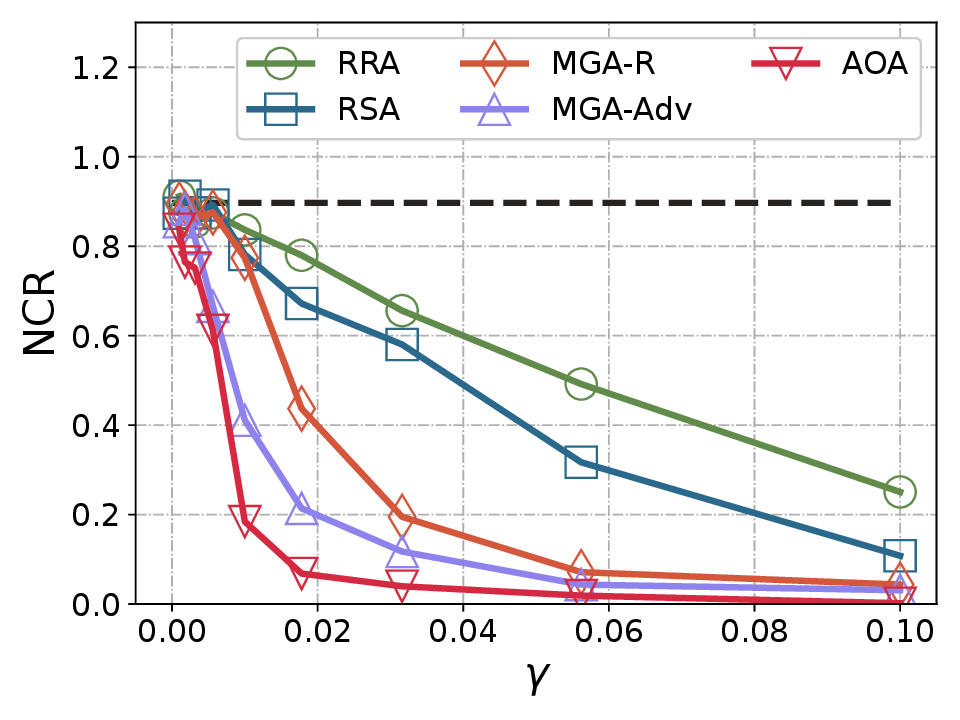}
		\end{minipage}%
    }
    
     \subfigure[Kosarak, $\mathrm{ACC}$]{
		\begin{minipage}[c]{0.2\textwidth}
		\centering
        \includegraphics[width=1\textwidth]{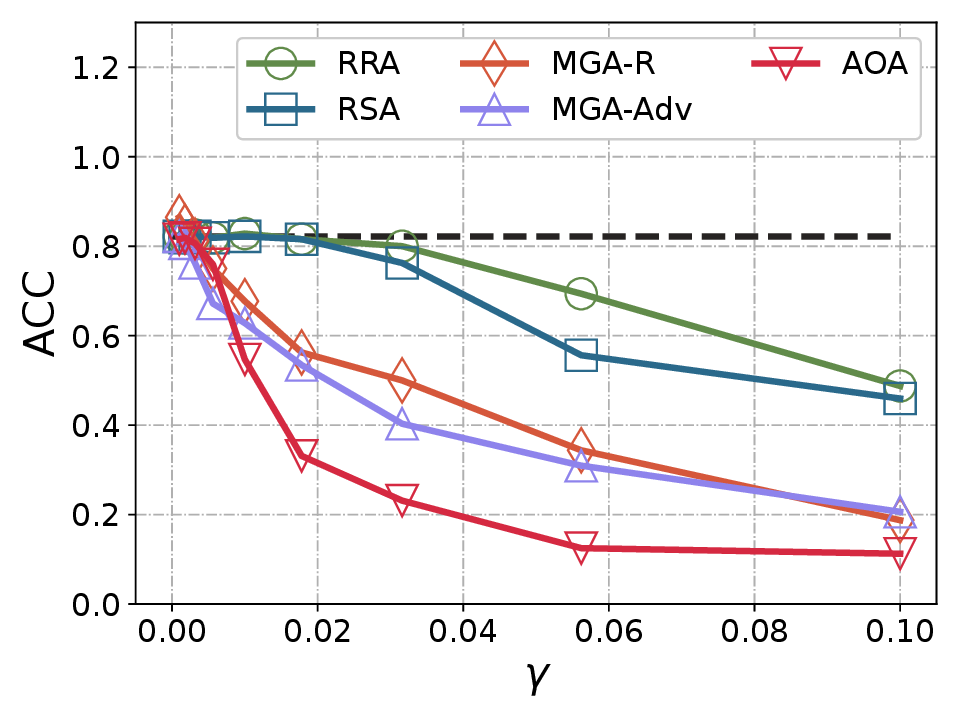}
		\end{minipage}%
	}
 	\subfigure[Kosarak, $\mathrm{NCR}$]{
		\begin{minipage}[c]{0.2\textwidth}
		\centering
        \includegraphics[width=1\textwidth]{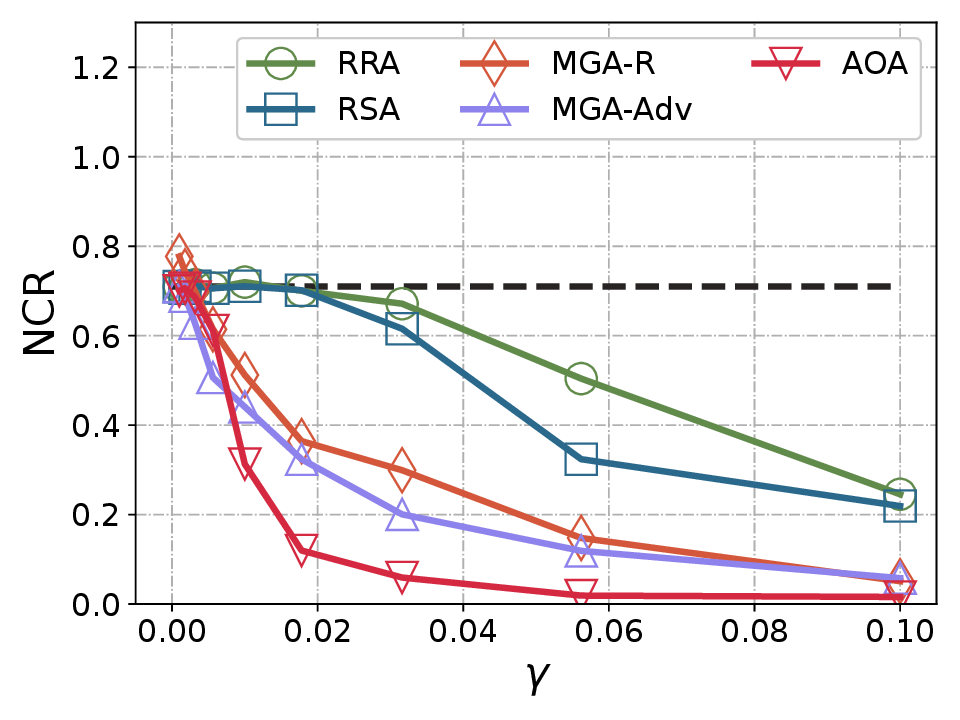}
		\end{minipage}%
    }
 
     \subfigure[BMS-POS, $\mathrm{ACC}$]{
		\begin{minipage}[c]{0.2\textwidth}
		\centering
        \includegraphics[width=1\textwidth]{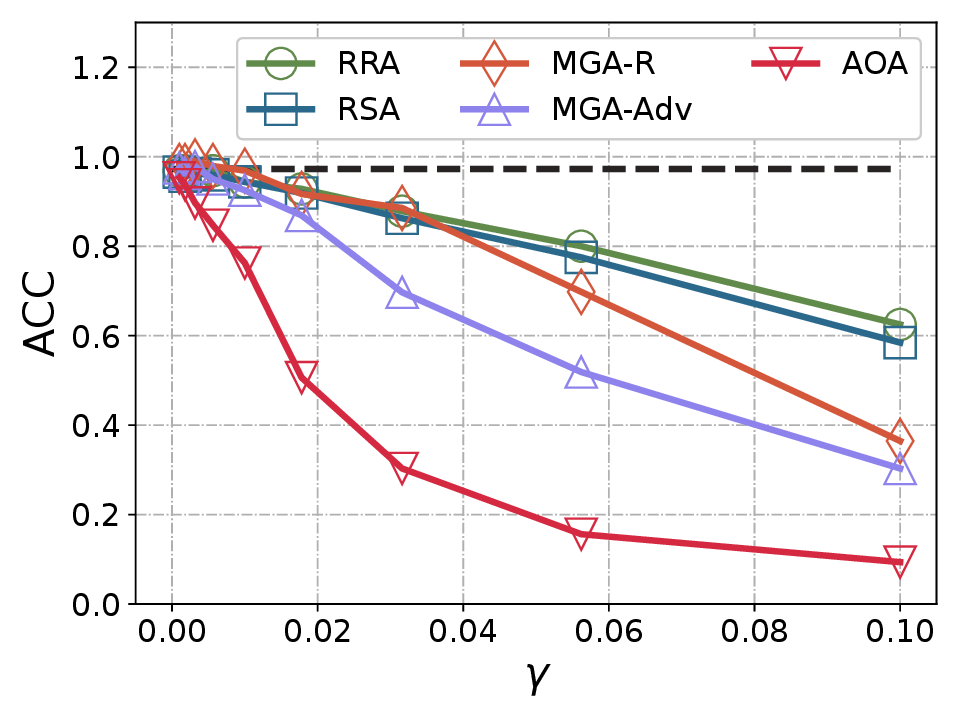}
		\end{minipage}%
	}
	\subfigure[BMS-POS, $\mathrm{NCR}$]{
		\begin{minipage}[c]{0.2\textwidth}
		\centering
        \includegraphics[width=1\textwidth]{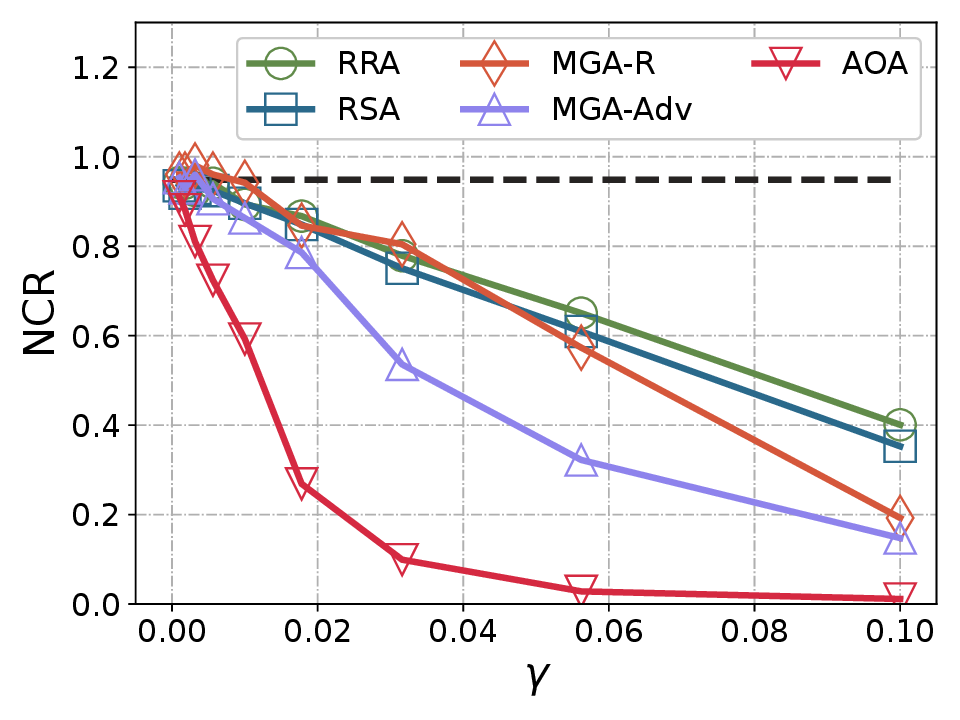}
		\end{minipage}%
    }
    \centering
	\caption{Attacking SVIM, with $k = 32$, $\epsilon=4.0$ (black dashed line - no attack).} 
 \Description[Attacking SVIM ($\epsilon=4.0$)]{Attacking SVIM, with $k = 32$, $\epsilon=4.0$ (black dashed line - no attack).}
	\label{fig:gamma:svim}
\end{figure}

\begin{figure}
\centering
	\subfigure[SVSM, $\epsilon$]{
		\begin{minipage}[c]{0.2\textwidth}
		\centering
        \includegraphics[width=1\textwidth]{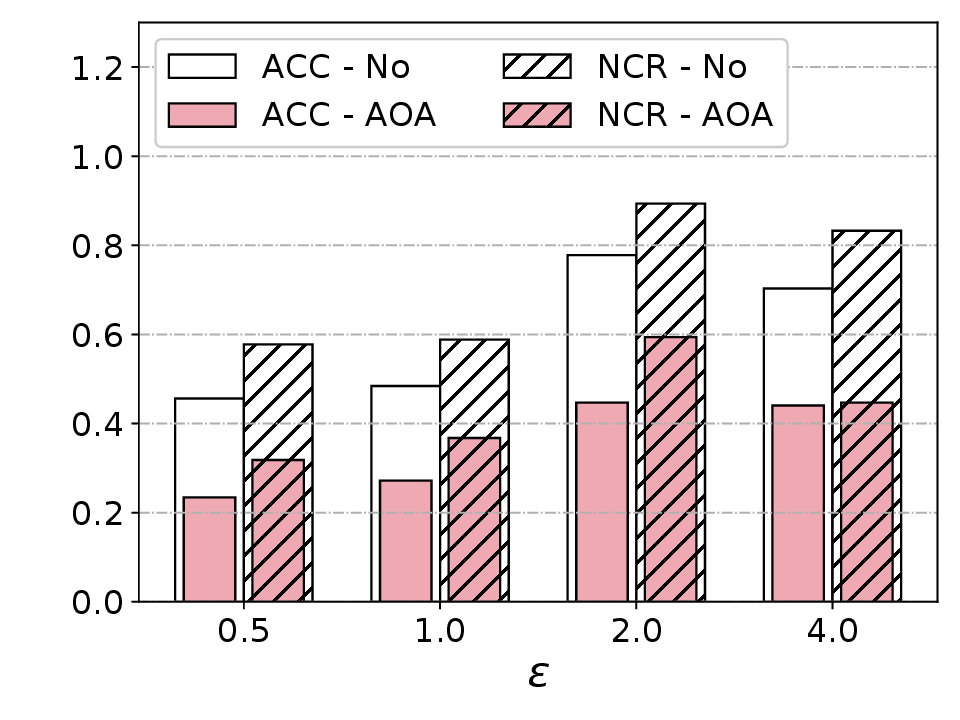}
		\end{minipage}%
	}
     	\subfigure[SVSM, k]{
		\begin{minipage}[c]{0.2\textwidth}
		\centering
        \includegraphics[width=1\textwidth]{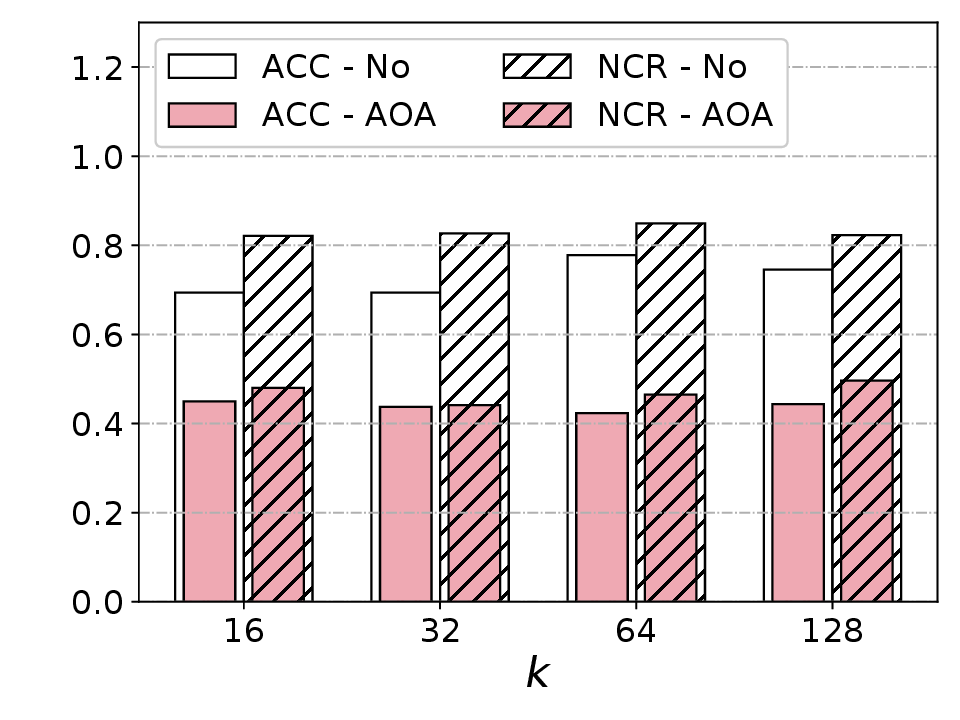}
		\end{minipage}%
	}
 
	\subfigure[LDPMiner, $\epsilon$]{
		\begin{minipage}[c]{0.2\textwidth}
		\centering
        \includegraphics[width=1\textwidth]{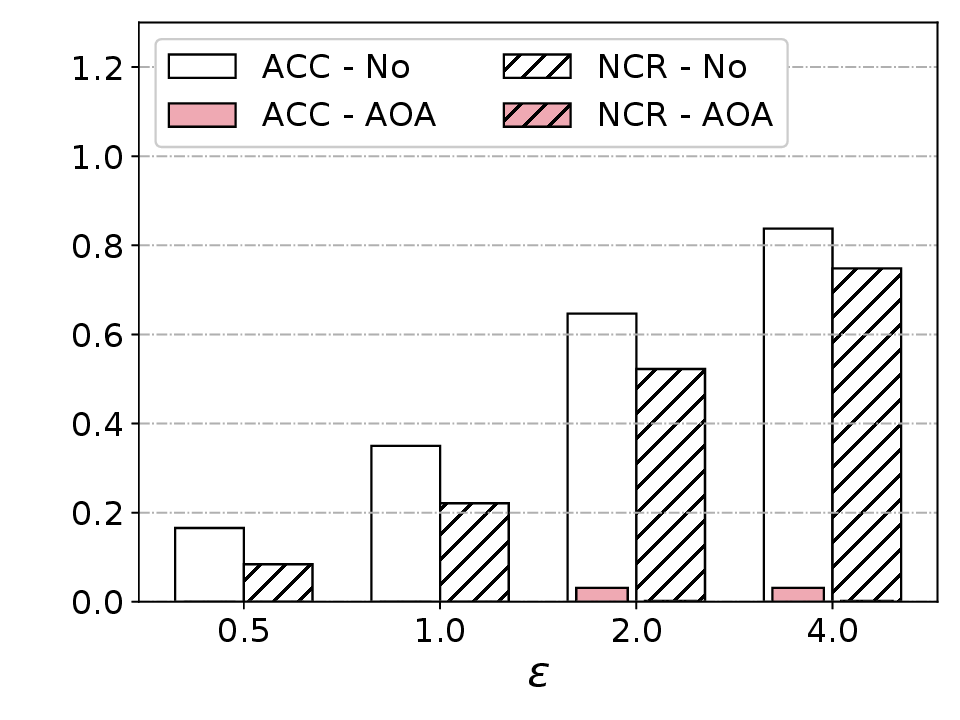}
		\end{minipage}%
    }
    	\subfigure[LDPMiner, k]{
		\begin{minipage}[c]{0.2\textwidth}
		\centering
        \includegraphics[width=1\textwidth]{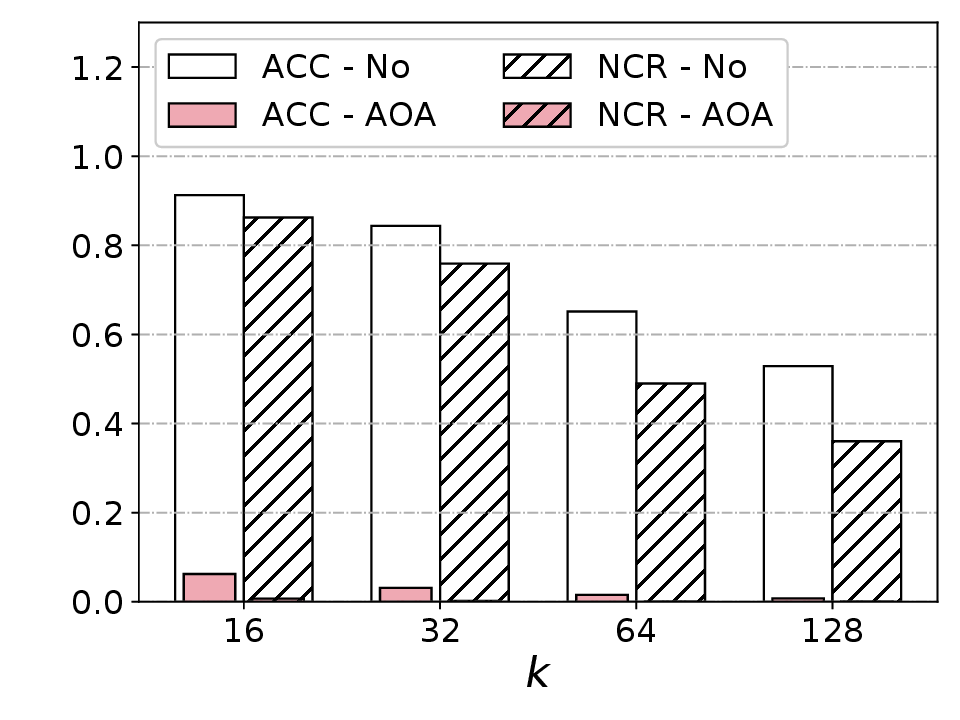}
		\end{minipage}%
    }
    
    \subfigure[SVIM, $\epsilon$]{
		\begin{minipage}[c]{0.2\textwidth}
		\centering
        \includegraphics[width=1\textwidth]{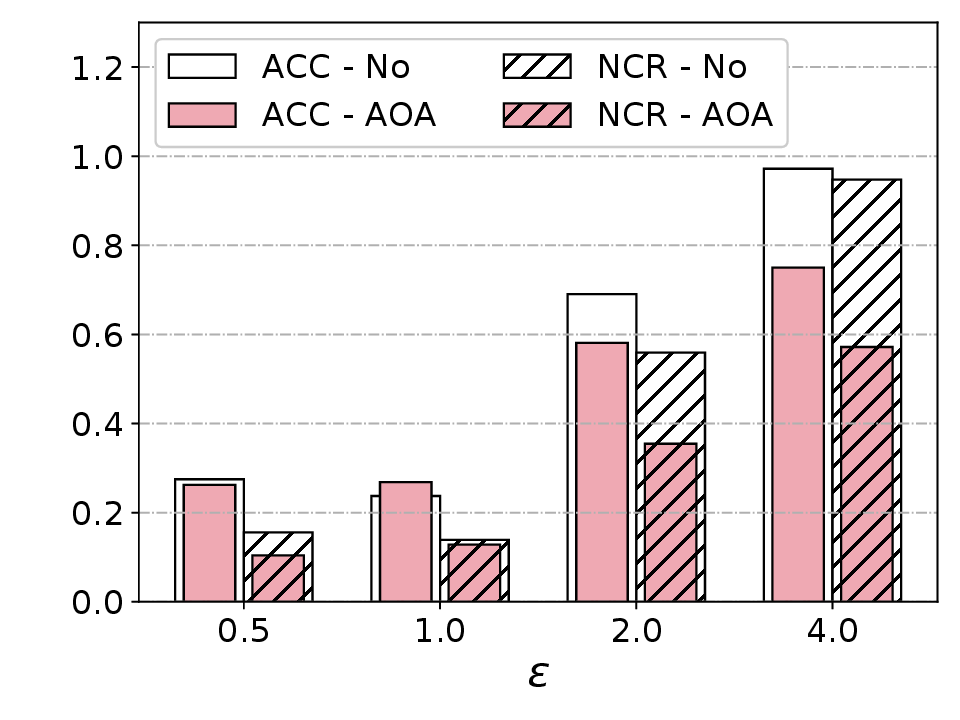}
		\end{minipage}%
    }
    \subfigure[SVIM, k]{
		\begin{minipage}[c]{0.2\textwidth}
		\centering
        \includegraphics[width=1\textwidth]{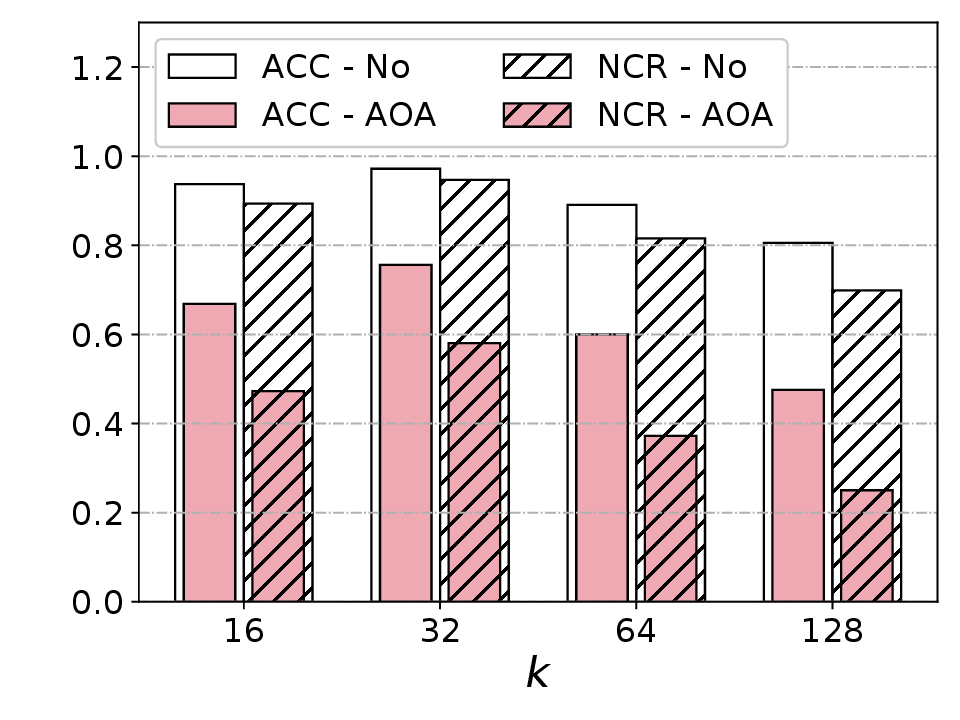}
		\end{minipage}%
    }
	\caption{AOA on BMS-POS with changing $\epsilon$ and $k$; Default: $\gamma = 0.01$, $k=32$, and $\epsilon=4.0$. \ccsrev{The $y$-axis is ACC/NCR.} }
  \Description[Varying $\epsilon$ and $k$]{Attacking LDP mechanisms on BMS-POS with changing $\epsilon$; AOA with $\gamma = 0.01$; $k=32$.}
	\label{fig:params:eps}
\end{figure}

\paragraphbe{Metrics} We adopt two metrics to measure the performance of the proposed attacks. 
\begin{itemize}[leftmargin=10pt]
    \item Accuracy ($\mathrm{ACC}$). Let $\mathcal{K}^*$ be the true top-$k$ items/itemsets, and $\mathcal{K}^\prime$ be the top-$k$ result mined by the LDP frequent itemset mining protocol under attack: $\mathrm{ACC} = |\mathcal{K}^* \cap \mathcal{K}^\prime|/k$. 
    \item Normalized Cumulative Rank ($\mathrm{NCR}$). Let $w(x)$ be the weight of an item/itemset $x$, for the item or itemset with the highest frequency its weight is $k$, and for others, the weight decreases by one according to their ranks. For the item/itemset with rank $k$, its weight is $1$, and for $x \notin \mathcal{K}^*$, $w(x) = 0$: $\mathrm{NCR} = \frac{\sum_{x\in \mathcal{K}^*} w(x)}{\sum_{x^\prime\in \mathcal{K}^\prime} w(x^\prime)}$. 
\end{itemize}
We note that an attacker tries to manipulate an LDP frequent itemset mining protocol to degrade its performance. Therefore, for the metrics in our experiments, \textbf{lower values of the evaluated metrics mean better attack performance.}

\subsection{Results for LDP Frequent Itemset Mining}
\label{sec:eval:itemset}
We first present the evaluation of the attack performance against the frequent itemset mining protocols, SVSM~\cite{DBLP:conf/sp/WangLJ18} \ccsrev{and FIML-IS~\cite{DBLP:conf/cikm/LiGGWY22}}. We set the privacy budget of both SVSM \ccsrev{and FIML-IS} as $\epsilon = 4.0$, and the protocol tries to find top-$k$ itemsets with $k = 32$. We change the fraction of malicious users $\gamma$ from $0.001$ to $0.1$. \ccsrev{More results with other settings will be included in Appendix~\ref{sec:app:addres}.}

\reffig{fig:gamma:svsm} and \reffig{fig:gamma:fimlitemset:eps4}  illustrate the effect of $\gamma$ on the attack performance on SVSM and FIML-IS, respectively. In general, we can observe that both the baseline attacks and our attack can degrade the performance of the LDP frequent itemset mining protocol to a lower level with more malicious users, and our attack, AOA, can cause the most serious damage to SVSM. Our analysis of the IBM Synthesize dataset reveals that MGA-R and MGA-Adv also exhibit favorable attack performance when $\gamma$ is sufficiently high compared with other two random baseline attacks, while AOA still exhibiting superior attack performance in various cases. 

For SVSM, \ccsrev{\reffig{fig:gamma:svsm}c and \reffig{fig:gamma:svsm}d} show the results on the Kosarak dataset, and \reffig{fig:gamma:svsm}e and \reffig{fig:gamma:svsm}f show the results on the BMS-POS dataset. We can observe that our attack AOA can significantly undermine the protocol, while the other four attacks fail to produce effective attacks. Additionally, it is worth noting that MGA-Adv demonstrates comparatively strong performance on the IBM Synthesize dataset, but performs inadequately on both the BMS-POS and Kosarak datasets. This disparity can be attributed to the fact that real-world datasets (\eg, BMS-POS and Kosarak) often contain itemsets with vastly differing frequencies, making it difficult for attacks on LDP frequency estimation to be successful against them. \ccsrev{For FIML-IS, it is observed from \reffig{fig:gamma:fimlitemset:eps4} that the advantage of AOA becomes less significant compared to other attacks. The reason for this is that the $\mathsf{SelectTop}$ phase of FIML-IS restricts the reports of users to membership tests only. While this approach reduces the attack surface, it also results in a limitation on the performance of the protocol when the domain of itemsets is large, e.g., for Kosarak and BMS-POS.}

\subsection{Results for LDP Frequent Item Mining}
\label{sec:eval:item}
We evaluate the attack performance against the set-valued frequent item mining protocols, LDPMiner~\cite{DBLP:conf/ccs/QinYYKXR16}, SVIM~\cite{DBLP:conf/sp/WangLJ18}, and FIML-I~\cite{DBLP:conf/cikm/LiGGWY22}. 
We set $\epsilon = 4.0$ and $k = 32$, and change the fraction of malicious users $\gamma$ from $0.001$ to $0.1$. \reffig{fig:gamma:ldpminer}, \reffig{fig:gamma:svim}, and \reffig{fig:gamma:fimlitem:eps4} show the impact of $\gamma$ on the attack performance. \ccsrev{More results with other settings will be included in Appendix~\ref{sec:app:addres}.}

For LDPMiner, we can find that both the $\mathrm{ACC}$ and $\mathrm{NCR}$ drop rapidly with $\gamma$ increasing when it is attacked by AOA, and metrics decrease to less than $0.1$ when $\gamma \geq 0.03$ for all three datasets. We can observe that both MGA-R and MGA-Adv also achieve a good attack performance against LDPMiner because LDPMiner uses basic RAPPOR~\cite{DBLP:conf/ccs/ErlingssonPK14} as the FO in the $\mathsf{SelectTop}$ phase. Since basic RAPPOR submits a $d$-bit array with each bit representing support for each item/itemset, thus MGA-Adv and AOA can both provide support for all items/itemsets in the target set, even without the attack resource estimation and allocation operations in our proposed attack.

For SVIM, AOA significantly outperforms other attacks when the fraction of malicious users is in a reasonable range. When the fraction of malicious users is too small ($\gamma < 0.01$), none of the attacks can significantly degrade the performance of SVIM. With $\gamma$ increasing, AOA can significantly degrade the performance of SVIM. For example, when $\gamma$ increases to about $0.03$, AOA can reduce $\mathrm{ACC}$ from $0.82$ to $0.23$ and $\mathrm{NCR}$ from $0.71$ to $0.06$ on the Kosarak dataset, and $\mathrm{ACC}$ from $0.97$ to $0.30$ and $\mathrm{NCR}$ from $0.95$ to $0.10$ on the BMS-POS dataset. For the same reason as that for SVSM, MGA-Adv can also achieve a satisfactory performance on the IBM dataset. Both AOA and MGA-Adv can degrade the metrics of SVIM on the IBM Synthesize dataset to less than $0.05$ when $\gamma \geq 0.05$. Nevertheless, AOA still outperforms all the baseline attacks for various cases. In addition, when the number of malicious users is large ($\gamma = 0.10$), AOA is able to degrade the NCR to almost $0$ for all the three datasets as we can observe in \reffig{fig:gamma:svim}b, \reffig{fig:gamma:svim}d, and \reffig{fig:gamma:svim}f. Another interesting observation from our experiments is that SVIM has a better performance compared to LDPMiner, and SVIM is also more robust than LDPMiner against the data poisoning attacks.

Similar results can be observed for FIML-I, as in the case of FIML-IS, due to the restricted attack surface resulting from the $\mathsf{SelectTop}$ phase of the protocol. However, it is worth noting that the proposed attacks still demonstrate the best attack performance in various cases. Due to the limit of space, the results on FIML-I will be included in Appendix~\ref{sec:app:addres}.

\begin{figure}
        \centering
	\subfigure[SVSM]{
		\begin{minipage}[c]{0.38\textwidth}
		\centering
        \includegraphics[width=1\textwidth]{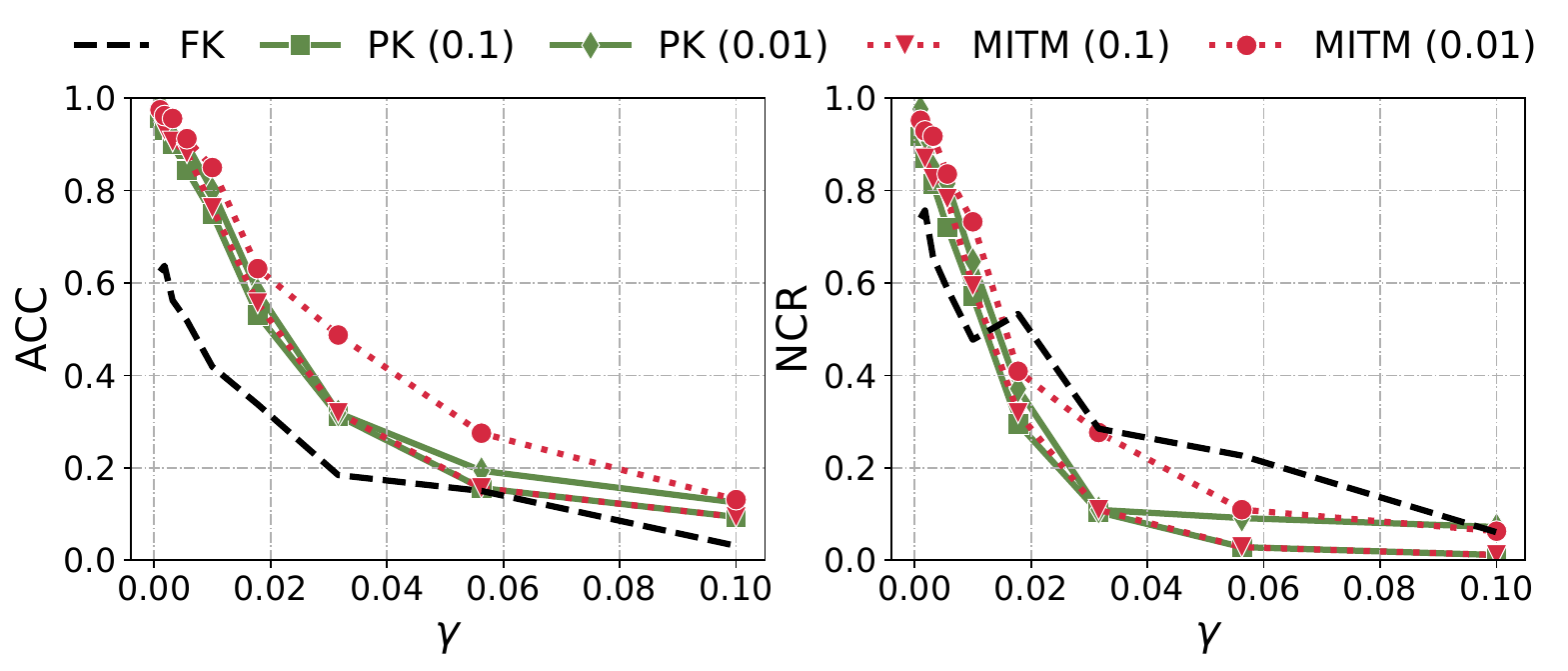}
		\end{minipage}%
	}

 	\subfigure[FIML-IS]{
		\begin{minipage}[c]{0.38\textwidth}
		\centering
        \includegraphics[width=1\textwidth]{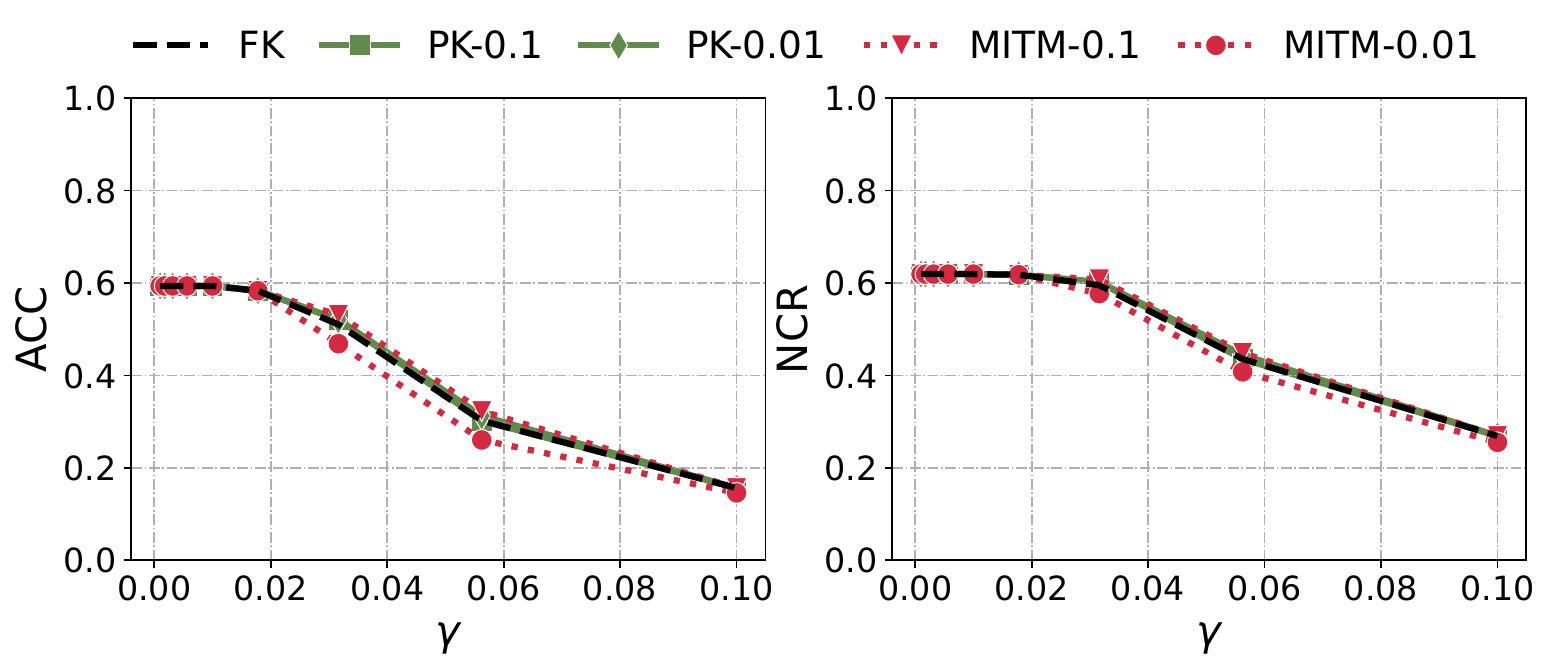}
		\end{minipage}%
	}
	\caption{Attacking the frequent itemset mining protocols with Limited Knowledge on BMS-POS.}
  \Description[Limited Knowledge on BMS-POS]{Attacking the frequent itemset mining protocols with Limited Knowledge on BMS-POS.}
	\label{fig:less:itemset}
\end{figure}

\subsection{Impact of Different Parameters} 
We investigate the impact of the privacy budget $\epsilon$ and the output size $k$ on the attack performance. Specifically, we aim to examine the robustness of the LDP frequent itemset mining protocols with different parameters against the proposed poisoning attack. We set $\gamma=0.01$ and $k=32$, and change $\epsilon$ from $0.5$ to $4.0$ to evaluate the impact of different privacy budgets. We set $\gamma=0.01$ and $\epsilon=4.0$, and change $k$ from $16$ to $128$ to evaluate the impact of different output sizes. The results of changing $\epsilon$ and $k$ are shown in~\reffig{fig:params:eps}. We define the drop ratio as $\frac{\mathrm{Met}^* - \mathrm{Met}^{\prime}}{\mathrm{Met}^*}$, where $\mathrm{Met}^*$ represents the value of the metric in the absence of an attack and $\mathrm{Met}^{\prime}$ is the value of the metric under attack. 

For SVIM and LDPMiner, in general, we observe that the drop ratios increase with the privacy budget increases. These findings suggest that protocols with stronger privacy protection (smaller privacy budget) may be more robust to data poisoning attacks. However, it is important to note that a smaller privacy budget does not guarantee satisfactory performance even without the attack. For SVSM, the drop ratios in $\mathrm{NCR}$ are found to be $0.45$, $0.38$, $0.34$, $0.46$, respectively, and the drop ratios are $0.49$, $0.44$, $0.43$, $0.37$ in $\mathrm{ACC}$, respectively. It is worth noting that in these experiments, we set the fraction of corrupted users to be $0.01$, which means only 1\% of the reports have been poisoned. This setting allows us to evaluate the impact of different parameters. While a stronger attack intensity may overwhelm the impact of parameters, we emphasize that a realistic setting could set $\gamma$ up to $0.1$, bringing a significant improvement in the drop ratios.

For SVSM, both the $\mathrm{NCR}$ and the $\mathrm{ACC}$ are not significantly affected by the change in $k$. For LDPMiner, the drop ratios in $\mathrm{NCR}$ and $\mathrm{ACC}$ are found to be larger than $0.95$ in most cases. As for SVIM, the drop ratios in $\mathrm{NCR}$ are $0.47$, $0.39$, $0.54$, and $0.64$ for $k = 16$, $32$, $64$, and $128$, respectively. The drop ratios in $\mathrm{ACC}$ exhibit a similar trend. Essentially, the results indicate that as $k$ increases, the attack becomes more powerful. A larger top-$k$ set of items or itemsets creates a wider attack surface, allowing the attacker more room to refine the target set and manipulate the results.

\begin{figure}
        \centering
    \subfigure[LDPMiner]{
		\begin{minipage}[c]{0.38\textwidth}
		\centering
        \includegraphics[width=1\textwidth]{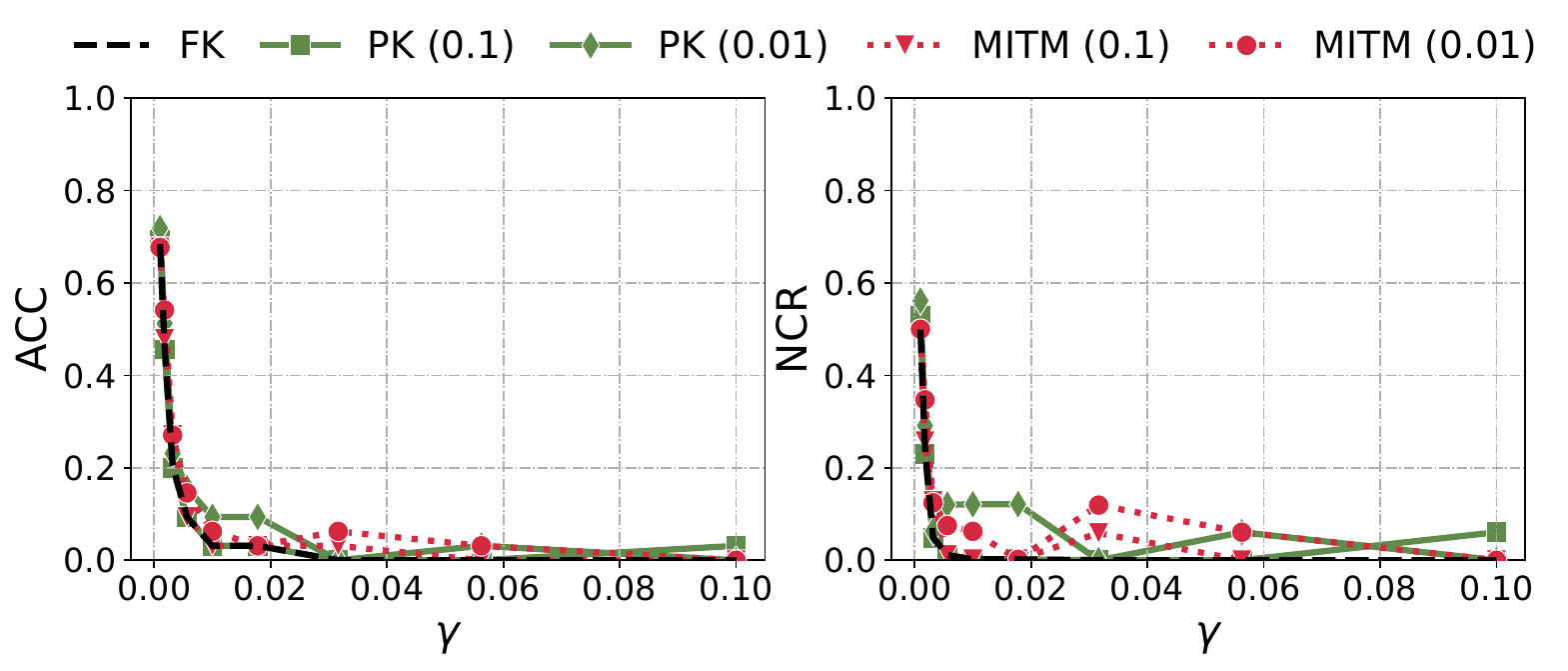}
		\end{minipage}%
	}

     \subfigure[SVIM]{
		\begin{minipage}[c]{0.38\textwidth}
		\centering
        \includegraphics[width=1\textwidth]{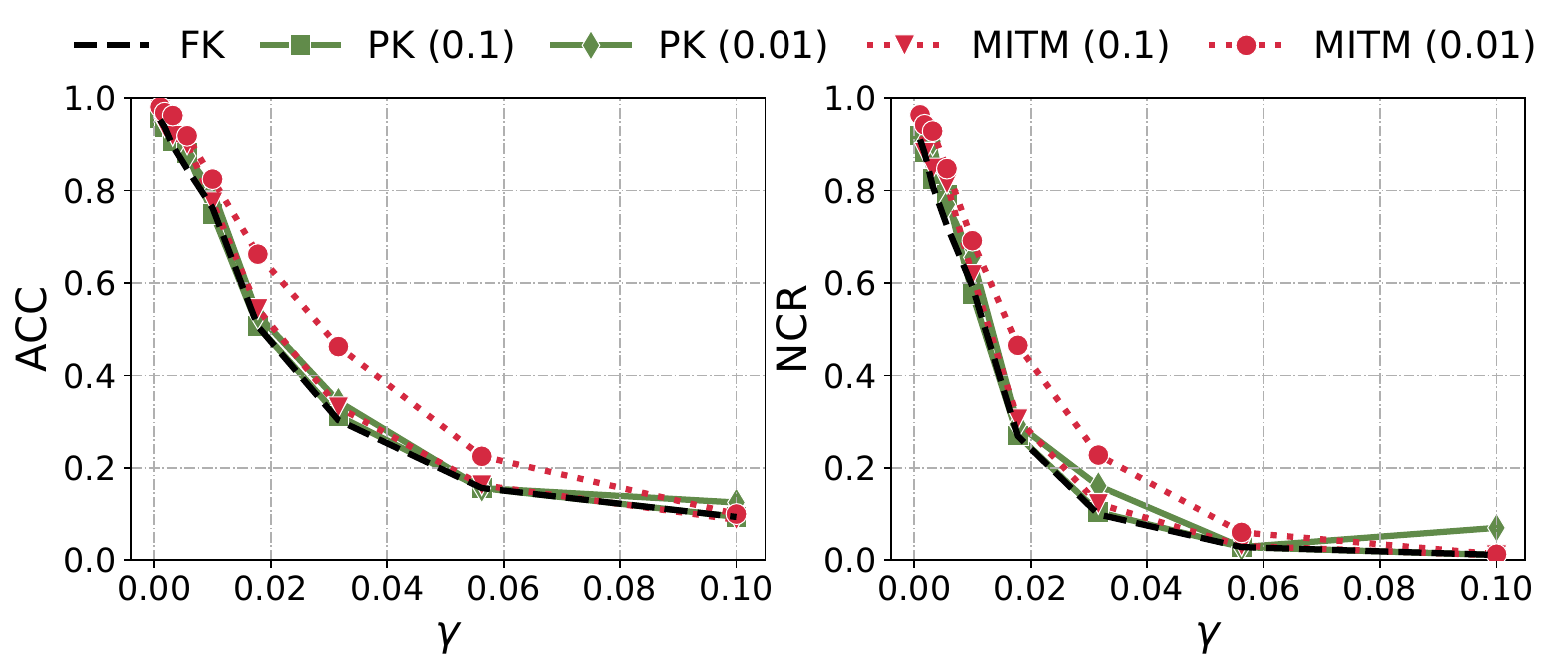}
		\end{minipage}%
	}
	\caption{Attacking the frequent item mining protocols with Limited Knowledge on BMS-POS.}
  \Description[Limited Knowledge on BMS-POS]{Attacking the frequent item mining protocols with Limited Knowledge on BMS-POS.}
	\label{fig:less:item}
\end{figure}

\subsection{Evaluating Attacks with Limited Knowledge} 
In this set of experiments, we evaluate attacks with limited knowledge. We set $\epsilon=4.0$ and $k=32$, comparing the performance of the partial-knowledge attack (PK), the man-in-the-middle attack (MITM), and the full-knowledge attack (FK) in two scenarios. In these scenarios, the PK attacker or the MITM attacker obtains 1\% or 10\% of the knowledge about the frequencies of itemsets, denoted by PK-0.01 and PK-0.1 in the figures, or intercepts 1\% or 10\% of the communication between the users and the data aggregator, denoted by MITM-0.01 and MITM-0.1 in the figures.

The results (\reffig{fig:less:itemset} and \reffig{fig:less:item}) show that both the PK attack and the MITM attack can achieve results comparable to those of the FK attack. This further demonstrates the feasibility of our data poisoning attacks against LDP frequent itemset mining protocols. Specifically, we observe that when 10\% of the knowledge or communication can be obtained or intercepted by the PK attacker or the MITM attacker, respectively, the attack performance of the attacks with limited knowledge is almost as good as that of the full-knowledge attack across various cases. However, when only 1\% of the knowledge or communication can be obtained or intercepted by the PK attacker or the MITM attacker, respectively, the attack performance degrades due to the inaccurate estimation of the frequencies under this setting. 
More results will be included in Appendix~\ref{sec:app}.

\section{Related Work}
\subsection{LDP Frequent Itemset Mining}
Differential privacy~\cite{DBLP:journals/jpc/DworkMNS16,DBLP:journals/fttcs/DworkR14} has become the de-facto approach for privacy-preserving data analysis. When the data is collected from distributed users and there is no trusted data collector, local differential privacy (LDP)~\cite{DBLP:conf/focs/KasiviswanathanLNRS08,DBLP:conf/ccs/ErlingssonPK14} has been widely adopted in various tasks, such as frequency estimation~\cite{DBLP:conf/uss/WangBLJ17,DBLP:conf/stoc/BassilyS15,DBLP:conf/ndss/0001LLSL20,DBLP:conf/ccs/ErlingssonPK14,DBLP:conf/sp/YeHMZ19,DBLP:journals/pvldb/WangQDYHX20}, heavy hitter identification~\cite{DBLP:conf/icalp/HsuKR12,DBLP:conf/nips/BassilyNST17,DBLP:journals/tdsc/0001LJ21,DBLP:journals/tifs/ZhuCXWZ24}, frequent itemset mining~\cite{DBLP:conf/sp/WangLJ18,DBLP:conf/ccs/QinYYKXR16}, and distributed learning~\cite{DBLP:conf/dasfaa/LiuCYC20,DBLP:journals/corr/abs-1807-04369,DBLP:conf/ijcai/SunQC21}.

In this paper, we focus on the LDP frequent itemset mining protocols. Qin et al. ~\cite{DBLP:conf/ccs/QinYYKXR16} propose LDPMiner, an LDP frequent item mining protocol, which is a special case of frequent itemset mining, and the mining result is the top-$k$ frequent items rather than itemsets. Frequent item mining is also a type of heavy hitter identification, but unlike the previous work, LDPMiner is performed on set-valued data (\i.e., each user has a set of items), while the typical LDP heavy hitter identification protocols~\cite{DBLP:conf/nips/BassilyNST17,DBLP:journals/tdsc/0001LJ21} are for the cases where each user has only a single value. Wang et al.~\cite{DBLP:conf/sp/WangLJ18} propose a more general protocol for LDP frequent item mining, named SVIM, which leverages a padding FO technique to improve the performance of the frequent item estimation. They also extend their work to propose SVSM, the first LDP frequent itemset mining protocol. Li et al.~\cite{DBLP:conf/cikm/LiGGWY22} also propose a frequent itemset mining protocl, named FIML, which incurs low communication and computation costs by using Random Response in the user response mechanism. Wang et al.~\cite{DBLP:conf/infocom/0003HNWXY18} propose a delicate Frequency Oracle called PrivSet, working on set-valued data to estimate user item distribution, using an exponential mechanism as the randomization mechanism. We wrap it into a single-phase frequent item mining protocol and apply our attack to further illustrate its versatility.

\subsection{Data Poisoning Attacks}
In recent years, data poisoning attacks have gained tremendous popularity in machine learning, \eg, ~\cite{DBLP:conf/icml/BiggioNL12,DBLP:conf/uss/FangCJG20,DBLP:conf/sp/JagielskiOBLNL18, DBLP:conf/nips/LiWSV16,DBLP:journals/csur/TianCLY23}, where the attacker contributes manipulated training data and attempts to influence the learned model.
Beyond the interest in poisoning attacks against machine learning, recent work has explored data poisoning attacks against LDP protocols, \eg, ~\cite{DBLP:conf/uss/CaoJG21,DBLP:conf/sp/CheuSU21,DBLP:journals/corr/abs-2111-11534,DBLP:journals/corr/abs-2205-11782,DBLP:journals/pvldb/ArcoleziGCP23}.

Cheu et al.~\cite{DBLP:conf/sp/CheuSU21} propose the attacks against LDP frequency estimation and heavy hitter identification protocols with the goal to degrade the aggregation performance, and the authors also explore the relationship between the attack bias and the injected data distribution. In~\cite{DBLP:conf/uss/CaoJG21}, Cao et al. propose an attack called maximal gain attack, with the aim of promoting the frequencies of target items by letting the fake users report crafted values against LDP frequency estimation and heavy hitter identification protocols. Although the data poisoning attacks on heavy hitter identification, which is similar to frequent item mining, have been studied in these two works, their attacks can only be applied to single-valued heavy hitter identification, e.g.,PEM~\cite{DBLP:journals/tdsc/0001LJ21}, and cannot apply well to set-valued heavy hitter identification (\ie, frequent item mining),  \eg, LDPMiner~\cite{DBLP:conf/ccs/QinYYKXR16}, SVIM~\cite{DBLP:conf/sp/WangLJ18}, which we have considered in this paper.
Besides the frequency estimation task over single-valued data, Wu et al.~\cite{DBLP:journals/corr/abs-2111-11534} investigate the data poisoning attack against frequency estimation and mean over key-value data, and in~\cite{DBLP:journals/corr/abs-2205-11782}, Li et al. investigate how to manipulate the estimation of the mean and the variance against LDP protocols.

In the face of the risks of data poisoning attacks in LDP protocols, prior research, \eg, ~\cite{DBLP:conf/pkc/AmbainisJL04,DBLP:conf/uss/CaoJG21,DBLP:conf/dbsec/Kato0Y21,DBLP:journals/tifs/SongXZ23}, also proposes some approaches to mitigate the attacks. Cao et al.~\cite{DBLP:conf/uss/CaoJG21} propose two types of defenses against attacks at the level of FOs, namely normalization and fake user detection, which can mitigate the attacks to some extent. Cryptographic techniques, such as Multi-Party Computation and Zero-Knowledge Proof, are also employed to ensure the honest behavior of users, \eg,
 \cite{DBLP:conf/dbsec/Kato0Y21,DBLP:journals/tifs/SongXZ23}. Kato et al.~\cite{DBLP:conf/dbsec/Kato0Y21} propose Cryptographic Randomized Response Technique (CRRT) and use it to build a cryptographic version of FOs, which is equivalent to employing a Trusted Third Party (TTP) to execute the FO.

In this work, we demonstrate that the LDP frequent itemset mining protocols are also vulnerable to data poisoning attacks. Compared with previous work, our work focus different tasks (\ie, frequent itemset mining) and different types of data (\ie, set-valued data), which make the previous work cannot apply well to the problem in this paper.

\section{Potential Defenses}
In this section, we delve into potential defenses against data poisoning attacks for locally differentially private frequent itemset mining. Additionally, we explore whether these potential defenses, along with existing ones against such attacks on LDP frequency estimation, can effectively mitigate the proposed attacks through extensive experimental evaluation.

\subsection{Securing the FOs}
To counteract the attack, one possible approach is to secure the frequency oracles used in frequent itemset mining protocols. While some defenses for LDP frequency estimation have been explored in~\cite{DBLP:conf/uss/CaoJG21}, such as normalizing estimated frequencies or detecting malicious users, the effectiveness of these defenses against GRR or SH remains unclear. Furthermore, the proposed defenses have been shown to lack satisfactory performance against OLH. An important observation is that malicious users often try to craft the LDP response to support more preimages. To mitigate the attack effect on FOs, one potential approach is to limit the number of items a response can support by clipping it, which would restrict the ability of malicious users.

Specifically, we examine a defense strategy called FilterFO, which filters user reports supporting items larger than the threshold. For OLH, we discard all reports with supports larger than $\theta$, where $\theta\in[1,d]$ is a parameter to trade off between utility and security. The false positive rate (FPR) of the filter, i.e., the tail distribution for the number of supports from a benign report $y$ is close to the tail of binomial ditribution $B(d,1/g)$:
   \[
    \Pr\left[S(y)\geq \theta\right]=\frac{I_{(1/g)}\left(\lceil \theta\rceil-1 ,d-\lceil \theta\rceil+1\right)+I_{(1/g)}\left(\lceil \theta\rceil ,d-\lceil \theta\rceil\right)}{2}
   \]
where $d$ represents the size of the domain of items/itemsets, $g$ represents the size of the outputs, $I$ is the regularized incomplete beta function.
For RAPPOR, the expectation of the number of support from a benign report $y$ for threshold $\theta$ is $\E[S(y)] = 1+(d-2)q$,
where $q=\frac{1}{\exp(\epsilon/2)+1}$ represents the probability to flip a bit of the reported $d$-bit array. The distribution of the number of support from a benign report $y$ is
    \[
    \Pr[S(y)\ge \theta] = q\cdot I_{q}(\lceil \theta\rceil ,d-\lceil \theta\rceil)+(1-q)\cdot I_{q}(\lceil \theta\rceil-1 ,d-\lceil \theta\rceil+1)
   \]
This defense strategy shares similar idea with the detect fake user method proposed in~\cite{DBLP:conf/uss/CaoJG21}, but is more neat and can be applied to various FOs.

Another type of defense aims to secure the FOs by using cryptographic techniques to verify the reports, e.g., \cite{DBLP:journals/tifs/SongXZ23,DBLP:conf/dbsec/Kato0Y21}, in order to prevent the manipulation of the output of $\Psi_{\rm{FO}(\epsilon)}(\cdot)$ for a given input item/itemset. However, this type of defense faces a major drawback: there is a significant computational overhead. The use of cryptographic techniques also weakens the motivation for deploying LDP because good performance can be achieved by incorporating cryptography with central differential privacy if more computational overheads are allowed, as shown in~\cite{DBLP:conf/uss/BohlerK20,DBLP:conf/ccs/BohlerK21}. Additionally, this type of defense cannot mitigate the manipulation of the inputs, i.e., changing the value of a user before using $\Psi_{\rm{FO}(\epsilon)}(\cdot)$ to perturb it. We denote this type of defense as CryptoFO in the following parts of this paper.

\begin{figure*}
        \centering
        \includegraphics[width=0.8\textwidth]{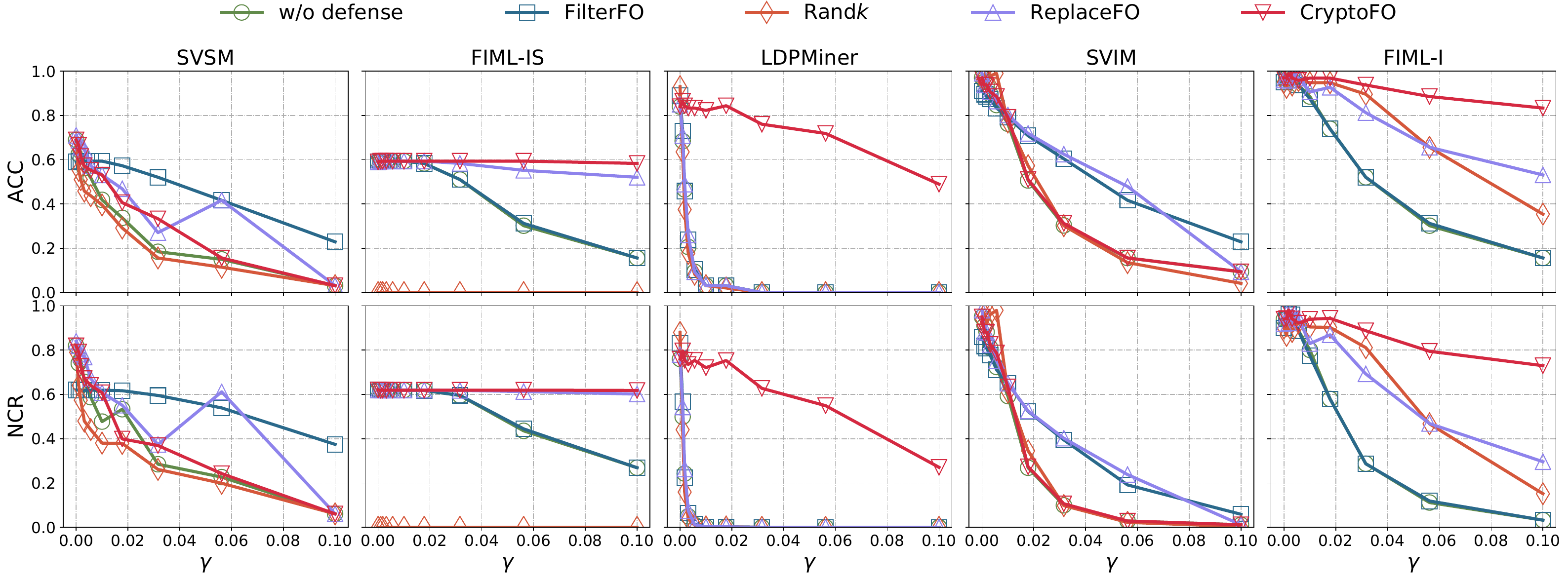}
	\caption{Defenses against AOA on BMS-POS dataset, with $k=32$, $\epsilon=4.0$.}
	\label{fig:defense:pos}
\end{figure*}

\subsection{Configuring the Protocols with Uncertainty}
Although securing the underlying FOs is a valid approach for defending against attacks on LDP frequent itemset mining,  the deployment of such a defense may face challenges. LDP frequent itemset mining protocols often use FOs as black-box building blocks, making it impractical to modify the details of FOs in some cases. We also explore an alternative strategy to defending against attacks by introducing randomness into the LDP frequent itemset mining. For example, the knowledge of the padding size and the number of candidates forms a key ability of the attacker to make the data poisoning attack successful. If the aggregator chooses the padding size and determines the number of candidates randomly, the attacks may be mitigated to a great extent. For example, in SVSM, LDPMiner, and SVIM, the size of the candidate set is set to $2k$, while it is set to $1.5k$ in FIML. The padding size will affect the choice of FOs in SVSM and SVIM.

Specifically, we consider two countermeasures in this part: (1) Rand$k$: using a randomly chosen $k^\prime > k$ to replace the original mining objective, aiming to disrupt the attacker's target set estimation step; (2) ReplaceFO: it is difficult to dynamically change the FO, but the aggregator can replace the FO used in the protocols. As aforementioned, OLH can achieve good performance when there is no attack, but it also leaves a larger attack surface for the attacker. Therefore, ReplaceFO replaces OLH with GRR in the protocols.

\subsection{Experimental Results}
We empirically evaluate the effectiveness of the four aforementioned defenses against the proposed attack. It is important to note that CryptoFO employs cryptographic techniques to prevent the poisoning of reports sent to the aggregator. In our evaluation, we let corrupted users to manipulate the mining results by crafting their original values (inputs to the perturbation function $\Psi_{\rm{FO}(\epsilon)}(\cdot)$).

We have conducted experiments with $k=32$, $\epsilon=4.0$, and $\gamma=0.001$, $0.01$, or $0.1$ to evaluate the effectiveness of various defense strategies on BMS-POS dataset. Figure~\ref{fig:defense:pos} shows the experimental results for scenarios with no defense, FilterFO, Rand$k$, ReplaceFO, and CryptoFO. In our experiments, the filter threshold of FilterFO is set as $\frac{2d}{g}$ for OLH and $\frac{2d}{\exp(\epsilon/2)+1}$ for RAPPOR.
Our observations indicate that the defense strategies generally show some effectiveness, but they cannot fully mitigate the impact of the attack. For FIML-IS, LDPMiner, and FIML-I, CryptoFO achieves better defense performance by preventing the manipulation of LDP mechanism outputs, thereby significantly limiting the attacker's capabilities. However, it is worth noting that this defense may not be practical as it weakens the motivation and restricts the deployment scenarios for LDP protocols.

Among the other defenses, ReplaceFO demonstrates better performance across a wider range of settings compared to the other two defenses for FIML-IS. By replacing OLH with GRR, ReplaceFO effectively restricts the attacker's ability to carry out the attack. FilterFO can achieve better performance for SVSM and SVIM because the OLH is used in these two protocols. It is worth noting that, in general, for attacks involving a large number of corrupted users (\ie, $\gamma=0.1$), none of the defenses can achieve satisfactory performance. Additionally, it may cause some performance degradation to LDP protocols when there is no attack. The results highlight the limited effectiveness of the current defenses, indicating the need for more systematic defenses against our attacks. Additional defense results on the IBM Synthesize and Kosarak datasets will be presented in Appendix~\ref{sec:app:addres}.

\section{Conclusion}
In this paper, we conduct the first systematic study on data poisoning attacks to LDP frequent itemset mining protocols. We propose a unified attack framework that can significantly degrade the performance of the essential LDP protocols for frequent itemset mining by carefully crafting the reports from malicious users. We evaluate the proposed attack with extensive experiments on a range of scenarios and the results demonstrate the effectiveness of our attack against LDP frequent itemset mining protocols, even with a limited proportion of malicious participants. We have also discussed and evaluated potential defenses against our attacks.
In conclusion, our work sheds light on the security of LDP frequent itemset mining protocols and highlights the need for further research in this area to ensure the robustness and reliability of these protocols in real-world scenarios.

\begin{acks}
We thank the anonymous reviewers for their comments. This work was supported by the National Natural Science Foundation of China (62372226, 62272215, 62002159) and the Leading-Edge Technology Program (BK20202001) of the Natural Science Foundation of Jiangsu Province, China.
\end{acks} 


\bibliographystyle{ACM-Reference-Format}
\bibliography{refs-full}


\begin{thebibliography}{42}


\ifx \showCODEN    \undefined \def \showCODEN     #1{\unskip}     \fi
\ifx \showDOI      \undefined \def \showDOI       #1{#1}\fi
\ifx \showISBNx    \undefined \def \showISBNx     #1{\unskip}     \fi
\ifx \showISBNxiii \undefined \def \showISBNxiii  #1{\unskip}     \fi
\ifx \showISSN     \undefined \def \showISSN      #1{\unskip}     \fi
\ifx \showLCCN     \undefined \def \showLCCN      #1{\unskip}     \fi
\ifx \shownote     \undefined \def \shownote      #1{#1}          \fi
\ifx \showarticletitle \undefined \def \showarticletitle #1{#1}   \fi
\ifx \showURL      \undefined \def \showURL       {\relax}        \fi
\providecommand\bibfield[2]{#2}
\providecommand\bibinfo[2]{#2}
\providecommand\natexlab[1]{#1}
\providecommand\showeprint[2][]{arXiv:#2}

\bibitem[Agrawal et~al\mbox{.}(1996)]%
        {DBLP:books/mit/fayyadPSU96/AgrawalMSTV96}
\bibfield{author}{\bibinfo{person}{Rakesh Agrawal}, \bibinfo{person}{Heikki
  Mannila}, \bibinfo{person}{Ramakrishnan Srikant}, \bibinfo{person}{Hannu
  Toivonen}, {and} \bibinfo{person}{A.~Inkeri Verkamo}.}
  \bibinfo{year}{1996}\natexlab{}.
\newblock \showarticletitle{Fast Discovery of Association Rules}.
\newblock In \bibinfo{booktitle}{\emph{Advances in Knowledge Discovery and Data
  Mining}}. \bibinfo{publisher}{{AAAI/MIT} Press}, \bibinfo{address}{{USA}},
  \bibinfo{pages}{307--328}.
\newblock


\bibitem[Ambainis et~al\mbox{.}(2004)]%
        {DBLP:conf/pkc/AmbainisJL04}
\bibfield{author}{\bibinfo{person}{Andris Ambainis}, \bibinfo{person}{Markus
  Jakobsson}, {and} \bibinfo{person}{Helger Lipmaa}.}
  \bibinfo{year}{2004}\natexlab{}.
\newblock \showarticletitle{Cryptographic Randomized Response Techniques}. In
  \bibinfo{booktitle}{\emph{Public Key Cryptography - {PKC} 2004, 7th
  International Workshop on Theory and Practice in Public Key Cryptography,
  Singapore, March 1-4, 2004}}, Vol.~\bibinfo{volume}{2947}.
  \bibinfo{publisher}{Springer}, \bibinfo{pages}{425--438}.
\newblock


\bibitem[{Apple}(2017)]%
        {iosdp}
\bibfield{author}{\bibinfo{person}{{Apple}}.} \bibinfo{year}{2017}\natexlab{}.
\newblock \bibinfo{title}{Apple Differential Privacy Technical Overview}.
\newblock
\newblock
\urldef\tempurl%
\url{https://www.apple.com/privacy/docs/Differential_Privacy_Overview.pdf}
\showURL{%
\tempurl}


\bibitem[Arcolezi et~al\mbox{.}(2023)]%
        {DBLP:journals/pvldb/ArcoleziGCP23}
\bibfield{author}{\bibinfo{person}{H{\'{e}}ber~Hwang Arcolezi},
  \bibinfo{person}{S{\'{e}}bastien Gambs},
  \bibinfo{person}{Jean{-}Fran{\c{c}}ois Couchot}, {and}
  \bibinfo{person}{Catuscia Palamidessi}.} \bibinfo{year}{2023}\natexlab{}.
\newblock \showarticletitle{On the Risks of Collecting Multidimensional Data
  Under Local Differential Privacy}.
\newblock \bibinfo{journal}{\emph{Proc. {VLDB} Endow.}} \bibinfo{volume}{16},
  \bibinfo{number}{5} (\bibinfo{year}{2023}), \bibinfo{pages}{1126--1139}.
\newblock


\bibitem[Bassily et~al\mbox{.}(2017)]%
        {DBLP:conf/nips/BassilyNST17}
\bibfield{author}{\bibinfo{person}{Raef Bassily}, \bibinfo{person}{Kobbi
  Nissim}, \bibinfo{person}{Uri Stemmer}, {and} \bibinfo{person}{Abhradeep~Guha
  Thakurta}.} \bibinfo{year}{2017}\natexlab{}.
\newblock \showarticletitle{Practical Locally Private Heavy Hitters}. In
  \bibinfo{booktitle}{\emph{Advances in Neural Information Processing Systems
  30: NeurIPS 2017, Long Beach, {CA}, {USA}, December 4-9, 2017}}.
  \bibinfo{publisher}{Neural Information Processing Systems Foundation},
  \bibinfo{address}{{USA}}, \bibinfo{pages}{2288--2296}.
\newblock


\bibitem[Bassily and Smith(2015)]%
        {DBLP:conf/stoc/BassilyS15}
\bibfield{author}{\bibinfo{person}{Raef Bassily} {and} \bibinfo{person}{Adam~D.
  Smith}.} \bibinfo{year}{2015}\natexlab{}.
\newblock \showarticletitle{Local, Private, Efficient Protocols for Succinct
  Histograms}. In \bibinfo{booktitle}{\emph{Proceedings of the Forty-Seventh
  Annual {ACM} on Symposium on Theory of Computing, {STOC} 2015, Portland,
  {OR}, {USA}, June 14-17, 2015}}. \bibinfo{publisher}{{ACM}},
  \bibinfo{address}{{USA}}, \bibinfo{pages}{127--135}.
\newblock


\bibitem[Benson et~al\mbox{.}(2018)]%
        {DBLP:conf/wsdm/Benson0T18}
\bibfield{author}{\bibinfo{person}{Austin~R. Benson}, \bibinfo{person}{Ravi
  Kumar}, {and} \bibinfo{person}{Andrew Tomkins}.}
  \bibinfo{year}{2018}\natexlab{}.
\newblock \showarticletitle{A Discrete Choice Model for Subset Selection}. In
  \bibinfo{booktitle}{\emph{Proceedings of the Eleventh {ACM} International
  Conference on Web Search and Data Mining, {WSDM} 2018, Marina Del Rey, {CA},
  {USA}, February 5-9, 2018}}. \bibinfo{publisher}{{ACM}},
  \bibinfo{address}{{USA}}, \bibinfo{pages}{37--45}.
\newblock


\bibitem[Biggio et~al\mbox{.}(2012)]%
        {DBLP:conf/icml/BiggioNL12}
\bibfield{author}{\bibinfo{person}{Battista Biggio}, \bibinfo{person}{Blaine
  Nelson}, {and} \bibinfo{person}{Pavel Laskov}.}
  \bibinfo{year}{2012}\natexlab{}.
\newblock \showarticletitle{Poisoning Attacks against Support Vector Machines}.
  In \bibinfo{booktitle}{\emph{Proceedings of the 29th International Conference
  on Machine Learning, {ICML} 2012, Edinburgh, Scotland, {UK}, June 26 - July
  1, 2012}}. \bibinfo{publisher}{icml.cc / Omnipress},
  \bibinfo{address}{{USA}}, \bibinfo{pages}{1467--1474}.
\newblock


\bibitem[B{\"{o}}hler and Kerschbaum(2020)]%
        {DBLP:conf/uss/BohlerK20}
\bibfield{author}{\bibinfo{person}{Jonas B{\"{o}}hler} {and}
  \bibinfo{person}{Florian Kerschbaum}.} \bibinfo{year}{2020}\natexlab{}.
\newblock \showarticletitle{Secure Multi-party Computation of Differentially
  Private Median}. In \bibinfo{booktitle}{\emph{29th {USENIX} Security
  Symposium, {USENIX} Security 2020, August 12-14, 2020}}.
  \bibinfo{publisher}{{USENIX} Association}, \bibinfo{pages}{2147--2164}.
\newblock


\bibitem[B{\"{o}}hler and Kerschbaum(2021)]%
        {DBLP:conf/ccs/BohlerK21}
\bibfield{author}{\bibinfo{person}{Jonas B{\"{o}}hler} {and}
  \bibinfo{person}{Florian Kerschbaum}.} \bibinfo{year}{2021}\natexlab{}.
\newblock \showarticletitle{Secure Multi-party Computation of Differentially
  Private Heavy Hitters}. In \bibinfo{booktitle}{\emph{{CCS} '21: 2021 {ACM}
  {SIGSAC} Conference on Computer and Communications Security, Virtual Event,
  Republic of Korea, November 15 - 19, 2021}}. \bibinfo{publisher}{{ACM}},
  \bibinfo{pages}{2361--2377}.
\newblock


\bibitem[Cao et~al\mbox{.}(2021)]%
        {DBLP:conf/uss/CaoJG21}
\bibfield{author}{\bibinfo{person}{Xiaoyu Cao}, \bibinfo{person}{Jinyuan Jia},
  {and} \bibinfo{person}{Neil~Zhenqiang Gong}.}
  \bibinfo{year}{2021}\natexlab{}.
\newblock \showarticletitle{Data Poisoning Attacks to Local Differential
  Privacy Protocols}. In \bibinfo{booktitle}{\emph{30th {USENIX} Security
  Symposium, {USENIX} Security 2021, August 11-13, 2021}}.
  \bibinfo{publisher}{{USENIX} Association}, \bibinfo{address}{{USA}},
  \bibinfo{pages}{947--964}.
\newblock


\bibitem[Cheu et~al\mbox{.}(2021)]%
        {DBLP:conf/sp/CheuSU21}
\bibfield{author}{\bibinfo{person}{Albert Cheu}, \bibinfo{person}{Adam~D.
  Smith}, {and} \bibinfo{person}{Jonathan~R. Ullman}.}
  \bibinfo{year}{2021}\natexlab{}.
\newblock \showarticletitle{Manipulation Attacks in Local Differential
  Privacy}. In \bibinfo{booktitle}{\emph{42nd {IEEE} Symposium on Security and
  Privacy, {SP} 2021, San Francisco, {CA}, {USA}, 24-27 May 2021}}.
  \bibinfo{publisher}{{IEEE}}, \bibinfo{address}{{USA}},
  \bibinfo{pages}{883--900}.
\newblock


\bibitem[Ding et~al\mbox{.}(2017)]%
        {DBLP:conf/nips/DingKY17}
\bibfield{author}{\bibinfo{person}{Bolin Ding}, \bibinfo{person}{Janardhan
  Kulkarni}, {and} \bibinfo{person}{Sergey Yekhanin}.}
  \bibinfo{year}{2017}\natexlab{}.
\newblock \showarticletitle{Collecting Telemetry Data Privately}. In
  \bibinfo{booktitle}{\emph{Advances in Neural Information Processing Systems
  30: NeurIPS 2017, Long Beach, {CA}, {USA}, December 4-9, 2017}}.
  \bibinfo{publisher}{Neural Information Processing Systems Foundation},
  \bibinfo{address}{{USA}}, \bibinfo{pages}{3571--3580}.
\newblock


\bibitem[Dwork et~al\mbox{.}(2006)]%
        {DBLP:conf/tcc/DworkMNS06}
\bibfield{author}{\bibinfo{person}{Cynthia Dwork}, \bibinfo{person}{Frank
  McSherry}, \bibinfo{person}{Kobbi Nissim}, {and} \bibinfo{person}{Adam~D.
  Smith}.} \bibinfo{year}{2006}\natexlab{}.
\newblock \showarticletitle{Calibrating Noise to Sensitivity in Private Data
  Analysis}. In \bibinfo{booktitle}{\emph{Theory of Cryptography, Third Theory
  of Cryptography Conference, {TCC} 2006, New York, {NY}, {USA}, March 4-7,
  2006, Proceedings}}, Vol.~\bibinfo{volume}{3876}.
  \bibinfo{publisher}{Springer}, \bibinfo{address}{Germany},
  \bibinfo{pages}{265--284}.
\newblock


\bibitem[Dwork et~al\mbox{.}(2016)]%
        {DBLP:journals/jpc/DworkMNS16}
\bibfield{author}{\bibinfo{person}{Cynthia Dwork}, \bibinfo{person}{Frank
  McSherry}, \bibinfo{person}{Kobbi Nissim}, {and} \bibinfo{person}{Adam~D.
  Smith}.} \bibinfo{year}{2016}\natexlab{}.
\newblock \showarticletitle{Calibrating Noise to Sensitivity in Private Data
  Analysis}.
\newblock \bibinfo{journal}{\emph{J. Priv. Confidentiality}}
  \bibinfo{volume}{7}, \bibinfo{number}{3} (\bibinfo{year}{2016}),
  \bibinfo{pages}{17--51}.
\newblock


\bibitem[Dwork and Roth(2014)]%
        {DBLP:journals/fttcs/DworkR14}
\bibfield{author}{\bibinfo{person}{Cynthia Dwork} {and} \bibinfo{person}{Aaron
  Roth}.} \bibinfo{year}{2014}\natexlab{}.
\newblock \showarticletitle{The Algorithmic Foundations of Differential
  Privacy}.
\newblock \bibinfo{journal}{\emph{Found. Trends Theor. Comput. Sci.}}
  \bibinfo{volume}{9}, \bibinfo{number}{3-4} (\bibinfo{year}{2014}),
  \bibinfo{pages}{211--407}.
\newblock


\bibitem[Erlingsson et~al\mbox{.}(2014)]%
        {DBLP:conf/ccs/ErlingssonPK14}
\bibfield{author}{\bibinfo{person}{{\'{U}}lfar Erlingsson},
  \bibinfo{person}{Vasyl Pihur}, {and} \bibinfo{person}{Aleksandra Korolova}.}
  \bibinfo{year}{2014}\natexlab{}.
\newblock \showarticletitle{{What Can We Learn Privately:} Randomized
  Aggregatable Privacy-Preserving Ordinal Response}. In
  \bibinfo{booktitle}{\emph{Proceedings of the 2014 {ACM} {SIGSAC} Conference
  on Computer and Communications Security, Scottsdale, {AZ}, {USA}, November
  3-7, 2014}}. \bibinfo{publisher}{{ACM}}, \bibinfo{address}{{USA}},
  \bibinfo{pages}{1054--1067}.
\newblock


\bibitem[Fang et~al\mbox{.}(2020)]%
        {DBLP:conf/uss/FangCJG20}
\bibfield{author}{\bibinfo{person}{Minghong Fang}, \bibinfo{person}{Xiaoyu
  Cao}, \bibinfo{person}{Jinyuan Jia}, {and} \bibinfo{person}{Neil~Zhenqiang
  Gong}.} \bibinfo{year}{2020}\natexlab{}.
\newblock \showarticletitle{Local Model Poisoning Attacks to Byzantine-Robust
  Federated Learning}. In \bibinfo{booktitle}{\emph{29th {USENIX} Security
  Symposium, {USENIX} Security 2020, August 12-14, 2020}}.
  \bibinfo{publisher}{{USENIX} Association}, \bibinfo{address}{{USA}},
  \bibinfo{pages}{1605--1622}.
\newblock


\bibitem[Fang et~al\mbox{.}(2021)]%
        {DBLP:conf/www/FangSLGT021}
\bibfield{author}{\bibinfo{person}{Minghong Fang}, \bibinfo{person}{Minghao
  Sun}, \bibinfo{person}{Qi Li}, \bibinfo{person}{Neil~Zhenqiang Gong},
  \bibinfo{person}{Jin Tian}, {and} \bibinfo{person}{Jia Liu}.}
  \bibinfo{year}{2021}\natexlab{}.
\newblock \showarticletitle{Data Poisoning Attacks and Defenses to
  Crowdsourcing Systems}. In \bibinfo{booktitle}{\emph{{WWW} '21: The Web
  Conference 2021, Virtual Event / Ljubljana, Slovenia, April 19-23, 2021}}.
  \bibinfo{publisher}{{ACM} / {IW3C2}}, \bibinfo{address}{{USA}},
  \bibinfo{pages}{969--980}.
\newblock


\bibitem[Hsu et~al\mbox{.}(2012)]%
        {DBLP:conf/icalp/HsuKR12}
\bibfield{author}{\bibinfo{person}{Justin Hsu}, \bibinfo{person}{Sanjeev
  Khanna}, {and} \bibinfo{person}{Aaron Roth}.}
  \bibinfo{year}{2012}\natexlab{}.
\newblock \showarticletitle{Distributed Private Heavy Hitters}. In
  \bibinfo{booktitle}{\emph{Automata, Languages, and Programming - 39th
  International Colloquium, {ICALP} 2012, Warwick, {UK}, July 9-13, 2012,
  Proceedings, Part {I}}}, Vol.~\bibinfo{volume}{7391}.
  \bibinfo{publisher}{Springer}, \bibinfo{address}{Germany},
  \bibinfo{pages}{461--472}.
\newblock


\bibitem[Jagielski et~al\mbox{.}(2018)]%
        {DBLP:conf/sp/JagielskiOBLNL18}
\bibfield{author}{\bibinfo{person}{Matthew Jagielski}, \bibinfo{person}{Alina
  Oprea}, \bibinfo{person}{Battista Biggio}, \bibinfo{person}{Chang Liu},
  \bibinfo{person}{Cristina Nita{-}Rotaru}, {and} \bibinfo{person}{Bo Li}.}
  \bibinfo{year}{2018}\natexlab{}.
\newblock \showarticletitle{Manipulating Machine Learning: Poisoning Attacks
  and Countermeasures for Regression Learning}. In
  \bibinfo{booktitle}{\emph{2018 {IEEE} Symposium on Security and Privacy, {SP}
  2018, Proceedings, San Francisco, {CA}, {USA}, 21-23 May 2018}}.
  \bibinfo{publisher}{{IEEE} Computer Society}, \bibinfo{address}{{USA}},
  \bibinfo{pages}{19--35}.
\newblock


\bibitem[Kasiviswanathan et~al\mbox{.}(2008)]%
        {DBLP:conf/focs/KasiviswanathanLNRS08}
\bibfield{author}{\bibinfo{person}{Shiva~Prasad Kasiviswanathan},
  \bibinfo{person}{Homin~K. Lee}, \bibinfo{person}{Kobbi Nissim},
  \bibinfo{person}{Sofya Raskhodnikova}, {and} \bibinfo{person}{Adam~D.
  Smith}.} \bibinfo{year}{2008}\natexlab{}.
\newblock \showarticletitle{What Can We Learn Privately?}. In
  \bibinfo{booktitle}{\emph{49th Annual {IEEE} Symposium on Foundations of
  Computer Science, {FOCS} 2008, Philadelphia, {PA}, {USA}, October 25-28,
  2008}}. \bibinfo{publisher}{{IEEE} Computer Society},
  \bibinfo{address}{{USA}}, \bibinfo{pages}{531--540}.
\newblock


\bibitem[Kato et~al\mbox{.}(2021)]%
        {DBLP:conf/dbsec/Kato0Y21}
\bibfield{author}{\bibinfo{person}{Fumiyuki Kato}, \bibinfo{person}{Yang Cao},
  {and} \bibinfo{person}{Masatoshi Yoshikawa}.}
  \bibinfo{year}{2021}\natexlab{}.
\newblock \showarticletitle{Preventing Manipulation Attack in Local
  Differential Privacy Using Verifiable Randomization Mechanism}. In
  \bibinfo{booktitle}{\emph{Data and Applications Security and Privacy {XXXV} -
  35th Annual {IFIP} {WG} 11.3 Conference, DBSec 2021, Calgary, Canada, July
  19-20, 2021}}, Vol.~\bibinfo{volume}{12840}. \bibinfo{publisher}{Springer},
  \bibinfo{pages}{43--60}.
\newblock


\bibitem[{KDD Cup 2000}(2000)]%
        {KDDCup-2000}
\bibfield{author}{\bibinfo{person}{{KDD Cup 2000}}.}
  \bibinfo{year}{2000}\natexlab{}.
\newblock \bibinfo{title}{{BMS-POS}: Online retailer website clickstream
  analysis}.
\newblock
\newblock
\urldef\tempurl%
\url{https://www.kdd.org/kdd-cup/view/kdd-cup-2000}
\showURL{%
\tempurl}


\bibitem[Li et~al\mbox{.}(2016)]%
        {DBLP:conf/nips/LiWSV16}
\bibfield{author}{\bibinfo{person}{Bo Li}, \bibinfo{person}{Yining Wang},
  \bibinfo{person}{Aarti Singh}, {and} \bibinfo{person}{Yevgeniy Vorobeychik}.}
  \bibinfo{year}{2016}\natexlab{}.
\newblock \showarticletitle{Data Poisoning Attacks on Factorization-Based
  Collaborative Filtering}. In \bibinfo{booktitle}{\emph{Annual Conference on
  Neural Information Processing Systems 2016, Barcelona, Spain, December 5-10,
  2016}}. \bibinfo{publisher}{Neural Information Processing Systems
  Foundation}, \bibinfo{address}{{USA}}, \bibinfo{pages}{1885--1893}.
\newblock


\bibitem[Li et~al\mbox{.}(2022)]%
        {DBLP:conf/cikm/LiGGWY22}
\bibfield{author}{\bibinfo{person}{Junhui Li}, \bibinfo{person}{Wensheng Gan},
  \bibinfo{person}{Yijie Gui}, \bibinfo{person}{Yongdong Wu}, {and}
  \bibinfo{person}{Philip~S. Yu}.} \bibinfo{year}{2022}\natexlab{}.
\newblock \showarticletitle{Frequent Itemset Mining with Local Differential
  Privacy}. In \bibinfo{booktitle}{\emph{Proceedings of the 31st {ACM}
  International Conference on Information {\&} Knowledge Management, Atlanta,
  GA, USA, October 17-21, 2022}}. \bibinfo{publisher}{{ACM}},
  \bibinfo{pages}{1146--1155}.
\newblock


\bibitem[Li et~al\mbox{.}(2023)]%
        {DBLP:journals/corr/abs-2205-11782}
\bibfield{author}{\bibinfo{person}{Xiaoguang Li}, \bibinfo{person}{Ninghui Li},
  \bibinfo{person}{Wenhai Sun}, \bibinfo{person}{Neil~Zhenqiang Gong}, {and}
  \bibinfo{person}{Hui Li}.} \bibinfo{year}{2023}\natexlab{}.
\newblock \showarticletitle{Fine-grained poisoning attack to local differential
  privacy protocols for mean and variance estimation}. In
  \bibinfo{booktitle}{\emph{Proceedings of the 32nd USENIX Conference on
  Security Symposium, Anaheim, {CA}, {USA}}}. \bibinfo{publisher}{USENIX
  Association}, \bibinfo{address}{{USA}}, Article \bibinfo{articleno}{98},
  \bibinfo{numpages}{18}~pages.
\newblock


\bibitem[Liu et~al\mbox{.}(2020)]%
        {DBLP:conf/dasfaa/LiuCYC20}
\bibfield{author}{\bibinfo{person}{Ruixuan Liu}, \bibinfo{person}{Yang Cao},
  \bibinfo{person}{Masatoshi Yoshikawa}, {and} \bibinfo{person}{Hong Chen}.}
  \bibinfo{year}{2020}\natexlab{}.
\newblock \showarticletitle{{FedSel}: Federated {SGD} Under Local Differential
  Privacy with Top-k Dimension Selection}. In
  \bibinfo{booktitle}{\emph{Database Systems for Advanced Applications - 25th
  International Conference, {DASFAA} 2020, Jeju, South Korea, September 24-27,
  2020, Proceedings, Part {I}}}, Vol.~\bibinfo{volume}{12112}.
  \bibinfo{publisher}{Springer}, \bibinfo{address}{Germany},
  \bibinfo{pages}{485--501}.
\newblock


\bibitem[Pihur et~al\mbox{.}(2018)]%
        {DBLP:journals/corr/abs-1807-04369}
\bibfield{author}{\bibinfo{person}{Vasyl Pihur}, \bibinfo{person}{Aleksandra
  Korolova}, \bibinfo{person}{Frederick Liu}, \bibinfo{person}{Subhash
  Sankuratripati}, \bibinfo{person}{Moti Yung}, \bibinfo{person}{Dachuan
  Huang}, {and} \bibinfo{person}{Ruogu Zeng}.} \bibinfo{year}{2018}\natexlab{}.
\newblock \showarticletitle{Differentially-Private `Discard' Machine Learning}.
\newblock \bibinfo{journal}{\emph{CoRR}}  \bibinfo{volume}{abs/1807.04369}
  (\bibinfo{year}{2018}).
\newblock


\bibitem[Qin et~al\mbox{.}(2016)]%
        {DBLP:conf/ccs/QinYYKXR16}
\bibfield{author}{\bibinfo{person}{Zhan Qin}, \bibinfo{person}{Yin Yang},
  \bibinfo{person}{Ting Yu}, \bibinfo{person}{Issa Khalil},
  \bibinfo{person}{Xiaokui Xiao}, {and} \bibinfo{person}{Kui Ren}.}
  \bibinfo{year}{2016}\natexlab{}.
\newblock \showarticletitle{Heavy Hitter Estimation over Set-Valued Data with
  Local Differential Privacy}. In \bibinfo{booktitle}{\emph{Proceedings of the
  2016 {ACM} {SIGSAC} Conference on Computer and Communications Security,
  Vienna, Austria, October 24-28, 2016}}. \bibinfo{publisher}{{ACM}},
  \bibinfo{address}{{USA}}, \bibinfo{pages}{192--203}.
\newblock


\bibitem[Song et~al\mbox{.}(2023)]%
        {DBLP:journals/tifs/SongXZ23}
\bibfield{author}{\bibinfo{person}{Shaorui Song}, \bibinfo{person}{Lei Xu},
  {and} \bibinfo{person}{Liehuang Zhu}.} \bibinfo{year}{2023}\natexlab{}.
\newblock \showarticletitle{Efficient Defenses Against Output Poisoning Attacks
  on Local Differential Privacy}.
\newblock \bibinfo{journal}{\emph{{IEEE} Trans. Inf. Forensics Secur.}}
  \bibinfo{volume}{18} (\bibinfo{year}{2023}), \bibinfo{pages}{5506--5521}.
\newblock


\bibitem[Sun et~al\mbox{.}(2021)]%
        {DBLP:conf/ijcai/SunQC21}
\bibfield{author}{\bibinfo{person}{Lichao Sun}, \bibinfo{person}{Jianwei Qian},
  {and} \bibinfo{person}{Xun Chen}.} \bibinfo{year}{2021}\natexlab{}.
\newblock \showarticletitle{{LDP-FL:} Practical Private Aggregation in
  Federated Learning with Local Differential Privacy}. In
  \bibinfo{booktitle}{\emph{Proceedings of the Thirtieth International Joint
  Conference on Artificial Intelligence, Virtual Event / Montreal, Canada,
  {IJCAI} 2021, 19-27 August 2021}}. \bibinfo{publisher}{ijcai.org},
  \bibinfo{address}{{USA}}, \bibinfo{pages}{1571--1578}.
\newblock


\bibitem[Tian et~al\mbox{.}(2023)]%
        {DBLP:journals/csur/TianCLY23}
\bibfield{author}{\bibinfo{person}{Zhiyi Tian}, \bibinfo{person}{Lei Cui},
  \bibinfo{person}{Jie Liang}, {and} \bibinfo{person}{Shui Yu}.}
  \bibinfo{year}{2023}\natexlab{}.
\newblock \showarticletitle{A Comprehensive Survey on Poisoning Attacks and
  Countermeasures in Machine Learning}.
\newblock \bibinfo{journal}{\emph{{ACM} Comput. Surv.}} \bibinfo{volume}{55},
  \bibinfo{number}{8} (\bibinfo{year}{2023}), \bibinfo{pages}{166:1--166:35}.
\newblock


\bibitem[Wang et~al\mbox{.}(2018a)]%
        {DBLP:conf/infocom/0003HNWXY18}
\bibfield{author}{\bibinfo{person}{Shaowei Wang}, \bibinfo{person}{Liusheng
  Huang}, \bibinfo{person}{Yiwen Nie}, \bibinfo{person}{Pengzhan Wang},
  \bibinfo{person}{Hongli Xu}, {and} \bibinfo{person}{Wei Yang}.}
  \bibinfo{year}{2018}\natexlab{a}.
\newblock \showarticletitle{PrivSet: Set-Valued Data Analyses with Locale
  Differential Privacy}. In \bibinfo{booktitle}{\emph{2018 {IEEE} Conference on
  Computer Communications, {INFOCOM} 2018, Honolulu, {HI}, {USA}, April 16-19,
  2018}}. \bibinfo{publisher}{{IEEE}}, \bibinfo{address}{{USA}},
  \bibinfo{pages}{1088--1096}.
\newblock


\bibitem[Wang et~al\mbox{.}(2020b)]%
        {DBLP:journals/pvldb/WangQDYHX20}
\bibfield{author}{\bibinfo{person}{Shaowei Wang}, \bibinfo{person}{Yuqiu Qian},
  \bibinfo{person}{Jiachun Du}, \bibinfo{person}{Wei Yang},
  \bibinfo{person}{Liusheng Huang}, {and} \bibinfo{person}{Hongli Xu}.}
  \bibinfo{year}{2020}\natexlab{b}.
\newblock \showarticletitle{Set-valued Data Publication with Local Privacy:
  Tight Error Bounds and Efficient Mechanisms}.
\newblock \bibinfo{journal}{\emph{Proc. {VLDB} Endow.}} \bibinfo{volume}{13},
  \bibinfo{number}{8} (\bibinfo{year}{2020}), \bibinfo{pages}{1234--1247}.
\newblock


\bibitem[Wang et~al\mbox{.}(2017)]%
        {DBLP:conf/uss/WangBLJ17}
\bibfield{author}{\bibinfo{person}{Tianhao Wang}, \bibinfo{person}{Jeremiah
  Blocki}, \bibinfo{person}{Ninghui Li}, {and} \bibinfo{person}{Somesh Jha}.}
  \bibinfo{year}{2017}\natexlab{}.
\newblock \showarticletitle{Locally Differentially Private Protocols for
  Frequency Estimation}. In \bibinfo{booktitle}{\emph{26th {USENIX} Security
  Symposium, {USENIX} Security 2017, Vancouver, {BC}, Canada, August 16-18,
  2017}}. \bibinfo{publisher}{{USENIX} Association}, \bibinfo{address}{{USA}},
  \bibinfo{pages}{729--745}.
\newblock


\bibitem[Wang et~al\mbox{.}(2018b)]%
        {DBLP:conf/sp/WangLJ18}
\bibfield{author}{\bibinfo{person}{Tianhao Wang}, \bibinfo{person}{Ninghui Li},
  {and} \bibinfo{person}{Somesh Jha}.} \bibinfo{year}{2018}\natexlab{b}.
\newblock \showarticletitle{Locally Differentially Private Frequent Itemset
  Mining}. In \bibinfo{booktitle}{\emph{2018 {IEEE} Symposium on Security and
  Privacy, {SP} 2018, Proceedings, San Francisco, {CA},{USA}, 21-23 May 2018}}.
  \bibinfo{publisher}{{IEEE} Computer Society}, \bibinfo{address}{{USA}},
  \bibinfo{pages}{127--143}.
\newblock


\bibitem[Wang et~al\mbox{.}(2021)]%
        {DBLP:journals/tdsc/0001LJ21}
\bibfield{author}{\bibinfo{person}{Tianhao Wang}, \bibinfo{person}{Ninghui Li},
  {and} \bibinfo{person}{Somesh Jha}.} \bibinfo{year}{2021}\natexlab{}.
\newblock \showarticletitle{Locally Differentially Private Heavy Hitter
  Identification}.
\newblock \bibinfo{journal}{\emph{{IEEE} Trans. Dependable Secur. Comput.}}
  \bibinfo{volume}{18}, \bibinfo{number}{2} (\bibinfo{year}{2021}),
  \bibinfo{pages}{982--993}.
\newblock


\bibitem[Wang et~al\mbox{.}(2020a)]%
        {DBLP:conf/ndss/0001LLSL20}
\bibfield{author}{\bibinfo{person}{Tianhao Wang}, \bibinfo{person}{Milan
  Lopuha{\"{a}}{-}Zwakenberg}, \bibinfo{person}{Zitao Li},
  \bibinfo{person}{Boris Skoric}, {and} \bibinfo{person}{Ninghui Li}.}
  \bibinfo{year}{2020}\natexlab{a}.
\newblock \showarticletitle{Locally Differentially Private Frequency Estimation
  with Consistency}. In \bibinfo{booktitle}{\emph{27th Annual Network and
  Distributed System Security Symposium, {NDSS} 2020, San Diego, {CA}, {USA},
  February 23-26, 2020}}. \bibinfo{publisher}{The Internet Society},
  \bibinfo{address}{{USA}}, \bibinfo{numpages}{16}~pages.
\newblock


\bibitem[Wu et~al\mbox{.}(2022)]%
        {DBLP:journals/corr/abs-2111-11534}
\bibfield{author}{\bibinfo{person}{Yongji Wu}, \bibinfo{person}{Xiaoyu Cao},
  \bibinfo{person}{Jinyuan Jia}, {and} \bibinfo{person}{Neil~Zhenqiang Gong}.}
  \bibinfo{year}{2022}\natexlab{}.
\newblock \showarticletitle{Poisoning Attacks to Local Differential Privacy
  Protocols for Key-Value Data}. In \bibinfo{booktitle}{\emph{31st {USENIX}
  Security Symposium, {USENIX} Security 2022, Boston, MA, USA, August 10-12,
  2022}}. \bibinfo{publisher}{{USENIX} Association}, \bibinfo{address}{{USA}},
  \bibinfo{pages}{519--536}.
\newblock


\bibitem[Ye et~al\mbox{.}(2019)]%
        {DBLP:conf/sp/YeHMZ19}
\bibfield{author}{\bibinfo{person}{Qingqing Ye}, \bibinfo{person}{Haibo Hu},
  \bibinfo{person}{Xiaofeng Meng}, {and} \bibinfo{person}{Huadi Zheng}.}
  \bibinfo{year}{2019}\natexlab{}.
\newblock \showarticletitle{PrivKV: Key-Value Data Collection with Local
  Differential Privacy}. In \bibinfo{booktitle}{\emph{2019 {IEEE} Symposium on
  Security and Privacy, {SP} 2019, San Francisco, {CA}, {USA}, May 19-23,
  2019}}. \bibinfo{publisher}{{IEEE}}, \bibinfo{address}{{USA}},
  \bibinfo{pages}{317--331}.
\newblock


\bibitem[Zhu et~al\mbox{.}(2024)]%
        {DBLP:journals/tifs/ZhuCXWZ24}
\bibfield{author}{\bibinfo{person}{Youwen Zhu}, \bibinfo{person}{Yiran Cao},
  \bibinfo{person}{Qiao Xue}, \bibinfo{person}{Qihui Wu}, {and}
  \bibinfo{person}{Yushu Zhang}.} \bibinfo{year}{2024}\natexlab{}.
\newblock \showarticletitle{Heavy Hitter Identification Over Large-Domain
  Set-Valued Data With Local Differential Privacy}.
\newblock \bibinfo{journal}{\emph{{IEEE} Trans. Inf. Forensics Secur.}}
  \bibinfo{volume}{19} (\bibinfo{year}{2024}), \bibinfo{pages}{414--426}.
\newblock


\end{thebibliography}

\clearpage
\appendix
\section{Appendix}
\label{sec:app}
\begin{figure}
	\centering
	\subfigure[IBM, $\mathrm{ACC}$]{
		\begin{minipage}[c]{0.2\textwidth}
		\centering
        \includegraphics[width=1\textwidth]{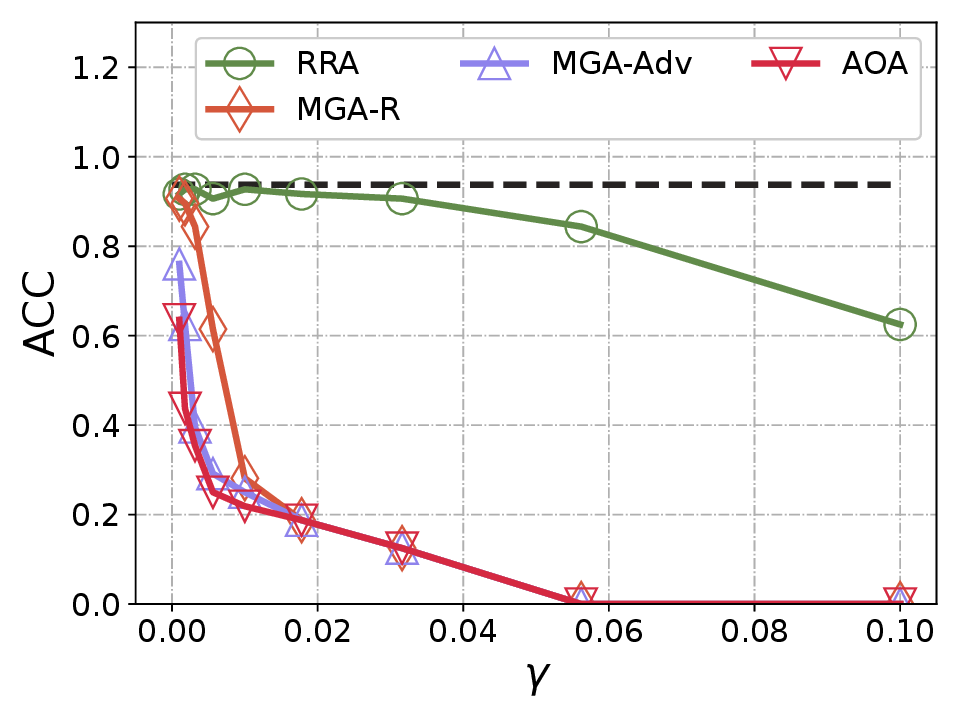}
		\end{minipage}%
	}
 	\subfigure[IBM, $\mathrm{NCR}$]{
		\begin{minipage}[c]{0.2\textwidth}
		\centering
        \includegraphics[width=1\textwidth]{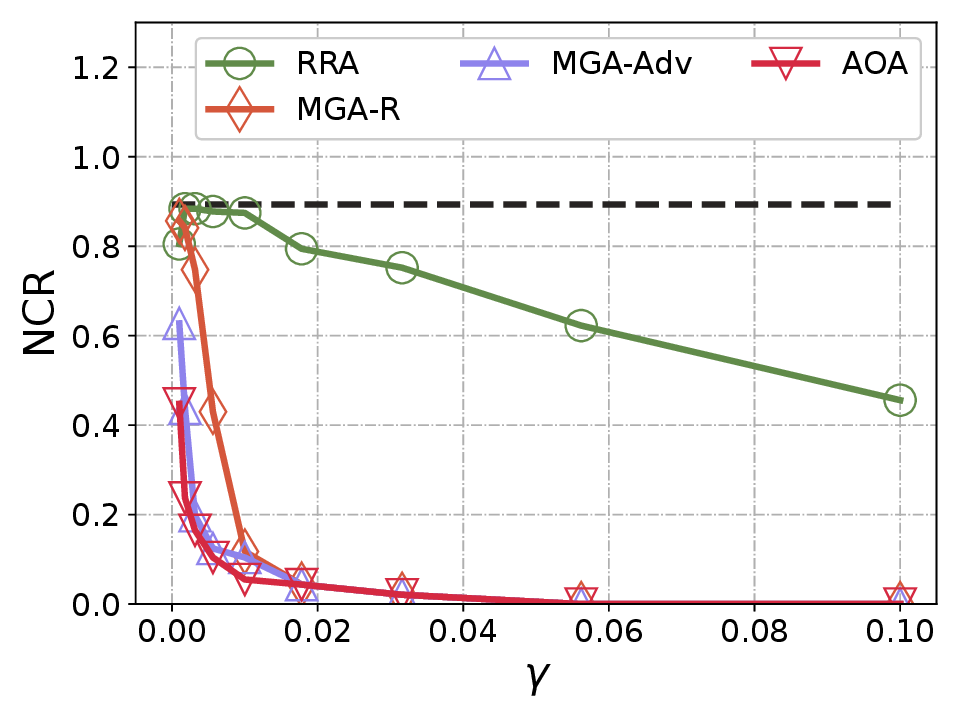}
		\end{minipage}%
    }
 
      \subfigure[BMS-POS, $\mathrm{ACC}$]{
		\begin{minipage}[c]{0.2\textwidth}
		\centering
        \includegraphics[width=1\textwidth]{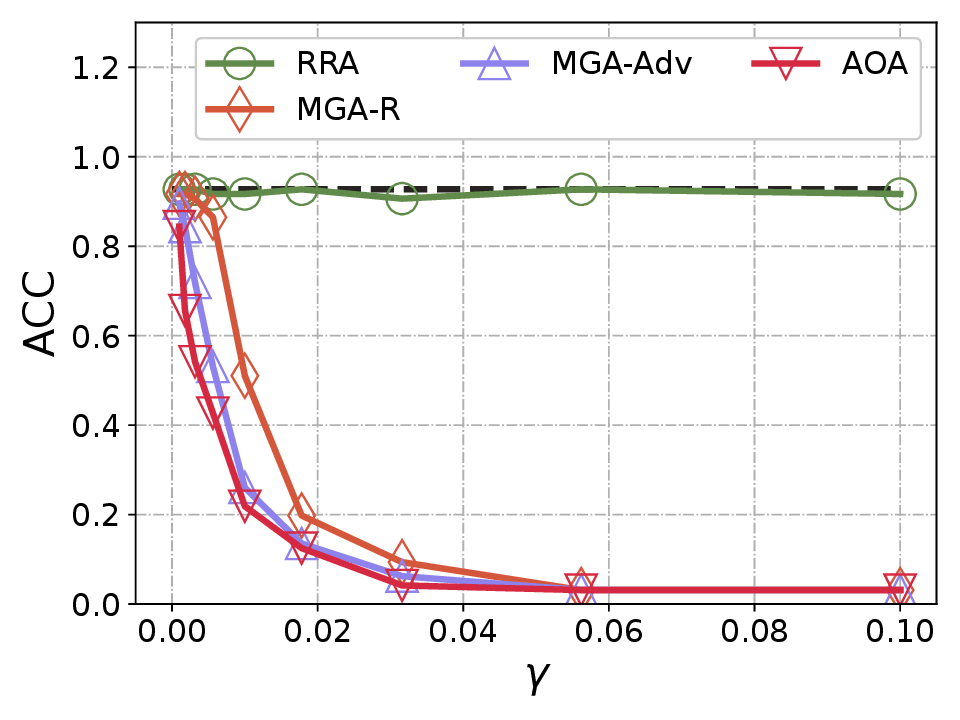}
		\end{minipage}%
	}
     	\subfigure[BMS-POS, $\mathrm{NCR}$]{
		\begin{minipage}[c]{0.2\textwidth}
		\centering
        \includegraphics[width=1\textwidth]{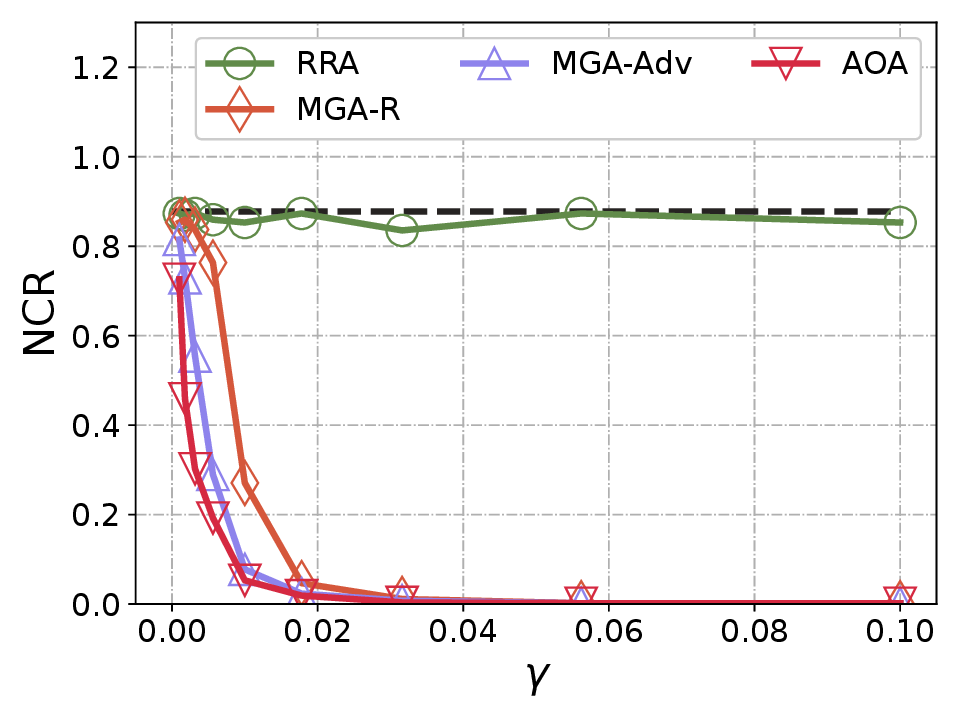}
		\end{minipage}%
    }
    \centering
	\caption{\ccsrev{Attacking PrivSet, with $k = 32$, $\epsilon=4.0$ (black dashed line - no attack).}}
	\label{fig:gamma:privset:eps4}
\end{figure}

\ccsrev{
\subsection{Attacking PrivSet}
\label{sec:attackprivset}
In the PrivSet protocol, each user sends a report with $\kappa$ items to the aggregator, where $\kappa$ is a parameter dependent on $\epsilon$, $d$, and the padding size $l$. This means that given a set of corrupted users $U^c$ with a size of $m$, the amount of resources allowed for manipulating PrivSet is $m \times \min(L, \kappa)$, where $L$ represents the size of the target set.

A user report can support multiple items in PrivSet. But when the report length $\kappa < L$, only part of the targets can be supported. In this case, we adopt a similar method to generate poisoning reports as in OLH, by always providing support for the top-$\kappa$ targets with the largest frequency gaps upon generating each poisoning report. Subsequently, we update the gaps and identify new top-$\kappa$ target items.

Since PrivSet only consists of the $\mathsf{SelectTop}$ phase, where it estimates the frequencies of items and identifies the top-$k$ items based on these estimates, attacks on PrivSet can only be launched against this phase. The attacker utilizes the poisoned data generation operation to carry out the attack.

The results on attacking PrivSet are shown in \reffig{fig:gamma:privset:eps4}. 
}

\subsection{Additional Results}
\label{sec:app:addres}
This section reports more results from the experiments. 

\begin{figure}
	\centering
	\subfigure[IBM, $\mathrm{ACC}$]{
		\begin{minipage}[c]{0.2\textwidth}
		\centering
        \includegraphics[width=1\textwidth]{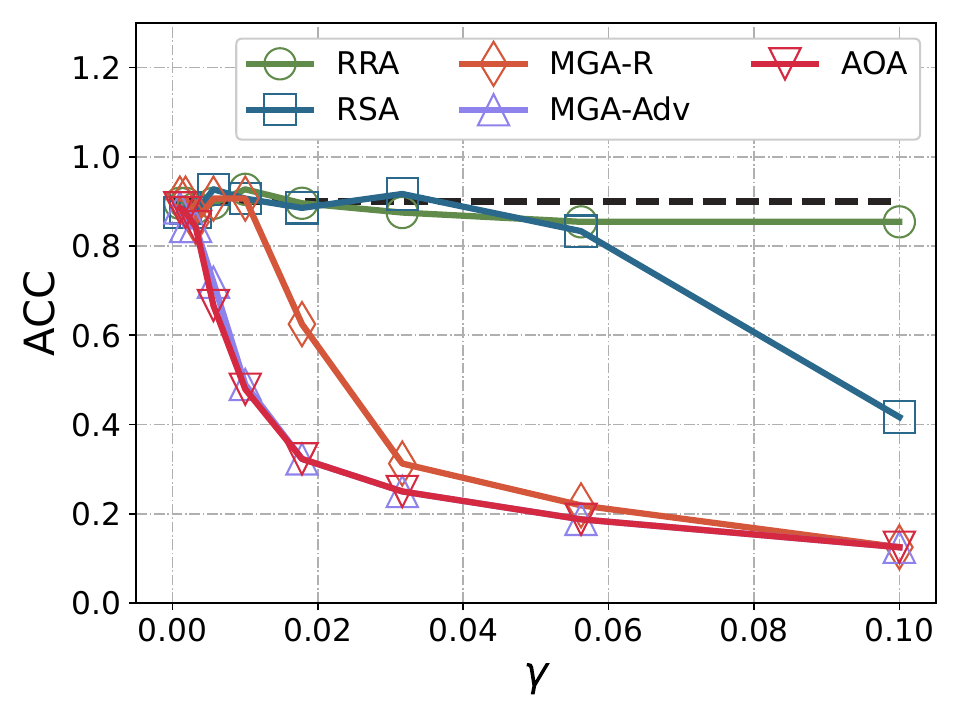}
		\end{minipage}%
	}
 	\subfigure[IBM, $\mathrm{NCR}$]{
		\begin{minipage}[c]{0.2\textwidth}
		\centering
        \includegraphics[width=1\textwidth]{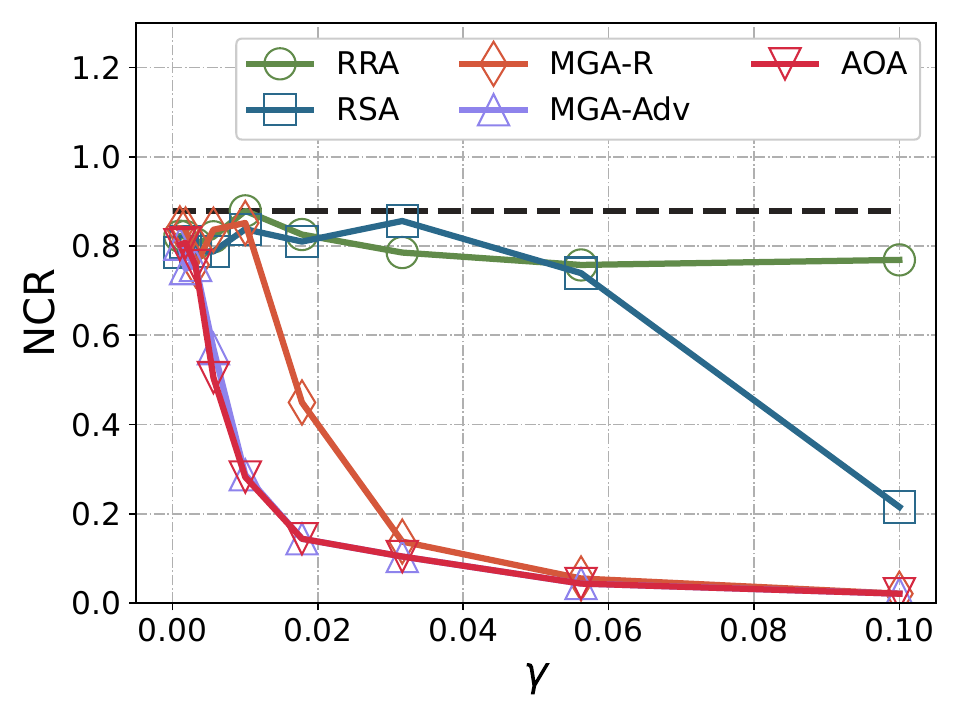}
		\end{minipage}%
    }
 
      \subfigure[Kosarak, $\mathrm{ACC}$]{
		\begin{minipage}[c]{0.2\textwidth}
		\centering
        \includegraphics[width=1\textwidth]{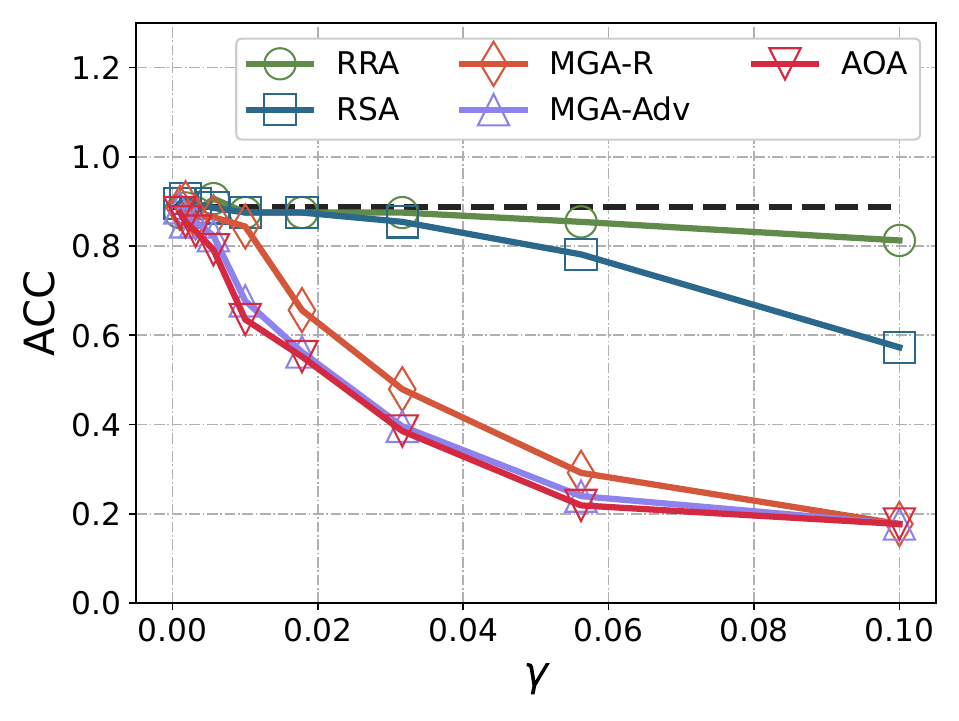}
		\end{minipage}%
	}
     	\subfigure[Kosarak, $\mathrm{NCR}$]{
		\begin{minipage}[c]{0.2\textwidth}
		\centering
        \includegraphics[width=1\textwidth]{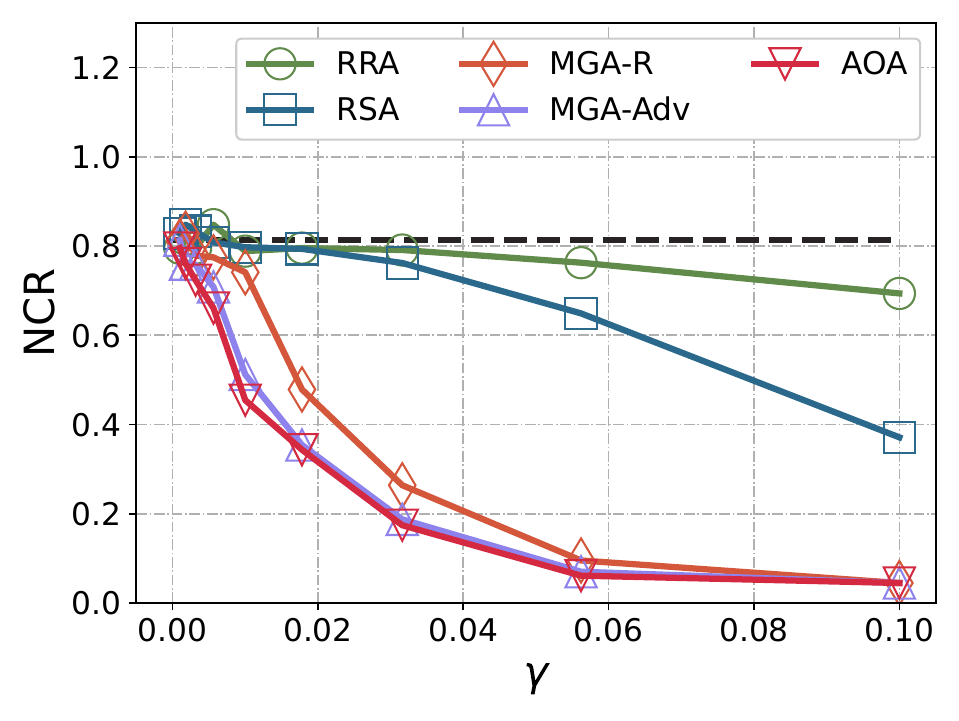}
		\end{minipage}%
    }
 
     \subfigure[BMS-POS, $\mathrm{ACC}$]{
		\begin{minipage}[c]{0.2\textwidth}
		\centering
        \includegraphics[width=1\textwidth]{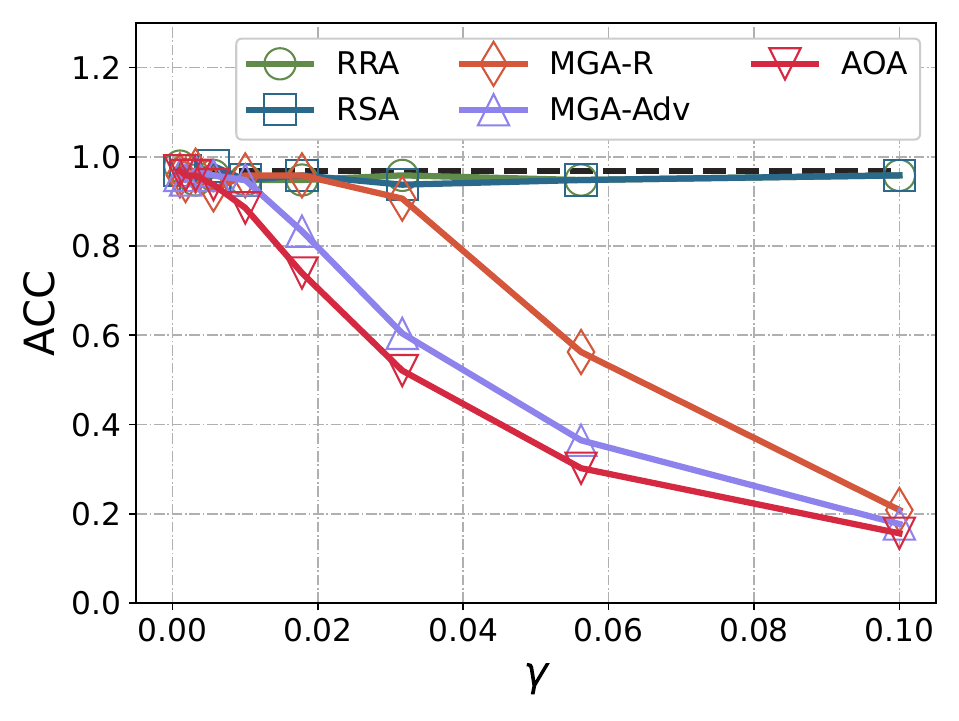}
		\end{minipage}%
	}
	\subfigure[BMS-POS, $\mathrm{NCR}$]{
		\begin{minipage}[c]{0.2\textwidth}
		\centering
        \includegraphics[width=1\textwidth]{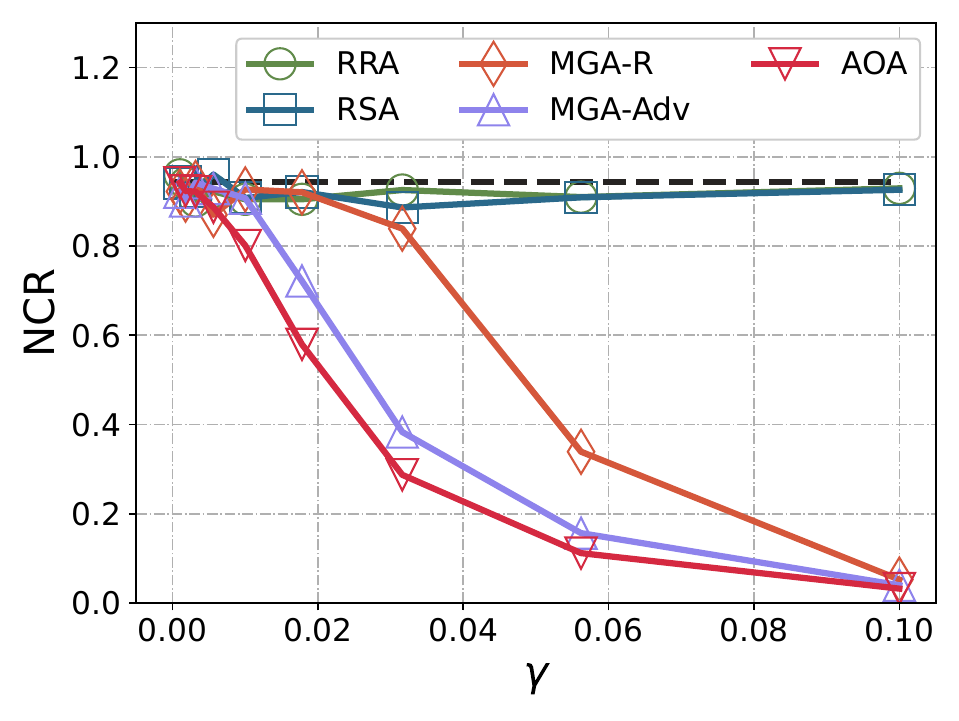}
		\end{minipage}%
    }

    \centering
	\caption{\ccsrev{Attacking FIML-I, with $k = 32$, $\epsilon=4.0$ (black dashed line - no attack).}}
 \Description[Attacking FIML-I ($\epsilon=4.0$)]{Attacking FIML-I, with $k = 32$, $\epsilon=4.0$ (black dashed line - no attack).}
	\label{fig:gamma:fimlitem:eps4}
\end{figure}

\begin{figure}
\centering
\includegraphics[width=0.4\textwidth]{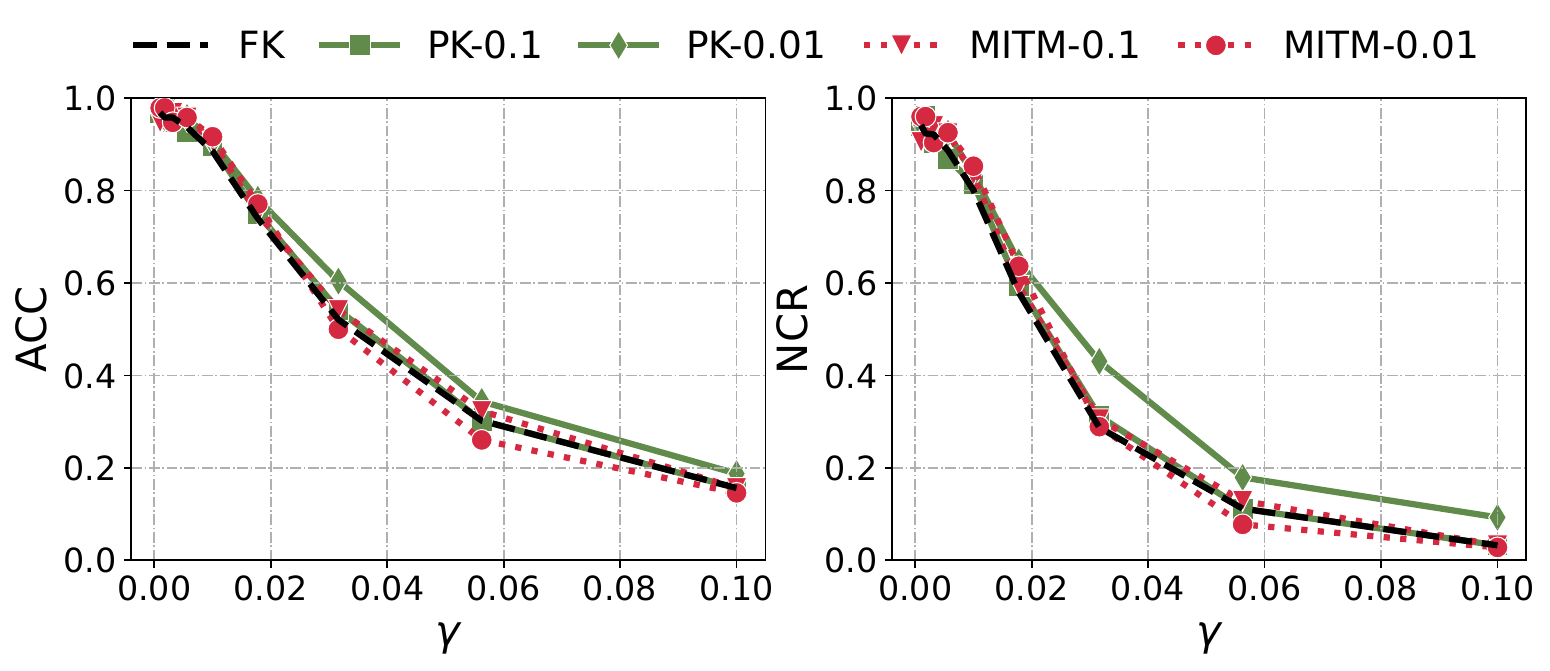}
 
\caption{\ccsrev{Attacking FIML-I with Limited Knowledge on BMS-POS dataset.}}
	\label{fig:less:fimli}
\end{figure}

\ccsrev{
\paragraphbe{Attacking FIML-I} The experimental results on attacking FIML-I are shown in \reffig{fig:gamma:fimlitem:eps4} and \reffig{fig:less:fimli}. The results demonstrate the effectiveness of the proposed attacks.
}

\begin{figure*}
\centering
    \subfigure[\ccsrev{IBM, SVSM}]{
		\begin{minipage}[c]{0.25\textwidth}
		\centering
        \includegraphics[width=1\textwidth]{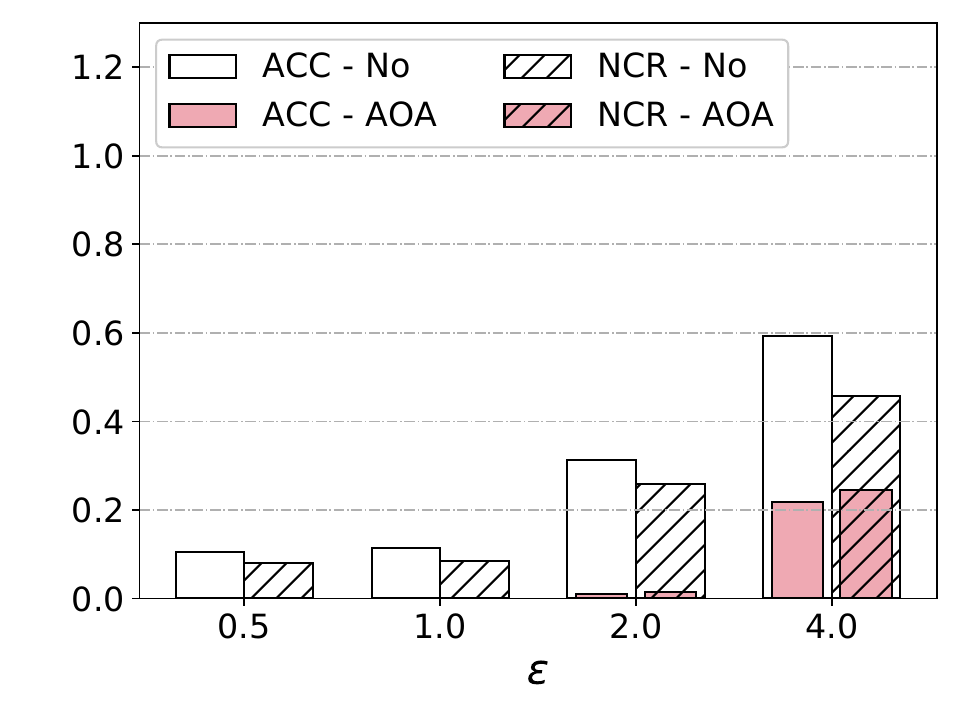}
		\end{minipage}%
	}	
	\subfigure[\ccsrev{IBM, LDPMiner}]{
		\begin{minipage}[c]{0.25\textwidth}
		\centering
        \includegraphics[width=1\textwidth]{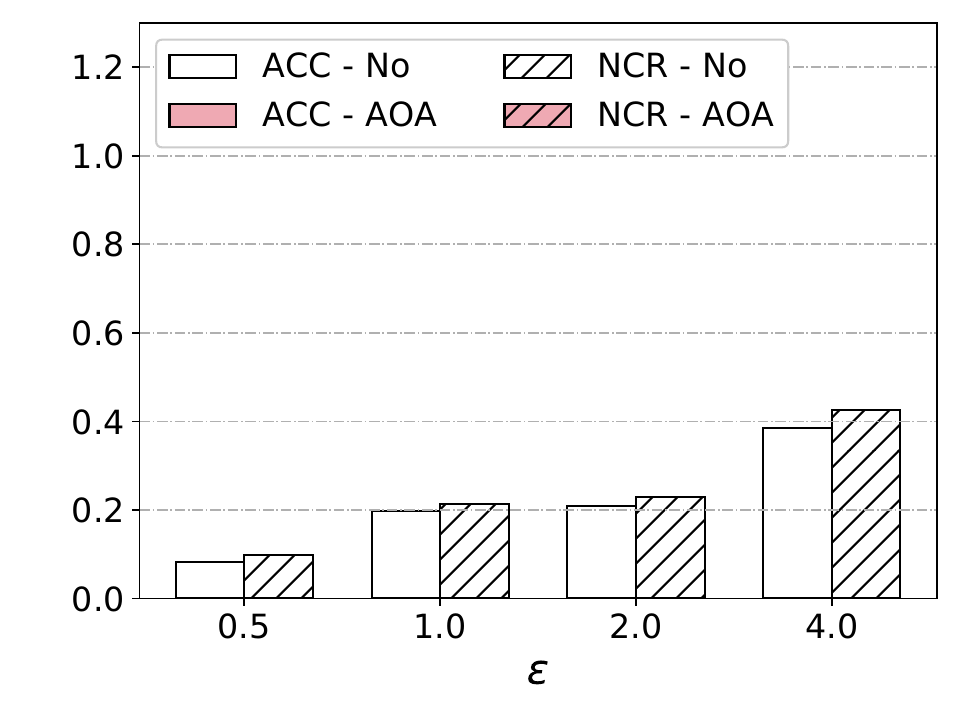}
		\end{minipage}%
    }
    \subfigure[\ccsrev{IBM, SVIM}]{
		\begin{minipage}[c]{0.25\textwidth}
		\centering
        \includegraphics[width=1\textwidth]{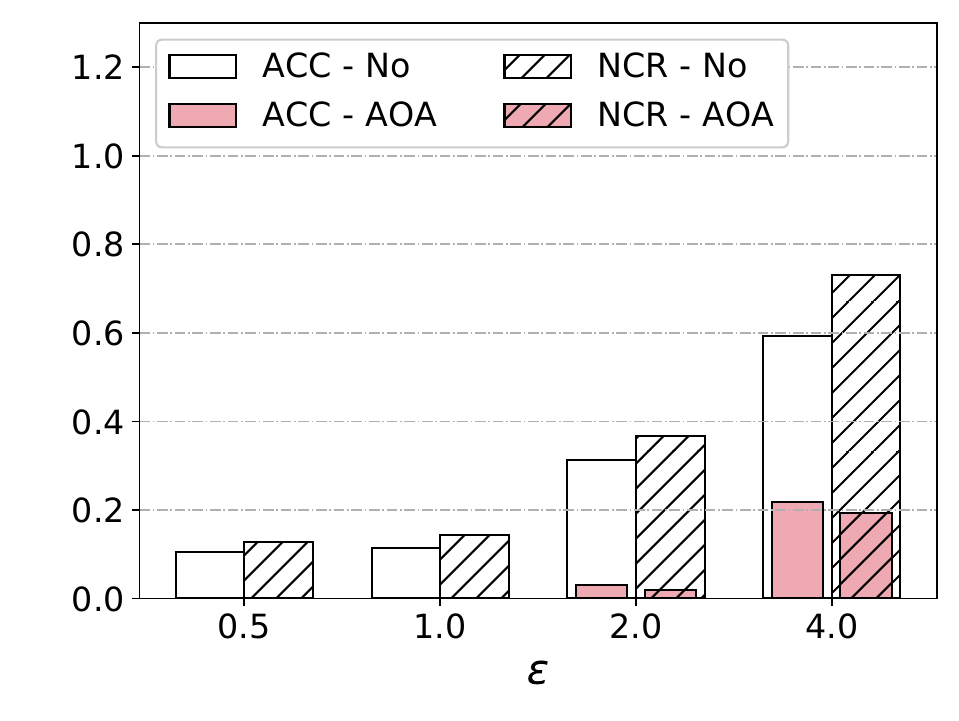}
		\end{minipage}%
    }
    
     \subfigure[Kosarak, SVSM]{
		\begin{minipage}[c]{0.25\textwidth}
		\centering
        \includegraphics[width=1\textwidth]{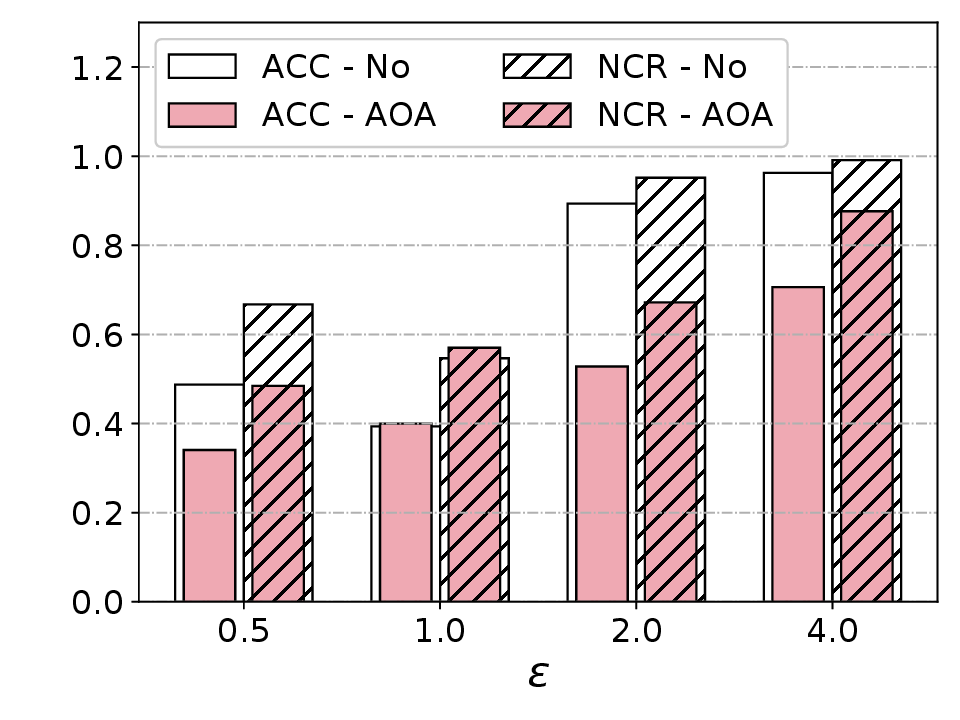}
		\end{minipage}%
	}
    \subfigure[Kosarak, LDPMiner]{
		\begin{minipage}[c]{0.25\textwidth}
		\centering
        \includegraphics[width=1\textwidth]{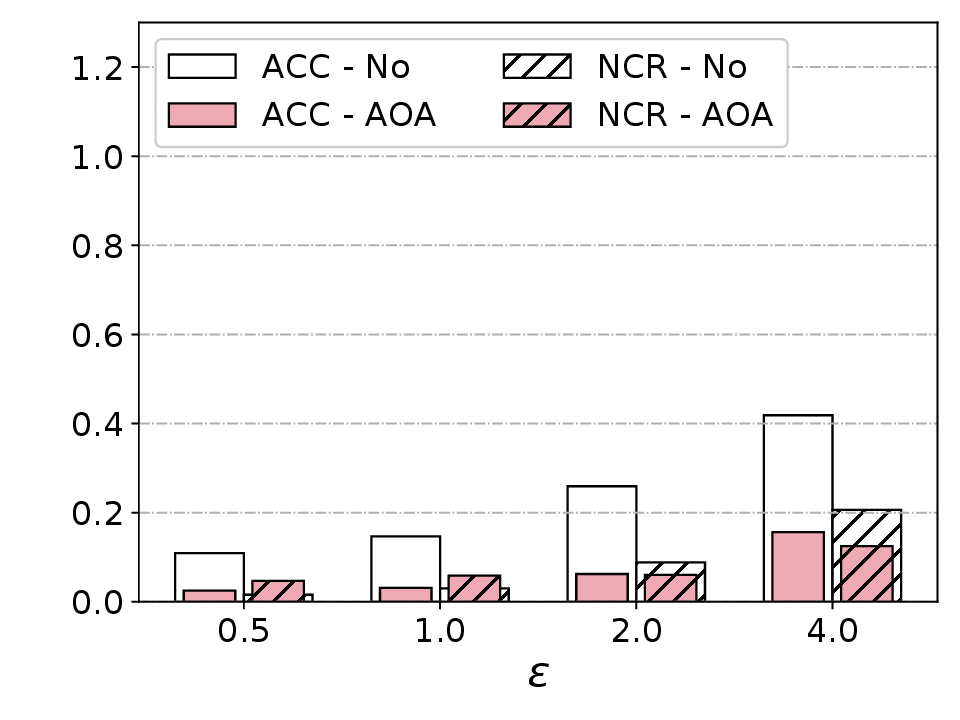}
		\end{minipage}%
    }
    \subfigure[Kosarak, SVIM]{
		\begin{minipage}[c]{0.25\textwidth}
		\centering
        \includegraphics[width=1\textwidth]{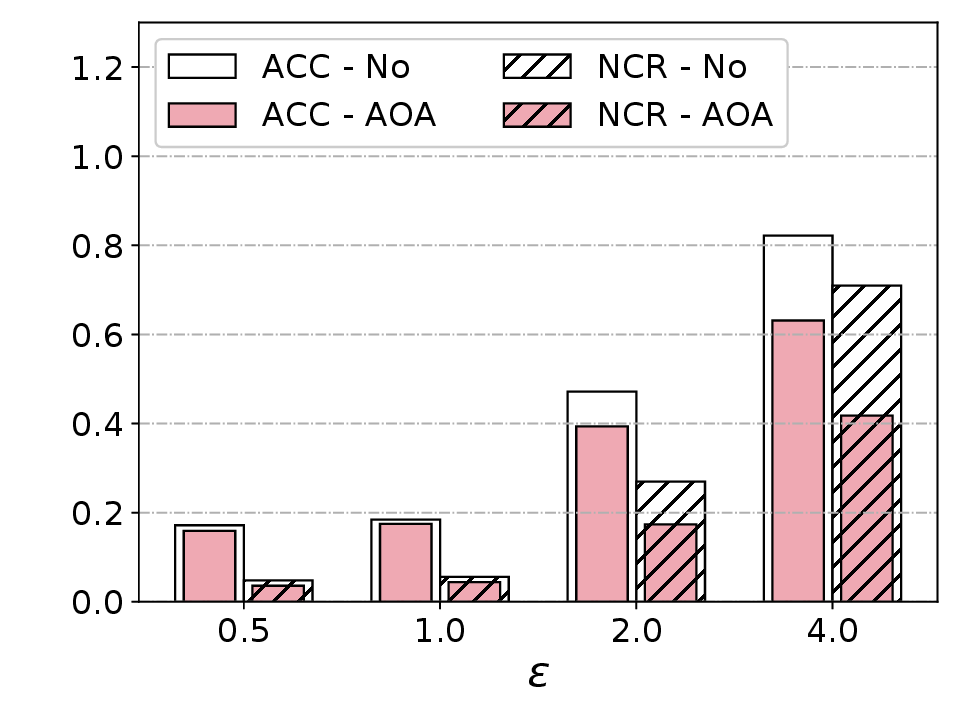}
		\end{minipage}%
    }
	\caption{Attacking LDP mechanisms on the datasets of IBM Synthesize and Kosarak with changing $\epsilon$; AOA with $\gamma = 0.01$; $k=32$. \ccsrev{The $y$-axis is ACC/NCR.}}
	\label{fig:app:params:eps}
\end{figure*}

\begin{figure*}
\centering
    \subfigure[\ccsrev{IBM, SVSM}]{
		\begin{minipage}[c]{0.25\textwidth}
		\centering
        \includegraphics[width=1\textwidth]{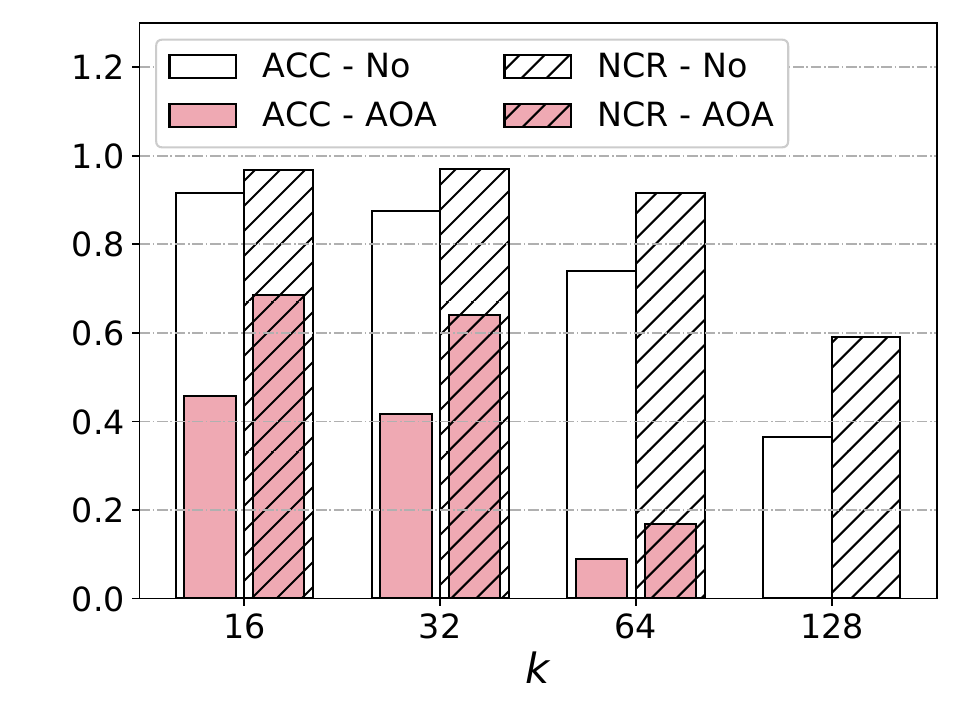}
		\end{minipage}%
	}
	\subfigure[\ccsrev{IBM, LDPMiner}]{
		\begin{minipage}[c]{0.25\textwidth}
		\centering
        \includegraphics[width=1\textwidth]{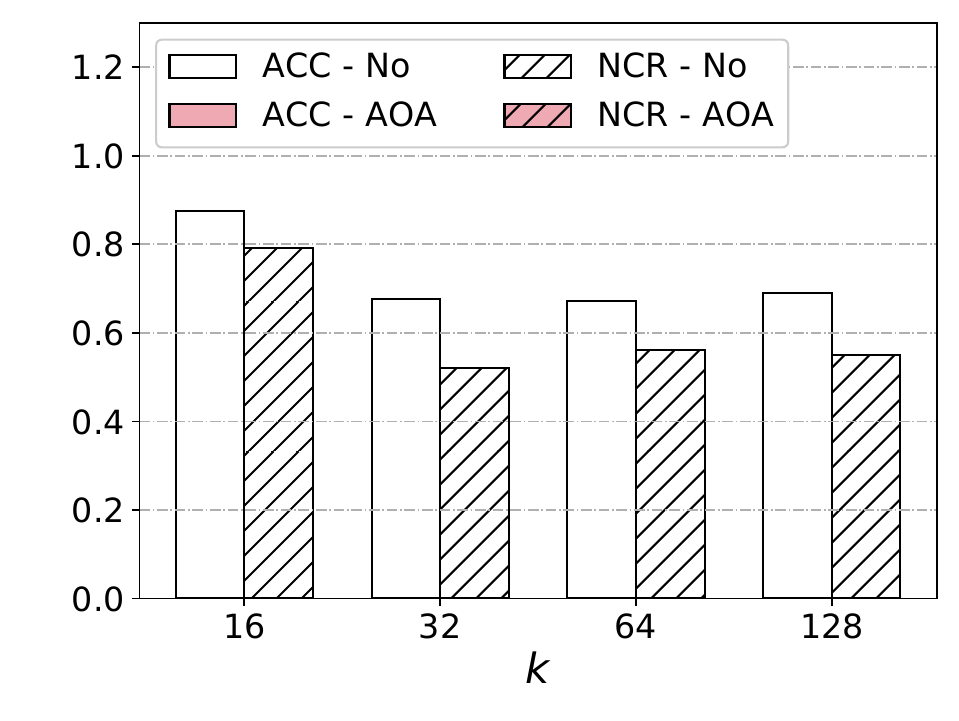}
		\end{minipage}%
    }
    \subfigure[\ccsrev{IBM, SVIM}]{
		\begin{minipage}[c]{0.25\textwidth}
		\centering
        \includegraphics[width=1\textwidth]{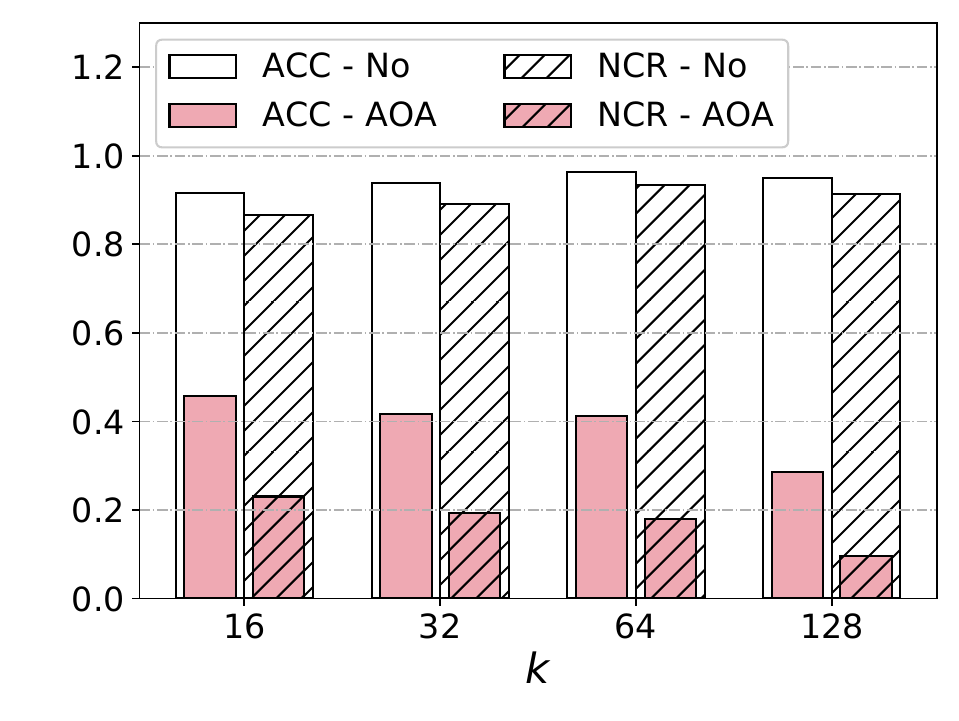}
		\end{minipage}%
    }

    	\subfigure[Kosarak, SVSM]{
		\begin{minipage}[c]{0.25\textwidth}
		\centering
        \includegraphics[width=1\textwidth]{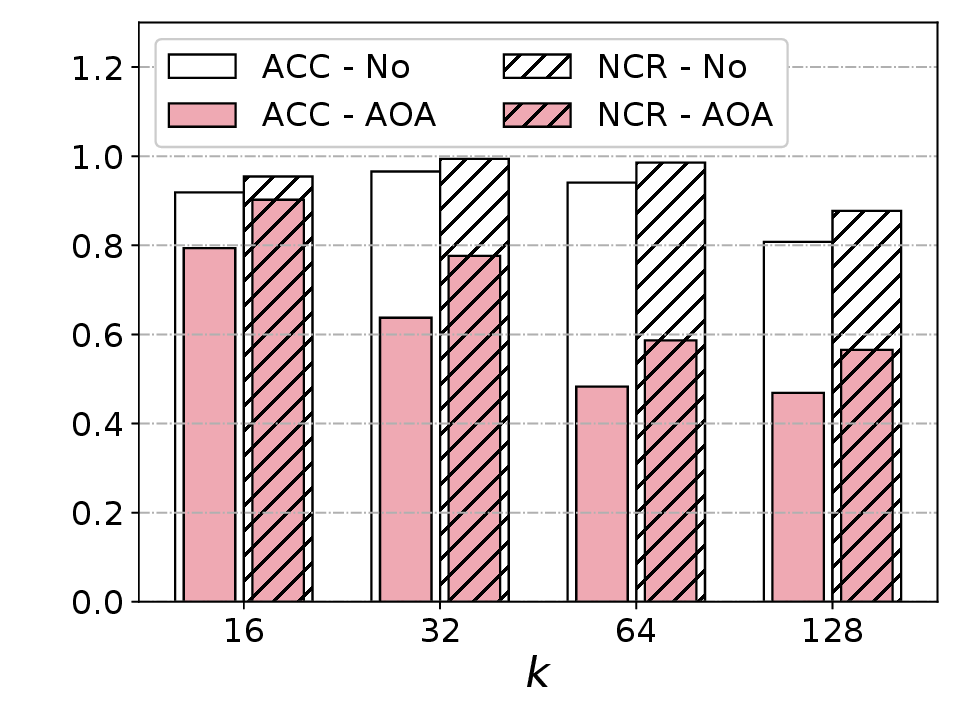}
		\end{minipage}%
	}
	\subfigure[Kosarak, LDPMiner]{
		\begin{minipage}[c]{0.25\textwidth}
		\centering
        \includegraphics[width=1\textwidth]{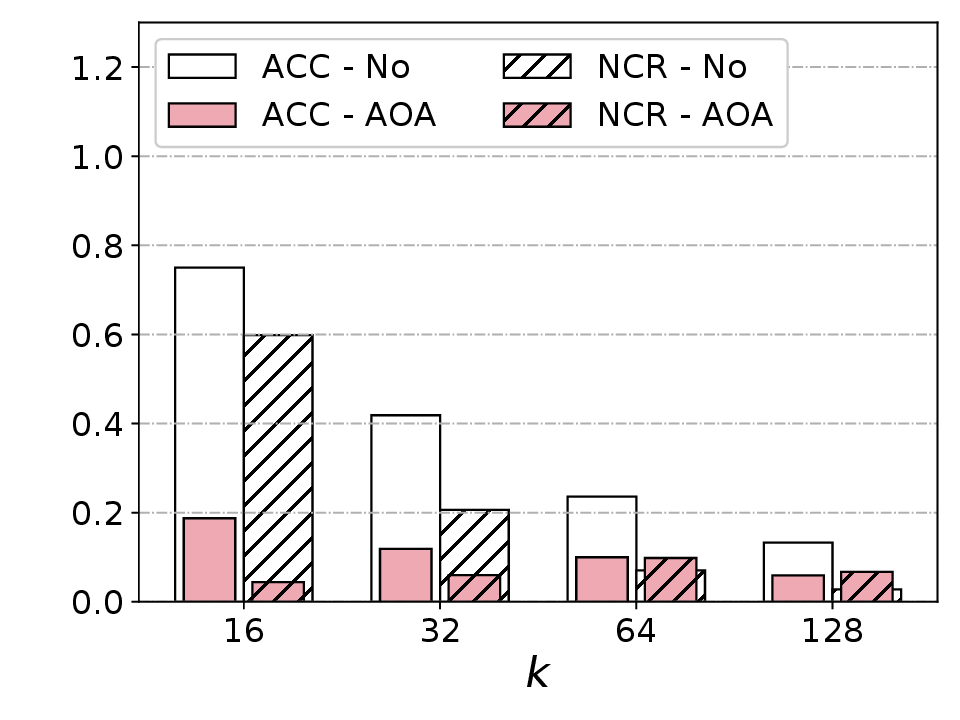}
		\end{minipage}%
    }
    \subfigure[Kosarak, SVIM]{
		\begin{minipage}[c]{0.25\textwidth}
		\centering
        \includegraphics[width=1\textwidth]{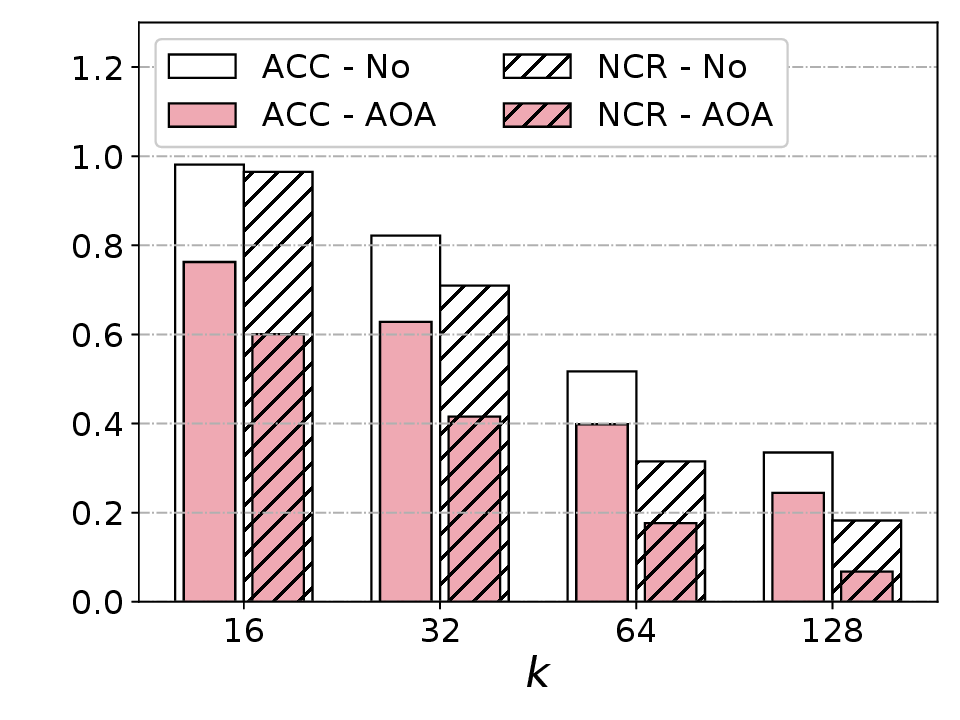}
		\end{minipage}%
    }
    
	\caption{Attacking LDP mechanisms on the datasets of IBM Synthesize and Kosarak with changing $k$; AOA with $\gamma = 0.01$; $\epsilon=4.0$. \ccsrev{The $y$-axis is ACC/NCR.}}
	\label{fig:app:params:k}
\end{figure*}

\paragraphbe{Impact of different parameters} We report experimental results on the impact of different parameters for the Kosarak dataset in \reffig{fig:app:params:eps}. Similar to the experiments conducted on the BMS-POS dataset, we set $\gamma=0.01$ and $k=32$ to evaluate the impact of $\epsilon$, and $\gamma=0.01$ and $\epsilon=4.0$ to evaluate the impact of $k$. We observe similar trends as those shown in \reffig{fig:params:eps}. An interesting observation can be made when attacking LDPMiner on the Kosarak dataset (\reffig{fig:app:params:eps}b and \reffig{fig:app:params:k}b). We can see that because the protocol is difficult to achieve good performance when $\epsilon$ is small or $k$ is large, even without an attack, the NCR may increase when an attack is conducted. Despite the insignificant attack performance of the proposed attack in this case, the unsatisfactory performance of the protocol in this setting makes it an unlikely valuable target for practical attacks. 

\begin{figure*}[h!]
    \centering
     	\subfigure[IBM, SVSM]{
		\begin{minipage}[c]{0.40\textwidth}
		\centering
        \includegraphics[width=1\textwidth]{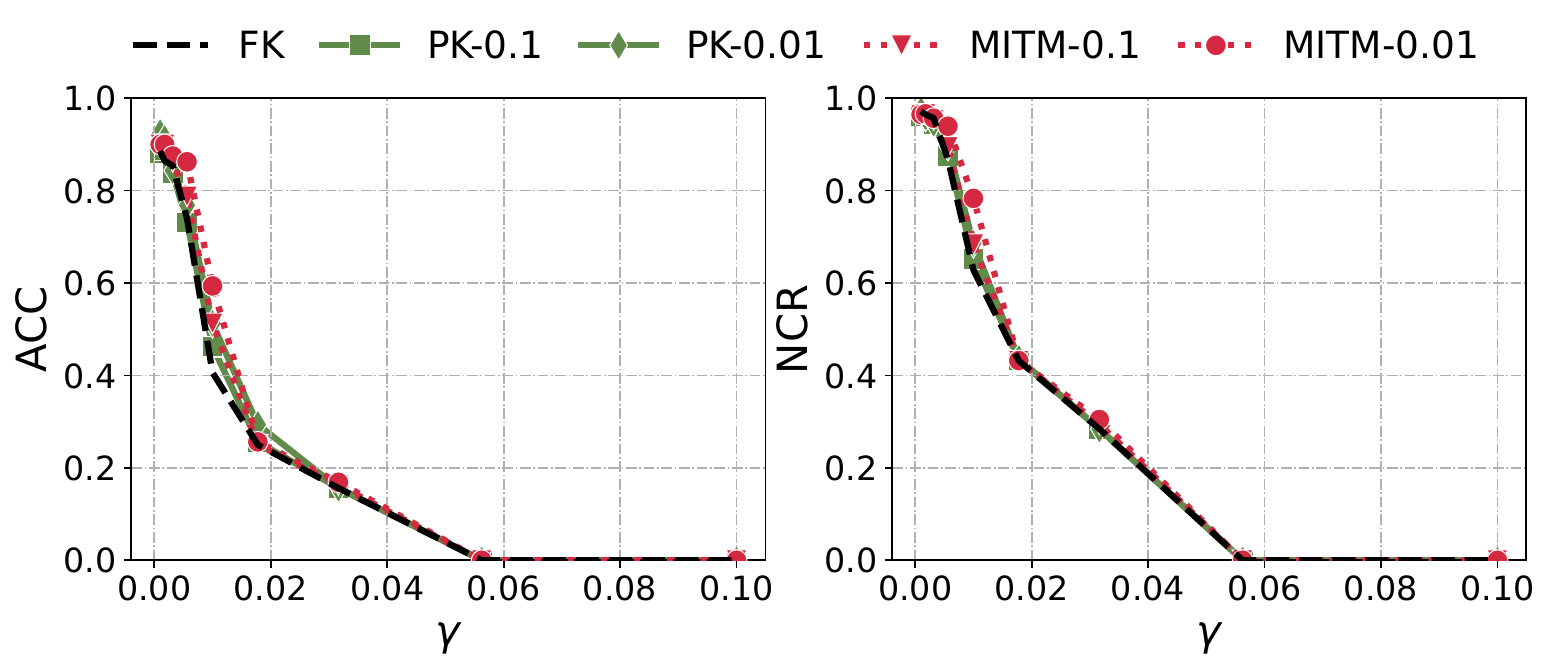}
		\end{minipage}%
	}
	\subfigure[Kosarak, SVSM]{
		\begin{minipage}[c]{0.40\textwidth}
		\centering
        \includegraphics[width=1\textwidth]{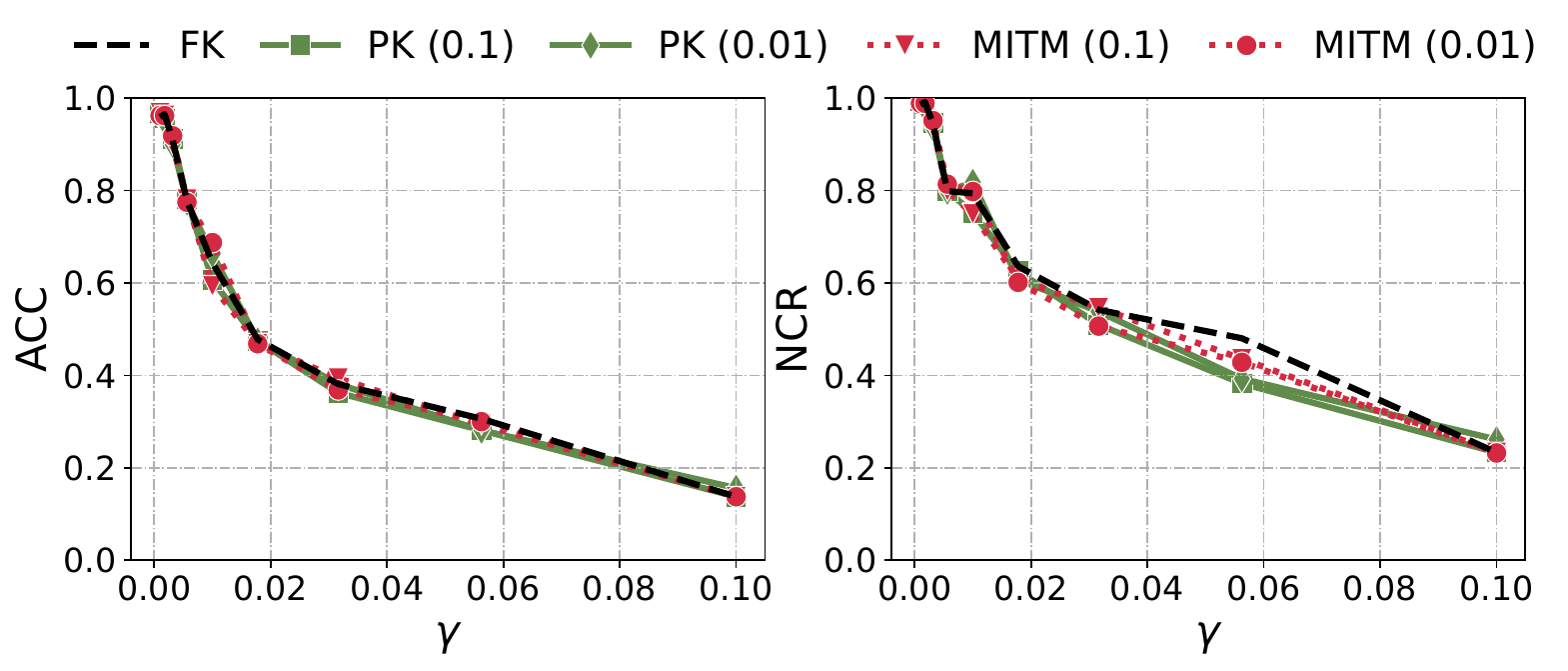}
		\end{minipage}%
	}

        \subfigure[\ccsrev{IBM, FIML-IS}]{
		\begin{minipage}[c]{0.40\textwidth}
		\centering
        \includegraphics[width=1\textwidth]{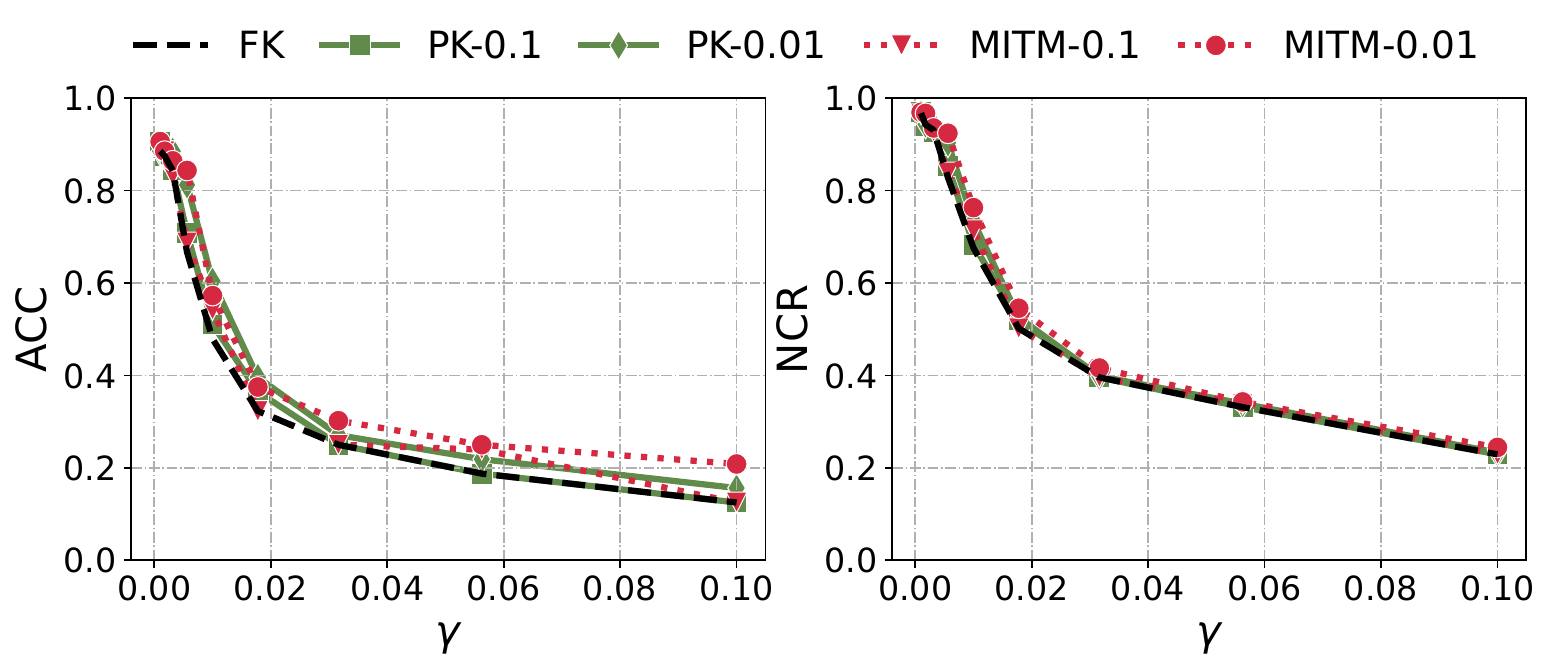}
		\end{minipage}%
	}
     \subfigure[\ccsrev{Kosarak, FIML-IS}]{
		\begin{minipage}[c]{0.40\textwidth}
		\centering
        \includegraphics[width=1\textwidth]{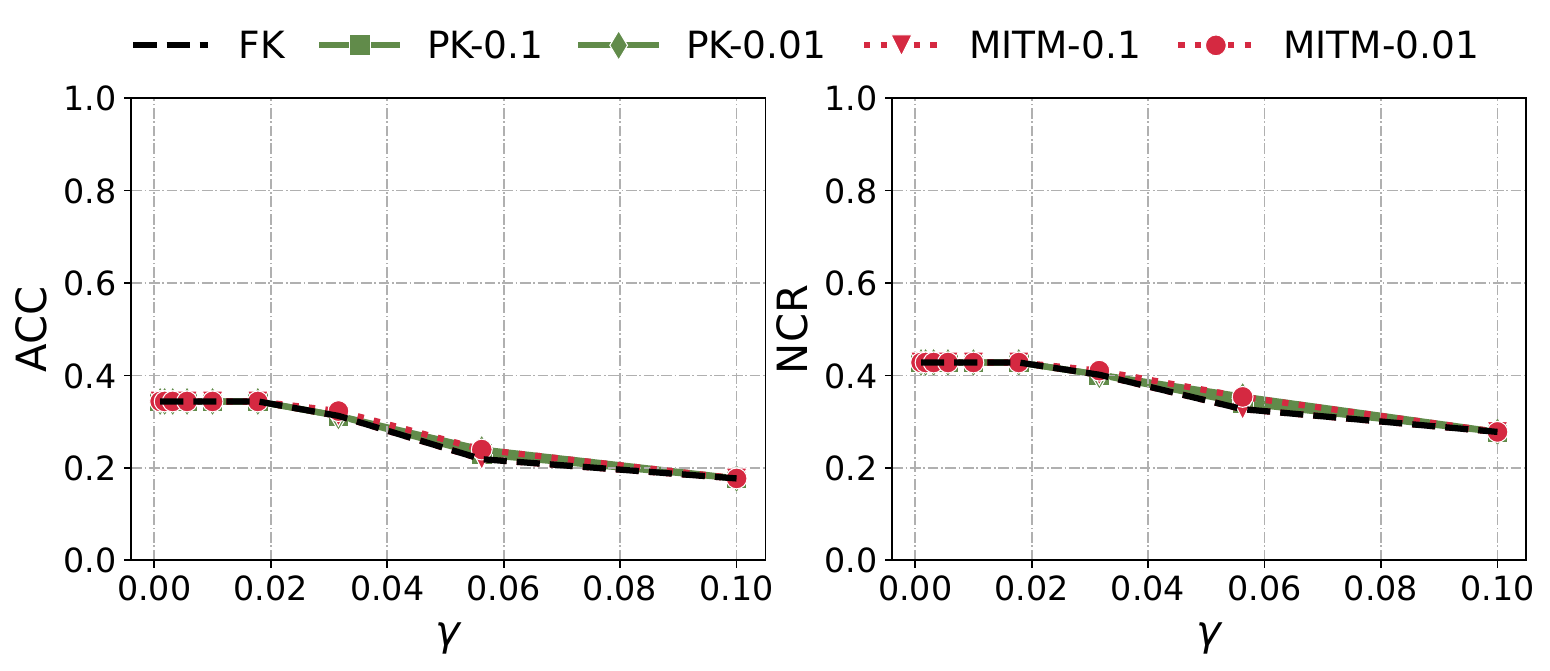}
		\end{minipage}%
	}

      \subfigure[IBM, LDPMiner]{
		\begin{minipage}[c]{0.40\textwidth}
		\centering
        \includegraphics[width=1\textwidth]{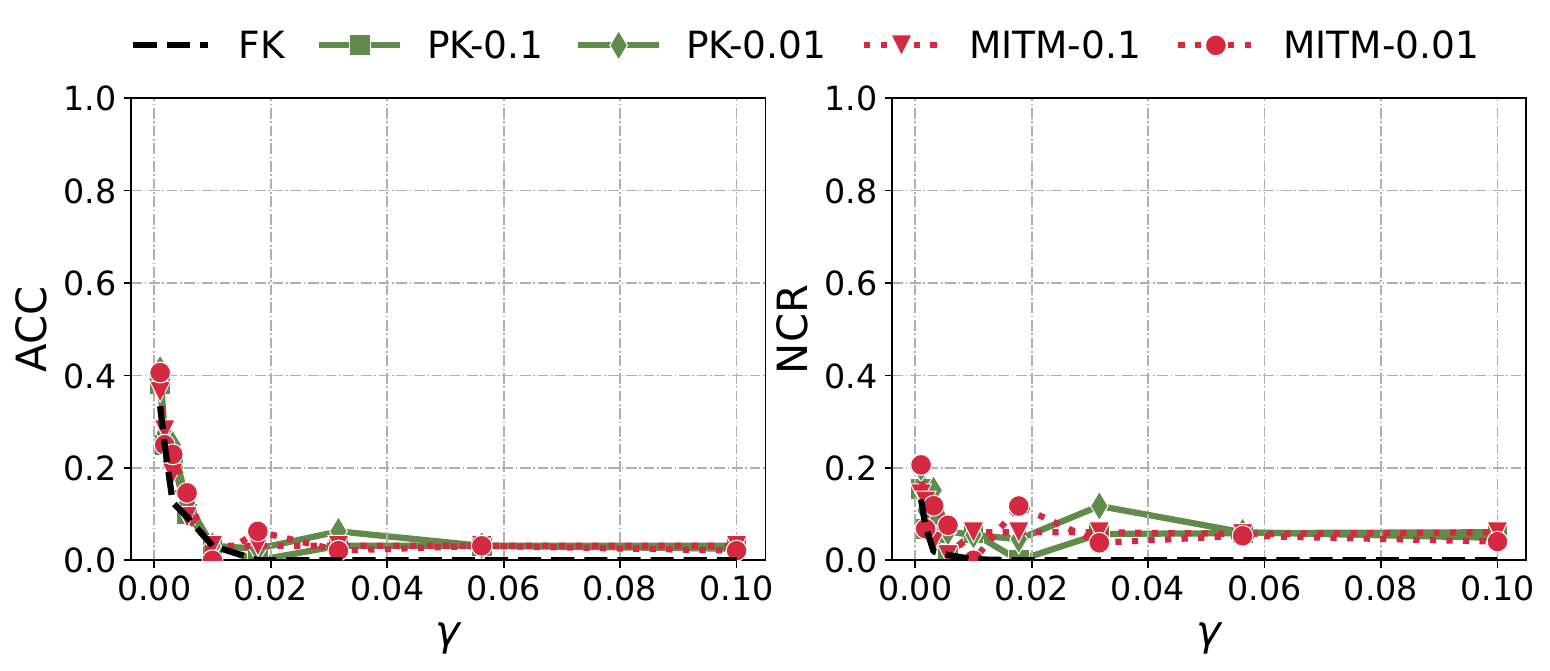}
		\end{minipage}%
	}
    \subfigure[Kosarak, LDPMiner]{
		\begin{minipage}[c]{0.40\textwidth}
		\centering
        \includegraphics[width=1\textwidth]{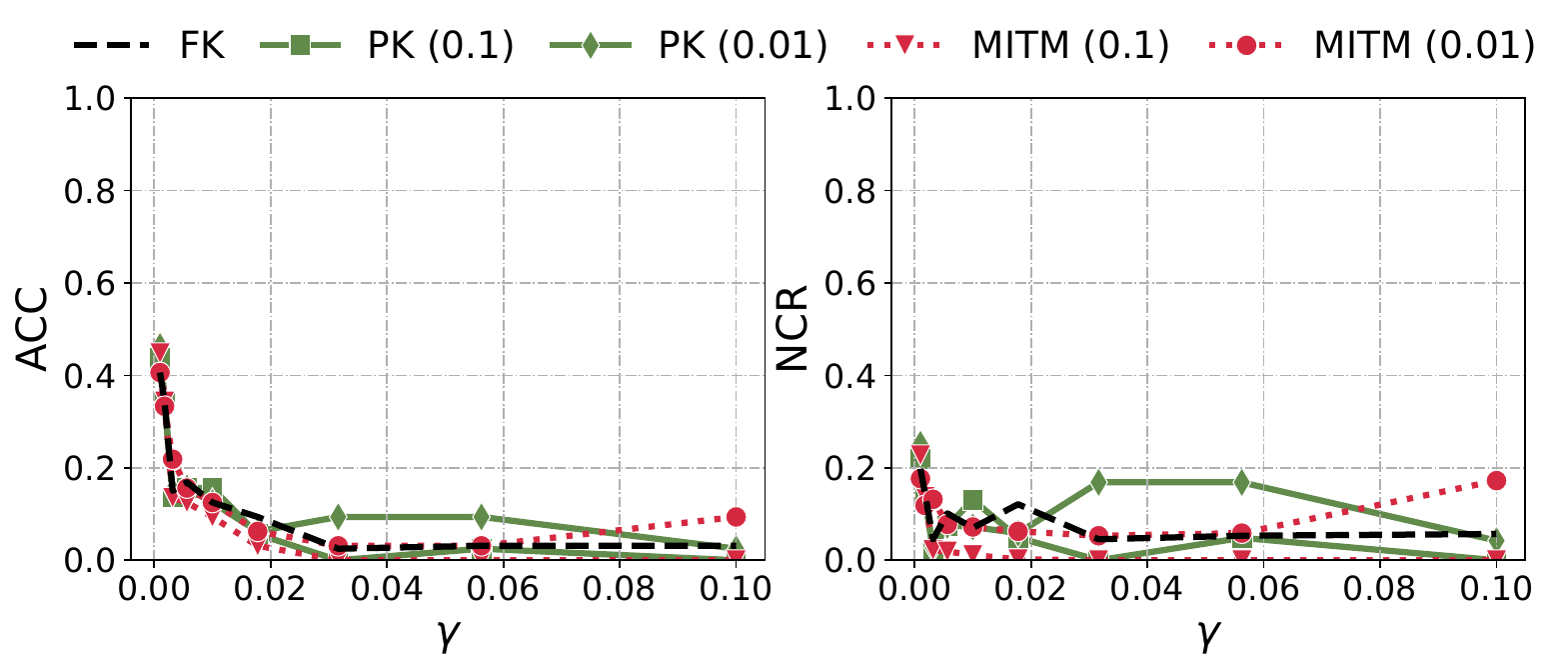}
		\end{minipage}%
	}

      \subfigure[IBM, SVIM]{
		\begin{minipage}[c]{0.40\textwidth}
		\centering
        \includegraphics[width=1\textwidth]{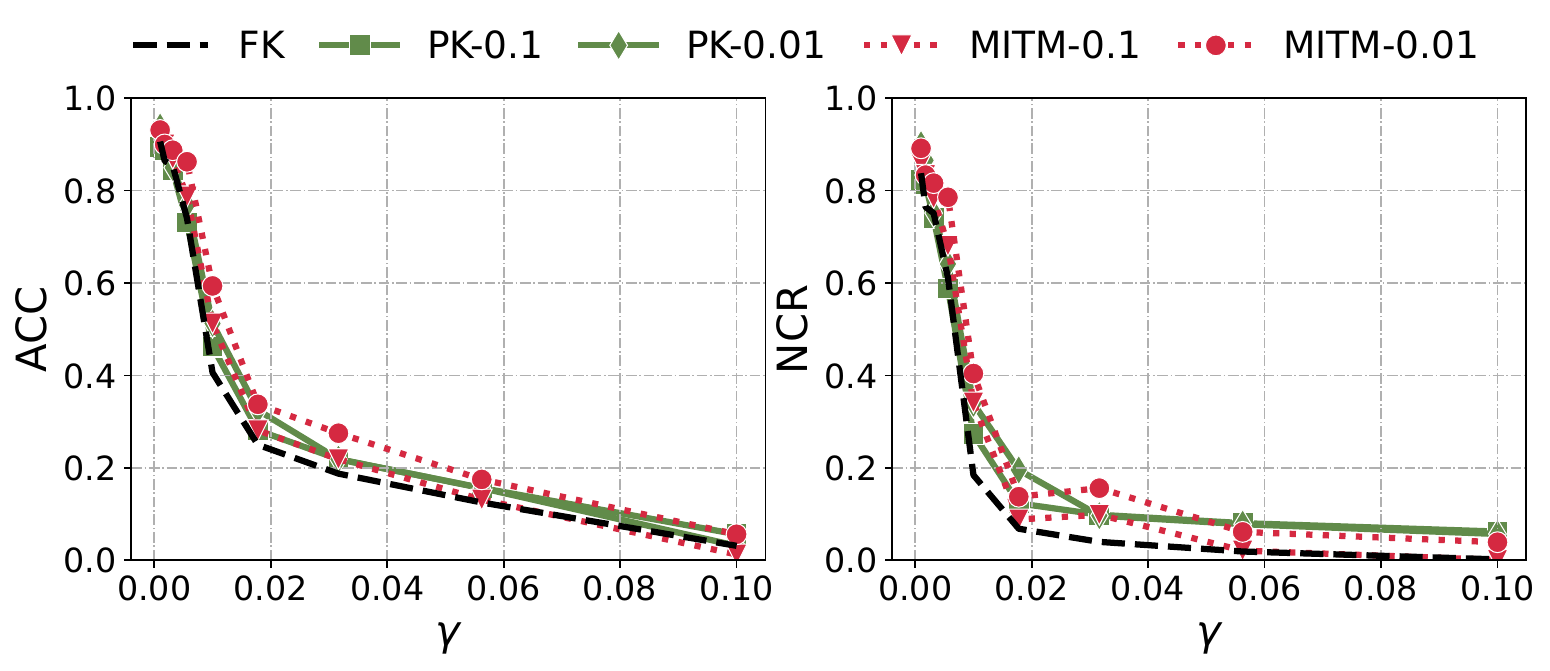}
		\end{minipage}%
	}
     \subfigure[Kosarak, SVIM]{
		\begin{minipage}[c]{0.40\textwidth}
		\centering
        \includegraphics[width=1\textwidth]{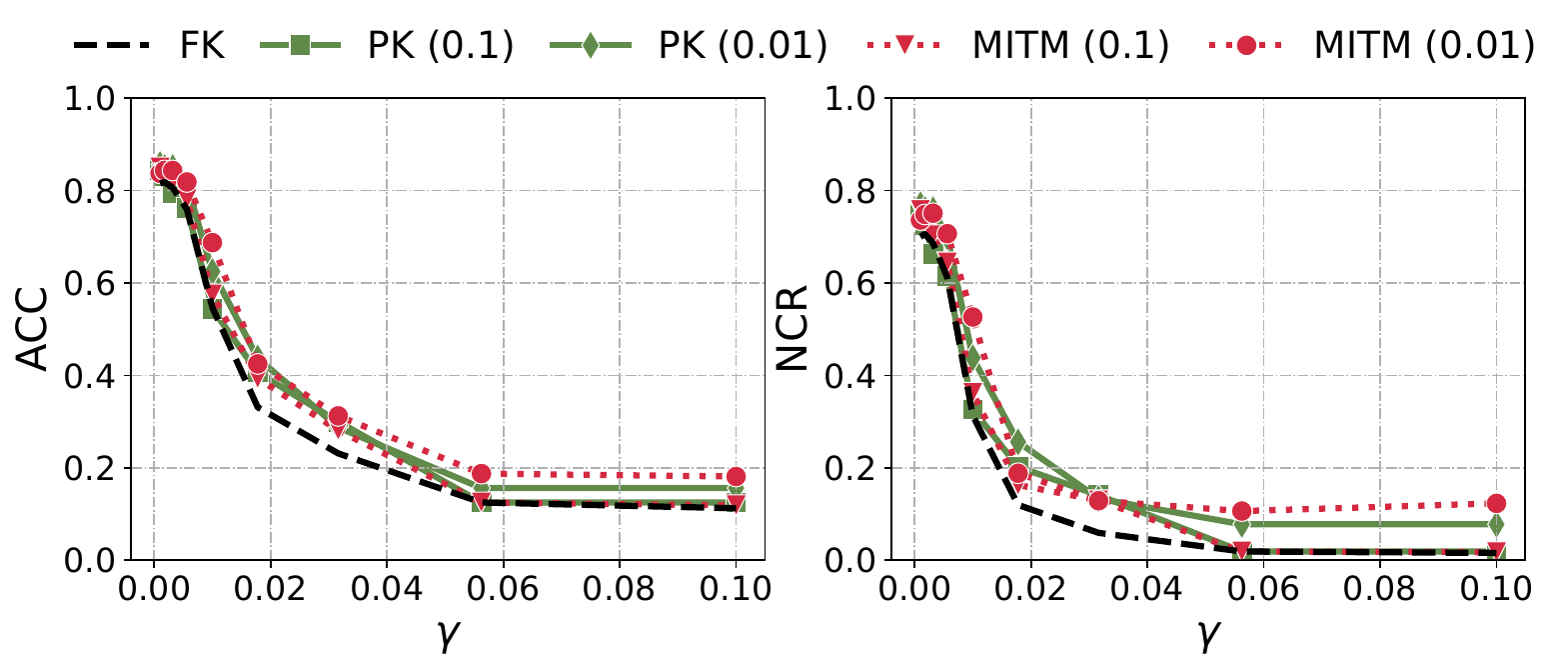}
		\end{minipage}%
	}
 
       \subfigure[\ccsrev{IBM, FIML-I}]{
		\begin{minipage}[c]{0.40\textwidth}
		\centering
        \includegraphics[width=1\textwidth]{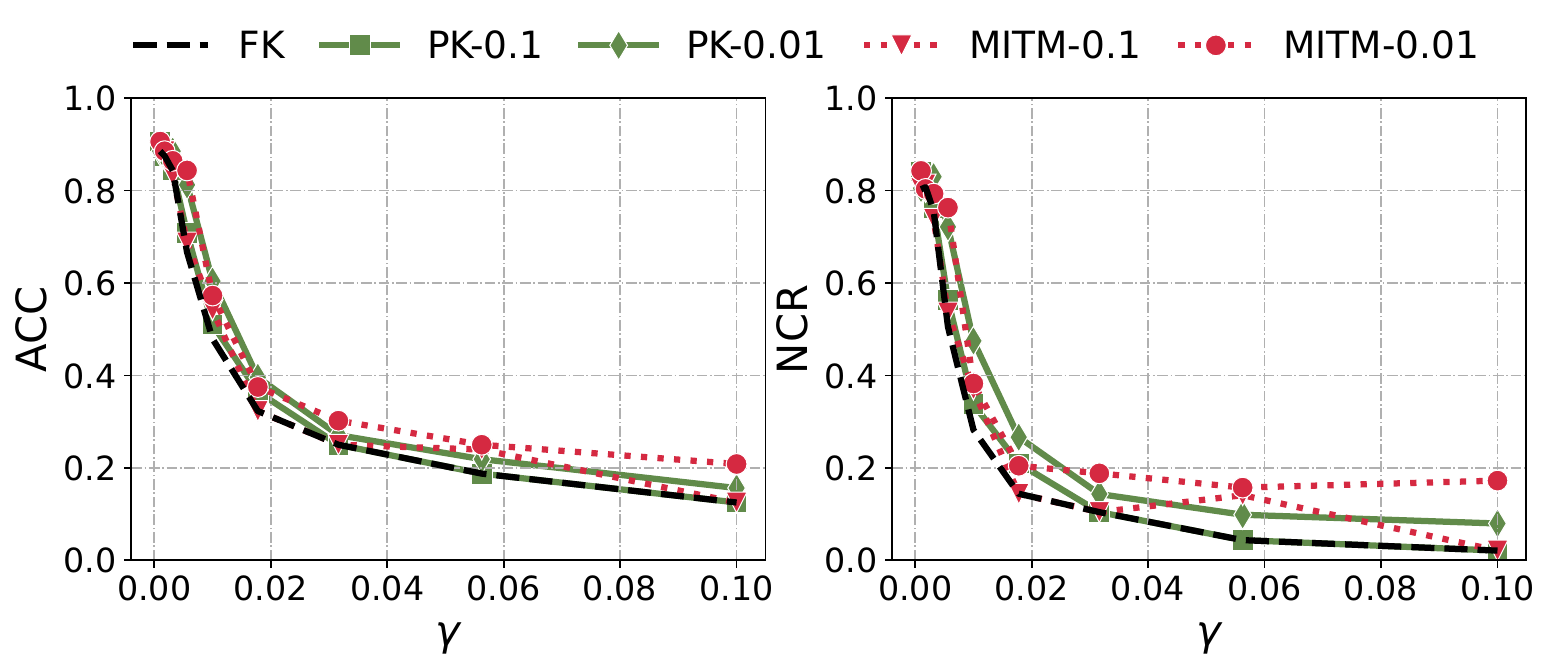}
		\end{minipage}%
	}
     \subfigure[\ccsrev{Kosarak, FIML-I}]{
		\begin{minipage}[c]{0.40\textwidth}
		\centering
        \includegraphics[width=1\textwidth]{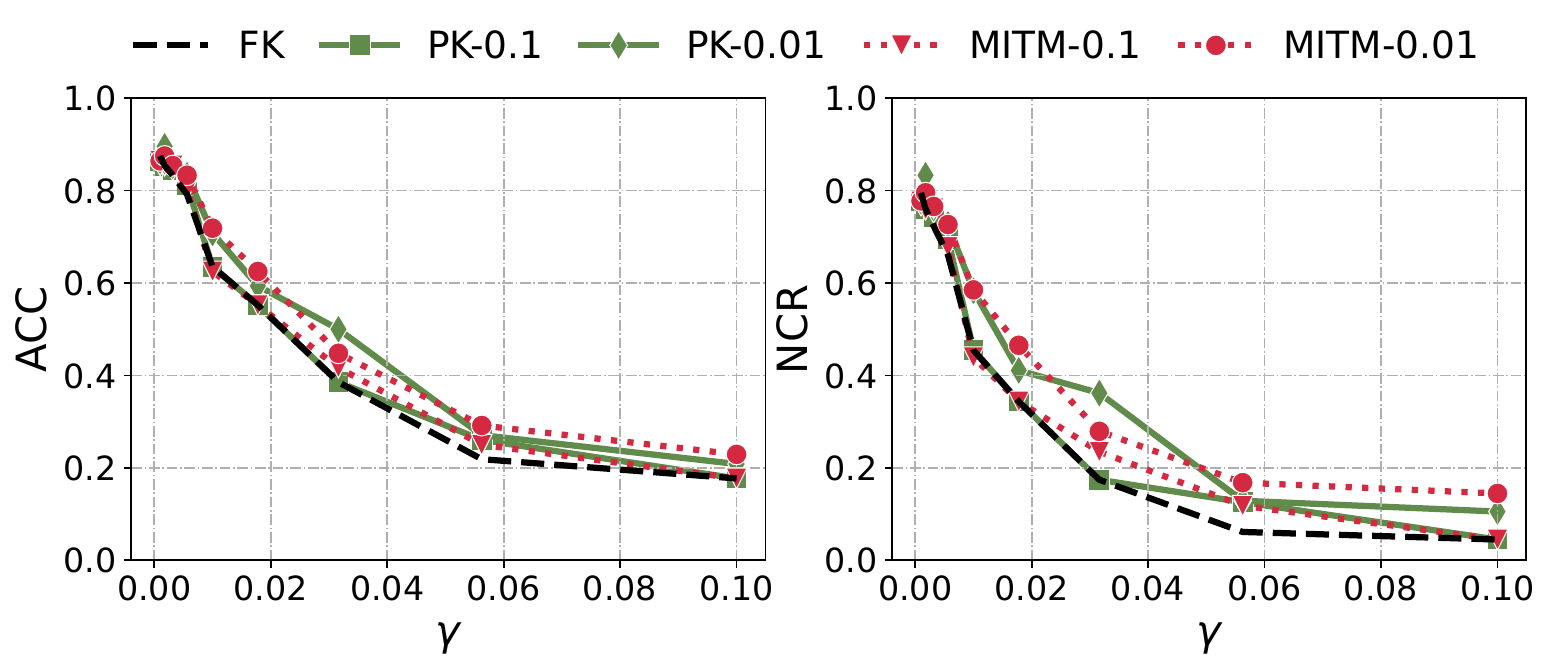}
		\end{minipage}%
	}
 
	\caption{Attacks with limited knowledge on the datasets of IBM Synthesize and Kosarak ($\epsilon = 4.0$).}
 \Description[Limited Knowledge - More (4.0)]{Attacks with limited knowledge on the datasets of IBM Synthesize and Kosarak ($\epsilon = 4.0$).}
	\label{fig:less:kosarak}
\end{figure*}

\paragraphbe{Attacks with limited knowledge} We report experimental results on attacks with limited knowledge for the datasets of IBM Synthesize and Kosarak in \reffig{fig:less:kosarak}. Similar to the experiments for the BMS-POS dataset, we set $\epsilon=4.0$ and $k=32$, and varied $\gamma$ from $0.001$ to $0.1$ to evaluate the attack with partial knowledge with 1\% or 10\% knowledge, as well as the man-in-the-middle attack with 1\% or 10\% intercepted communication. We can observe a similar trend as shown in \reffig{fig:less:item} and \reffig{fig:less:itemset}. The proposed attacks with limited knowledge also demonstrate comparable attack performance, and as more knowledge is provided, the attack performance improves. 

\begin{figure*}
	\centering
	\subfigure[\ccsrev{IBM, $\mathrm{ACC}$}]{
		\begin{minipage}[c]{0.25\textwidth}
		\centering
        \includegraphics[width=1\textwidth]{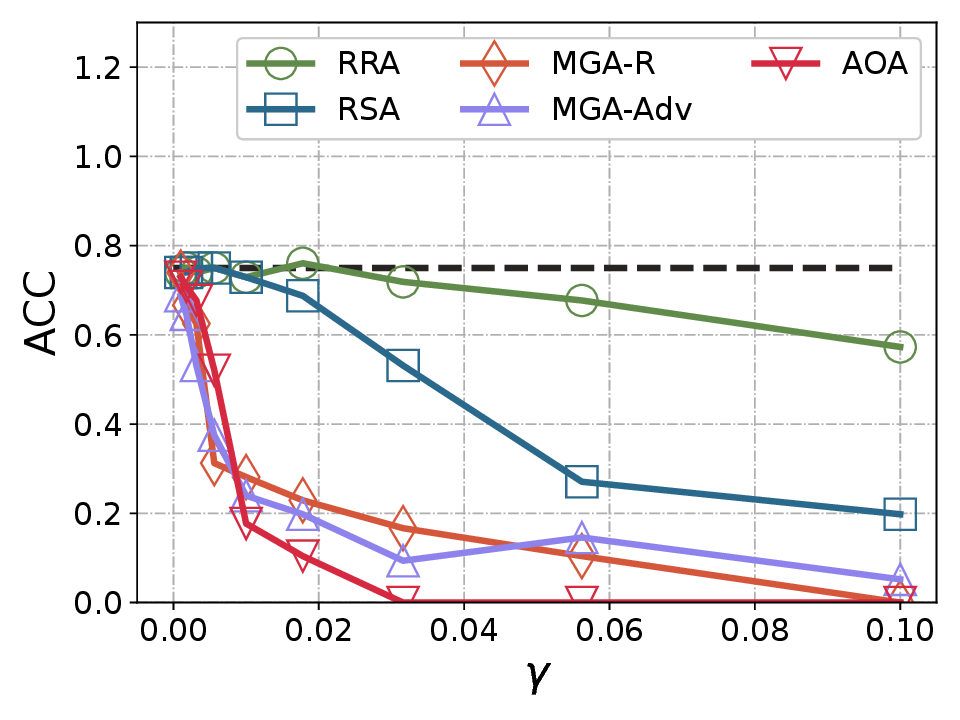}
		\end{minipage}%
	}
      \subfigure[\ccsrev{Kosarak, $\mathrm{ACC}$}]{
		\begin{minipage}[c]{0.25\textwidth}
		\centering
        \includegraphics[width=1\textwidth]{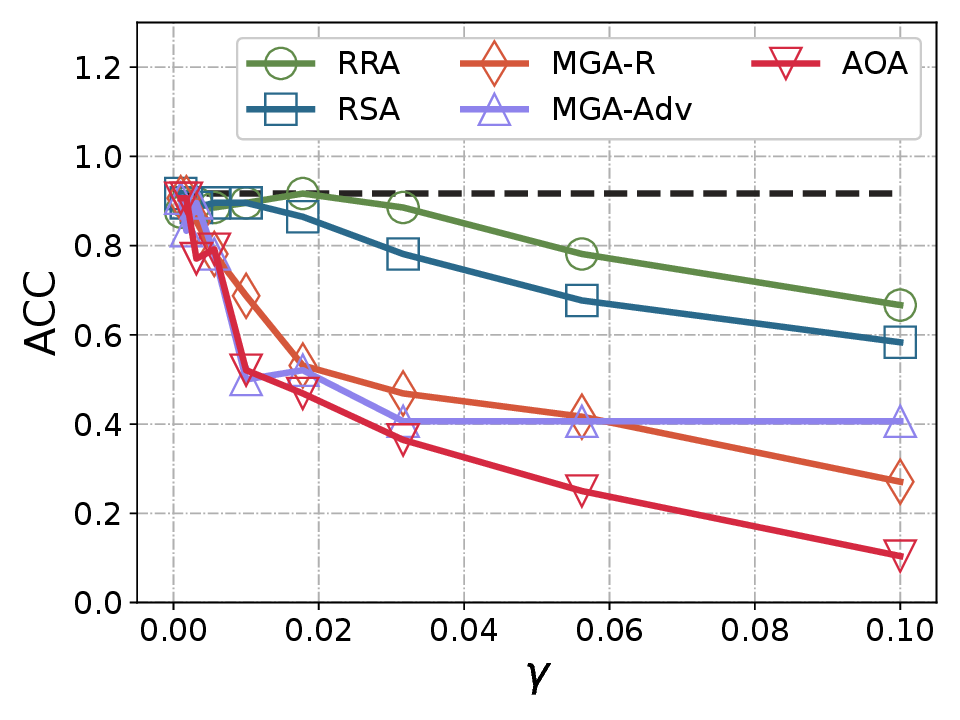}
		\end{minipage}%
	}
     \subfigure[\ccsrev{BMS-POS, $\mathrm{ACC}$}]{
		\begin{minipage}[c]{0.25\textwidth}
		\centering
        \includegraphics[width=1\textwidth]{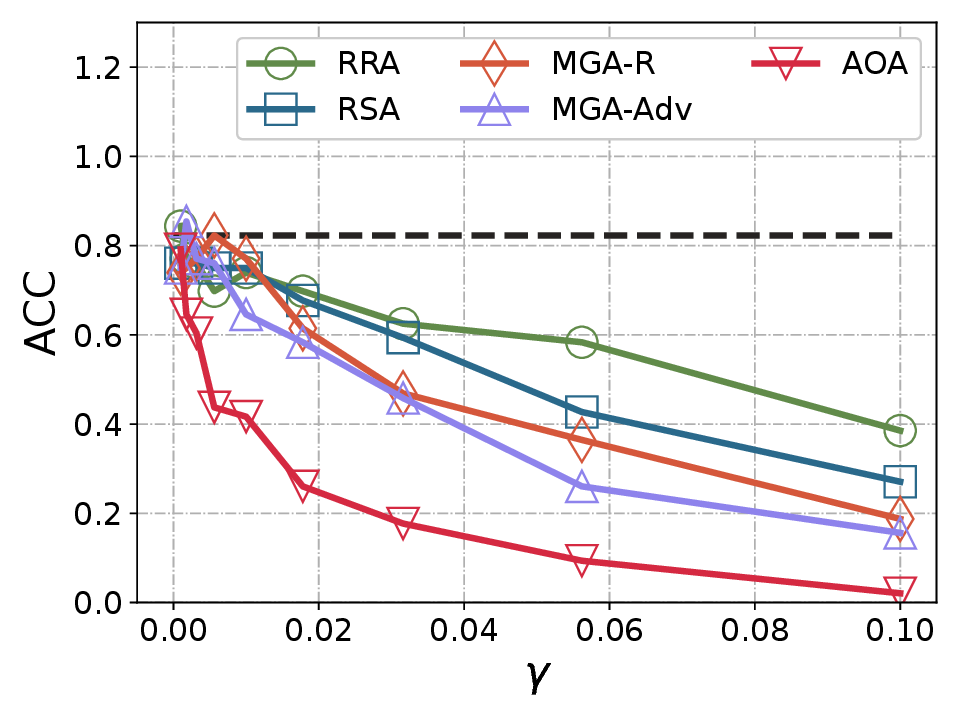}
		\end{minipage}%
	}

	\subfigure[\ccsrev{IBM, $\mathrm{NCR}$}]{
		\begin{minipage}[c]{0.25\textwidth}
		\centering
        \includegraphics[width=1\textwidth]{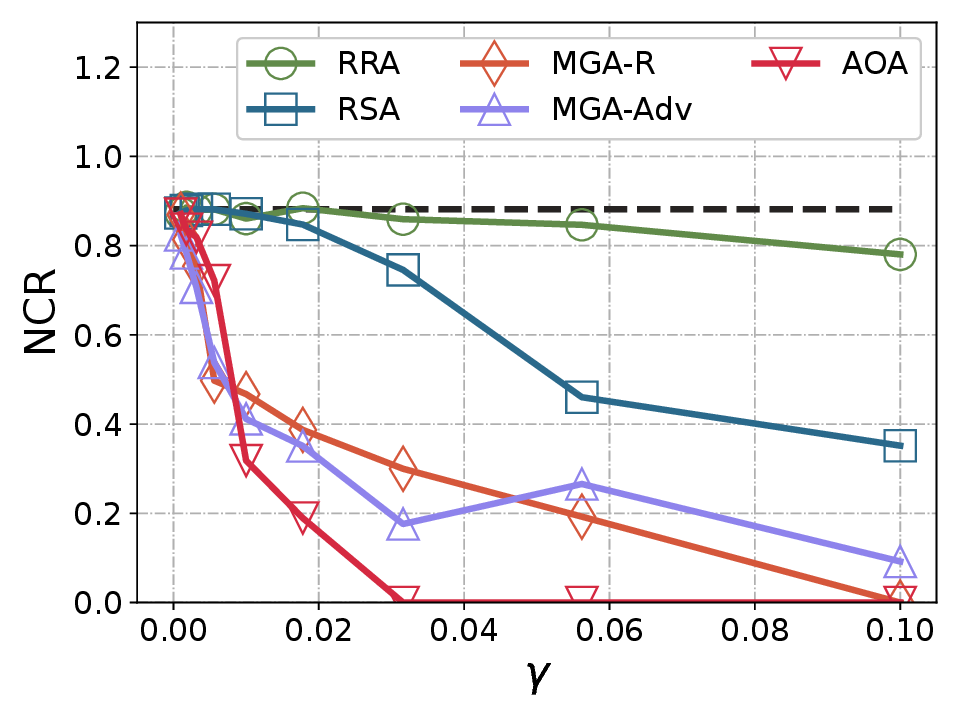}
		\end{minipage}%
    }
    	\subfigure[\ccsrev{Kosarak, $\mathrm{NCR}$}]{
		\begin{minipage}[c]{0.25\textwidth}
		\centering
        \includegraphics[width=1\textwidth]{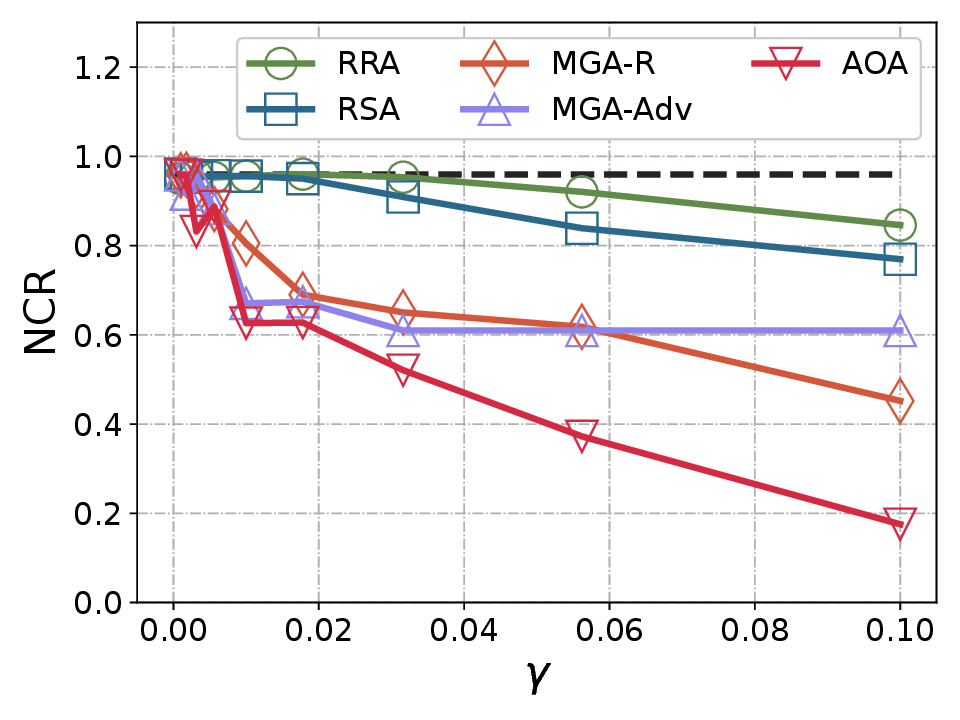}
		\end{minipage}%
    }
	\subfigure[\ccsrev{BMS-POS, $\mathrm{NCR}$}]{
		\begin{minipage}[c]{0.25\textwidth}
		\centering
        \includegraphics[width=1\textwidth]{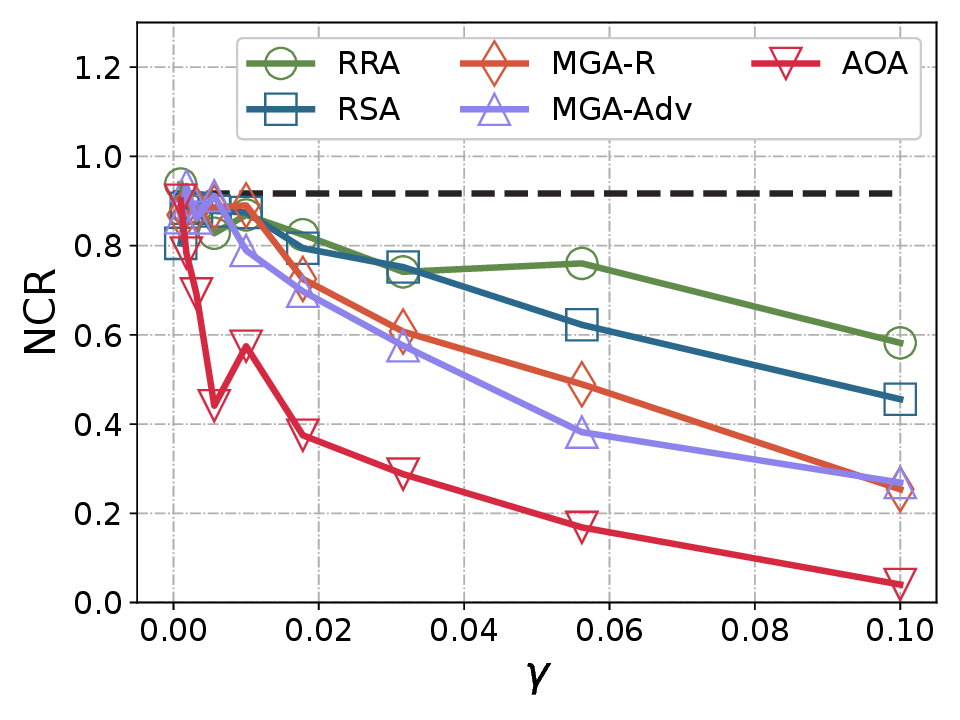}
		\end{minipage}%
    }

    \centering
	\caption{\ccsrev{Attacking SVSM, with $k = 32$, $\epsilon=2.0$ (black dashed line - no attack).}}
	\label{fig:gamma:svsm:eps2}
\end{figure*}

\begin{figure*}
	\centering
	\subfigure[IBM, $\mathrm{ACC}$]{
		\begin{minipage}[c]{0.25\textwidth}
		\centering
        \includegraphics[width=1\textwidth]{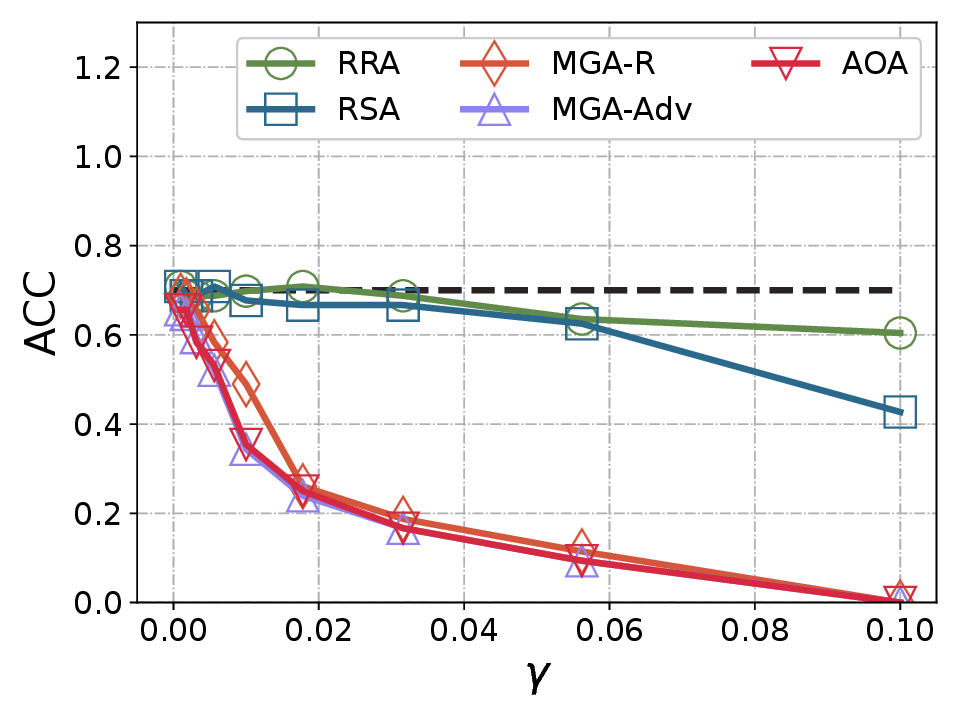}
		\end{minipage}%
	}
     \subfigure[Kosarak, $\mathrm{ACC}$]{
		\begin{minipage}[c]{0.25\textwidth}
		\centering
        \includegraphics[width=1\textwidth]{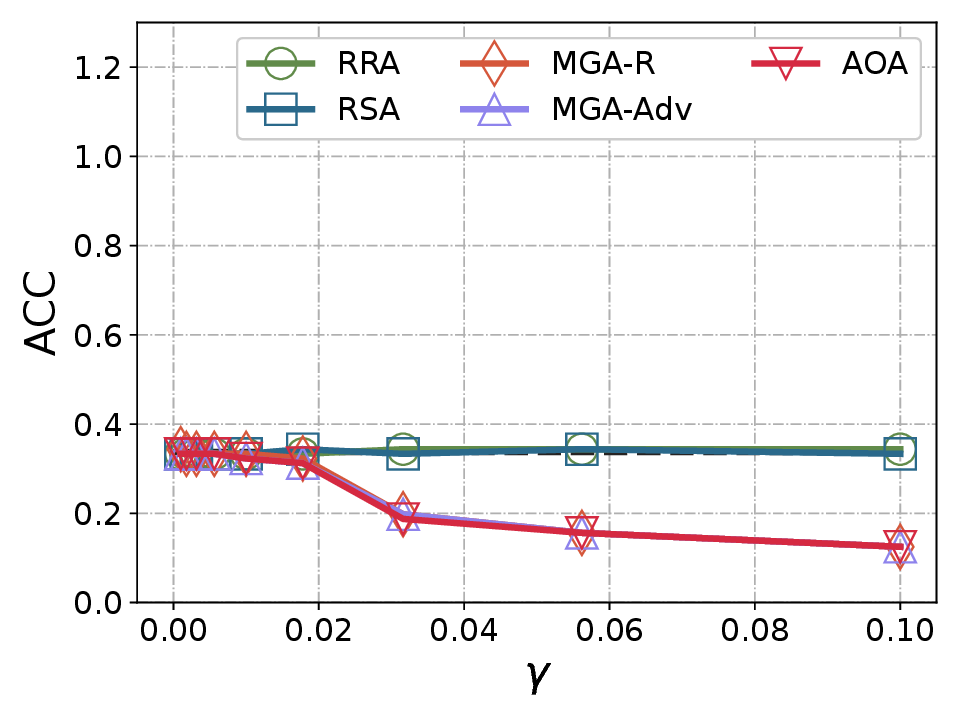}
		\end{minipage}%
	}
     \subfigure[BMS-POS, $\mathrm{ACC}$]{
		\begin{minipage}[c]{0.25\textwidth}
		\centering
        \includegraphics[width=1\textwidth]{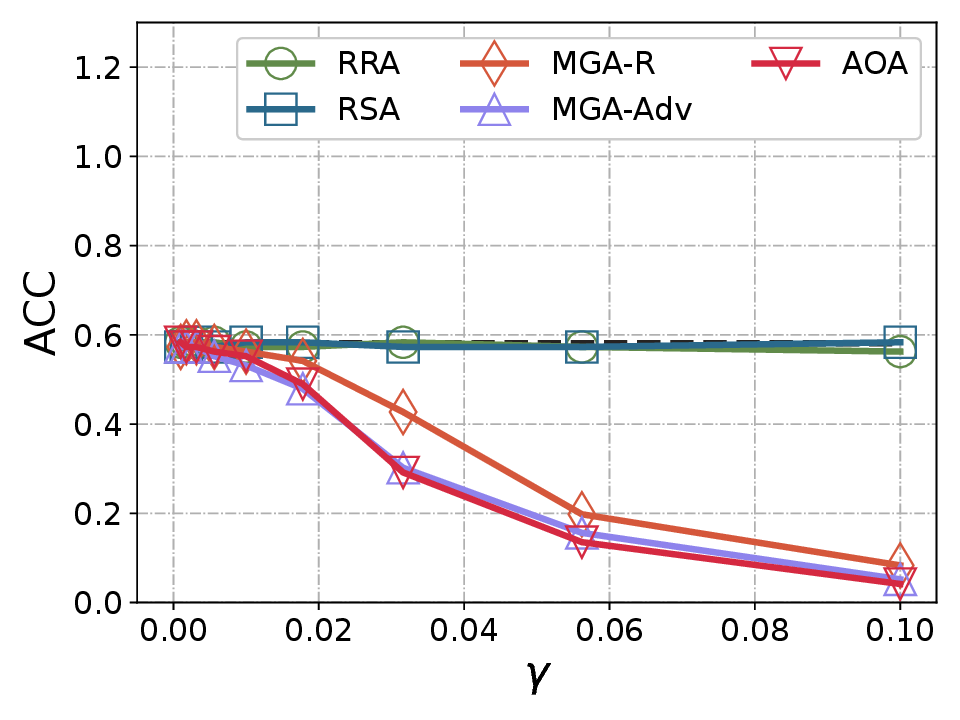}
		\end{minipage}%
	}
 
	\subfigure[IBM, $\mathrm{NCR}$]{
		\begin{minipage}[c]{0.25\textwidth}
		\centering
        \includegraphics[width=1\textwidth]{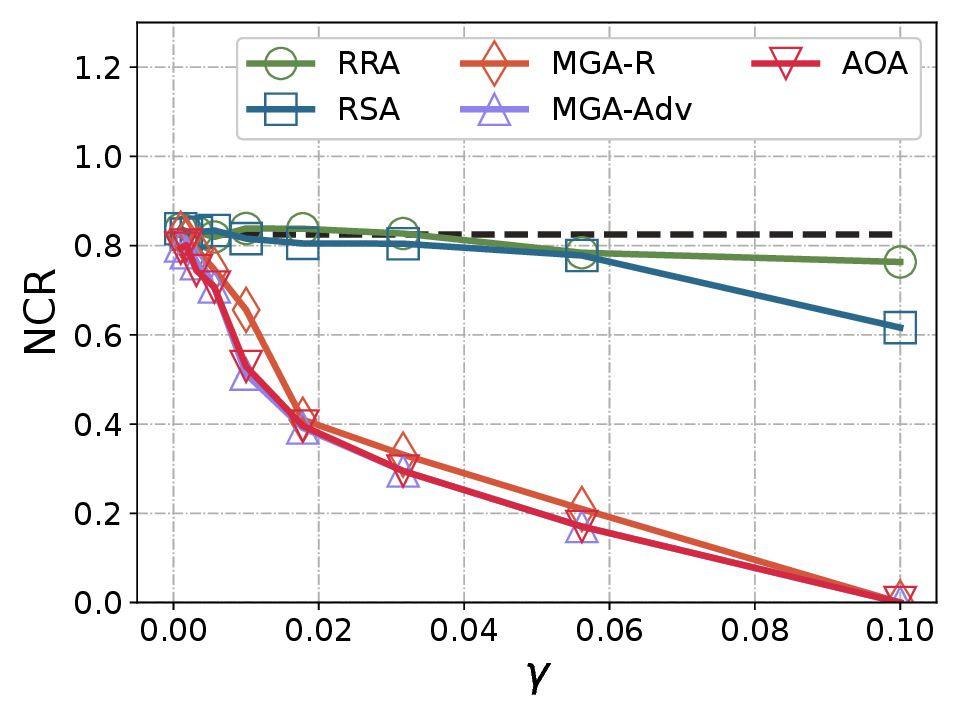}
		\end{minipage}%
    }
	\subfigure[Kosarak, $\mathrm{NCR}$]{
		\begin{minipage}[c]{0.25\textwidth}
		\centering
        \includegraphics[width=1\textwidth]{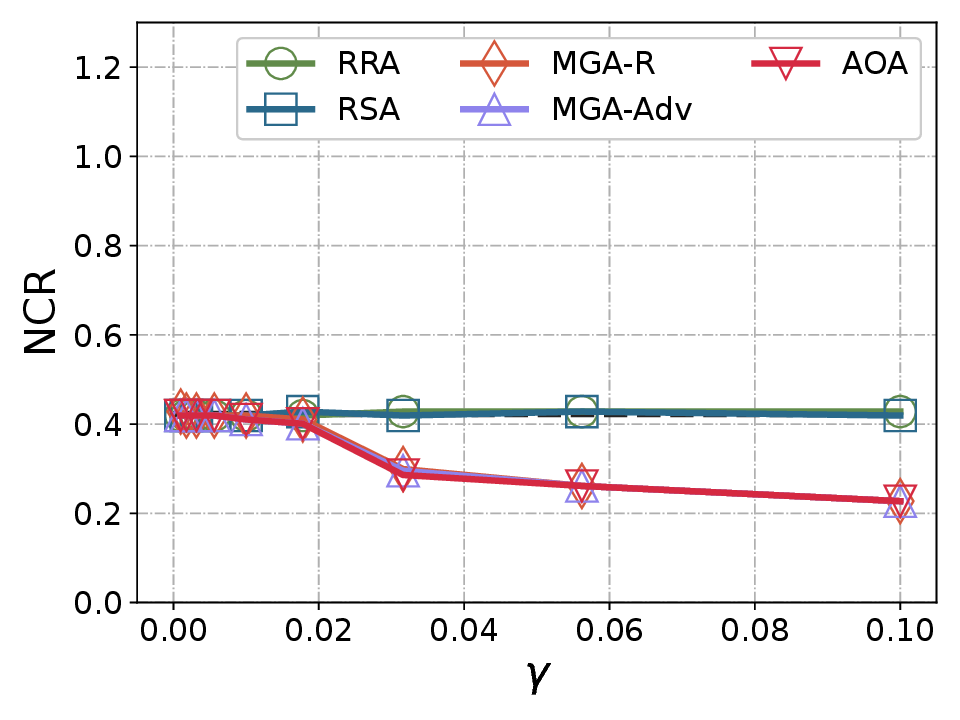}
		\end{minipage}%
    }
	\subfigure[BMS-POS, $\mathrm{NCR}$]{
		\begin{minipage}[c]{0.25\textwidth}
		\centering
        \includegraphics[width=1\textwidth]{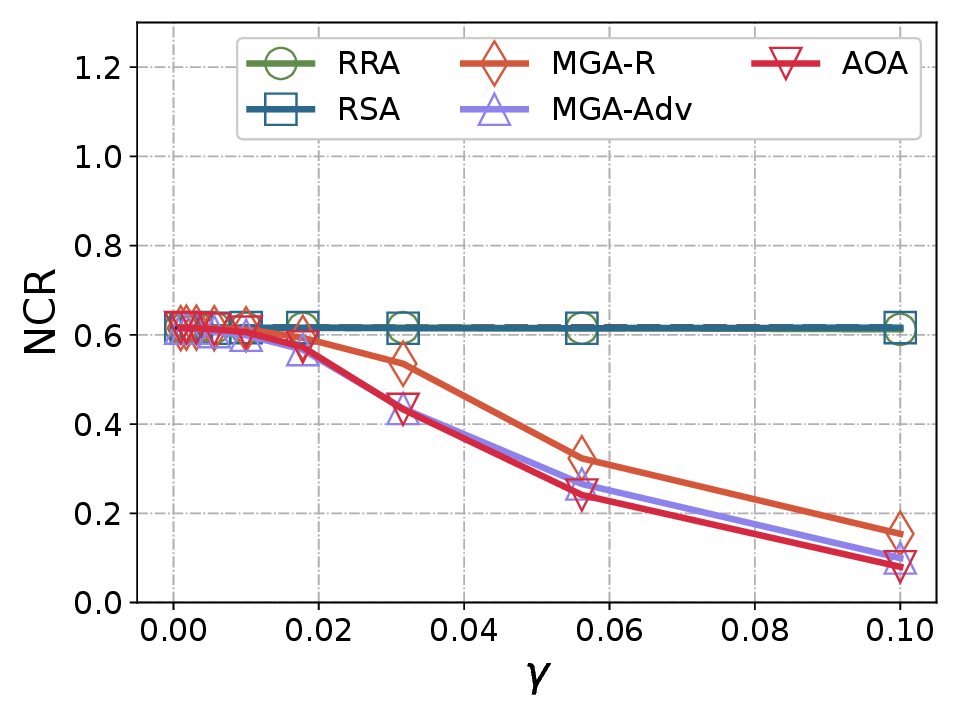}
		\end{minipage}%
    }
    \centering
	\caption{\ccsrev{Attacking FIML-IS, with $k = 32$, $\epsilon=2.0$ (black dashed line - no attack).}}
	\label{fig:gamma:fimlitemset:eps2}
\end{figure*}

\begin{figure*}
	\centering
	\subfigure[IBM, $\mathrm{ACC}$]{
		\begin{minipage}[c]{0.25\textwidth}
		\centering
        \includegraphics[width=1\textwidth]{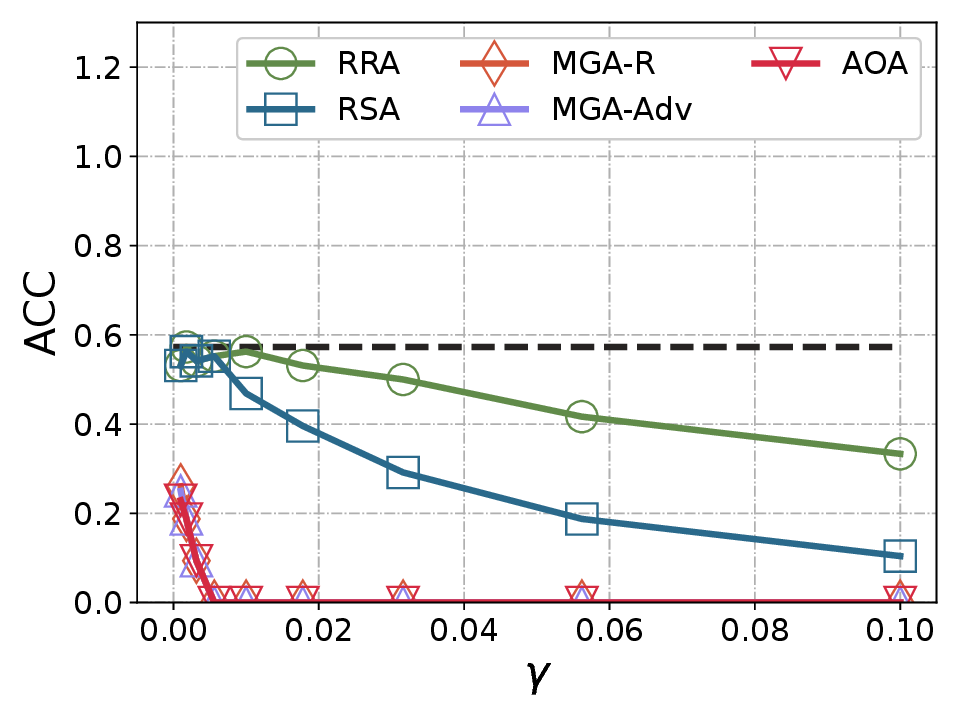}
		\end{minipage}%
	}
     \subfigure[Kosarak, $\mathrm{ACC}$]{
		\begin{minipage}[c]{0.25\textwidth}
		\centering
        \includegraphics[width=1\textwidth]{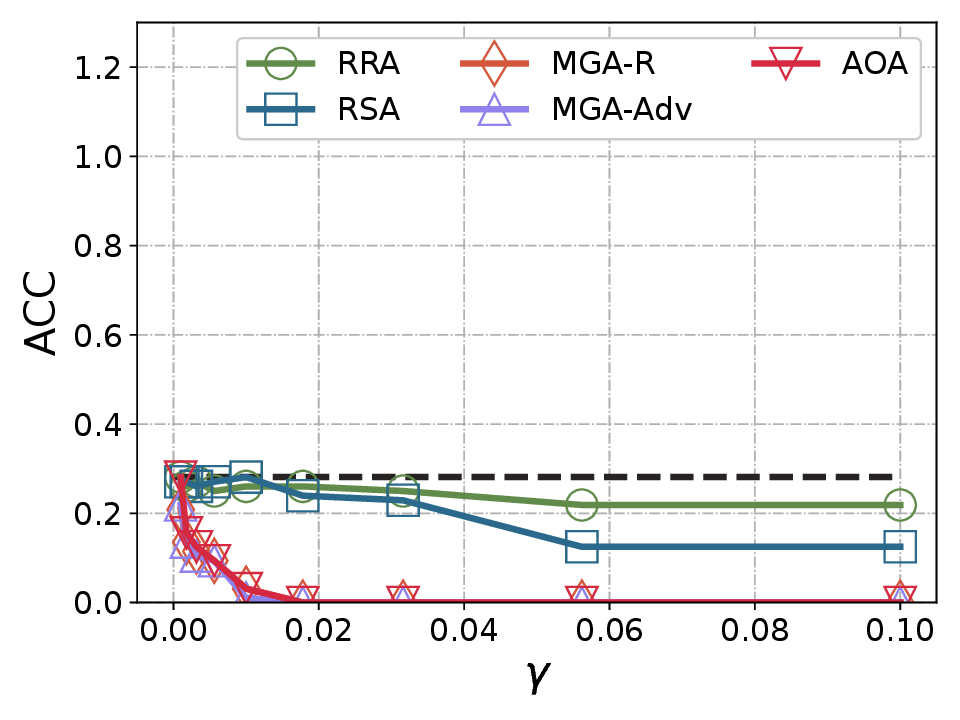}
		\end{minipage}%
	}
     \subfigure[BMS-POS, $\mathrm{ACC}$]{
		\begin{minipage}[c]{0.25\textwidth}
		\centering
        \includegraphics[width=1\textwidth]{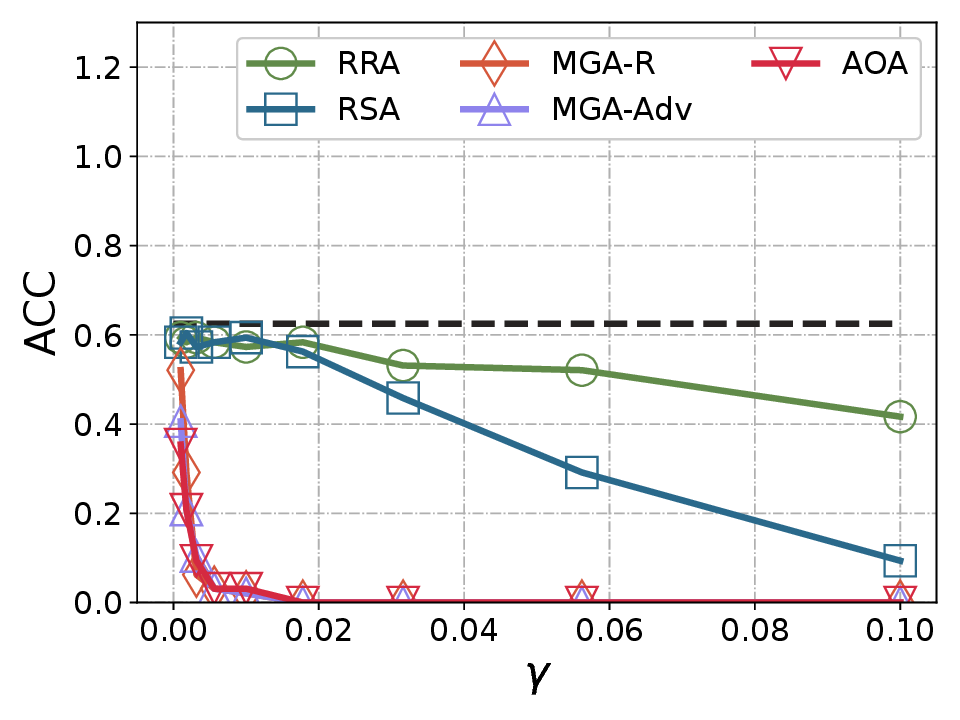}
		\end{minipage}%
	}
 
	\subfigure[IBM, $\mathrm{NCR}$]{
		\begin{minipage}[c]{0.25\textwidth}
		\centering
        \includegraphics[width=1\textwidth]{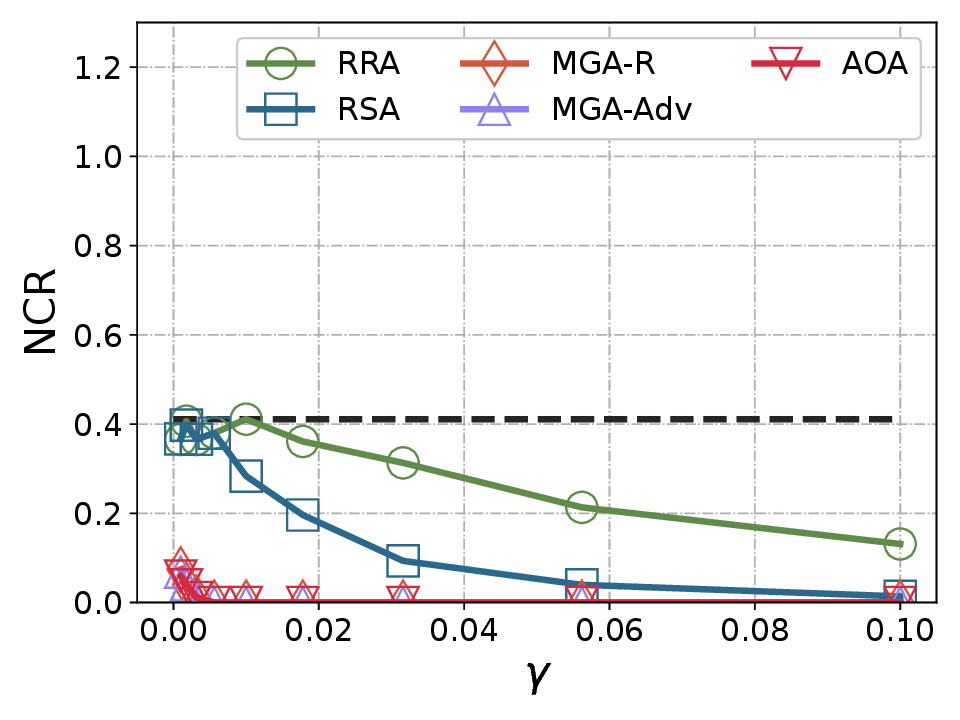}
		\end{minipage}%
    }
	\subfigure[Kosarak, $\mathrm{NCR}$]{
		\begin{minipage}[c]{0.25\textwidth}
		\centering
        \includegraphics[width=1\textwidth]{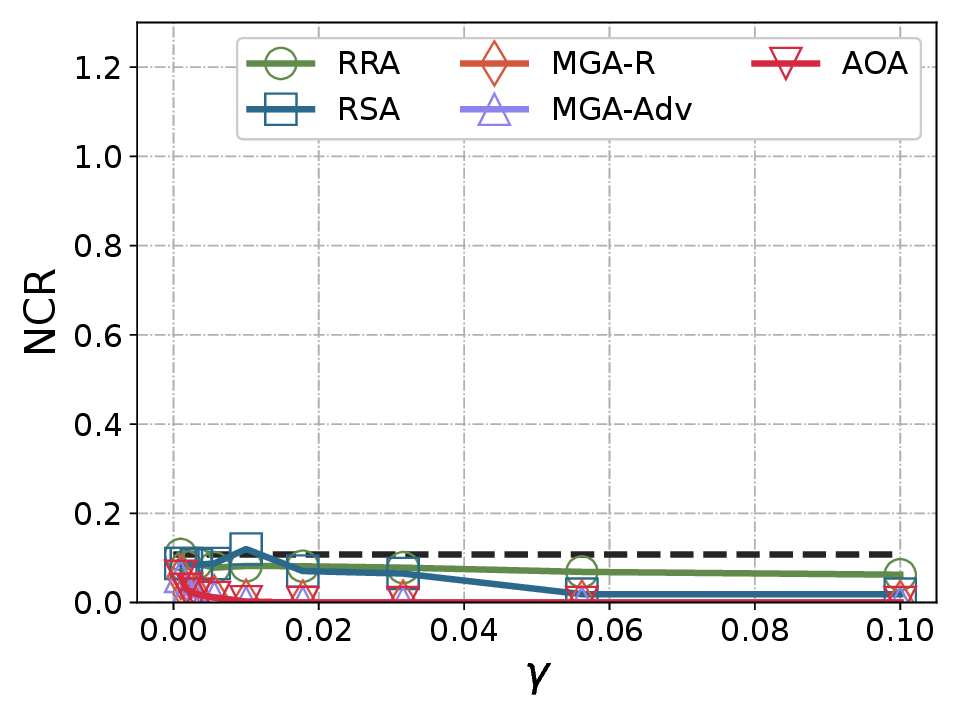}
		\end{minipage}%
    }
	\subfigure[BMS-POS, $\mathrm{NCR}$]{
		\begin{minipage}[c]{0.25\textwidth}
		\centering
        \includegraphics[width=1\textwidth]{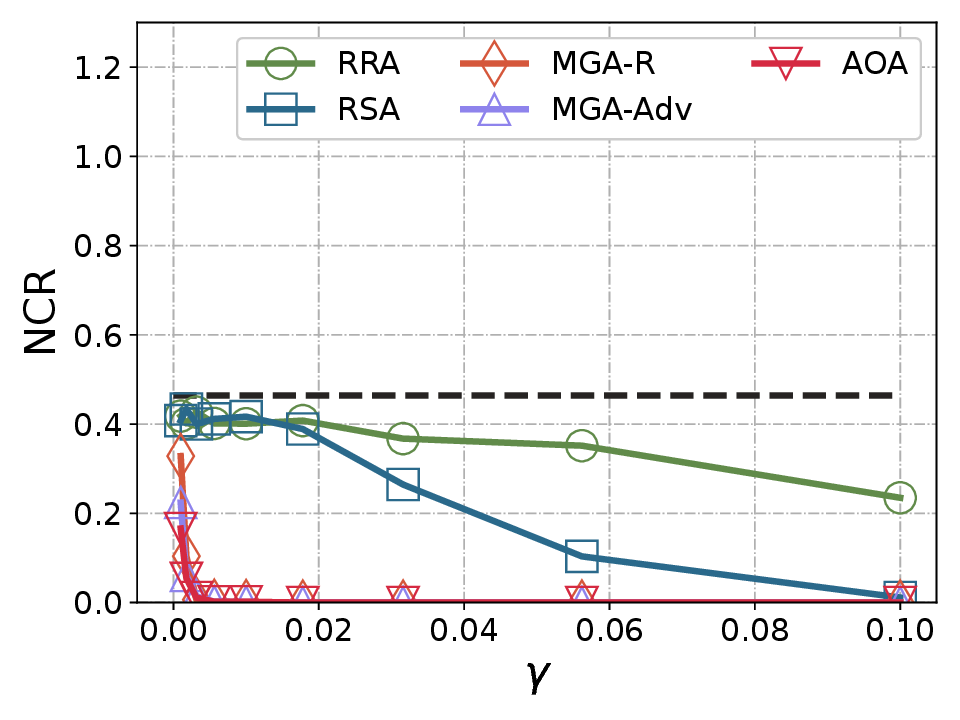}
		\end{minipage}%
    }
    \centering
	\caption{\ccsrev{Attacking LDPMiner, with $k = 32$, $\epsilon=2.0$ (black dashed line - no attack).}}
	\label{fig:gamma:ldpminer:eps2}
\end{figure*}

\begin{figure*}
	\centering
	\subfigure[IBM, $\mathrm{ACC}$]{
		\begin{minipage}[c]{0.25\textwidth}
		\centering
        \includegraphics[width=1\textwidth]{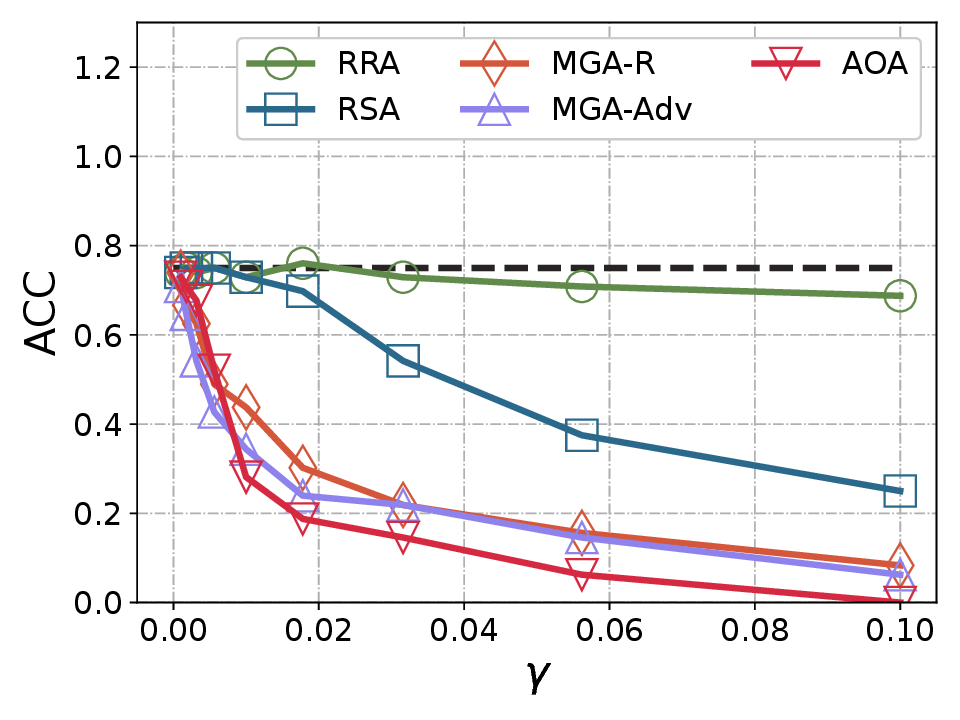}
		\end{minipage}%
	}
     \subfigure[Kosarak, $\mathrm{ACC}$]{
		\begin{minipage}[c]{0.25\textwidth}
		\centering
        \includegraphics[width=1\textwidth]{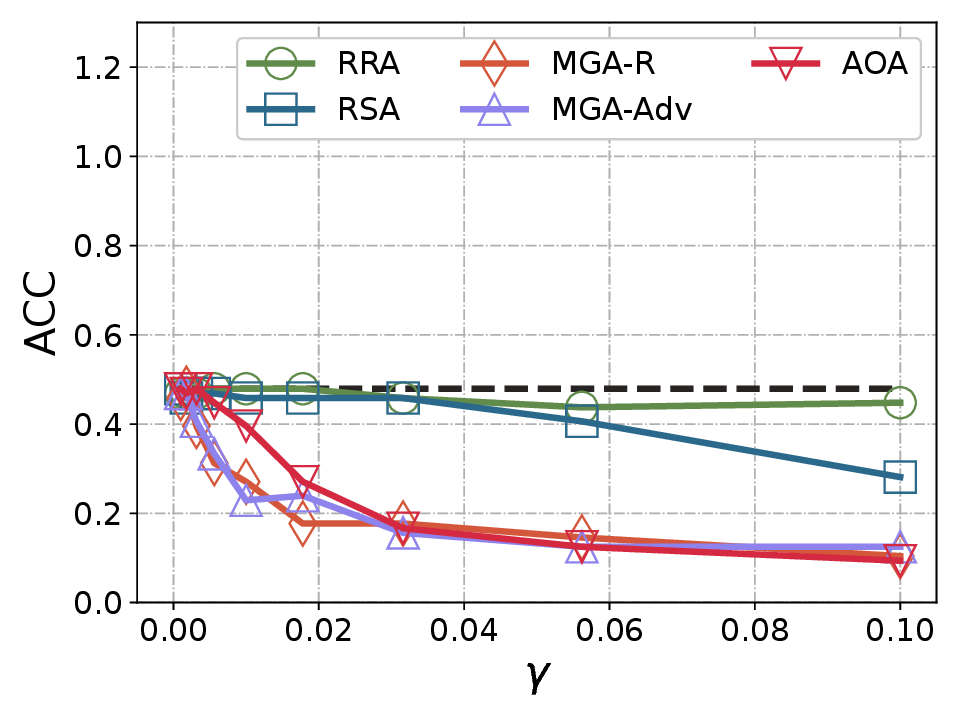}
		\end{minipage}%
	}
     \subfigure[BMS-POS, $\mathrm{ACC}$]{
		\begin{minipage}[c]{0.25\textwidth}
		\centering
        \includegraphics[width=1\textwidth]{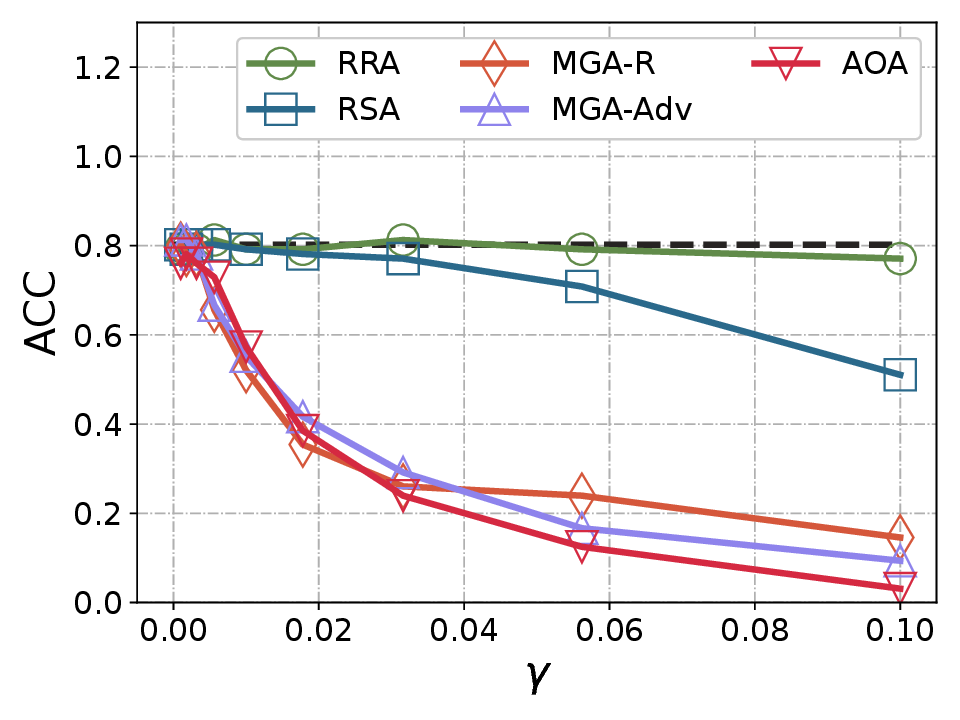}
		\end{minipage}%
	}
 
	\subfigure[IBM, $\mathrm{NCR}$]{
		\begin{minipage}[c]{0.25\textwidth}
		\centering
        \includegraphics[width=1\textwidth]{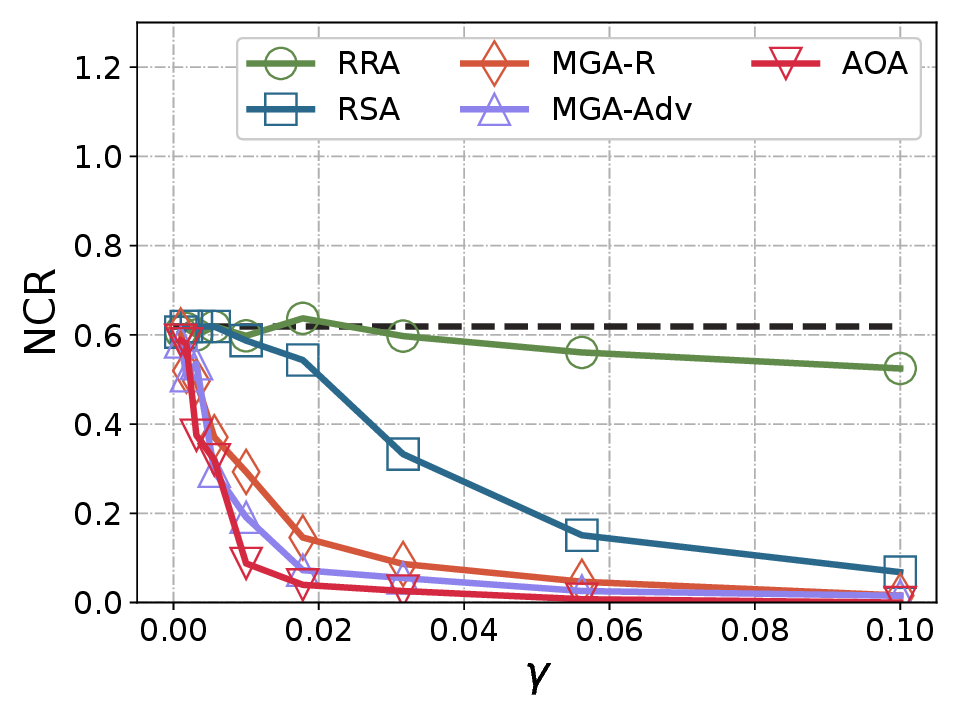}
		\end{minipage}%
    }
	\subfigure[Kosarak, $\mathrm{NCR}$]{
		\begin{minipage}[c]{0.25\textwidth}
		\centering
        \includegraphics[width=1\textwidth]{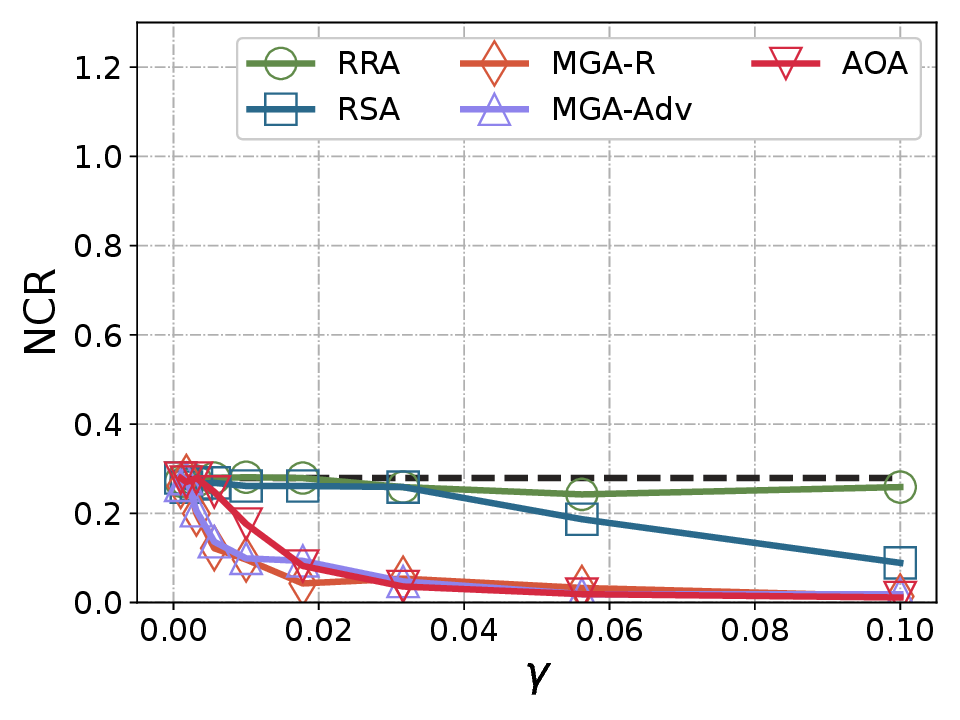}
		\end{minipage}%
    }
	\subfigure[BMS-POS, $\mathrm{NCR}$]{
		\begin{minipage}[c]{0.25\textwidth}
		\centering
        \includegraphics[width=1\textwidth]{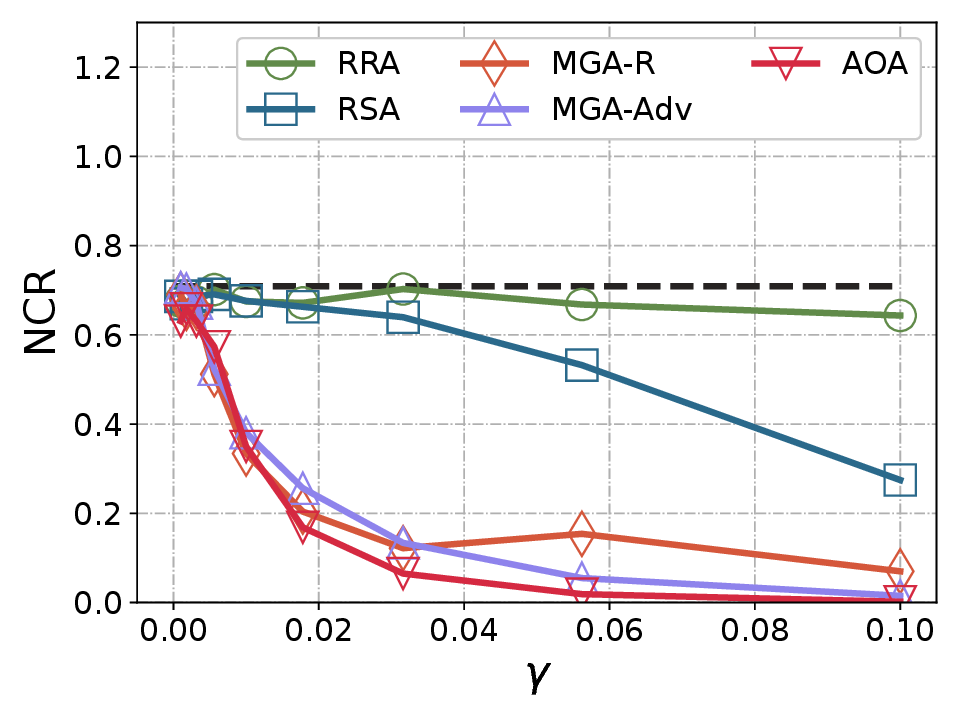}
		\end{minipage}%
    }
    \centering
	\caption{\ccsrev{Attacking SVIM, with $k = 32$, $\epsilon=2.0$ (black dashed line - no attack).}}
	\label{fig:gamma:svim:eps2}
\end{figure*}

\begin{figure*}
	\centering
	\subfigure[IBM, $\mathrm{ACC}$]{
		\begin{minipage}[c]{0.25\textwidth}
		\centering
        \includegraphics[width=1\textwidth]{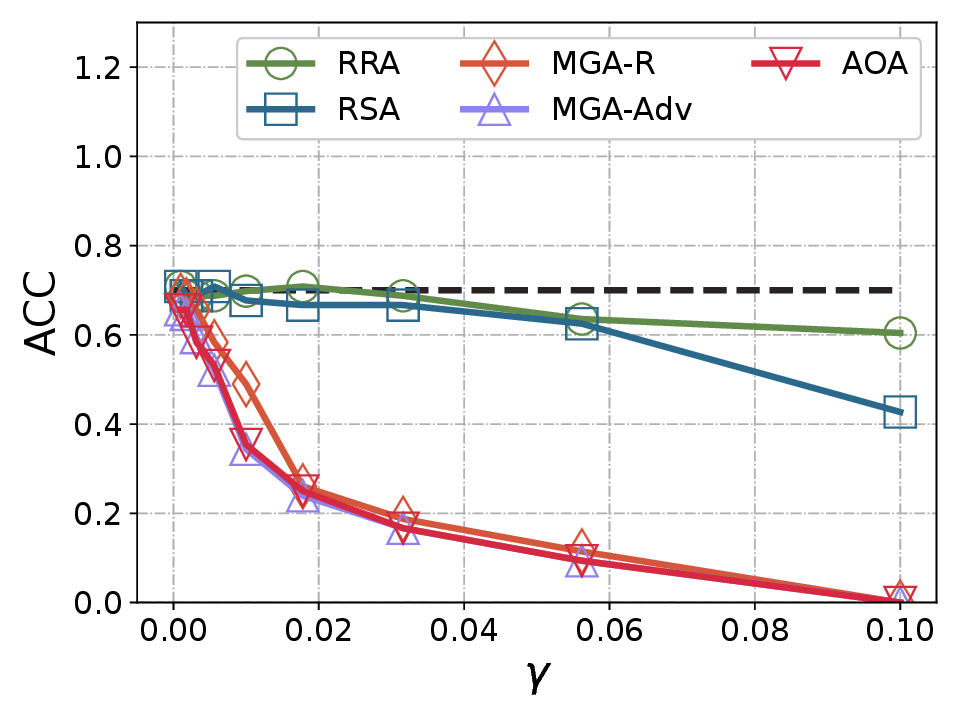}
		\end{minipage}%
	}
     \subfigure[Kosarak, $\mathrm{ACC}$]{
		\begin{minipage}[c]{0.25\textwidth}
		\centering
        \includegraphics[width=1\textwidth]{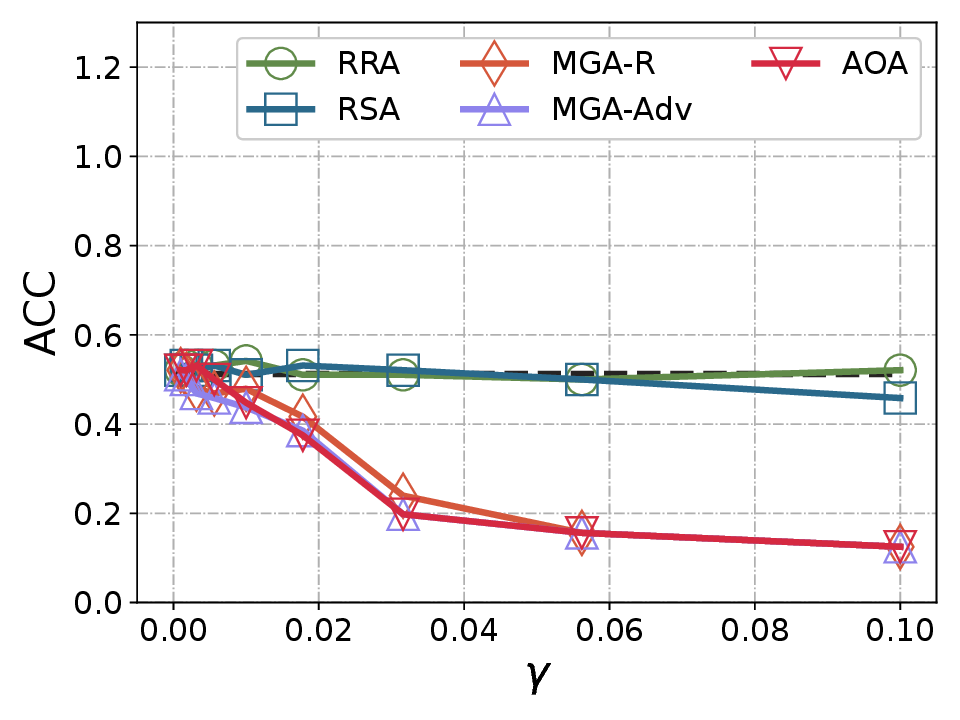}
		\end{minipage}%
	}
     \subfigure[BMS-POS, $\mathrm{ACC}$]{
		\begin{minipage}[c]{0.25\textwidth}
		\centering
        \includegraphics[width=1\textwidth]{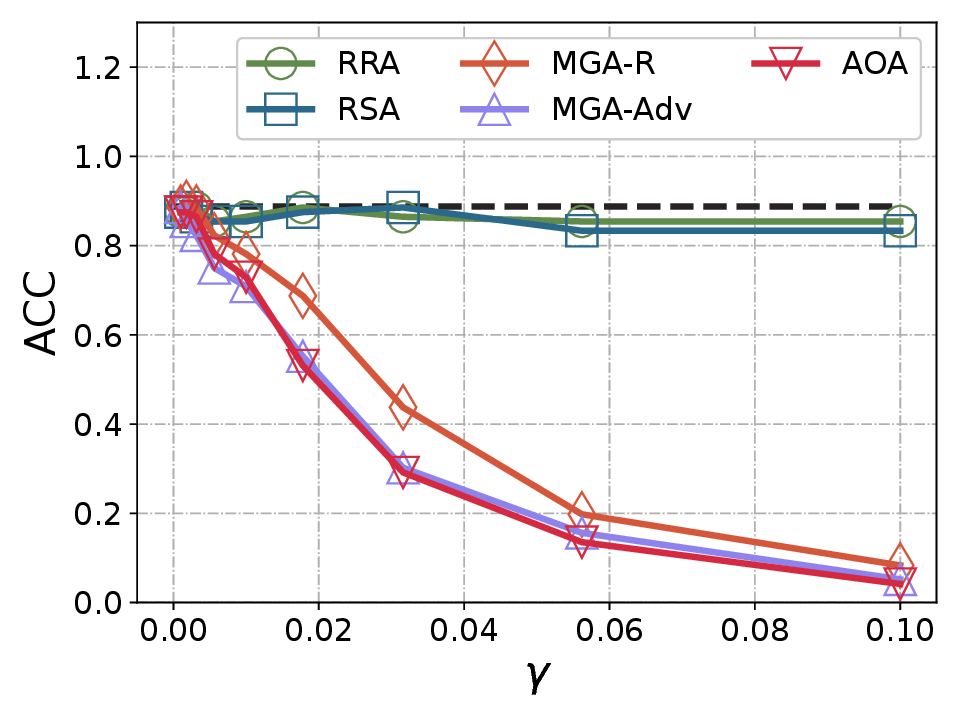}
		\end{minipage}%
	}
 
	\subfigure[IBM, $\mathrm{NCR}$]{
		\begin{minipage}[c]{0.25\textwidth}
		\centering
        \includegraphics[width=1\textwidth]{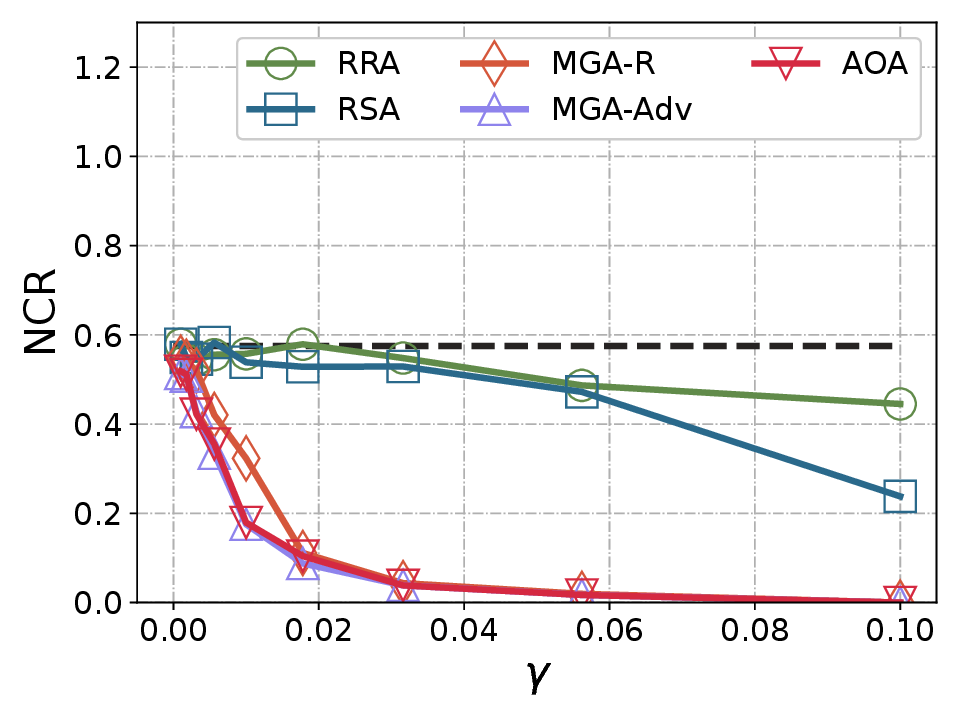}
		\end{minipage}%
    }
	\subfigure[Kosarak, $\mathrm{NCR}$]{
		\begin{minipage}[c]{0.25\textwidth}
		\centering
        \includegraphics[width=1\textwidth]{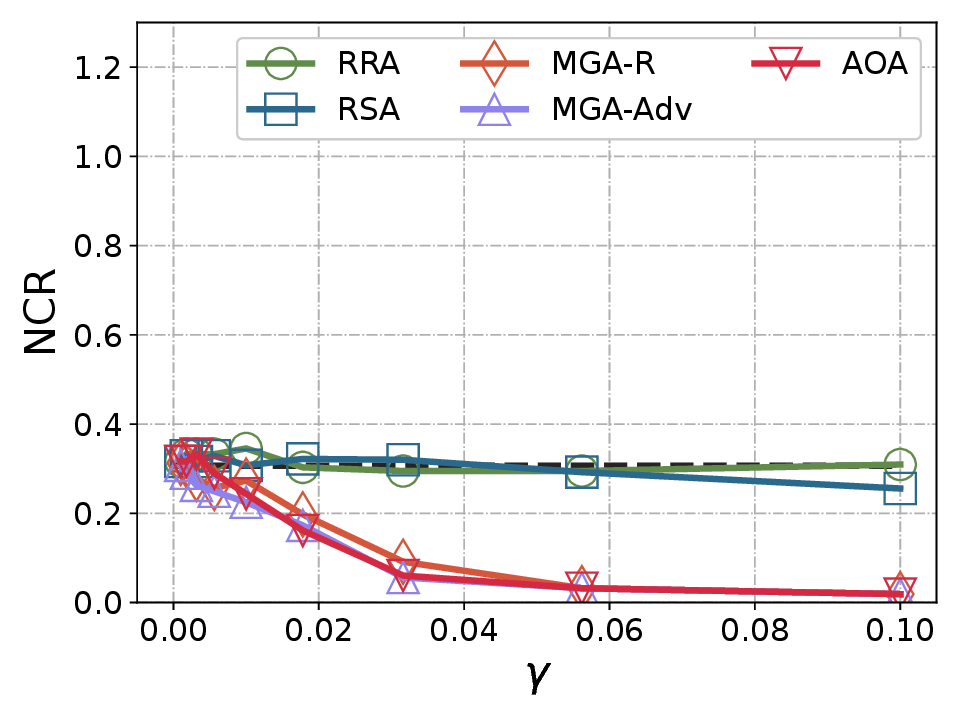}
		\end{minipage}%
    }
	\subfigure[BMS-POS, $\mathrm{NCR}$]{
		\begin{minipage}[c]{0.25\textwidth}
		\centering
        \includegraphics[width=1\textwidth]{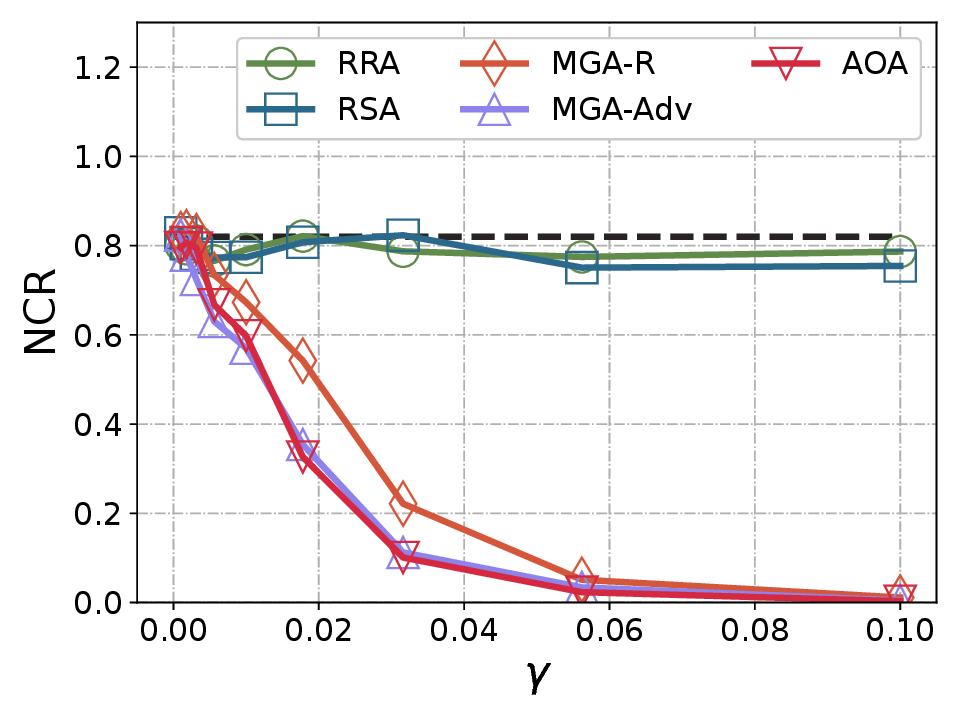}
		\end{minipage}%
    }
    \centering
	\caption{\ccsrev{Attacking FIML-I, with $k = 32$, $\epsilon=2.0$ (black dashed line - no attack).}}
	\label{fig:gamma:fimlitem:eps2}
\end{figure*}

\begin{figure*}[h!]
    \centering
     	\subfigure[\ccsrev{IBM, SVSM}]{
		\begin{minipage}[c]{0.40\textwidth}
		\centering
        \includegraphics[width=1\textwidth]{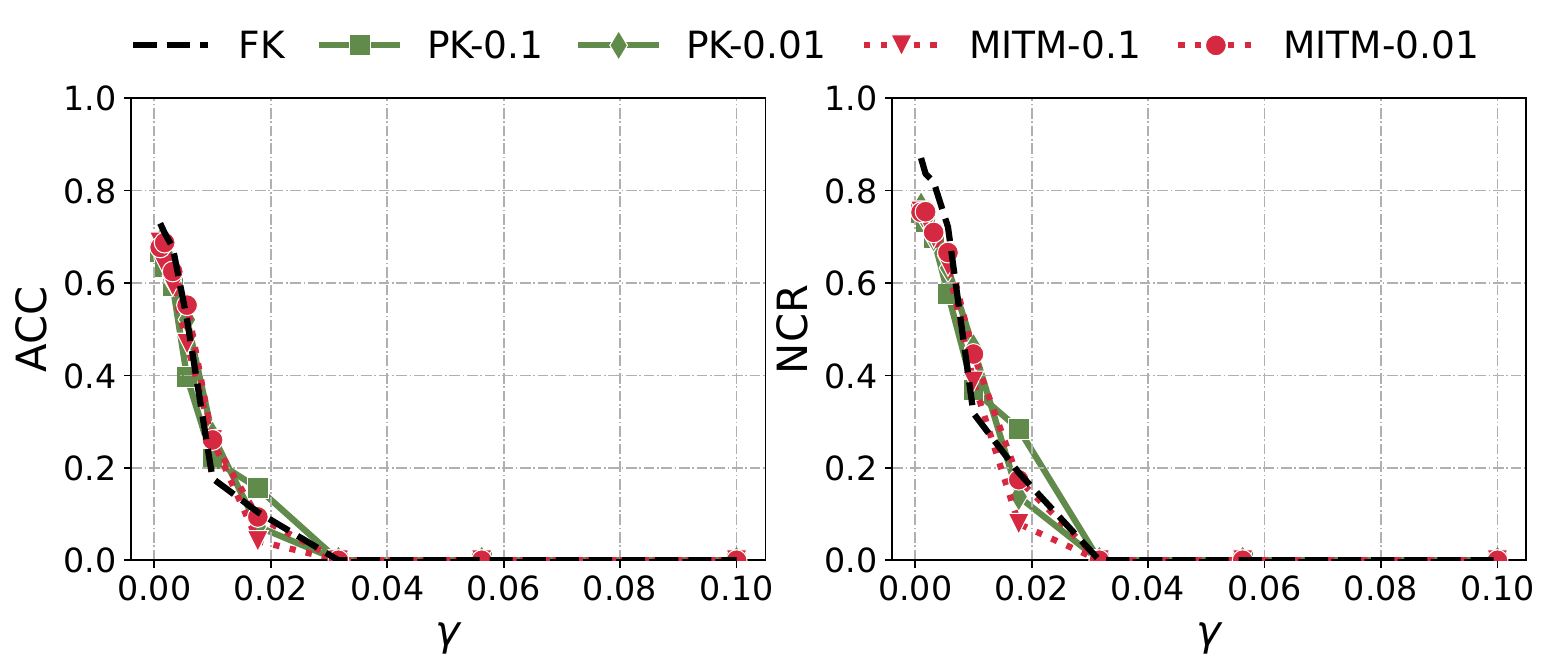}
		\end{minipage}%
	}
        \subfigure[\ccsrev{IBM, FIML-IS}]{
		\begin{minipage}[c]{0.40\textwidth}
		\centering
        \includegraphics[width=1\textwidth]{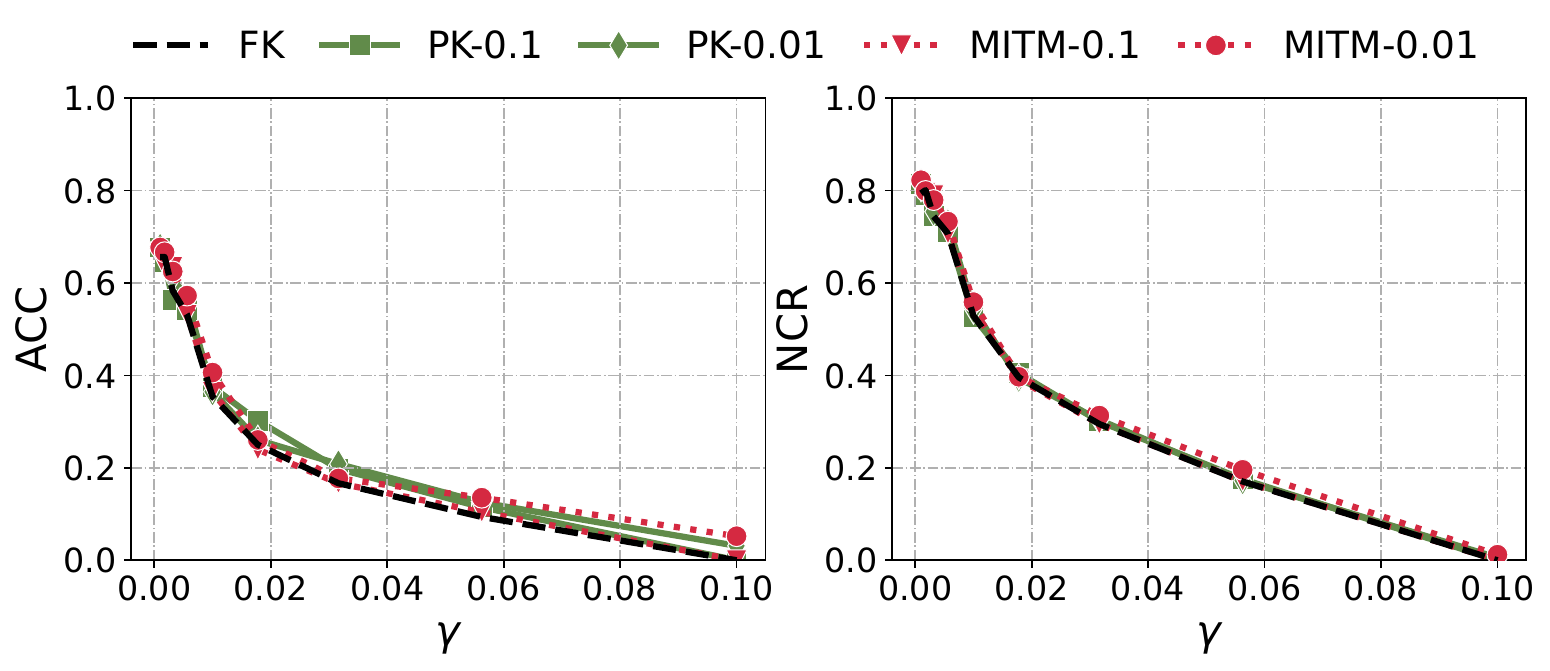}
		\end{minipage}%
	}

      \subfigure[\ccsrev{IBM, LDPMiner}]{
		\begin{minipage}[c]{0.40\textwidth}
		\centering
        \includegraphics[width=1\textwidth]{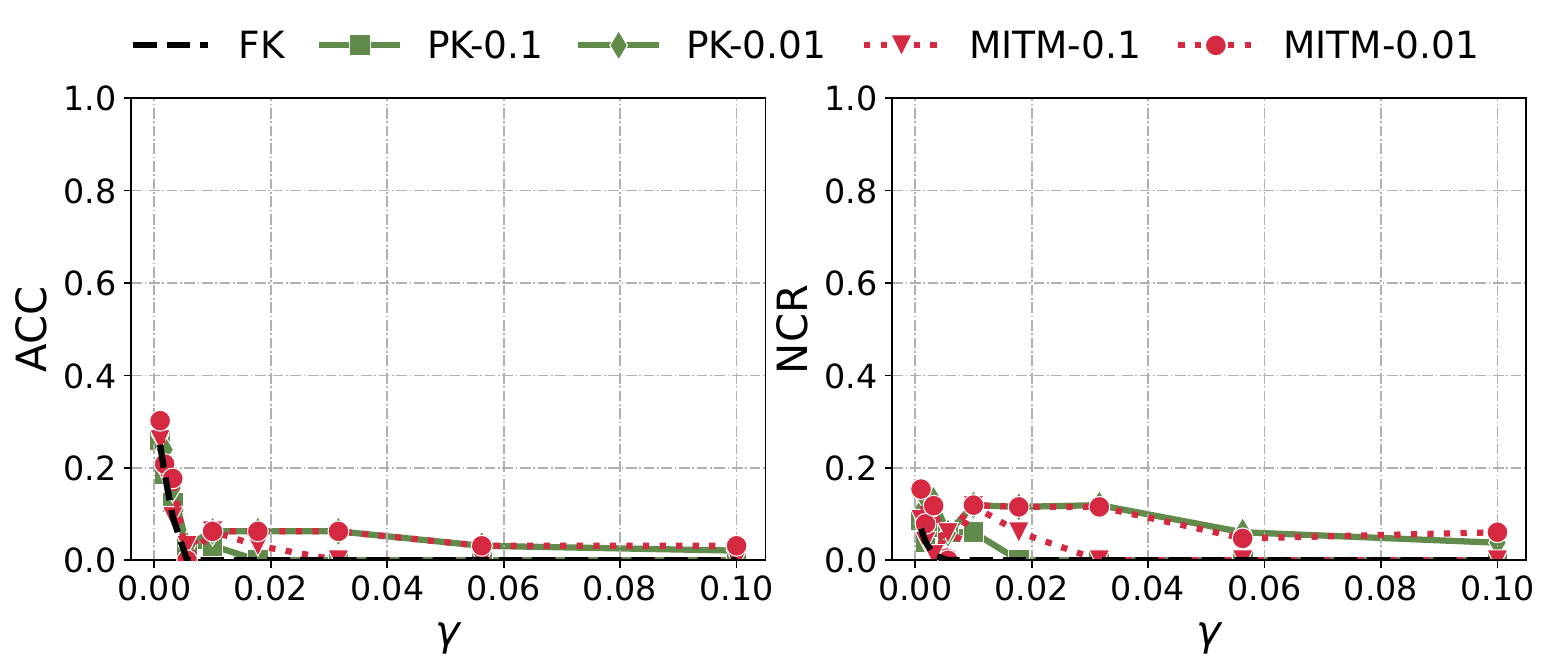}
		\end{minipage}%
	}
      \subfigure[\ccsrev{IBM, SVIM}]{
		\begin{minipage}[c]{0.40\textwidth}
		\centering
        \includegraphics[width=1\textwidth]{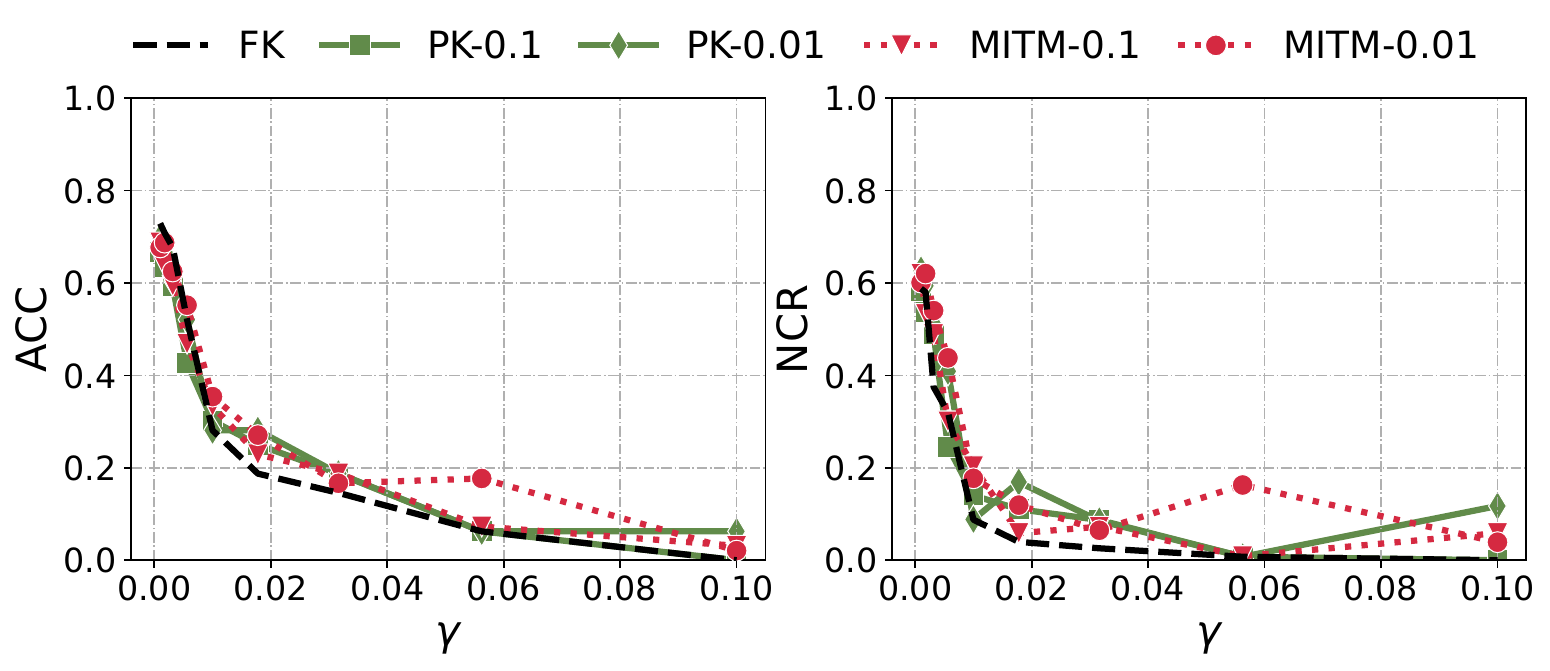}
		\end{minipage}%
	}

       \subfigure[\ccsrev{IBM, FIML-I}]{
		\begin{minipage}[c]{0.40\textwidth}
		\centering
        \includegraphics[width=1\textwidth]{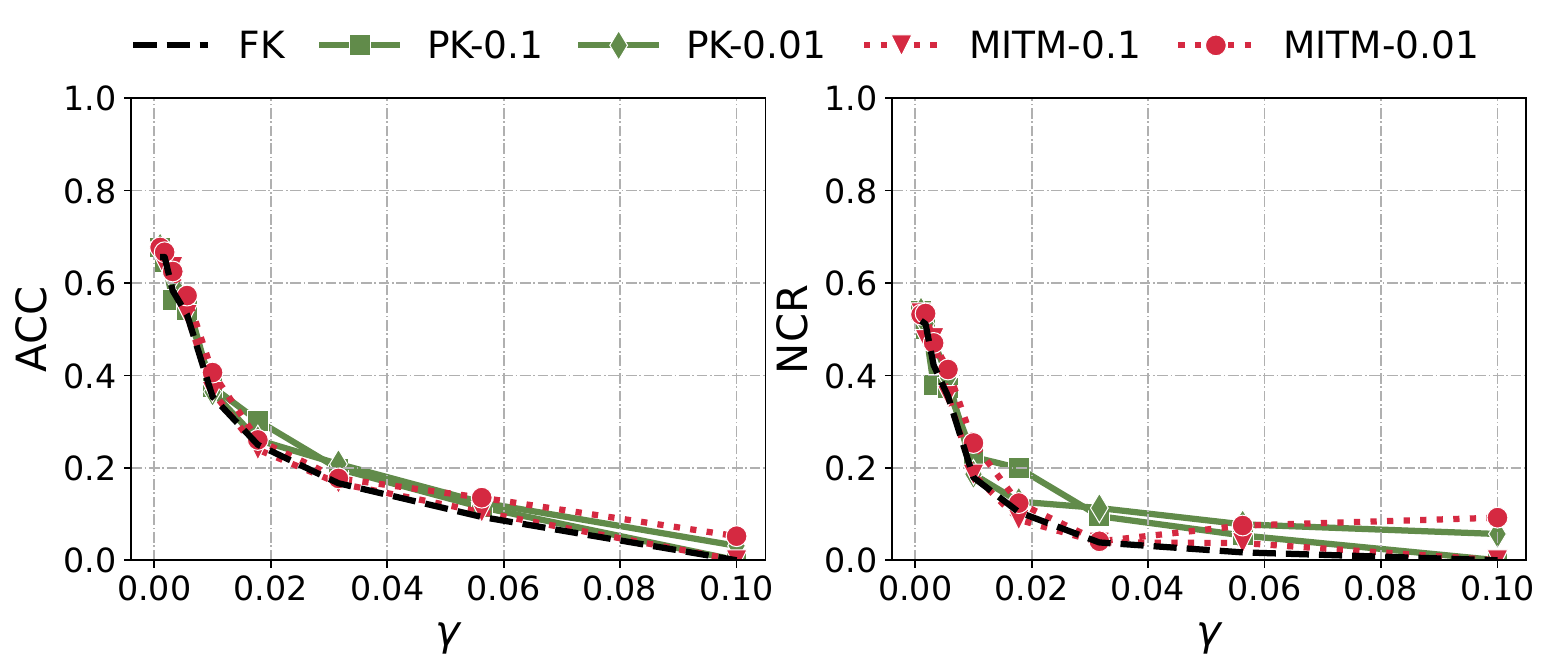}
		\end{minipage}%
	}
 
	\caption{\ccsrev{Attacks with limited knowledge on the IBM Synthesize dataset ($\epsilon = 2.0$).}}
	\label{fig:less:ibm:eps2}
\end{figure*}

\begin{figure*}[h!]
    \centering
	\subfigure[\ccsrev{Kosarak, SVSM}]{
		\begin{minipage}[c]{0.40\textwidth}
		\centering
        \includegraphics[width=1\textwidth]{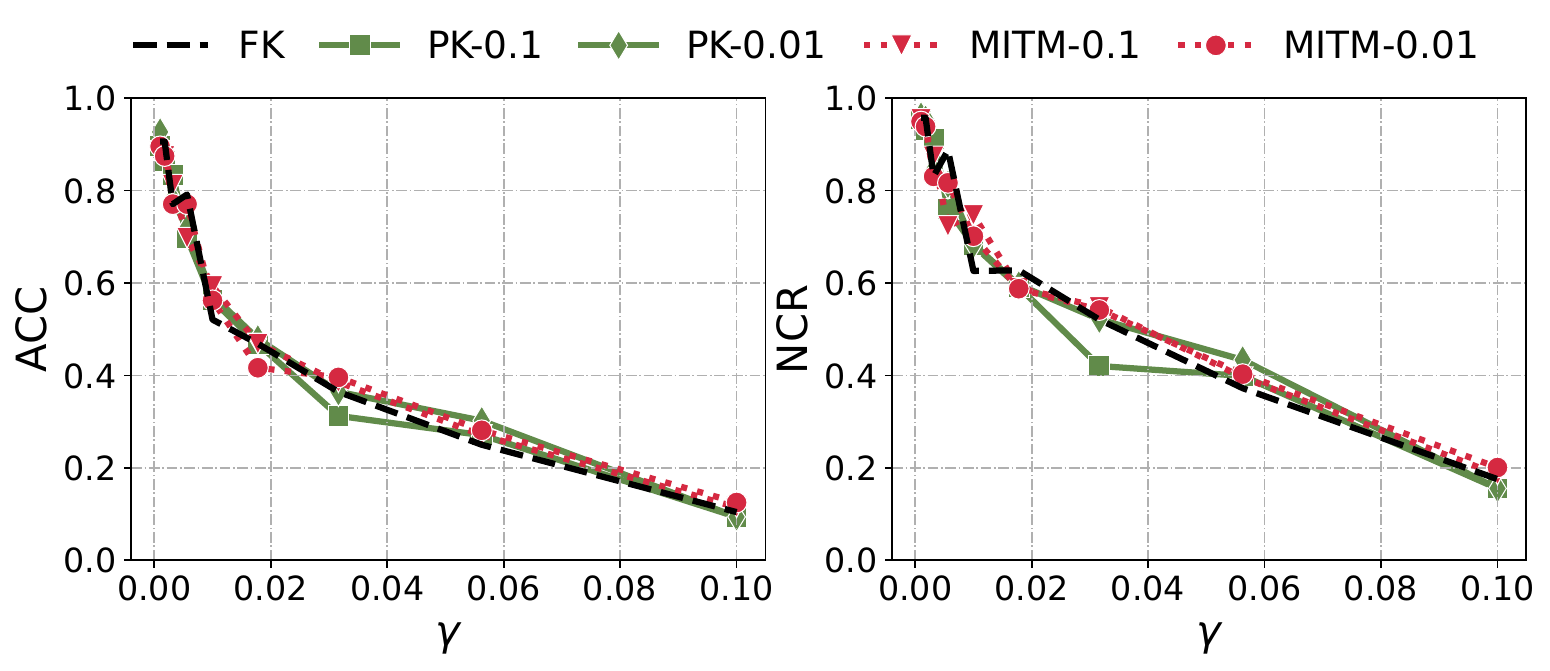}
		\end{minipage}%
	}
     \subfigure[\ccsrev{Kosarak, FIML-IS}]{
		\begin{minipage}[c]{0.40\textwidth}
		\centering
        \includegraphics[width=1\textwidth]{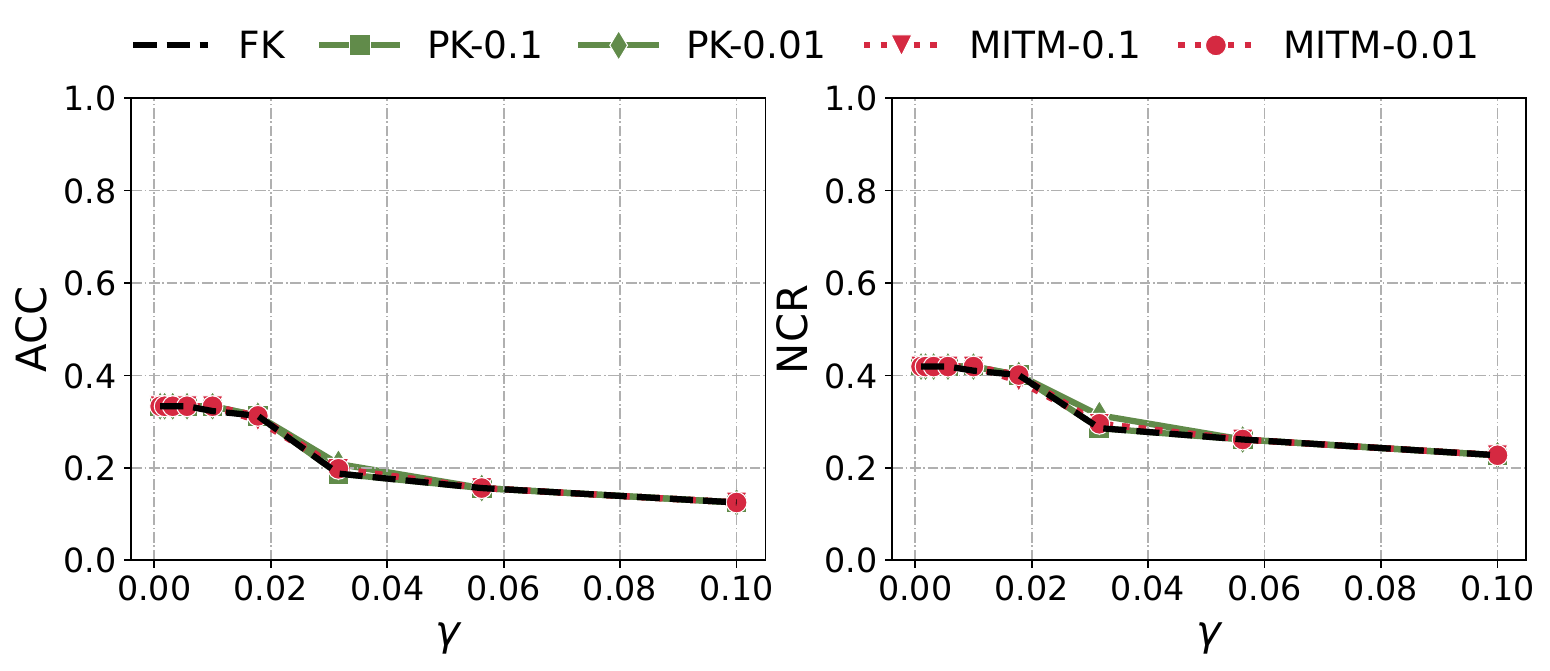}
		\end{minipage}%
	}
 
    \subfigure[\ccsrev{Kosarak, LDPMiner}]{
		\begin{minipage}[c]{0.40\textwidth}
		\centering
        \includegraphics[width=1\textwidth]{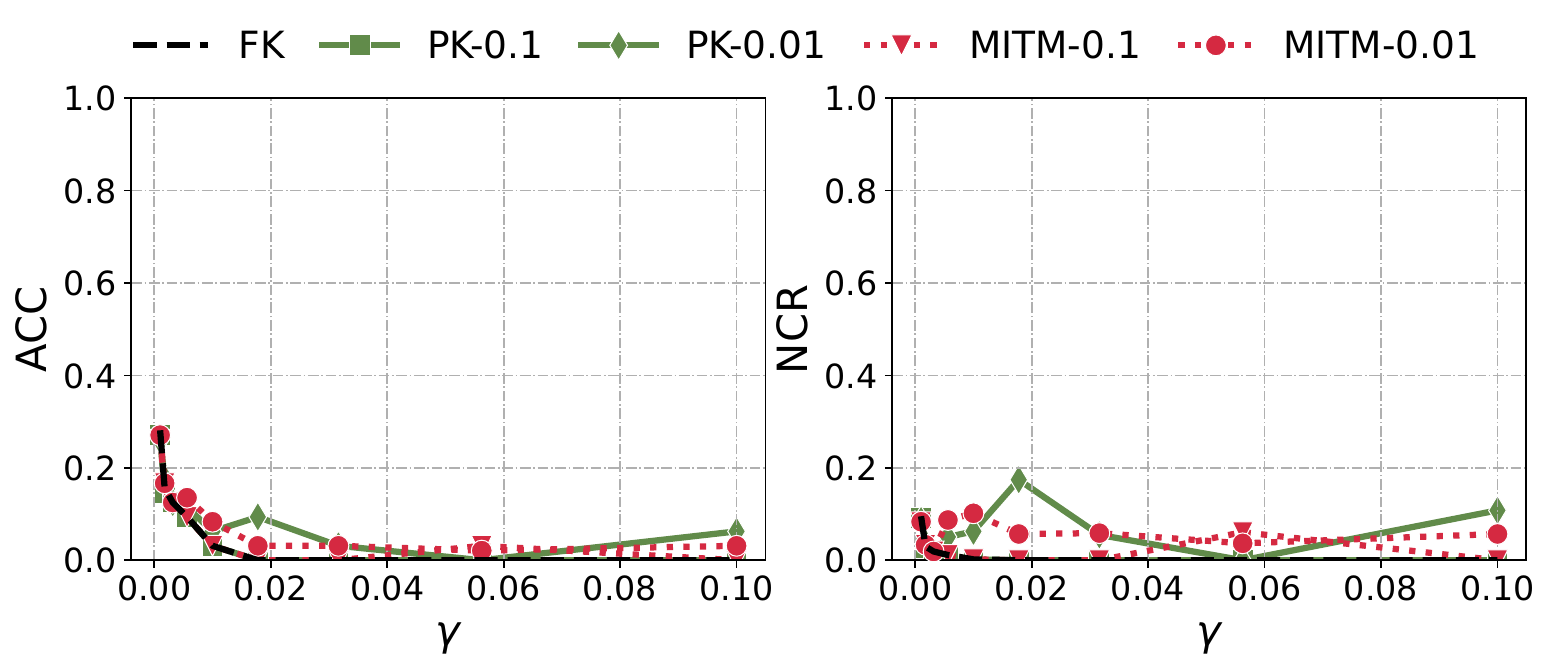}
		\end{minipage}%
	}
     \subfigure[\ccsrev{Kosarak, SVIM}]{
		\begin{minipage}[c]{0.40\textwidth}
		\centering
        \includegraphics[width=1\textwidth]{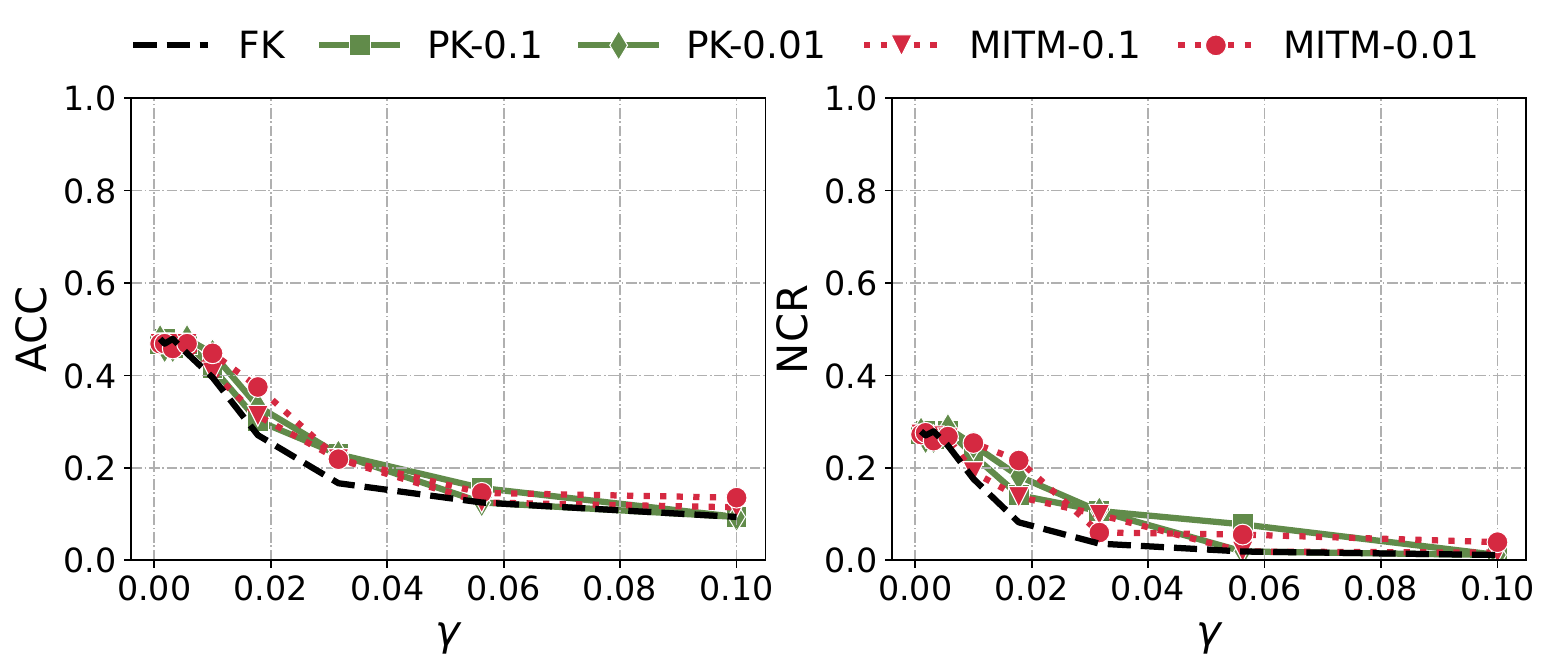}
		\end{minipage}%
	}
     \subfigure[\ccsrev{Kosarak, FIML-I}]{
		\begin{minipage}[c]{0.40\textwidth}
		\centering
        \includegraphics[width=1\textwidth]{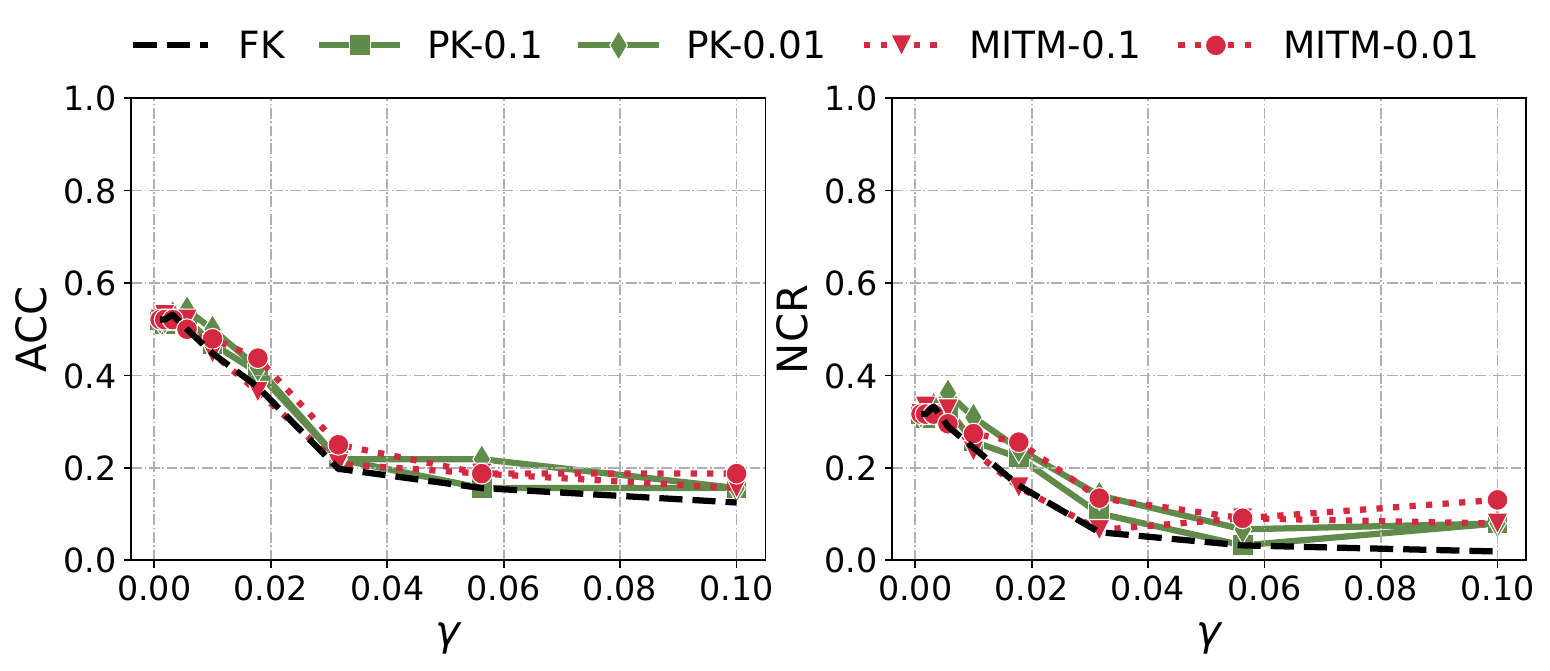}
		\end{minipage}%
	}
 
	\caption{\ccsrev{Attacks with limited knowledge on the Kosarak dataset($\epsilon = 2.0$).}}
	\label{fig:less:kosarak:eps2}
\end{figure*}

\begin{figure*}[h!]
    \centering
	\subfigure[\ccsrev{BMS-POS, SVSM}]{
		\begin{minipage}[c]{0.40\textwidth}
		\centering
        \includegraphics[width=1\textwidth]{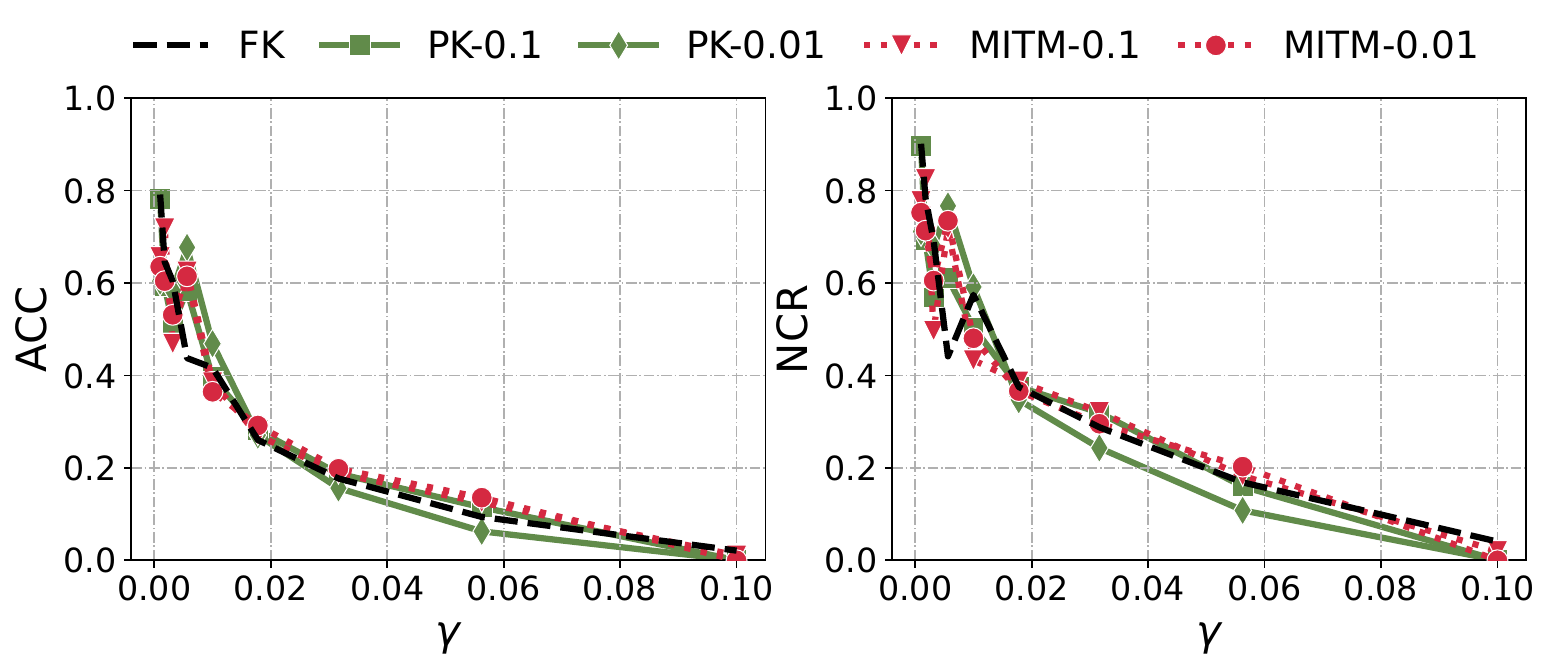}
		\end{minipage}%
	}
     \subfigure[\ccsrev{BMS-POS, FIML-IS}]{
		\begin{minipage}[c]{0.40\textwidth}
		\centering
        \includegraphics[width=1\textwidth]{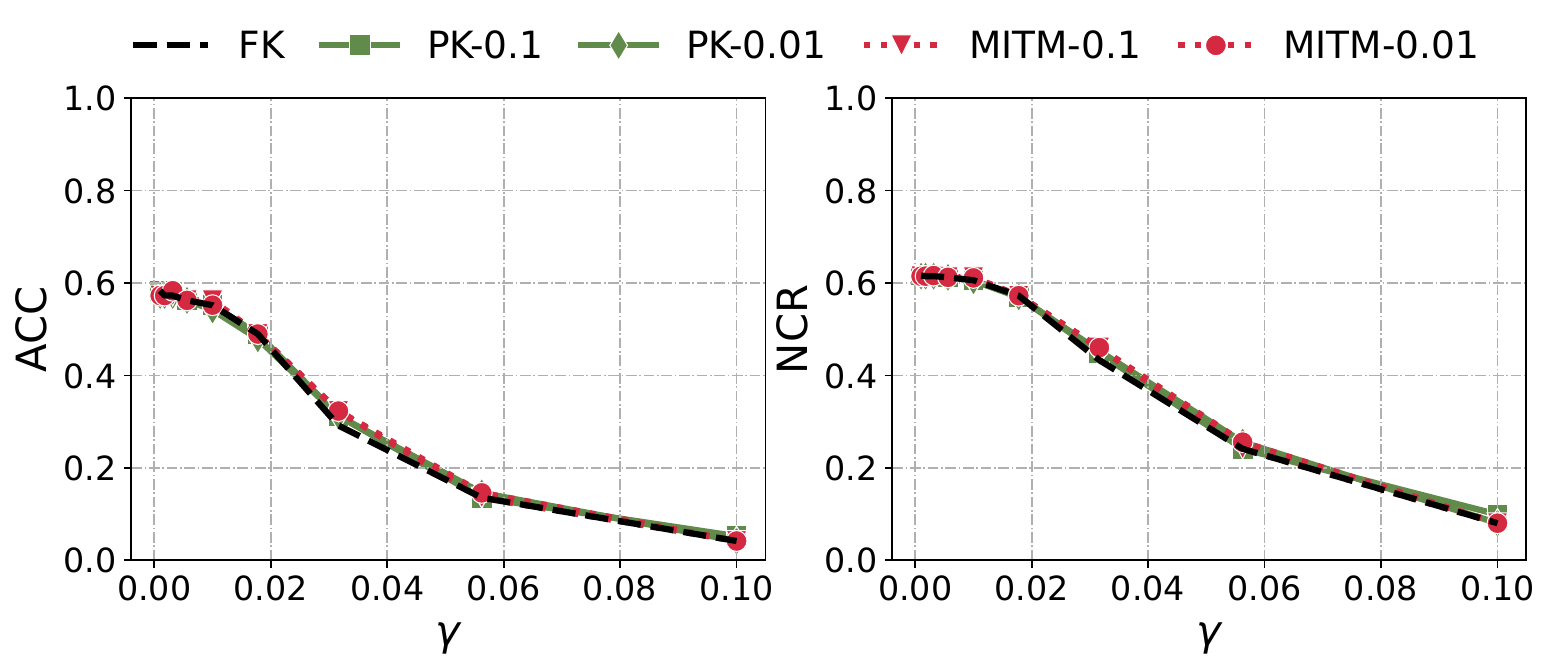}
		\end{minipage}%
	}
 
    \subfigure[\ccsrev{BMS-POS, LDPMiner}]{
		\begin{minipage}[c]{0.40\textwidth}
		\centering
        \includegraphics[width=1\textwidth]{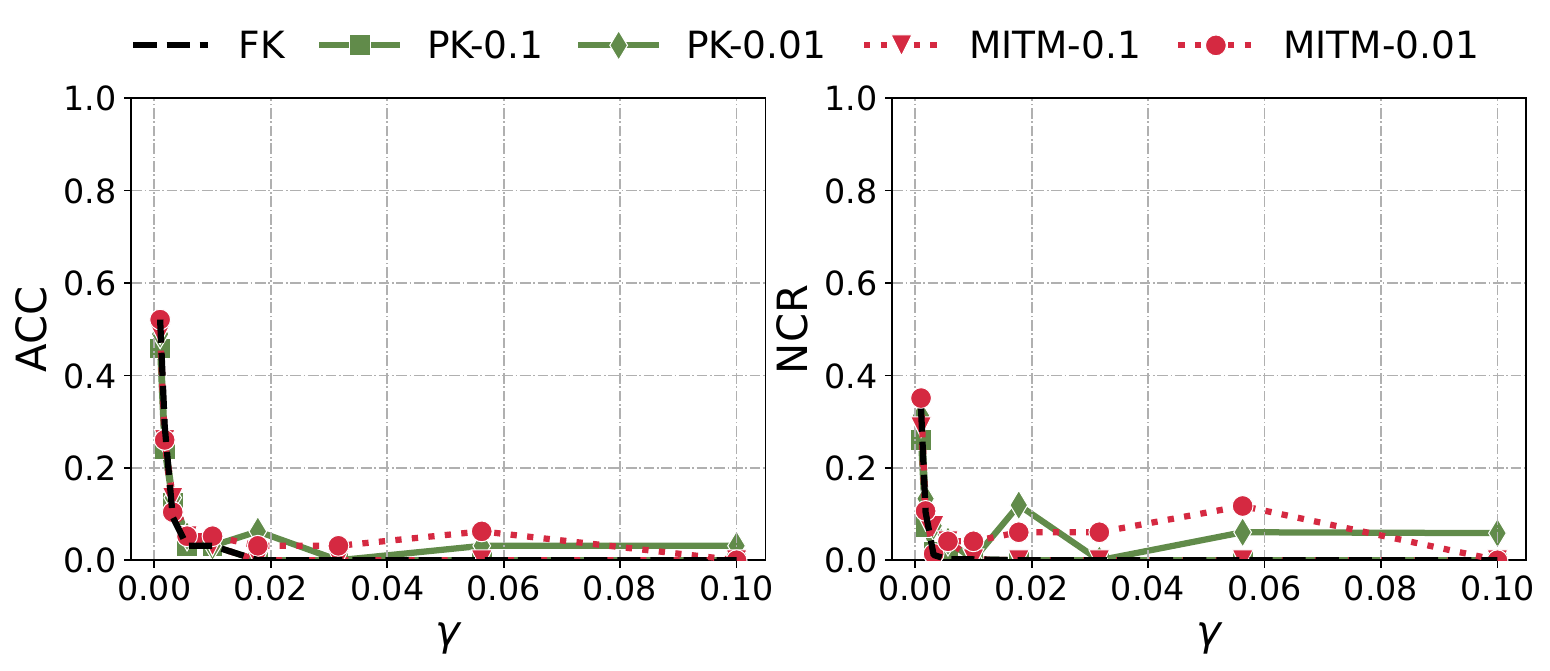}
		\end{minipage}%
	}
     \subfigure[\ccsrev{BMS-POS, SVIM}]{
		\begin{minipage}[c]{0.40\textwidth}
		\centering
        \includegraphics[width=1\textwidth]{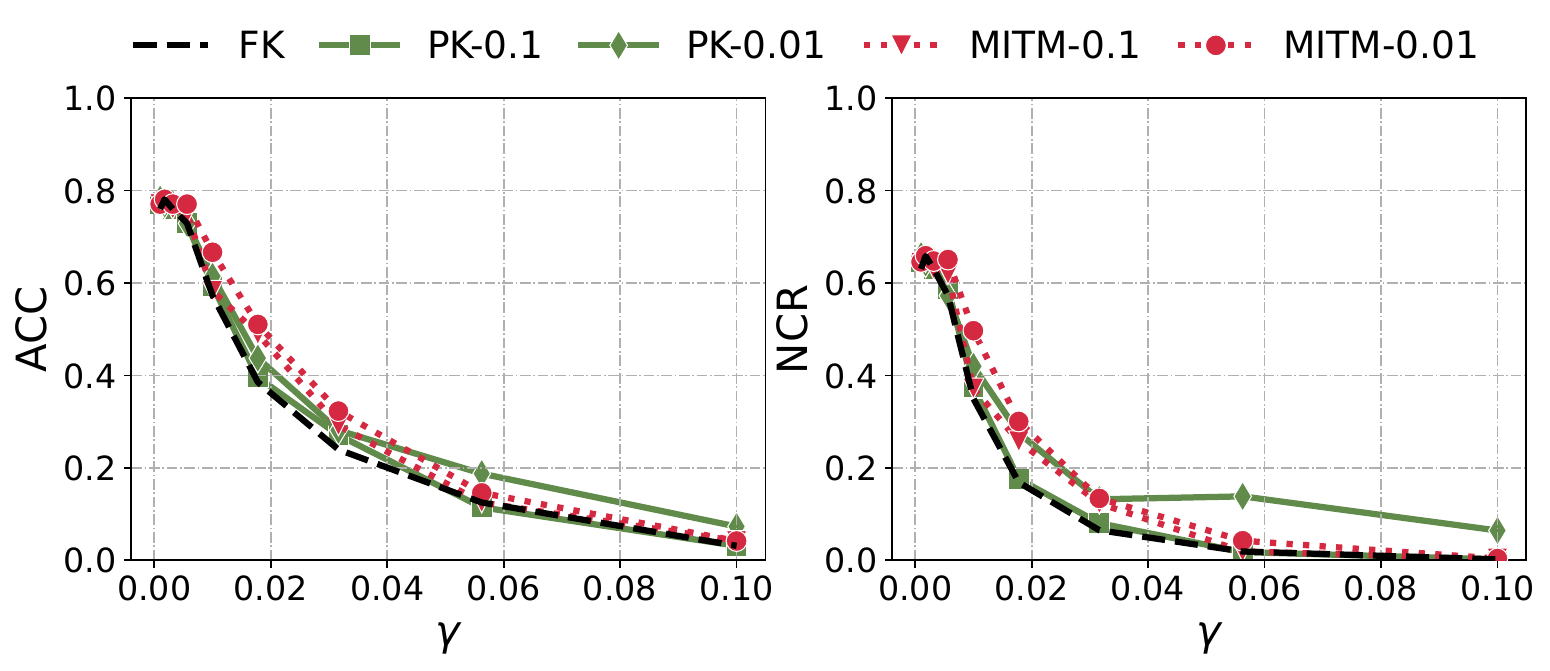}
		\end{minipage}%
	}
     \subfigure[\ccsrev{BMS-POS, FIML-I}]{
		\begin{minipage}[c]{0.40\textwidth}
		\centering
        \includegraphics[width=1\textwidth]{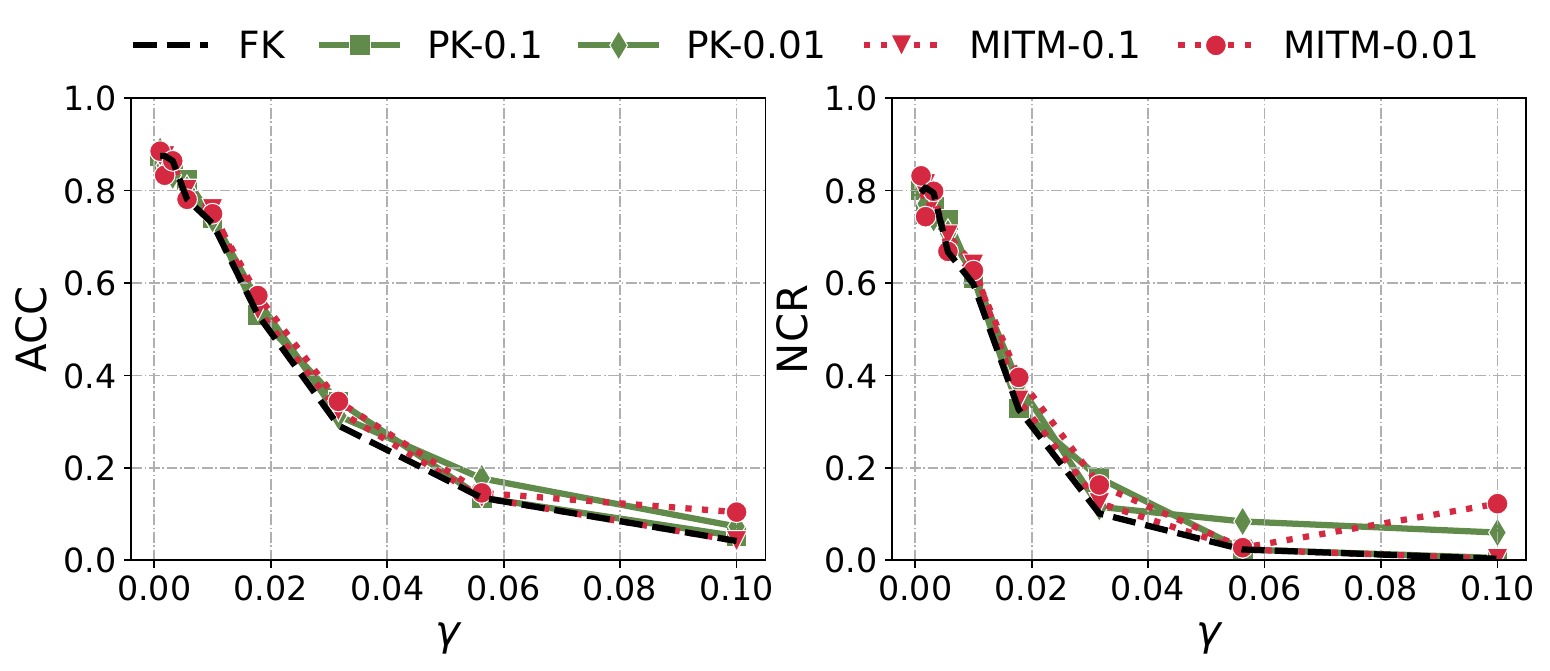}
		\end{minipage}%
	}
 
	\caption{\ccsrev{Attacks with limited knowledge on the BMS-POS dataset($\epsilon = 2.0$).}}
	\label{fig:less:pos:eps2}
\end{figure*}

\begin{figure*}
        \centering
        \includegraphics[width=0.8\textwidth]{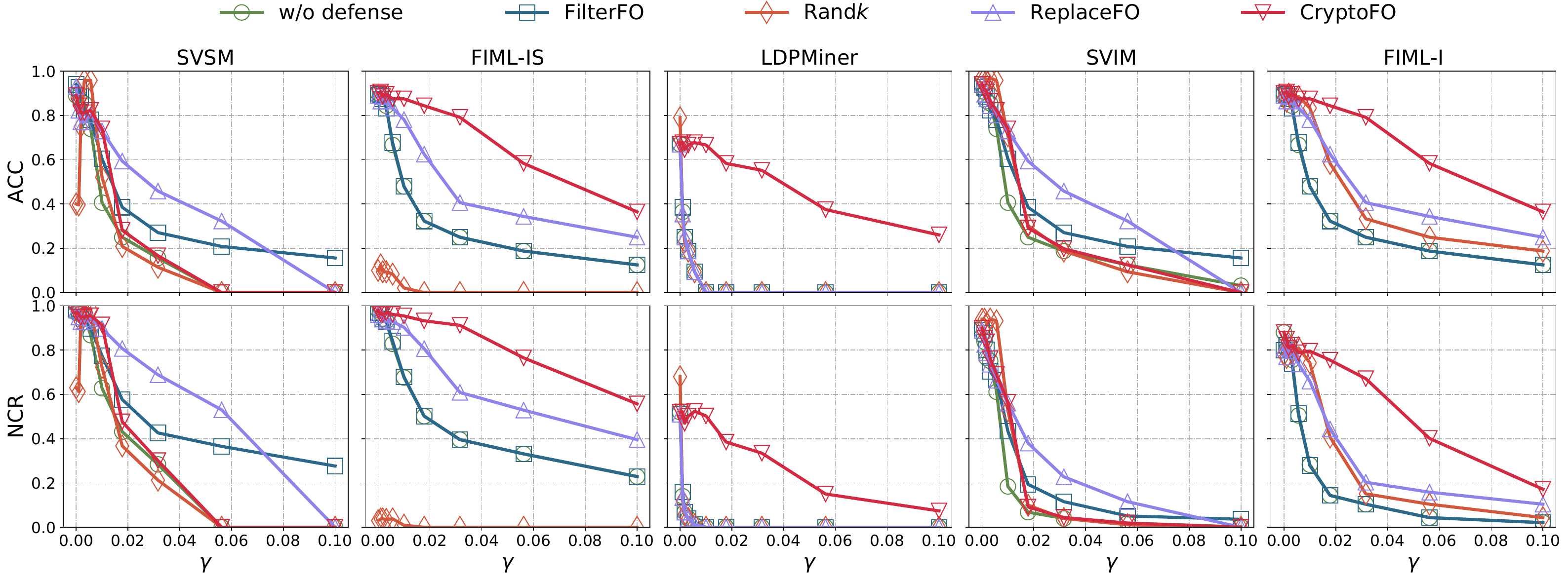}
	\caption{Defenses against AOA on IBM  Synthesize dataset, with $k=32$, $\epsilon=4.0$.}
	\label{fig:defense:ibm}
\end{figure*}

\begin{figure*}
        \centering
        \includegraphics[width=0.8\textwidth]{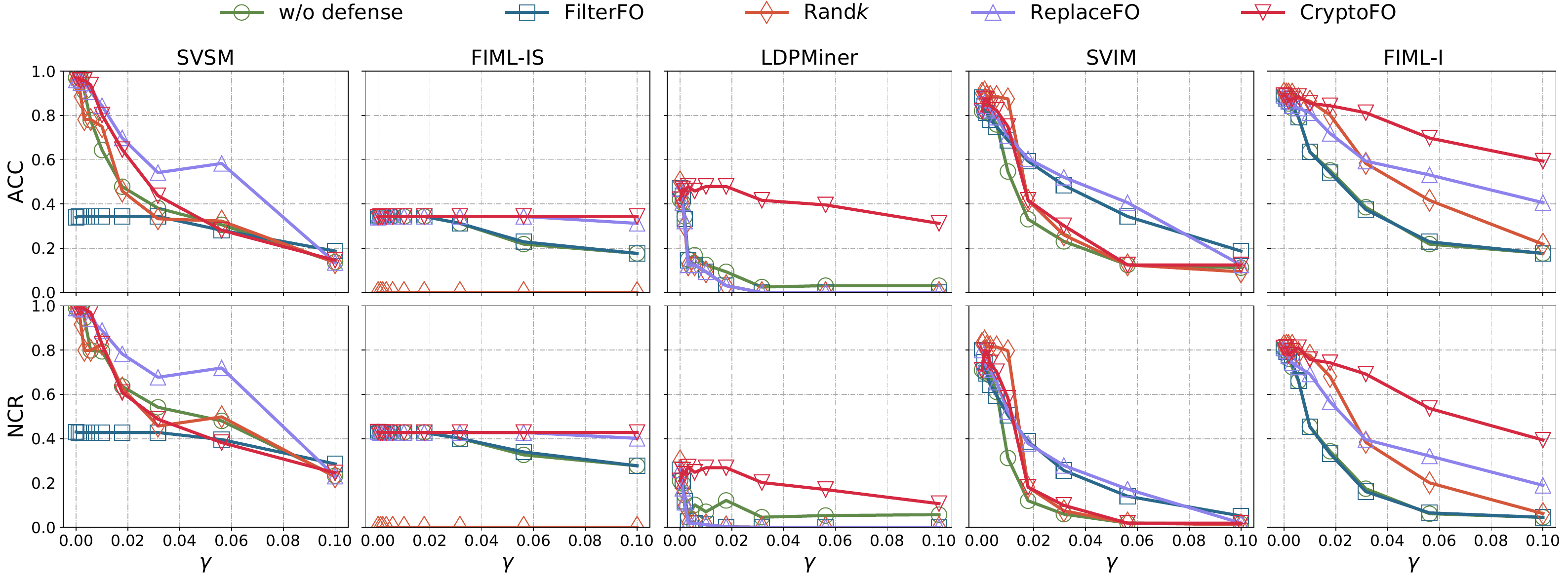}
	\caption{Defenses against AOA on Kosarak dataset, with $k=32$, $\epsilon=4.0$.}
	\label{fig:defense:kosarak}
\end{figure*}

\ccsrev{
\paragraphbe{Evaluating the attacks in different privacy regimes} We have also conducted experiments with $\epsilon=2.0$. The results of attacking SVSM, FIML-IS, LDPMiner, SVIM, and FIML-I are presented in \reffig{fig:gamma:svsm:eps2} to \reffig{fig:gamma:fimlitem:eps2}, respectively. Additionally, we report the results of attacks with limited knowledge in \reffig{fig:less:ibm:eps2}, \reffig{fig:less:kosarak:eps2}, and \reffig{fig:less:pos:eps2} for the IBM Synthesize, Kosarak, and BMS-POS datasets, respectively.

\paragraphbe{Defenses} More results on the evaluation of defenses against AOA are shown in \reffig{fig:defense:ibm} and \reffig{fig:defense:kosarak} for IBM Synthesize and Kosarak datasets, respectively. 
}

\end{document}